\documentstyle[12pt,epsf]{book}

\advance \topmargin by -\headheight

\advance \topmargin by -\headsep

\evensidemargin \oddsidemargin

 \def\mod{{\hbox{\rm mod}}}

\def\ie{{\em i.e.}}

\def\ie{\hbox{\it i.e.}}

\def\CC{{\mathchoice
{\rm C\mkern-8mu\vrule height1.45ex depth-.05ex
width.05em\mkern9mu\kern-.05em}
{\rm C\mkern-8mu\vrule height1.45ex depth-.05ex
width.05em\mkern9mu\kern-.05em}
{\rm C\mkern-8mu\vrule height1ex depth-.07ex
width.035em\mkern9mu\kern-.035em}
{\rm C\mkern-8mu\vrule height.65ex depth-.1ex
width.025em\mkern8mu\kern-.025em}}}

\def\RR{{\rm I\kern-1.6pt {\rm R}}}

\def\ZZ{{\rm Z}\kern-3.8pt {\rm Z} \kern2pt}

\def\IB{\relax{\rm I\kern-.18em B}}

\def\ID{\relax{\rm I\kern-.18em D}}

\def\II{\relax{\rm I\kern-.18em I}}

\def\IP{\relax{\rm I\kern-.18em P}}

\def\G{\Gamma}

\def\np{Nucl. Phys.}

\def\pl{Phys. Lett.}

\def\prl{Phys. Rev. Lett.}

\def\pr{Phys. Rev.}

\def\jhep{J. High Energy Phys.}

\newcommand{\beq}{\begin{equation}}

\newcommand{\eeq}{\end{equation}}

\newcommand{\rc}{\nonumber\\}

\newcommand{\bear}{\begin{eqnarray}}

\newcommand{\eear}{\end{eqnarray}}

\newcommand{\bo}{w}

\newcommand{\unbrw}{\underline{\breve{w}}}

\def\to{\rightarrow}

\def\to{\rightarrow}

\def\om{{\underline{w}}}

\def\CL{{\cal L}}

\def\half{{1 \over 2}}

\def\la{\lambda}

\def\us{\underline{\sigma}}

\newfont{\namefont}{cmr10}

\newfont{\addfont}{cmti7 scaled 1440}

\newfont{\boldmathfont}{cmbx10}

\newfont{\headfontb}{cmbx10 scaled 1728}

\begin{document}

\begin{titlepage}

\begin{center} \Large \bf Supersymmetric
solutions of supergravity from wrapped branes

\end{center}

\vskip 0.3truein
\begin{center}
\'Angel Paredes
\footnote{Ph.D. thesis, Universidade de Santiago de
Compostela, Spain (june, 2004). \\
Advisor: Alfonso V. Ramallo}

\vspace{0.3in}

Departamento de F\'\i sica de Part\'\i culas, 
Universidade de Santiago de Compostela \\
E-15782 Santiago de Compostela, Spain\\
angel@fpaxp1.usc.es

\vspace{0.3in}

\end{center}

\vskip 1truein

\begin{center}
\bf ABSTRACT
\end{center}

We consider several solutions of supergravity 
 with reduced supersymmetry which are related to
wrapped branes, and elaborate
on their geometrical and physical interpretation.
The Killing spinors are computed for each configuration.
In particular, all the known metrics on the conifold
and all $G_2$ holonomy metrics  with cohomogeneity
one and $S^3\times S^3$ principal orbits are constructed
from D=8 gauged supergravity in a unified formalism. 
The addition of 4-form fluxes piercing the unwrapped
directions is also considered. We also study the problem
of finding kappa-symmetric D5-probes in the 
so-called Maldacena-N\'u\~nez model. Some of these
solutions are related to the addition of flavor to the
dual gauge theory. We match our results with some known
features of ${\cal N}=1$ SQCD with a small number of
flavors and compute its meson mass spectrum.
Moreover, the gravity solution dual to three dimensional
${\cal N}=1$ gauge theory, solutions related to 
branes wrapping hyperbolic spaces, $Spin(7)$ holonomy
metrics and $SO(4)$ twistings in D=7 gauged sugra
are studied in the last chapter.

\smallskip

\vskip2.6truecm
\leftline{hep-th/0407013}
\smallskip

\end{titlepage}
\setcounter{footnote}{0}

%****************************************

\pagestyle{empty}

\

\vspace{9cm}

\hspace{8cm}
\textsl{
What Immortal hand or eye}

\hspace{8cm}
\textsl{
Dare frame thy fearful symmetry?}

\

\hspace{9cm} William Blake, ``The Tyger".

\setcounter{page}{0}

%****************************************

\tableofcontents

%****************************************

\pagestyle{headings}

\chapter*{Motivation}
\addcontentsline{toc}{chapter}{Motivation}

\medskip

There are four kinds of interactions known to exist in nature:
gravitational, electromagnetic, weak and strong.
The first one, much weaker than the rest, is described
by Einstein's General Theory of Relativity. The other three are
accurately explained by the successful Standard Model, based on quantum
gauge theories. Unfortunately, it is known that both theories are
incompatible at very small distances or very high energy scales, of the
order of the Planck mass, about $10^{19}$ GeV. This means that some new
physics must happen when approaching the Planck scale.

 Superstring theory, despite being originally formulated as an attempt 
to explain strong interactions, is, nowadays, the most promising
candidate for solving this puzzle. Its basic idea is to suppose that
particles, instead of being points, have some natural extension and
are, in fact, vibrational modes of some fundamental strings.

 By studying the spectrum of excitations
of a closed string, one finds a massless spin-two field, which can
be identified with the graviton. On the other hand, the infinite
tower of massive string modes can cure the non-renormalizability of
General Relativity, thus yielding a consistent theory of quantum
gravity. Since the eighties, lots of theoretical physicists have
hoped that string theory can lead to a 
``Theory of Everything", that should describe consistently all the
measured phenomena. Standard texts on string theory are
\cite{stringbooks}.

The fact that  five consistent string theories can be formulated
was a puzzle: how should nature choose one among several
possibilities? This question was nicely solved when the 
existence of a web of dualities relating all of them  was discovered.
The so-called M-theory lives in eleven dimensions and the different
string theories are different perturbative regimes. Besides, eleven
dimensional supergravity also appears as a low energy limit.

The objects called branes play an important r\^ole in this picture.
D-branes are non-perturbative solitonic objects that can be identified
with hyperplanes where open strings can end (an introduction on branes
can be found in
\cite{johnson}). The dynamics of the D-branes can be described by the
physics of the open strings, thus giving rise to a gauge theory living on
the worldvolume of the brane (see \cite{givkuta} for a review of the
interplay between brane dynamics and gauge theory). However, there is
another way of thinking about D-branes, as sources of closed strings.
From this point of view, branes are objects that modify the gravitational
background, {\it i.e.} the geometry of space-time. Therefore, this
open/closed string duality leads to a gauge/gravity duality.
This  notion has opened new and amazing possibilities. Besides
addressing the problem of unification,   string theory can give an
insight in the search of  duals of different gauge theories.

This fact is related to a much older proposal by t'Hooft \cite{Hooft}.
He pointed out that the Feynman diagrams of a $U(N)$ gauge theory
can be rearranged as a sum over the genus of the surfaces in
which the diagrams can be drawn. This is pretty similar to the
computation of   string amplitudes, where there is a sum over
the genus of the possible worldsheets. Then, there is a gauge/string
duality, at least in some regime of  parameters. The problem is that no
hints are given of which should be the string theory that corresponds to
each gauge theory.

In 1997, Maldacena formulated an astonishing conjecture along these
lines \cite{jm}. The statement is that type IIB string theory
living on $AdS_5\times S^5$ is exactly dual to four dimensional ${\cal
N}=4$ super Yang-Mills theory with $SU(N)$ gauge group (which is called
AdS/CFT
 duality since the gauge theory is conformal). 
Although a strict proof
has not been given, the duality has overcome a large number of tests (the
standard review on these topics is
\cite{magoo}). The duality between two such different theories was
reached by looking at the dual open/closed string descriptions of the
near horizon limit of a stack of $N$ D3-branes.
The low energy limit of string theory yields a supergravity theory, and
we find that type IIB sugra on $AdS_5\times S^5$ is dual to ${\cal N}=4$
SYM in its non-perturbative regime.

A remarkable fact is that the relation between the two theories 
that are supposed to be equivalent is holographic \cite{hologram}. This
means that the number of dimensions in which they live is different and,
that, somehow, the physics on the boundary of a space encodes all
the bulk information.

Following these ideas, a lot of work has been devoted to the research
on other possible dualities involving more realistic gauge theories.
In particular, one would like to have less supersymmetry and break
conformal invariance. The final goal is to find a gravity dual of QCD, at
least for the limit with large number of colors.

The motivation of this work is this amazing interplay between strings,
gravity, geometry and gauge theory. When branes are wrapped, the
amount of supersymmetry of the solution gets reduced and, in 
fact, conformal symmetry gets broken. However, in order to have some
supersymmetry, the branes must be wrapped along certain supersymmetric
cycles which are embedded in non-trivial spaces.
With these ingredients, one can engineer several setups, such that
different gauge theories live on the worldvolume of the wrapped branes.
Supergravity techniques will be used in order to get geometrical results
about these spaces with reduced supersymmetry, and also 
to obtain features of the dual gauge theories.

\subsubsection{About this thesis}

This Ph.D. thesis is mainly based on papers
\cite{epr,eprii,epriii,flavoring}, although a few unpublished 
results are also discussed. Throughout the work, technical
details are described thoroughly at many 
points (mainly the way of obtaining and solving BPS systems of equations).
However, sometimes it is possible to skip them while keeping
a comprehensive reading of the text. The plan for the rest
of the thesis is the following:

In chapter \ref{introsusy}, there is a brief introduction to the ideas
of supersymmetry and supergravity.
Then, some general strategies for computing supersymmetric solutions
of supergravity are presented. Finally, the degrees of freedom,
lagrangians and susy transformations of several supergravity theories
are reviewed. This chapter provides the  basic prerequisites needed and
sets up  the notation
for the computations of the following.

In chapter \ref{conichapter}, we use eight-dimensional supergravity
to study the geometry of the so-called conifold. The metric and
Killing spinors are found. In the  process, we will find an
important technical point: the need of having a rotated projection
on the spinor. In chapter \ref{g2chapter}, 8d sugra is used again,
this time for computing metrics of seven-dimensional $G_2$ holonomy
manifolds. Assuming again a rotated Killing spinor, all the
complete metrics of cohomogeneity one and $G_2$ holonomy with
$S^3\times S^3$ principal orbits are constructed. It will be
shown how asymptotically locally conical metrics are obtained
from this approach. Then, in chapter \ref{fluxes}, we find
a general procedure to incorporate RR fluxes in the unwrapped
directions of configurations of the type of those mentioned above.
From an M-theory point of view, this amounts to adding
M2-branes to the solution.

Then, in chapter \ref{MNchapter}, we turn to the so-called
Maldacena-N\'u\~nez model. In this scenario, the gravity
solution corresponding to D5-branes wrapping a supersymmetric two-cycle
inside a Calabi-Yau is dual to
${\cal N}=1$ super Yang-Mills theory in four dimensions. A brief review
of the model is presented and the supergravity solution is computed from
an analysis of supersymmetry. The Killing spinors are obtained in the
process. In chapter
\ref{flavoring}, we address a concrete problem within this model. We look 
for surfaces where supersymmetric brane probes can be placed. We argue
that some of these  brane probes introduce fundamental quarks in the dual
gauge theory, which are represented by fundamental strings stretching
from the  brane probe to the gauge theory brane. Some known features of
the ${\cal N}=1$ gauge theory are recovered from the gravity viewpoint and
a prediction is made about the meson mass spectrum. 

Finally, in chapter \ref{chothers}, the supersymmetry of a few more
supergravity solutions is studied.

%****************************************

\chapter{Some notes on supergravity}
\label{introsusy}
\medskip

The main goal of this thesis is
to find (bosonic) supersymmetric configurations which are
solutions of some supergravity equations of motion, and to
deepen in their physical and geometrical interpretation.
This chapter is aimed to be the basis of the analysis carried
out in the rest of this work.

First of all, a brief (and surely incomplete) introduction to
what is supergravity will be given. Then, the general strategy
that will be used to find the different solutions 
will be 
established. After that, several supergravities that will appear
throughout the rest of the thesis, and the
relation between them, will be presented. This is
useful to fix notation for the following chapters.
Notice that some of the actions and supersymmetry transformations
written here are not the most general ones, but 
only truncations
where some fields have been set to zero.
An important concept for finding supersymmetric solutions in
gauged supergravities is the twisting (which amounts to exciting the
gauge field). It will be introduced in section \ref{twistsec}.

%*******************************************************

\setcounter{equation}{0}
\section{What is Supersymmetry?}
\label{secsu}

Supersymmetry (susy)
can be defined as a Fermi-Bose symmetry, {\it i.e.}
as a transformation mixing bosonic and fermionic degrees of
freedom which leaves the physics (the equations of motion)
invariant.

It was first discovered by studying the interplay between
the space-time Poincar\'e symmetry 
(Lorentz group plus translations)
and internal 
symmetry groups. A theorem by Coleman and Mandula stated
that if both Poincar\'e
and internal symmetry are present (subject to a few
hypothesis), they do not have  non-trivial mixing, {\it i.e.},
the full symmetry group should be a direct product of both.

One of the hypothesis was that the internal symmetry was 
described by a Lie group based on commutators, but it was
realized in the seventies that the no-go theorem could be
avoided by taking a Lie algebra based on anticommutators.

The supersymmetry algebra can be written in a completely 
schematic fashion (for a rigorous discussion, see, for example,
\cite{susybooks}):

\bear
&&[P,P]=0 ;\qquad\quad\quad \, \, [P,M]=P ; \rc  
&&[M,M]=M; \qquad\, \quad  [P,Q^I]=0 ; \rc
&&[M,Q^I]=Q^I ; \  \ \quad \quad\,\{Q^I,\bar Q^J\}=P \, \delta^{IJ} ; \rc
 &&\{Q^I,Q^J\}=Z^{IJ} ;  \ \,\quad
\,\{\bar Q^I,\bar Q^J\}=Z^{IJ} ; \label{susyalg}
\eear
where the $P$ stands for translations, $M$ for Lorentz
generators (spatial rotations and boosts), $Q^I$,
$\bar Q^I$ for the
supersymmetry generators, and $Z^{IJ}$  are the
central charges. All space-time and spinor
 indices have been omitted. The
indices $I,J=1, ..., 
\cal N$ label different sets of supersymmetry generators.

For consistency, the supersymmetry generators must transform
as 1/2 spinors under Lorentz transformations (this fact could
have been intuitively anticipated because their algebra is
based on anticommutators). This has the immediate consequence
that under a susy transformation, bosons turn into fermions and
vice versa. So an irreducible representation of the susy algebra
will correspond to several particles, forming what is called
a supermultiplet. It can be proved that a supermultiplet always
contains the same number of bosonic and fermionic degrees of
freedom. Furthermore, as susy transformations commute with
momemtum generators, we have, in particular $[P^2,Q]=0$,
and therefore all particles in the same supermultiplet have the
same mass.

At this point, one may think that, although possibly interesting
from a theoretical point of view, supersymmetry could be far
from reality because, if there were supersymmetric particles
with the same mass as the usual ones, they would have been
certainly observed by now. The only way out to this problem
is to say that, 
if supersymmetry exists, it must be broken at a scale
of energy at least as high as the energies probed in
accelerators. Anyway, experimentally there is not even a hint
on the existence of susy particles, so the next question to
answer is: why physicists have been (and still are) so 
interested in supersymmetry during the last decades?

First of all, supersymmetric theories are the most natural extension of
the usual quantum field theories and they have the advantage
with respect to them of a better UV behavior because the
bosonic and fermionic loops cancel one against each other.

Maybe the fact that gives strongest support 
to the idea of susy really existing
in nature is Grand Unification. If the standard model gauge
group comes from the breaking of a larger gauge group at
some high mass scale, the three couplings (electromagnetic,
weak and strong) should get unified at that scale. 
Following the renormalization group flows using the standard
model spectrum of particles, the couplings fail to converge
at a point. But using a supersymmetric extension of the model,
this problem can be overcome. Another argument in favor of susy
is based on 
the hierarchy problem. The big difference between the Planck
scale and the electroweak scale suggests that susy should be
restored at a scale comparable to the Higgs mass. A third
physical puzzle which may be solved by the existence of 
supersymmetric particles is the dark matter. The WIMPS
(weakly interacting massive particles) that are thought to
form it, could be some of the yet unobserved superpartners.
All these three arguments tend to signal to the same value
for the mass of the lightest susy particles: about a few TeV.
If this is true, the Large Hadron Collider which will be soon
operative at CERN should confirm the existence of 
superpartners.

Yet another reason to believe in supersymmetry is 
that it appears
in a very natural way in string
theory, and it is required to keep
the theory free of tachyons (and therefore inconsistencies).

Even if susy turned out not to be real, it would continue to
be quite interesting for some reasons.
Susy theories are in general simpler than non-susy ones because
of the constraints imposed by the symmetry. Then, they can be 
used as toy models that could hopefully capture features of
more realistic theories and help to understand difficult 
problems like confinement. Susy can also give an insight into
mathematical problems (specially, in geometry) as it will become
clear in the following chapters, and  has led to great developments
like mirror symmetry. And finally, even if superstring theories
are not the correct description of quantum gravity, they would
still have a major physical interest because of the 
gauge/string duality, that can be explored along the lines of 
Maldacena's conjecture.

\subsubsection{Spinors in arbitrary dimensions}

As in the following we will deal with supersymmetric 
theories in different number of dimensions, it is
convenient to take a brief look at  spinor
representations in any number of dimensions. 
For a nice review on this topic, see \cite{tanii}.

A spinor
representation of the Lorentz group
is associated to a Clifford algebra:
\beq
\{\G_\mu,\G_\nu\}=2 \; g_{\mu\nu}\,\,,
\eeq
where $\mu,\nu=1,...,D$ are space-time indices (and so $D$ is
the dimension of space-time). It can be proved that,
in order to have a representation of this algebra, the Dirac
gamma matrices can be written as $2^{[D/2]}\times 2^{[D/2]}$
complex square matrices ($[D/2]$ being the integer part of
$D/2$). Then, a spinor has $2^{[D/2]}$ complex components,
whose degrees of freedom are halved because they must 
satisfy the Dirac equation. In conclusion, a Dirac spinor
in $D$ dimensions has $2^{[D/2]}$ real degrees of freedom
(but notice that this halving does not apply for the spinors
used for the susy transformations, as they are arbitrary and
do not need to satisfy any equation of motion). 
Furthermore, it is important to know whether it is possible
to impose some condition that consistently
reduces the number of degrees
of freedom of the spinor, as it turns out that supergravity
multiplets are constructed in each dimension with these reduced
spinors.

There are two types of such conditions. Each of them halves
the number of degrees of freedom: 
\begin{itemize}
\item[-] Imposing reality of the spinors
gives rise to the so-called (pseudo) Majorana 
spinors\footnote{For simplicity, no distinction will be made
between Majorana and pseudo-Majorana spinors in this
introduction.}. This 
can be done when $D$=0,1,2,3,4 mod 8 (assuming that there is just
one time-like dimension).
\item[-] Imposing that the spinors have a definite chirality
gives rise to the so-called Weyl spinors. This is possible
when the dimension of space-time is even.
\item[-] Both conditions can be imposed simultaneously 
when $D$=2 mod 8, getting (pseudo) Majorana-Weyl spinors.

\end{itemize}

This is
useful to know how many supercharges there exist in a susy theory. For
example, in an
${\cal N}=1$, $D=11$ theory, the supersymmetry is generated by 
one Majorana spinor that has 32 degrees of freedom and therefore there
are 32 associated real supercharges, while in an ${\cal N}=1$,
$D=4$ theory, there are only 4 supercharges.

%*******************************************************

\setcounter{equation}{0}
\section{What is Supergravity?}

Supergravity (sugra) is the gauge theory of supersymmetry.

The basic idea is to formulate a supersymmetric theory
including Einstein's general relativity. General 
relativity is a theory whose basic
field is a spin 2 particle,
the graviton. The supermultiplet of the graviton must,
at least, contain a spin 3/2 particle (a Rarita-Schwinger
field), the so-called gravitino\footnote{Particles with
spins bigger than 2 are generally problematic when coupling
to other particles. In particular, this is why
spin 5/2 particles are not considered as superpartners of
the graviton.}. In general relativity, there is a gauge symmetry 
that consists in
reparametrizations of space-time,
which are generated by the momentum operator. As supersymmetry
transformations are related to the momentum operator in eq.
(\ref{susyalg}), we conclude that they must also be local.

This fact is quite restrictive for the construction of 
supergravity theories. In particular, the maximum 
number of dimensions where a consistent supergravity
can be formulated is $D=11$ with ${\cal N}= 1$. In 
dimensions lower than 11, a bunch of supergravities have
been constructed in the last decades. Most of them are
obtainable by Kaluza-Klein
reducing the $D=11$ sugra in some compact
space. A set of new fields will always appear upon
dimensional reduction, as space-time indices along
the directions where compactification has been 
performed turn into internal indices. If one knows
the compactification ansatz that relates two 
supergravities of different dimension, a solution
of one of them can be easily reduced (or uplifted)
to the other one (provided the solution 
somehow respects the
symmetries of the compact space). 
This procedure is quite useful
in the search of solutions, as will become clear in 
the following chapters.

Historically, supergravity was born as a good candidate
to solve the problem of unifying gravity with the rest
of interactions. The boson-fermion loop cancellation
was expected to cure the non-renormalizability of
gravity, while the gauge fields and interactions
could come from the
Kaluza-Klein reduction. Today, it is apparent that this
is not the whole story. String/M-theory is now the main
candidate for unification. However, supergravities appear
as low energy limits of string theories (when the massive
string oscillations have been frozen). In particular, $D=11$
supergravity is the low energy limit of M-theory 
(it cannot be a coincidence when the maximal supergravity
lives in the same number of dimensions as the postulated
``Theory of Everything"!).

%*******************************************************

\setcounter{equation}{0}
\section{Looking for Supersymmetric Solutions: General Strategies}

The aim of this section is to describe the methods that will
be later used in the search of sugra solutions. The problem
in finding solutions is that the supergravity equations of
motion are, in general, complicated systems of second order
equations. The idea is to find somehow, systems of
first order equations (much simpler to deal with), which
automatically solve the second order problem. Supersymmetric
solutions will always satisfy such a first order system 
(of course, there  exist non-susy solutions that cannot
 be found following these strategies).

Another notion we should keep in mind is the possibility of
uplifting a solution obtained in a low dimensional sugra
to ten or eleven dimensions where its physical interpretation
is clearer. We will use gauged supergravity theories that can 
be formulated by compactifying another supergravity in higher
dimension. 
In section
\ref{sugras}, some expressions that facilitate the task of finding the
high dimensional solutions from the low dimensional ones can be found.

%*******************************************************

\subsection{Vanishing of fermion field variations}
\label{vffv}

We will look for classical configurations, so the expectation
value of the fermionic fields should be zero.
As explained in section \ref{secsu}, supercharges
are spin 1/2 fields and they turn fermions into bosons
and vice versa. Schematically:
\bear
\delta F&=&f(B)\,\,, \cr
\delta B&=&g(F)\,\,,
\eear
where $f(B)$ and $g(B)$ are some functions of the bosonic 
and fermionic fields respectively and $\delta$ means susy transformation. 
As the fermions are zero ($\Rightarrow g(F)=0$), the invariance
of the bosonic fields describing the solutions is guaranteed.
In order to preserve susy, the fermionic fields should also not
vary, hence:
\beq
f(B)\,=\,0\,\,,
\label{fB}
\eeq
which gives the desired system of equations, first order in
derivatives.

There are some points worth to comment about this:

The only way of having a manageable system of equations
is to start with an ansatz for the bosonic fields. 
Then, (\ref{fB}) leads to a system from which the functions
in the ansatz can be computed. Needless to say that, to 
find interesting solutions, it is a requisite to begin
with the correct ansatz.

That a configuration is supersymmetric does not necessarily
imply that it is a  solution of the supergravity equations of motion.
However, as susy configurations are related to BPS states, which saturate
some energy bound, they usually are, actually, solutions
of the sugra equations of motion. Anyway, the correct way
of proceeding is to first find the configurations by
imposing supersymmetry and then to directly check the
second order equations of motion.

Usually, for eqs. (\ref{fB}) to be solvable, one must impose
some projections on the spinor that parameterizes the 
transformation. When this happens, not all the supercharges present
in the supergravity theory are preserved by the solution.
These projections are of the type $P
\epsilon=\epsilon$ ($P$ being some function of the gamma matrices,
which will be explicitly showed for each solution.
Notice that all the projectors should commute among
themselves). An
important point is that each independent projection halves
the number of preserved supercharges.

%*******************************************************

\subsection{Superpotential method}
\label{superpotential}

By plugging an ansatz for the fields in terms of some functions
$\alpha^i$ depending on a single coordinate $\eta$ (which in all
the cases studied in this work will be a radial coordinate),
one gets a one-dimensional action. If the action satisfies
certain conditions, there is a direct way of getting a first-order
system which solves the equations of motion.

Let us consider a lagrangian of the type ($S=\int L\,d\eta$):
\beq
L\,=\,\frac{1}{2}\,g_{ij}\,\frac{d\alpha^i}{d\eta}
\,\frac{d\alpha^j}{d\eta}\,-\,V\,\,,
\label{lagr}
\eeq
where $g_{ij}$ is a symmetric matrix (that might depend on the
functions $\alpha^i$) and $V\equiv V(\alpha^i)$. 
The second order Euler-Lagrange 
equations are:
\beq
\frac{d}{d\eta}\left[g_{ij}\frac{d\alpha^j}{d\eta}\right]\,=\,
\frac{1}{2}
\left(\frac{\partial}{\partial\alpha^i}\,g_{jk}\right)
\frac{d\alpha^j}{d\eta}\frac{d\alpha^k}{d\eta}
-\frac{\partial V}{\partial\alpha^i}\,\,.
\label{order2}
\eeq
Let us assume that the potential can be written as:
\beq
V=-\,\frac{1}{2}\,g^{ij}\,\frac{\partial W}{\partial \alpha^i}
\frac{\partial W}{\partial \alpha^j}\,\,,
\label{defsuperp}
\eeq
for some function $W(\alpha^i)$, which we will call 
superpotential, and where $g^{ij}$ has been defined as
the inverse of $g_{ij}$: $g_{ij}g^{jk}=\delta_i^{\ k}$.
Then, it can be straightforwardly proved that the 
first order system:
\beq
\frac{d\alpha^i}{d\eta}\,=\,\mp\,g^{ij}\frac{\partial W}
{\partial \alpha^j}\,\,,
\label{order1}
\eeq
automatically solves (\ref{order2}). Moreover, on this 
solution of the equations of motion,
the hamiltonian identically vanishes:
\beq
H\,=\,\frac{\partial\alpha^i}{\partial \eta}
\frac{\partial L}{\partial \frac{d\alpha^i}{d\eta}}\,-\,L\,=\,0
\,\,.
\label{Hcero}
\eeq
Conversely, it can be proved that  any classical system
whose hamiltonian does not explicitly depend on time, and whose energy
is zero,
can be solved this way. In order to prove this assertion, let us use
Hamilton-Jacobi's formalism. The  Hamilton-Jacobi equation reads:
\beq
{\partial S \over \partial t}\,+\,
H\left(\alpha_i\,,{\partial S \over \partial \alpha_i}\right)\,=\,0\,\,,
\label{hjacobi}
\eeq
where the non-explicit dependence of $H$ on $t$ has been taken 
into account. The Hamilton's principal function $S$ is the 
generating function of a canonical transformation to a system
where coordinates and momenta are constant.
The solution to (\ref{hjacobi}) is $S=-E t+W(\alpha_i)$,
where the constant $E$ is the energy and $W$ is Hamilton's characteristic
function. The equations of motion written in terms of $W$ are
$p_i={\partial W \over
\partial \alpha^i}$, the equivalent of (\ref{order1}). Moreover, by
using (\ref{hjacobi}), the energy can be written:
\beq
E=-{\partial S\over \partial t}=
H=\frac{1}{2}\,g_{ij}\,\frac{d\alpha^i}{d\eta}
\,\frac{d\alpha^j}{d\eta}\,+V=\half g^{ij} p_i \,p_j+V=
\half g^{ij}\frac{\partial W}
{\partial \alpha^i}\frac{\partial W}
{\partial \alpha^j}+V\,\,.
\label{extr1}
\eeq
Now, if $E=0$, eq. (\ref{defsuperp}) is obtained from eq. (\ref{extr1}).
This reasoning shows that the function $W$ which has been
called superpotential is nothing else than Hamilton's characteristic
function.

A supersymmetric configuration can always be obtained as the
solution of a first order system, so this construction is
somehow related to supersymmetry in some cases, although 
by no means it is a proof of it.

It will be useful to apply this method to  lagrangians of the
type:
\beq
L\,=\,e^{c_1 A}\left[c_2 \,(\partial_\eta A)^2\,-\,
\frac{1}{2}\,G_{ab}(\varphi)\,\partial_\eta \varphi^a\,
\partial_\eta \varphi^b\,-\,\tilde{V}(\varphi)\right]\,\,,
\label{newlagr}
\eeq
where the fields $\alpha^i(\eta)$ have been split
$\alpha^i=(A,\varphi^a)$, and $c_1$ and $c_2$ are numbers.
This is of the form (\ref{lagr}) with the identifications:
\beq
g_{AA}=2\,c_2\,e^{c_1 A},\,\,\,\,\,\,\,\,\,\,\,\,\,\,\,\,
g_{ab}=-e^{c_1 A}\,G_{ab},\,\,\,\,\,\,\,\,\,\,\,\,\,\,\,\,
V=e^{c_1 A}\, \tilde{V}(\varphi)\,\,.
\eeq
Suppose it is 
possible to find a function $\tilde{W}(\varphi)$ such
that:
\beq
\tilde{V}(\varphi)\,=\,{c_3^2 \over 2}\,\,G^{ab}\,
\frac{\partial\tilde{W}}{\partial \varphi^a}\,
\frac{\partial\tilde{W}}{\partial \varphi^b}\,\,-\,\,
{c_1^2\,c_3^2 \over 4\, c_2}\,\tilde{W}^2\,\,\,,
\label{newdefsuperp}
\eeq
where $c_3$ is any constant (notice that it is only an
irrelevant rescaling in $\tilde{W}$). Then, equation (\ref{newdefsuperp})
is equivalent to (\ref{defsuperp}) with superpotential:
\beq
W\,=\,c_3\,e^{c_1\,A}\,\tilde{W}\,\,.
\eeq
And so, equations (\ref{order1}) read:
\bear
{dA \over d\eta}&=&\mp\,{c_1\,c_3 \over 2\,c_2}\,
\tilde{W}(\varphi)\,\,\,,\cr\cr
{d\varphi^a \over d\eta}&=&\pm\,c_3\,G^{ab}\,
{\partial \tilde{W}(\varphi) \over  \partial\varphi^b}
\,\,\,,
\label{neworder1}
\eear
which is the sought system of first-order equations.

Let us summarize all the above reasoning: Having a 
lagrangian of the form (\ref{newlagr}), if one manages
to find a function $\tilde{W}(\varphi)$ such that 
(\ref{newdefsuperp}), then the system (\ref{neworder1})
solves the equations of motion and, on this solution,
condition (\ref{Hcero}) holds.

\vskip.1in

To finish the section, let us make a brief comparison 
between the two methods presented. The superpotential method is much less
straightforward, in the sense that, even if a superpotential
(\ref{newdefsuperp}) exists, there is no direct way of finding it, and the
task may be quite difficult if $\tilde{W}(\varphi)$ is a complicated
function. The only problem with the susy variation method is that
one has to deal with the Dirac matrices algebra and it may not
be simple to find the correct projections that must be imposed
on the spinor. But this method has the additional advantage
that one finds the amount of supersymmetry preserved
and the Killing spinors.

%*******************************************************

%*******************************************************

\setcounter{equation}{0}
\section{Supergravity actions and supersymmetry transformations}
\label{sugras}

The aim of this section is to compile the expressions of the
different supergravity theories that will be used throughout 
the thesis. As  we will look for bosonic supersymmetric
configurations, the only sector of the action we will
need is the bosonic one
 (the configurations must be solution of the 
Euler-Lagrange equations derived from it). As also explained in
the previous section, the supersymmetry transformation of the
fermionic fields must be set to zero, so their explicit
expression is needed.
The relation between supergravities in different number of 
dimensions is given.

Notice that not all the bosonic fields are excited in the
solutions that will be explored, so for the sake of simplicity
the fields that will not be used are neglected in the 
expressions of this section. Anyway, references to the 
original papers where the full equations can be found will
be provided in each subsection.

%*******************************************************

\subsubsection{A note on notation}

Both the formalism of differential forms and the formalism
where indices are written explicitly will be used. A differential
$p$-form is defined as:
\beq
\omega_{(p)}\,=\,{1 \over p!}\omega_{{\mu_1}\dots{\mu_p}}
dx^{\mu_1}\wedge\dots\wedge dx^{\mu_p}\,\,.
\eeq
Here, ${\mu_1}\dots{\mu_p}$ are curved indices, referred to
the coordinate basis. We will often use the tangent space
basis, and therefore flat indices:
\beq
ds^2\,=\,g_{\mu\nu}\,dx^\mu dx^\nu\,=\,\eta_{ab}\,e^a e^b\,\,,
\eeq
where $\eta_{ab}={\rm diag}(-1,+1,\dots,+1)$. The $e^a$ one-forms
are the components of the so-called vielbein. They can be expressed in
components: $e^a=e^a_\mu(x) dx^\mu$, so the $e^a_\mu(x)$ transform
between flat and curved indices. The spin connection of a
metric can be found by solving the so-called Cartan's structure
equations:
\beq
0=de^a+ \omega^a_{\ b} \wedge e^b\,\,.
\label{structure}
\eeq
The spin connection is a one-form which in components
reads: $\omega^a_{\ b}=\omega^a_{\ b\mu} dx^\mu$.
We need to define the covariant derivative, whose
action on a spinor is given by:
\beq
D_{\mu} \epsilon = (\partial_{\mu} + {1\over 4}\,\omega_{\mu}^{ab}
\G_{ab})\,\epsilon\,.
\label{covariant}
\eeq
As we will deal with gauged supergravities, we will need at
some point to introduce gauge covariant derivatives, {\it i.e.}
that take into account the gauge connection besides the spin
connection. They will be denoted by the symbol ${\cal D}$ and
its precise definition will be given in each case.

In (\ref{covariant}), the $\G_{ab}$ are Dirac matrices with indices
referring to the vielbein basis. They satisfy the algebra:
\beq
\{\G_a,\G_b\}\,=\,2\,\eta_{ab}\,\,.
\eeq
The symbol with several indices in a single gamma
$\G_{\mu_1\cdots\mu_n}$ will denote an antisymmetrized product
of Dirac matrices.

%*******************************************************

\subsection{D=11, ${\cal N}=1$ supergravity}

Eleven is the maximal dimension where a supergravity can exist
\cite{Nahm}. The action and supersymmetry transformation laws
were first constructed in \cite{Cremmer}. The number of
supercharges is 32, corresponding to one Majorana spinor.

The bosonic content of the theory includes only the metric
and a 3-form potential (with a 4-form field strength
$F_{(4)}=dC_{(3)}$). The action for these fields is:

\beq
{\cal L}=\sqrt{-g} \left[ R-\frac{1}{48}
F_{\mu\nu\rho\sigma}F^{\mu\nu\rho\sigma}\right]
+\frac{1}{144^2}\epsilon^{\alpha_1\dots\alpha_4\beta_1\dots\beta_4
\mu\nu\rho}F_{\alpha_1\dots\alpha_4}F_{\beta_1\dots\beta_4}
C_{\mu\nu\rho}\,\,.
\eeq

The only fermionic degrees of freedom are those corresponding to a
Rarita-Schwinger field $\psi_\mu$, the gravitino. Its  supersymmetry
variation is given by:
\beq
\delta\psi_{\mu}\,=\,D_{\mu}\,\epsilon\,+\,{1\over 288}\,
F^{(4)}_{\mu_1\dots \mu_4}\,\,
\Bigg(\,\G_{\mu}^{\,\,\mu_1\dots \mu_4}\,-\,8\,\delta_{\mu}^{\mu_1}\,
\G^{\,\,\mu_2\dots \mu_4}\,\Bigg)\,\epsilon\,\,,
\label{susy11}
\eeq

%*******************************************************

\subsection{D=10 type IIA supergravity}
\label{typeIIA}

Eleven dimensional supergravity can be dimensionally reduced yielding
maximal ({\it i.e.} with 32 supercharges) non-chiral supergravity in ten
dimensions \cite{type2a}. The resulting theory is called type IIA
supergravity and it is a low energy limit of type IIA string theory. The
Kaluza-Klein reduction ansatz for the metric is:
\beq
ds^2_{11}\,=\,e^{-{2\over 3}\,\phi}\,ds^2_{10}\,+\,
e^{{4\over 3}\,\phi}\,(\,dz\,+\,C_{(1)}\,)^2\,\,.
\label{KKred}
\eeq
Furthermore, one has to reduce the eleven dimensional three-form, which
generates in ten dimensions a three-form and a two-form, depending on
whether or not the reduction direction is comprised among the indices of
the original form. Therefore, the bosonic content of this theory consists
of a  metric $g_{\mu\nu}$, a dilaton $\phi$ and a Ramond-Ramond one-form
$C_{(1)}$ coming from the reduction of the metric, besides a
Neveu-Schwarz two-form $B_{(2)}$ and an RR three-form $C_{(3)}$ coming
from the reduction of the three-form\footnote{It is worth pointing out
the meaning of each potential in terms of the branes of the corresponding
string theory. Fundamental strings are electrically charged with respect
to $B_{(2)}$, while their duals NS5-branes couple magnetically to this
potential. In the same vein, D0 and D6-branes couple to $C_{(1)}$ and D2
and D4-branes to $C_{(3)}$. Notice that, as $C_{(1)}$ comes from the
eleven dimensional metric, D0 and D6-brane configurations must uplift to
pure geometry in eleven dimensions.}. The bosonic action (in Einstein
frame\footnote{In the so-called Einstein frame, the lagrangian includes
an Einstein-like gravitational term $\sqrt{-g}\,R$. On the other hand,
in the so-called string frame, which is natural from the string sigma
model point of view, the corresponding term of the lagrangian reads
$\sqrt{-g}\,e^{-2\phi}\,R$. Both frames are related by a rescaling of
the metric by some power (which depends on the number of dimensions) of
the dilaton. See \cite{librortin} for a description of the relation
between the actions and equations of motion in both frames.}) of this
theory is:
\bear
S&=&\int d^{10}x \sqrt{-g}\,\left[R-\half
\partial_{\mu}\phi\,\partial^{\mu}\phi\,-{1 \over 12}e^{-\phi}
H_{(3)}^2-{1\over 4}e^{{3\over 2}\phi}F_{(2)}^2
-{1\over 48}e^{\half \phi}F_{(4)}^2\right]+\rc
&&+\,\half \int B_{(2)}\wedge dC_{(3)} \wedge dC_{(3)}\,\,,
\eear
where   the field strengths have been defined: $F_{(2)}=dC_{(1)}$,
$H_{(3)}=dB_{(2)}$ and $F_{(4)}=dC_{(3)}+C_{(1)}\wedge H_{(3)}\,$.
On the other hand, the fermionic content of the theory comprises two
Majorana spinors: a gravitino $\psi_\mu$ and a dilatino $\la$, each
decomposable into two Majorana-Weyl components. Their supersymmetry
variations read (Einstein frame):
\bear
\delta\la&=&{1\over 4}\sqrt{2}\,D_\mu\phi\,\G^\mu\G^{11}\epsilon
+{3\over 16}{1\over
\sqrt{2}}\,e^{{3\phi\over 4}}F^{(2)}_{\mu_1\mu_2}
\G^{\mu_1\mu_2}\epsilon+\rc
&&+{1\over 24}{i\over
\sqrt{2}}\,e^{-{\phi\over 2}}H^{(3)}_{\mu_1\mu_2\mu_3}
\G^{\mu_1\mu_2\mu_3}\epsilon
-{1\over 192}{i\over \sqrt{2}}\,e^{{\phi
\over 4}}F^{(4)}_{\mu_1\mu_2\mu_3\mu_4}
\G^{\mu_1\mu_2\mu_3\mu_4}\epsilon\,\,,\rc\rc
\delta\psi_\mu&=&D_\mu\epsilon+{1\over 64} \,e^{{3\phi \over 4}}\,
F_{\mu_1\mu_2}^{(2)}\,
\left(\G_\mu^{\ \mu_1\mu_2}-14\delta_\mu^{\mu_1}\G^{\mu_2}\right)\G^{11}
\,\epsilon\,+\rc
&&+\,{1\over 96}\,e^{-{\phi \over 2}}\,H_{\mu_1\mu_2\mu_3}^{(3)}\,
\left(\G_\mu^{\
\mu_1\mu_2\mu_3}-9\delta_\mu^{\mu_1}\G^{\mu_2\mu_3}\right)\G^{11}
\,\epsilon\,+\rc
&&+\,{i\over 256}\,e^{{\phi \over 4}}\,F_{\mu_1\mu_2\mu_3\mu_4}^{(4)}\,
\left(\G_\mu^{\
\mu_1\mu_2\mu_3\mu_4}-
{20 \over 3}\delta_\mu^{\mu_1}\G^{\mu_2\mu_3\mu_4}\right)\G^{11}
\,\epsilon\,\,.
\eear
The chirality operator $\G^{11}$ is defined as 
$\G^{11}=i\G^0\G^1\dots\G^9\,$.

%*******************************************************

\subsection{D=10 type IIB supergravity}
\label{typeIIB}

There is another maximal supergravity that can be constructed in ten
dimensions \cite{type2b}. This type IIB theory is chiral and cannot be
obtained by dimensional reduction from eleven dimensions. Nevertheless,
it is related to type IIA sugra by T-duality. The bosonic degrees of
freedom are the metric $g_{\mu\nu}$, the dilaton $\phi$, a NSNS two-form
$B_{(2)}$, a Ramond-Ramond scalar $\chi$, an RR two-form $C_{(2)}$ and
an RR four-form $C_{(4)}$. The action for these fields reads
(in Einstein frame):
\bear
S&=&\int d^{10}x \sqrt{-g}\Big[R-\half
\partial_{\mu}\phi\,\partial^{\mu}\phi-{1\over 12}e^{-\phi}H_{(3)}^2-
\half e^{2\phi}\partial_{\mu}\chi\,\partial^{\mu}\chi
-{1\over 12}e^\phi F_{(3)}^2-\rc
&&-{1\over 240} F_{(5)}^2\,\Big]+\int C_{(4)}\wedge F_{(3)}
\wedge H_{(3)}\,\,,
\eear
where the following definitions have been used: $F_{(3)}=dC_{(2)}-\chi
H_{(3)}\ $ and $F_{(5)}=dC_{(4)}+C_{(2)}\wedge H_{(3)}$. Apart from the
equations of motions that arise from this action, one has additionally to
impose the self-duality condition $F_{(5)}={}^*F_{(5)}\ $.

Let us now consider the susy variations of the fermionic fields, a
dilatino $\la$ and a gravitino $\psi_\mu\ $.
In the type IIB theory the spinor $\epsilon$ is actually composed
by two Majorana-Weyl spinors $\epsilon_L$ and $\epsilon_R$ of well defined
ten-dimensional chirality, which can be arranged as a two-component vector in
the form:
\beq
\epsilon\,=\,\pmatrix{\epsilon_L\cr\epsilon_R}\,\,.
\label{striibsp}
\eeq
We can use complex spinors instead of working with the real 
two-component spinor written in eq.
(\ref{striibsp}).  If 
$\epsilon_R$ and $\epsilon_L$ are the two components of the real spinor written
in eq. (\ref{striibsp}), the complex spinor is simply:
\beq
\epsilon\,=\,\epsilon_L\,+\,i\,\epsilon_R\,\,.
\eeq
We have the following rules to pass 
from one notation to the other:
\beq
\epsilon^*\,\leftrightarrow\,\sigma_3\,\epsilon\,\,,
\,\,\,\,\,\,\,\,\,\,\,\,\,\,\,\,\,\,\,
i\,\epsilon^*\,\leftrightarrow\,\sigma_1\,\epsilon\,\,,
\,\,\,\,\,\,\,\,\,\,\,\,\,\,\,\,\,\,\,
i\,\epsilon\,\leftrightarrow\,-i\,\sigma_2\,\epsilon\,\,,
\label{striibcn}
\eeq
where the $\sigma$'s are Pauli matrices.
Using complex spinors,  
the supersymmetry transformations of the dilatino $\lambda$ and gravitino 
$\psi_\mu$ in type IIB supergravity are (Einstein frame):
\bear
\delta\la&=&i\,P_{\mu}\,\G^{\mu}\epsilon^{*}\,-\,{i\over 24}
\,F_{\mu_1\mu_2\mu_3}\,
\G^{\mu_1\mu_2\mu_3}\,\epsilon\,\,,\rc\rc
\delta\psi_{\mu}&=&D_{\mu}\epsilon\,-\,{i\over 1920}
\,F^{(5)}_{\mu_1\dots\mu_5}\,\G^{\mu_1\dots\mu_5}
\G_{\mu}\,\epsilon\,+\rc\rc
&+&{1\over 96}\,
F_{\mu_1\mu_2\mu_3}\,\Big(\,\G_{\mu}^{\,\,\,\mu_1\mu_2\mu_3}\,-\,9\,
\delta_{\mu}^{\mu_1}\,\G^{\mu_2\mu_3}\,\Big)\,\epsilon^*\,\,,
\label{SUSYIIB}
\eear
where
$P_{\mu}$ and $F_{\mu_1\mu_2\mu_3}$ are given by:
\bear
P_{\mu}&=&{1\over 2}\,\big[\,\partial_\mu\phi\,+\,ie^{\phi}\,
\partial_{\mu}\chi\,\big]\,\,,\rc\rc
F_{\mu_1\mu_2\mu_3}&=&e^{-{\phi\over 2}}\,H^{(3)}_{\mu_1\mu_2\mu_3}\,+\,
ie^{{\phi\over 2}}\,F^{(3)}_{\mu_1\mu_2\mu_3}\,\,.
\label{ncapaccinco}
\eear
A  thorough review on eleven and ten dimensional supergravities, the
relations among them (Kaluza-Klein reduction, T-duality), solutions
from branes and many other topics on gravity and its relation with
strings can be found in \cite{librortin}.

%*******************************************************

\subsection{D=8, ${\cal N}=2$ $SU(2)$ gauged supergravity} 
\label{ss8d}

The easiest way to dimensionally reduce a supergravity theory
is to impose that nothing depends on the coordinates of the
dimensions where the reduction is made. However, Scherk and
Schwarz proved (\cite{ssch}, see also \cite{ssch2}) that one
can allow the fields and transformation laws to depend on
the internal coordinates in a well defined fashion, satisfying 
some criteria. The idea is to reduce in a Lie group ($G$) 
 manifold such
that the dependence of the fields and transformation laws on
the internal coordinates appears in a simple factorizable form. It turns
out that the vector fields coming from the reduction
of the metric are gauge fields with gauge group $G$ in the lower
dimensional theory.

\vskip.1in

The maximal eight dimensional gauged supergravity was constructed by 
Salam and
Sezgin in ref. \cite{ss} by means of a Scherk-Schwarz
compactification of D=11
supergravity
on a
$SU(2)$ group manifold. The total number of supercharges is
32 (two Weyl spinors).

The bosonic field content of this
theory can be truncated to include the metric $g_{\mu\nu}$,
a dilatonic scalar $\phi$, five scalars parametrized by a $3\times 3$
unimodular matrix $L_{\alpha}^i$ which lives in the coset
$SL(3,\RR)/SO(3)$, an $SU(2)$ gauge potential $A_{\mu}^i$ and a 
three-form
potential $B_{(3)}$. The kinetic energy of the coset scalars
$L_{\alpha}^i$  is given
in terms of the symmetric traceless  matrix $P_{\mu\,ij}$ defined by
 means of
the expression:
\beq
\left(P_\mu\right)_{(ij)}+\left(Q_\mu\right)_{[ij]}\,\,=\,L_i^{\alpha}\,
(\,\partial_{\mu}\,\delta_{\alpha}^{\beta}\,-\,\epsilon_{\alpha\beta\gamma}\,A_{\mu}^{\gamma}\,)\,L_{\beta
j}\,\,,
\label{apauno}
\eeq
where $Q_{\mu\,ij}$ is, by definition, the antisymmetric part of the right-hand
side of eq. (\ref{apauno}). Furthermore, the potential energy of the coset
scalars is written in terms of the so-called $T$-tensor, $T^{ij}$, and of its
trace, $T$, defined as:
\beq
T^{ij}\,=\,L^i_{\alpha}\,L^j_{\beta}\,\delta^{\alpha\beta}\,\,,
\,\,\,\,\,\,\,\,\,\,\,\,\,\,\,\,\,\,\,\,\,\,\,\,
T\,=\,\delta_{ij}\,T^{ij}\,\,.
\label{apados}
\eeq
The field strength $F_{\mu\nu}^{\alpha}$ of
the $SU(2)$ gauge field
$A_{\mu}^\alpha$ reads:
\beq
F^\alpha=dA^\alpha+\frac{1}{2}\epsilon_{\alpha\beta\gamma}
A^\beta\wedge A^\gamma,
\label{gaugefstr}
\eeq
where the $SU(2)$ gauge coupling constant has been set to 1.
If $G_{\mu\nu\rho\sigma}$ denotes the components of
$dB_{(3)}$, the bosonic lagrangian for this truncation of D=8 gauged
supergravity is:
\bear
{\cal L}&=&\sqrt{-g_{(8)}}\,\,\Big[\,
{1\over 4}\,R\,-\,{1\over 4}\,e^{2\phi}\,
F_{\mu\nu}^{i}\,F^{\mu\nu\,\,i}\,-\,{1\over 4}\,
P_{\mu\,ij}\,P^{\mu\,ij}\,-\,{1\over 2}\,
\partial_{\mu}\,\phi\partial^{\mu}\,\phi\,-\cr\cr
&&-{1\over 16}\,e^{-2\phi}\,(\,T_{ij}\,T^{ij}\,-\,{1\over 2}T^2\,)\,-\,
{1\over 48}\,e^{2\phi}\,G_{\mu\nu\rho\sigma}\,
G^{\mu\nu\rho\sigma}\,\,\Big]\,\,.
\label{apatres}
\eear
This truncation is not consistent in general, as some of these
fields act as sources for the other fields present in the 
full Salam-Sezgin supergravity which have been ignored.
For these sources to vanish, the following conditions must be
imposed:
\beq
G \wedge G = \ast G \wedge F^i = 0 \,,
\label{consist}
\eeq
where $\ast G$ is the Hodge dual
of $G$ in eight dimensions\footnote{This can be immediately
obtained by looking at the equations of motion written in
\cite{ss}. Clearly, if $G=0$, the consistency is trivial.}.

 The eleven
dimensional reduction anstaz,
that can be readily used to uplift eight dimensional solutions is,
for the metric:
\beq
ds^2_{11}\,=\,e^{-{2\over 3}\,\phi}\,ds^2_8\,+\,4\,e^{{4\over 3}\phi}\,
(\,A^i\,+\,{1\over 2}\,L^i\,)^2\,\,,
\label{uplift8d}
\eeq
where $L^i$ is defined as:
\beq
L^i\,=\,2\, \tilde{w}^{\alpha}\,L_{\alpha}^i\,\,,
\label{apaseis}
\eeq
with $\tilde{w}^{i}$ being left-invariant forms on the $SU(2)$ group
manifold, satisfying:
\beq
d\tilde{w}^i\,=\,{1\over
2}\,\epsilon_{ijk}\,\tilde{w}^j\,\wedge\,\tilde{w}^k\,\,.
\label{tres}
\eeq
In terms of the angles parameterizing the $S^3$:
\bear
\tilde{w}^1&=& \cos\tilde{\psi}\,
d\tilde\theta\,+\,\sin\tilde{\psi}\,\sin\tilde\theta
\,d\tilde\varphi\,\,,\rc
\tilde{w}^2&=&\sin\tilde{\psi}\,
d\tilde\theta\,-\,\cos\tilde{\psi}\,\sin\tilde\theta 
\,d\tilde\varphi\,\,,\rc
\tilde{w}^3&=&d\tilde{\psi}\,+\,\cos\tilde\theta \,d\tilde\varphi\,\,.
\label{omegas}
\eear
The three angles $\tilde\varphi$, $\tilde\theta$ and $\tilde{\psi}$ 
take values in the
rank $0\le\tilde\varphi< 2\pi$, $0\le\tilde\theta\le\pi$ and
$0\le\tilde{\psi}< 4\pi$.

Finally, the 4-form $G$ comes directly from the reduction of the
four form field strength $F$ of D=11 sugra, when no index is along
a reduced dimension. The relation between both is\footnote{The factor
of two is needed to pass from the Salam--Sezgin conventions of eleven
dimensional supergravity to the more standard ones.}:
\beq
F_{\underline{\mu\nu\rho\sigma}}\,=\,2\,e^{{4\phi\over 3}}\,
G_{\underline{\mu\nu\rho\sigma}}\,\,,
\label{cseissalam}
\eeq
with underlined indices referring to the tangent space basis.

The fermionic fields are
two pseudo-Majorana spinors $\psi_{\lambda}$
and $\chi_i$ and their supersymmetry transformations are:
\bear
\delta\psi_{\lambda}&=&{\cal D}_{\lambda}\,\epsilon\,+\,{1\over
24}\,e^{\phi}\,
F_{\mu\nu}^{i}\,\hat\Gamma_i\,(\,\Gamma_{\lambda}^{\mu\nu}\,-\,
10\,\delta_{\lambda}^{\mu}\,\Gamma^{\nu}\,)\,\epsilon\,-\,{1\over 288}\,
e^{-\phi}\epsilon_{ijk}\,\hat\Gamma^{ijk}\Gamma_{\lambda}\,T\epsilon\,-\rc\rc
&&-{1\over 96}\,e^{\phi}\,G_{\mu\nu\rho\sigma}\,(\,
\Gamma_{\lambda}^{\mu\nu\rho\sigma}\,
-\,4\delta_{\lambda}^{\mu}\,\Gamma^{\nu\rho\sigma}\,)\,\epsilon\,\,,\rc\rc
\delta\chi_i&=&{1\over 2}\,(P_{\mu ij}\,+\,{2\over 3}\,\delta_{ij}\,
\partial_{\mu}\phi\,)\,\hat\Gamma^j\,\Gamma^{\mu}\,\epsilon\,-\,
{1\over 4}\,e^{\phi}\,F_{\mu\nu i}\,\Gamma^{\mu\nu}\,\epsilon\,-\,
{1\over 8}\,e^{-\phi}\,(\,T_{ij}\,-\,{1\over 2}\,\delta_{ij}\,T\,)\,
\epsilon^{jkl}\hat\Gamma_{kl}\epsilon\,-\,\rc\rc
&&-{1\over 144}\,e^{\phi}\,G_{\mu\nu\rho\sigma}\,\hat\Gamma_i\,
\Gamma^{\mu\nu\rho\sigma}\,\epsilon\,\,,
\label{susy8}
\eear
where the symbol ${\cal D}$ stands for the full gauge covariant
derivative. Its explicit definition is:
\beq
{\cal D}_\mu\,\epsilon\,=\,\left(
\partial_\mu\,+\,\frac{1}{4}\,\omega^{ab}_\mu\,\G_{ab}\,+\,
\frac{1}{4}\,Q_{\mu ij}\,\hat\G^{ij}\right)\,\epsilon\,\,.
\eeq
The following representation of the Clifford algebra can be used:
\beq
\Gamma^{a}\,=\,\gamma^{\underline a}\,\,\otimes\,\II\,\,,
\,\,\,\,\,\,\,\,\,\,\,\,\,\,\,\,\,\,\,\,\,\,
\hat\Gamma^{i}\,=\,\gamma_9\,\,\otimes\,\sigma^i\,\,,
\label{gammasalam}
\eeq
where $\gamma^{\underline a}$ are eight dimensional Dirac matrices,
$\sigma^i$ are Pauli matrices and  $\gamma_9\,=\,i \gamma^{\underline 0}\,
\gamma^{\underline 1}\,\cdots\,\gamma^{\underline 7}$ ($\gamma_9^2\,=\,1$).
From this representation of the gamma matrices, it is 
immediate to find the useful expression:
\beq
\Gamma_{01\cdots 7} 
\hat\Gamma_{123} \,=\,-1\,\,\,.
\label{gammaprod}
\eeq

The way of writing the Dirac matrices is a remnant from the eleven
dimensional theory. Upon uplifting, the unhatted gammas would
become the 11d gammas along the directions present in the 8d
solution, while the hatted gammas would directly
correspond to 11d Dirac matrices
along the three directions of the $SU(2)$
group manifold, using the vielbein
that naturally arises from (\ref{uplift8d}).

%*******************************************************

\subsection{D=7, ${\cal N}=2$ $SU(2)$ gauged supergravity}
\label{TVN}

This supergravity was first constructed by Townsend and van
Nieuwenhuizen \cite{TVN} by directly gauging simple ${\cal N}=2$
supergravity in  seven 
dimensions\footnote{An $SO(4)$ ${\cal N}=2$ gauged sugra in
D=7 was constructed in \cite{ss2}.}. 
Much later, it was reobtained as an $S^4$
reduction of D=11 sugra \cite{d7d11} and as
a Scherk-Schwarz compactification of D=10  
sugra in an $SU(2)$ group
manifold \cite{sabra} (see also \cite{kksphere}).
There are only 16 supercharges. Therefore, it is not the
most extended sugra that can be formulated in D=7. In fact, it
can be obtained as a truncation of the theory that will be
presented in the next subsection. The advantages in
looking for solutions of the reduced theory, instead of dealing with the
maximal one,
are that it is quite simpler and that uplifting  is much more
trivial.

The bosonic field content consists of the metric $g_{\mu\nu}$,
the dilaton $\phi$, a 3-form potential $B_{(3)}$ (which can be, 
equivalently, dualized into a 2-form) and the $SU(2)$ gauge fields
$A_\mu^i$\ . Once again, we will set to one the gauge coupling constant.

The action for these fields in string frame \cite{acharya}
is\footnote{The action can be further generalized by the 
inclusion of a ``topological mass term" $h$ \cite{TVN},
which here will be set to zero.}:
\bear
{\cal L}&=&
\sqrt{-g_{(7)}}\,e^{-2\phi}\,\left[
R-\frac{1}{8}F_{\mu\nu}^{i}\,F^{\mu\nu\,\,i}\,+
\,4\partial_{\mu}\,\phi\partial^{\mu}\,\phi\,+4\,\right]
\,-\cr
&-&\frac{1}{2} e^{2\phi}\ \,{}^*G_{(4)}\wedge
G_{(4)}+\frac{1}{4}\,F^a\wedge F^a\wedge B_{(3)}\,\,,
\label{lagrTVN}
\eear
where $G_{(4)}=dB_{(3)}$ and the field strength is defined 
as in (\ref{gaugefstr}). From this action, it is immediate to
see that the 3-form $B_{(3)}$ can be consistently taken to
vanish only if $F \wedge F =0$ because, otherwise, it 
acts as a source because of the last term.

The fermionic fields are a dilatino $\lambda$ and a gravitino
$\psi_\mu$. Their supersymmetric variations 
are\footnote{Different conventions used in the literature may
lead to confusion. In order to maintain the definition
(\ref{gaugefstr}), the gauge field and its field strength must
be defined with the opposite sign to \cite{acharya}.
The action, being quadratic in $F$ does not get modified,
but the susy variations do.
This is the convention used in \cite{kksphere}, so the
uplifting equations written there can be used without changes in
what refers to the gauge field.} (string frame):
\bear
\delta\lambda&=&\left[\,\G^\mu\,\partial_\mu\phi\,+\,
\frac{i}{8}\,\G^{\mu\nu}\,F^i_{\mu\nu}\,\sigma^i\,+\,
\frac{1}{48}\,\G^{\mu\nu\rho\tau}\,G_{\mu\nu\rho\tau}\,+\,1
\,\right]\ \epsilon \,\,,\cr
\delta\psi_\mu&=&\left[D_\mu\,-\,\frac{i}{2}A^i_\mu\,\sigma^i\,+
\,\frac{i}{4}\,F_{\mu\nu}^i\,\G^\nu\,\sigma^i\,+\,
{1 \over 96}\,e^{\phi}\,\G_\mu^{\,\,\,\nu\rho\tau\delta}\,
G_{\nu\rho\tau\delta}\,\right]\ \epsilon\,\,\,,
\label{susyTVN}
\eear
where $\sigma^i$ are the Pauli matrices rotating the 
$SU(2)$ internal space and $D_\mu$ is as
in (\ref{covariant}).

The relation to higher dimensional fields can be read from
\cite{kksphere}. Adapting notations, we see that from
a seven dimensional solution, the corresponding ten
dimensional metric and NSNS three-form are (in Einstein 
frame)\footnote{It should be noticed that in \cite{kksphere}
(eqs. (35)-(38)), the low dimensional theory is also in
Einstein frame and the metric $ds_7^2$ must be multiplied by a factor
of $e^{-\frac{4\phi}{5}}$ in order to match notations.}:
\bear
ds_{10}^2&=&e^{-\frac{\phi}{2}}\,\left[\,ds_7^2\,+\,
\frac{1}{4}\,\sum_i\,(\om^i-A^i)^2\,\right] \,\,, \cr
H_{(3)}&=&-e^{2\phi}\ast G_{(4)}
-\frac{1}{4}(\om^1-A^1)\wedge(\om^2-A^2)\wedge(\om^3-A^3)\,+\,
\frac{1}{4}\sum_i\,F^i\wedge\,(\om^i-A^i)\,,\,\,\,\,\,\,\,\,\,\,\,\,
\,\,\,\,\,\,\,\,\,
\label{upliftTVN}
\eear
and the dilaton stays the same. The Hodge dual is calculated
with the 7d string frame metric. The $\om^i$ are left-invariant
$SU(2)$ one-forms as in (\ref{tres}), but $\om^2$ is defined
with the opposite sign to $\tilde{w}^2$, 
so their algebra is:
\beq
d\om^i\,=\,-\,{1\over 2}\,\epsilon_{ijk}\,\om^j\,\wedge\,\om^k\,\,,
\label{othertres}
\eeq
and their explicit expression:
\bear
\om^1&=& \cos\tilde{\psi}\,
d\tilde\theta\,+\,\sin\tilde{\psi}\,\sin\tilde\theta
\,d\tilde\varphi\,\,,\rc
\om^2&=&-\sin\tilde{\psi}\,
d\tilde\theta\,+\,\cos\tilde{\psi}\,\sin\tilde\theta 
\,d\tilde\varphi\,\,,\rc
\om^3&=&d\tilde{\psi}\,+\,\cos\tilde\theta\, d\tilde\varphi\,\,.
\label{otherom}
\eear
The underlining has been introduced with the purpose of minimizing
the degree of confusion introduced by notation. Hopefully, it
will be clear when we are using one-forms satisfying (\ref{tres})
and when they are of the kind (\ref{othertres}). Certainly, everything can
be defined using just one expression for all the one-forms, but
that would make more intricate the relation of the equations with those in
the cited literature.

%*******************************************************

\subsection{D=7, ${\cal N}=4$ SO(5) gauged supergravity}
\label{PPvN}

This maximal (32 supercharges) sugra was found by
Pernici, Pilch and van Nieuwenhuizen \cite{PPVN}. It
comes from compactification of D=11 sugra on an
$S^4$ \cite{vaman}. The seven dimensional supergravity of the previous
section is just a truncation of this one. Here, by allowing a
larger gauge group and more degrees of freedom, a more general
situation is taken into account.

The bosonic content of the theory includes the metric,
14 scalar degrees of freedom parametrizing the coset space
$SL(5,\RR)/SO(5)$ that will be denoted by $V_I^i$,  3-form
potentials $C^I_{(3)}$ and the SO(5) gauge field $A_\mu^{IJ}$.
The indices $i$, $j$, $I$, $J$ run from 1 to 5. The bosonic
lagrangian takes the form (the notation of \cite{liumin} is
used, mainly).
\bear
{\cal L}\,&=&\,\half\sqrt{-g_{(7)}}\left[R+\half (T^2-2T_{ij}
T^{ij})-{\rm Tr}(P_\mu P^\mu)-\half(V_I^iV_J^jF_{\mu\nu}^{IJ})^2
+\left((V^{-1})^I_iC^I_{\mu\nu\rho}\right)^2\right]+\rc
&&+{1 \over 4}\delta^{IJ}(C_{(3)})_I\wedge (dC_{(3)})_J+
\half \epsilon_{IJKLM}(C_{(3)})_I \wedge F^{JK}\wedge F^{LM}+
\half p_2(A,F)\,\,.
\label{PPvNlagr}
\eear
The gauge coupling $m$ and gravitational coupling $\kappa$ have been taken
to one.
$p_2(A,F)$ is a  Chern-Simons term that vanishes for all the cases
considered in this work. Moreover, the gauge field strength is obtained
from the gauge field as:
\beq
F\,=\,dA\,+\,2\,[A,A]\,\,,
\eeq
and the $P$ and $Q$ matrices are defined as:
\beq
(V^{-1})_i^I {\cal D}_\mu V_I^j\,=\,(Q_\mu)_{[ij]}\,+\,
(P_\mu)_{(ij)}\,\,.
\eeq
${\cal D}_\mu$ is a gauge covariant derivative, and its action
on the scalars and on the spinors reads:
\beq
{\cal D}_\mu V_I^j\,=\,\partial_\mu
V_I^j\,+\,2(A_\mu)_I^{\,\,\,J}V_J^j\,\,,
\quad\quad
{\cal D}_\mu \psi\,=\,\left(\partial_\mu+{1\over 4}Q_{\mu ij}\G^{ij}
+{1\over 4}\omega_\mu^{ab}\gamma_{ab}\right)\psi\,\,.
\eeq
The $T$ tensor, coming from the scalar fields is defined by:
\beq
T_{ij}\,=\,(V^{-1})_i^I\,(V^{-1})_j^J\,\delta_{IJ}\,\,,
\quad\quad\quad
T\,=\,T_{ij}\,\delta^{ij}\,\,.
\eeq
The fermionic fields comprise the gravitino and a set of 
spin-$\half$ fermions. In the following, $\gamma_\mu$ will be 
the seven dimensional space-time Dirac matrices, while the
$\G_i$ will be a set of five dimensional Dirac matrices
living in the internal space, with signature $(+++++)$.
The spin-$\half$ fermions must fulfil the irreducibility
condition $\G^i\la_i=0$. The supersymmetry transformations
of the fermionic fields are:
\bear
\delta\psi_\mu&=&\Big[\,{\cal D}_\mu\,+\,{1\over 20}\,T
\,\gamma_\mu\,-\,{1\over 40}\,
(\gamma_\mu^{\
\nu\la}\,-\,8\,\delta_\mu^\nu\,\gamma^\la)\,\G^{ij}\,V_I^i\,V_J^j
\,F_{\nu\la}^{IJ}\,+\rc
&&+\,{1\over 10\sqrt{3}}\,(\gamma_\mu^{\ \nu\la\sigma}\,
-\,\frac{9}{2}\,\delta_\mu^\nu\,\gamma^{\la\sigma})\,\G^i\,(V^{-1})_i^I
\,C_{\nu\la\sigma}^I\,\Big]\,\epsilon\,\,,\rc\rc
\delta\la_i&=&\Big[\,\half\,(T_{ij}\,-\,\frac{1}{5}\,\delta_{ij}\,T)
\,\G^j\,+\,
\frac{1}{16}\,\gamma^{\mu\nu}\,(\G^{kl}\G^i\,-\,\frac{1}{5}\G^i\G^{kl})
V_K^k \,V_L^l\, F_{\mu\nu}^{KL}\,+\rc
&&+\,\half\,\gamma^\mu \,P_{\mu ij}\,\G^j\,+\,
{1\over 20\sqrt{3}}\,\gamma^{\mu\nu\la}\,(\G^{ij}-4\delta^{ij})
(V^{-1})_j^J\,C_{\mu\nu\la}^J\,\Big]\,\epsilon\,\,.
\label{PPvNsusy}
\eear
The formulae relating the seven dimensional fields to the 
eleven dimensional ones can be found in \cite{vaman}, see also
\cite{tran}.

%*******************************************************

\setcounter{equation}{0}
\section{The twist}
\label{twistsec}

Throughout this work we are going to consider non-trivial
supergravity solutions corresponding to branes which have
part of their worldvolume wrapped along some cycles. 
Such a curved worldvolume does not support, in general, a
covariantly constant spinor. This seems to contradict the
fact that D-branes are 1/2-supersymmetric objects. What happens
is that supersymmetry is not realized in the usual way, but
involves a twisted definition of the supercharges \cite{bvs}.

Let us think about this from the perspective of low dimensional gauged
supergravity, along the lines of \cite{MNfirst}. We start with a geometry
including the cycle where the brane is wrapped. Then, in general,
one cannot fulfil the condition
$D_\mu\epsilon=(\partial_\mu+\omega_\mu)\epsilon=0$. However, one
can couple the theory to the gauge field, in order to satisfy
the following schematic equations:
\beq
A_\mu=\omega_\mu\qquad \Rightarrow\qquad {\cal D}_\mu\epsilon=
(\partial_\mu+\omega_\mu-A_\mu)\epsilon=\partial_\mu\epsilon=0\,\,,
\label{twisting}
\eeq
which can be immediately solved by taking a constant spinor.
Therefore, the way of getting supersymmetric solutions related to
wrapped branes is by appropriately identifying the spin connection
with the gauge connection related to the R-symmetry group. This
coupling to the gauge field changes the spins of all fields, resulting in
what is called a twisted field theory. However, when one takes into
account that the cycle where the brane is wrapped is small and
one decouples the corresponding Kaluza-Klein modes, it is
possible to end up with an ordinary (not twisted) field theory living in
the unwrapped worldvolume of the brane.

We now consider the uplifting to eleven dimensions of a solution of this
kind. If there are only type IIA D6-branes, the eleven dimensional
solution must be pure geometry. Therefore, we get a Ricci flat
manifold with reduced supersymmetry ${\cal Y}_p$ (which implies reduced
holonomy). The gauge connection of the low dimensional theory becomes spin
connection upon the uplifting. Hence, the twisting can help us in
looking for such non-trivial metrics.
We now turn to the above mentioned type IIA solutions with
D6-branes corresponding to this eleven dimensional solution. The
D6-branes must be wrapping a supersymmetric cycle inside a different
manifold 
${\cal X}_{p-1}$ (or ${\cal X}_{p-2}$ in some cases). This manifold ${\cal
X}$ preserves the double of supersymmetries  than ${\cal Y}_p$, so
the total number of supercharges is the same, as the D6-branes half them
\cite{jaume}. For instance, D6-branes wrapping a supersymmetric
two-cycle inside an $SU(2)$ holonomy manifold uplift to an 
$SU(3)$ holonomy manifold, and D6-branes wrapping a SLag three-cycle
inside an $SU(3)$ holonomy manifold uplift to a $G_2$ holonomy
manifold. These cases will be considered in the following chapters.

%*******************************************************

\chapter{Supersymmetry and metrics on the conifold}
\label{conichapter}

\setcounter{equation}{0}
\section{Introducing the conifold}

The so-called conifold (see \cite{cdlo}) is  a Calabi-Yau
manifold with six (real) dimensions. Notably, it is one of
the few Calabi-Yau three-folds in which a Ricci-flat K\"ahler
metric is known. Its great physical importance comes from the
fact that it allows to construct string theory vacua with
reduced supersymmetry. This is very useful in the search for
gravity duals of four dimensional gauge theories with
${\cal N}=1$ supersymmetry \cite{KS,MN,civ}. Moreover, the
study of singularities and the ways in which they can be smoothed
provides a framework in which some non-trivial phenomena can
be studied. The conifold is also archetypical in the study of
geometric transitions \cite{gopavafa,vafa}. 
 
Let us start by defining the (singular) conifold as the
six-dimensional surface  embedded in $\CC^4$ according to:
\beq
\sum_{A=1}^4\,(z^A)^2\,=\,0\,\,,
\label{cf1}
\eeq
where the $z^A$ are complex numbers. Let us separate the real and
imaginary parts of the $z^A$'s:
\beq
z^A\,=\,x^A\,+\,i\,y^A\,\,,\qquad\qquad A=1,\dots ,4\,\,.
\label{xyconi}
\eeq
Notice that if $z^A$ solves eq. (\ref{cf1}), so it does $\la z^A$
for any $\la$. Therefore, the surface is made up of complex lines
through the origin, and thus it is a cone. The apex of the cone $z^A=0$
is the only singular point of the manifold.

The base of the cone can be described by the intersection of the
quadric with a sphere in $\CC^4$, which is given by:
\beq
\sum_{A=1}^4\,|z^A|^2\,=\,\rho^2\,\,.
\label{cf2}
\eeq
Eqs. (\ref{cf1}) and (\ref{cf2}) are better expressed in terms
of the real quantities $x^A$, $y^A$ of eq. (\ref{xyconi}). Using a
notation where they are four-dimensional vectors:
\beq
\vec{x}\cdot\vec{x}=\half \rho^2\,\,,\qquad\qquad
\vec{y}\cdot\vec{y}=\half \rho^2\,\,,\qquad\qquad
\vec{x}\cdot\vec{y}=0\,\,.
\eeq	
The first equation defines an $S^3$ while the other two define
an $S^2$ fiber over $S^3$. All such bundles are trivial, so
the topology of the base of the cone is $S^2\times S^3$.

\begin{figure}[h]
\centerline{\hskip -.8in \epsffile{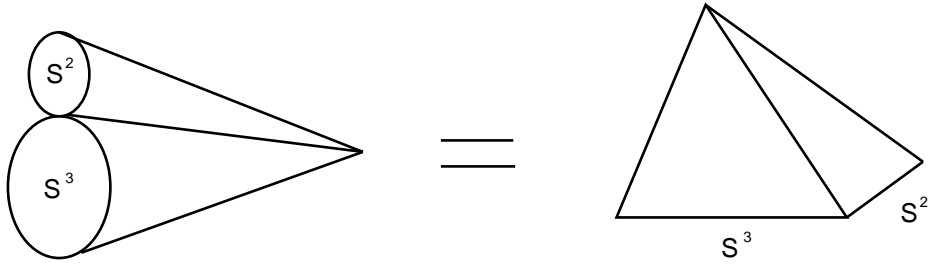}}
\caption{A pictorial representation of the singular conifold: it is
a cone whose base is topologically $S^2\times S^3$.}
\label{conifig}
\end{figure}

We now want to look for Ricci-flat K\"ahler metrics on the
conifold. Ricci flatness implies that the base of the cone
admits an Einstein metric. There are two possible metrics which represent
different geometries on $S^2\times S^3$ that fulfil this requirement.
However, by further imposing the K\"ahler 
condition\footnote{The K\"ahler condition implies that there
exists a function ${\cal F}$ (the so-called
K\"ahler potential) such that 
$g_{\mu\bar\nu}=\partial_\mu\partial_{\bar\nu}{\cal F}$,
where the $g_{\mu\bar\nu}$ is the metric of the six dimensional
space: $ds_6^2=g_{\mu\bar\nu}dz^\mu d\bar z^{\bar \nu}$.}
one is only left with the $T^{1,1}$ metric for the base of the
cone \cite{cdlo}:
\bear
ds_5^2\,(T^{1,1})&=&\,
\frac{1}{9}\,\left(d\tilde{\psi}\,+
\,\cos\tilde{\theta}\,d\tilde{\varphi}\,+\,
\cos\theta\,d\varphi\right)^2\,+\rc
&&+\,\frac{1}{6}\,(d\tilde{\theta}^2\,+
\,\sin^2\tilde{\theta}\,d\tilde{\varphi}^2)\,+\,
\frac{1}{6}\,(d\theta^2\,+
\,\sin^2\theta\, d\varphi^2)\,\,.
\label{T11}
\eear
The angles take values in the range: $0\leq\theta,\tilde\theta<\pi$ and
$0\leq\varphi,\tilde\varphi<2\pi$ and $0\leq\tilde\psi<4\pi$.
Notice that it is manifest that this metric is a $U(1)$ fibration
over $S^2\times S^2$. This compact homogeneous space $T^{1,1}$ can also
be defined as a coset space:
\beq
T^{1,1}\,=\,{SU(2)\times SU(2) \over U(1)}\,\,,
\eeq
and its volume is ${\mathrm Vol}(T^{1,1})={16\pi^3\over 27}$. The
(singular) conifold metric is:
\beq
ds_6^2\,=\,d\rho^2\,+\,\rho^2\,ds_5^2\,(T^{1,1})\,\,.
\label{singconi}
\eeq

A natural question to ask is how one can define a related manifold
where the singularity at the apex is avoided.
The most natural way seems to modify eq. (\ref{cf1}):
\beq
\sum_{A=1}^4\,(z^A)^2\,=\,\mu^2\,\,.
\eeq
This is the so-called  {\it deformation} of the conifold. At the 
apex there is a finite $S^3$, while the $S^2$ shrinks to zero.

There is also another way of getting rid of the singular point. By
defining:
\bear
&&X={1 \over \sqrt{2}}\,(z^3+i\,z^4)\,\,,\qquad\qquad
Y={1 \over \sqrt{2}}\,(-z^3+i\,z^4)\,\,,\rc
&&U={1 \over \sqrt{2}}\,(z^1-i\,z^2)\,\,,\qquad\qquad
V={1 \over \sqrt{2}}\,(z^1+i\,z^2)\,\,,
\eear
eq. (\ref{cf1}) can be reexpressed as:
\beq
X\,Y\,-\,U\,V\,=\,0\,\,.
\eeq
Now, replace this equation by:
\beq
\pmatrix{X&&U\cr V&&Y}\,\pmatrix{\la_1 \cr\la_2}\,=\,0\,\,,
\eeq
where $\la_1,\la_2 \in \CC$ are not both zero. Except at the apex,
the system gives a value to $\la_1/\la_2$. But at the apex,
 $\la_1/\la_2$ is not constrained, so one has an entire $\IP^1=S^2$.
This defines the so-called  {\it small resolution} of the 
conifold (this manifold will be called resolved conifold from now on). At
the  apex there is a finite $S^2$, while the $S^3$ shrinks to zero.

\begin{figure}[h]
\setlength{\unitlength}{1.1em}
\begin{center}
\begin{picture}(25,10)

\put(10,2){\line(1,0){6}}
%\put(10,2){\line(1,1){2}}
\put(16,2){\line(1,1){2}}
%\put(12,4){\line(1,0){6}}
\put(10,2){\line(1,2){3}}
\put(16,2){\line(-1,2){3}}
\put(18,4){\line(-5,4){5}}
\put(13,0.4){${\bf S}^3$}
\put(17,2){${\bf S}^2$}
\put(12,10){\small{conifold}}
\put(12,9){\small{singularity}}

\put(6.2,6){$\Longleftarrow$}
\put(5,7){\small{deformation}}

\put(-2,2){\line(1,0){6}}
%\put(-2,2){\line(1,1){2}}
\put(4,2){\line(1,1){2}}
%\put(0,4){\line(1,0){6}}
\put(-2,2){\line(1,3){2}}
\put(4,2){\line(-1,3){2}}
\put(0,8){\line(1,0){2}}
\put(6,4){\line(-1,1){4}}
\put(1,0.4){${\bf S}^3$}
\put(5,2){${\bf S}^2$}
\put(1,8.5){$\mu$}

\put(20,6){$\Longrightarrow$}
\put(19,7){\small{resolution}}

\put(22,2){\line(1,0){6}}
%\put(22,2){\line(1,1){2}}
\put(28,2){\line(1,1){2}}
%\put(24,4){\line(1,0){6}}
\put(22,2){\line(3,5){3}}
\put(28,2){\line(-3,5){3}}
\put(25,7){\line(1,1){2}}
\put(30,4){\line(-3,5){3}}
\put(25,0.4){${\bf S}^3$}
\put(29,2){${\bf S}^2$}
\put(25,8){$a$}

\end{picture}\end{center}
\caption{The two possible ways to smooth  the conifold singularity
are the deformation, replacing the node by an $S^3$ and the
resolution, replacing the node by an $S^2$. $\mu$ and $a$ are the
corresponding deformation and resolution parameters.}
\label{conifig2}
\end{figure}
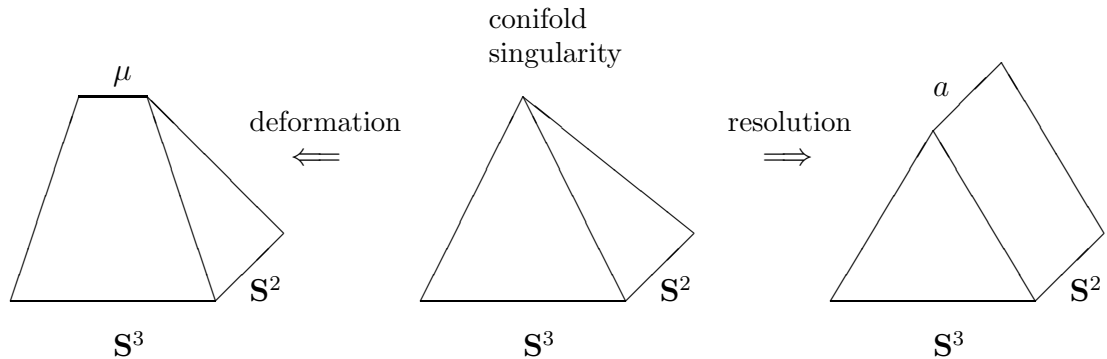

A schematic picture of the ways of repairing the singularity
is shown in figure \ref{conifig2}.

Homogeneous Ricci-flat K\"ahler metrics on the deformed and
resolved conifold where computed in \cite{cdlo}.  It was shown that  when
the resolving parameters are taken to zero, both metrics tend to 
(\ref{singconi}), in agreement with the fact that there is only one
Ricci-flat K\"ahler metric on the singular conifold. All these  metrics
have the same asymptotic behavior for large radial coordinate, far from
the singularity. These results will
be rederived below from a very different perspective.

\subsubsection{Using gauged supergravity}

In this chapter, it will be shown how gauged supergravity
provides a nice framework to study the conifold metrics. The
metrics on the singular, deformed and resolved conifolds
(and their  generalizations with one additional parameter
\cite{pt,pztdos}) can be found in a unified formalism. The different
metrics are different solutions of the same system of equations.
Furthermore, the Killing spinors will be obtained. Two independent
projections must be imposed on them, which is a direct check of the fact 
that the conifold is a 1/4-supersymmetric manifold.

Another lesson we will learn is that the excitation of new 
gauged sugra degrees
of freedom can  smooth singularities. The same will
happen in chapter \ref{g2chapter} in a different scenario.

In the rest of this chapter we will use Salam-Sezgin
gauged supergravity (see section \ref{ss8d}). By considering an
eight dimensional ansatz corresponding to D6-branes wrapped
on an $S^2$ sphere, we can find an 8d supersymmetric solution.
Then, one can easily uplift the solution to eleven 
dimensions (remember that this 8d sugra comes from compactification
of 11d sugra on $S^3$).

The uplifting formulae render a  fibration of the
$S^2$ over the $S^3$, due to the twisting performed in 
eight dimensions.
Then, the resulting 11d metric is ${\cal M}_{1,4}\times {\cal Y}_6$ 
where ${\cal M}$ stands for Minkowski space and ${\cal Y}_6$ is a cone
with a base topologically
$S^2\times S^3$ (the radial coordinate of the cone appears as the
distance to the D6, which are domain walls in 8d). Moreover, ${\cal
Y}_6$ must be Ricci-flat because the
D6 uplifts to pure geometry in eleven dimensions.
This six dimensional non-trivial metric turns out
to describe the conifold. The solution corresponds in ten dimensions
to D6-branes wrapping an $S^2$ inside a $K3$.

%*******************************************************

\setcounter{equation}{0}
\section{D6-brane wrapped on $S^2$}
\label{wrappingS2}

So let us consider a stack of D6-branes wrapping an $S^2$.
As explained above, the natural framework to deal with this
problem is D=8 gauged supergravity where they are domain walls.
The expressions (\ref{apauno})-(\ref{gammaprod}) will be used.
The ansatz for the metric is:
\beq
ds^2_8 = e^{2f} dx_{1,4}^2 + e^{2h} d\Omega_2^2 +
dr^2 ~,
\label{d6s2metric}
\eeq 
where $h\equiv h(r)$, $f\equiv f(r)$
and $d\Omega_2^2 = d\theta^2 + \sin^2\theta d\varphi^2$
is the metric of the unit $S^2$. Moreover, because of
the symmetry of the setup, it is enough to excite one scalar         
in the coset $SL(3,\RR)/SO(3)$. Accordingly, the $L_{\alpha}^i$
matrix will be taken as:
\beq
L_{\alpha}^i={\rm diag}(e^{\lambda},e^{\lambda},e^{-2\lambda})
\,\,\,.
\label{coniscalars}
\eeq
For the moment, we will consider $G_{(4)}$ to vanish
(see chapter \ref{fluxes} for its inclusion). Apart from
the metric and the scalar $\lambda$, the dilaton $\phi$
and the $SU(2)$ gauge potential $A^i$ are also present.
The lagrangian density (\ref{apatres}) becomes:

\beq
 \CL =\sqrt{-g}\,\left[ {1 \over 4} R - {1 \over
4} e^{2\phi} (F_{\mu\nu}^{i})^2 - {1 \over 4} (P_{\mu ij})^2 -
\half (\partial_\mu\phi)^2 - {1 \over 32} e^{-2\phi} (e^{-8\lambda} - 4
e^{-2\lambda}) \right]~,
\label{conilagr}
\eeq
where  $F^i$ is defined in (\ref{gaugefstr}) and the
$P$ and $Q$ matrices are (\ref{apauno}). 
\bear
P_{ij}&=&
\pmatrix{\la'\,dr&&0&&-A^2\,\sinh 3\la\cr\cr
0&&\la'\,dr&&A^1\,\sinh 3\la\cr\cr
-A^2\,\sinh 3\la&&A^1\,\sinh 3\la&&-2\la'\,dr}\,\,,
\rc\rc\rc
Q_{ij}& =&
\pmatrix{0&&-A^3&&A^2\,\cosh 3\la\cr\cr
A^3&&0&&-A^1\,\cosh 3\la\cr\cr
-A^2\,\cosh 3\la&&A^1\,\cosh 3\la&&0}.
\label{conipq}
\eear
Throughout the whole thesis, the prime will always denote derivative with
respect to $r$.

The ansatz for the gauge field is
better presented in terms of the triplet of Maurer--Cartan 1-forms
\footnote{These forms are defined as underlined sigmas 
to avoid confusion with Pauli matrices.} on $S^2$:
\beq\us^1 = d\theta ~, ~~~~~ \us^2 = \sin\theta
d\varphi ~, ~~~~~ \us^3 = \cos\theta d\varphi ~,
\label{MaurerCartanS2.}
\eeq
that obey the
conditions $d\us^i = - \half\, \epsilon_{ijk} \us^j\wedge \us^j$. The
gauge field $A$ will be taken as:
\beq
A^{1} = g(r) ~\us^1 ~, ~~~~~ \,\,\, A^{2} = g(r) ~\us^2
~, ~~~~~ \,\,\, A^{3} = \us^3 ~.
\label{conigauge}
\eeq
Notice that the form of the $A^3$ component is dictated by
 the schematic equation (\ref{twisting})
and therefore provides the appropriate twisting needed to preserve
some supersymmetry \cite{en}.
The other components of $A$ can also be switched on
\cite{epriii},
much in the spirit of the t'Hooft-Polyakov monopole, where
non-abelian gauge field degrees of freedom appear at short
distances and smooth the Dirac singularity.
 The field strength (\ref{gaugefstr}), reads (remember that
to get to flat indices on the internal group manifold
one must further multiply by the matrix of scalars $L^i_\alpha$):
\beq
F^{1} = g' ~dr \wedge \us^1 ~, ~~~~~ \,\,\, F^{2} =
g' ~dr \wedge \us^2 ~, ~~~~~ \,\,\, F^{3} = \big(g^2-1) ~\us^1 \wedge
\us^2 ~.
\label{conifstr}
\eeq
The uplifted eleven dimensional metric is (\ref{uplift8d}): 
\bear
ds^2_{11} &= & dx_{1,4}^2\,+\,
e^{2h-{2\phi\over 3}}\,d\Omega_2^2\,+\,e^{-{2\phi\over
3}}dr^2\,+\, 4\,e^{{4\phi\over 3}+2\lambda}\,\big(\,
\tilde{w}^1+\,g\,\us^1\,\big)^2 \,+\,\cr
 && + \,4\,e^{{4\phi\over
3}+2\lambda}\, \big(\, \tilde{w}^2+\,g\,\us^2
\,\big)^2\,+\, 4\,e^{{4\phi\over 3}-4\lambda}
\big(\, \tilde{w}^3+\,\us^3\,\big)^2 ~,
\label{coniuplift}
\eear
where 
\beq
f=\phi/3\,\,,
\label{fphi3}
\eeq
 has been imposed in order to have 
flat five dimensional Minkowski space-time in the 
unwrapped part of the metric (this condition can be
also obtained from consistency of the susy variation
equations). The $\tilde{w}^i$ were defined in (\ref{tres}).

Let us use the method described in section \ref{vffv}
to look for a solution. We have to impose:
\beq
\delta\psi_\mu=\delta\chi_i\,=0\,\,,\,\,\,\,\,\,\,\,\,\,\,
\,\,\,{\rm for \ all}\ \mu, \,i
\,\,\,,
\eeq
with the expressions of the variations given in (\ref{susy8}).
In the following, $\G_\theta$, $\G_\varphi$ and $\G_r$ are
Dirac matrices with flat indices, referred to the
vielbein which is natural from (\ref{d6s2metric}).
In order to seek for  solutions to the system, we
start by subjecting the spinor to the following angular projection
\beq
\Gamma_{\theta\varphi} \epsilon = -\hat\Gamma_{12}
\epsilon ~,
\label{coniangproj}
\eeq
which is imposed by the K\"ahler structure of the
ambient $K3$ manifold in which the two--cycle 
lives. By explicitly writing eqs. (\ref{susy8}) for 
this ansatz,
it can be seen that (\ref{coniangproj}) is needed.
The equations
$\delta\chi_1\,=\delta\chi_2=0$ give: 
\beq
\Big(\lambda' +
{2\over 3} \phi' \Big) \epsilon = g e^{-h} \sinh 3\lambda\,\,
\hat\Gamma_1 \Gamma_{\theta} \Gamma_{r} \hat\Gamma_{123} \epsilon
- e^{\phi+\lambda-h} g' \hat\Gamma_1\Gamma_{\theta} \epsilon  -
{1\over 4} e^{-\phi-4\lambda}\,\Gamma_{r} \hat\Gamma_{123}\epsilon
~,
\label{coniuno}
\eeq
while $\delta\chi_3\,=0$ reads: 
\bear
\Big(\,2\lambda'\,-\,{2\over 3}\,\phi'\,\Big)\,\epsilon &=& 
\Big[\,e^{\phi-2\lambda-2h}\,\big(g^2-1)\,-\, {1\over
4}\,e^{-\phi}\,\big(\,e^{-4\lambda}\,-\,2e^{2\lambda}\,\big)\,
\Big]\,\Gamma_{r}\,\hat\Gamma_{123}\,\epsilon \cr &&+
2g\,e^{-h}\,\sinh 3\lambda\,\,\hat\Gamma_1\,\Gamma_{\theta}\,
\Gamma_{r}\,\hat\Gamma_{123}\,\epsilon ~.
\label{conidos}
\eear
One can combine these
two equations to eliminate $\lambda'$: 
\beq
\phi'\epsilon +
e^{\phi+\lambda-h} g' \hat\Gamma_1\Gamma_{\theta} \epsilon
\,+\,\Big[\,{1\over 2}\,e^{\phi-2\lambda-2h} \big(g^2-1)\,+\,
{1\over 8}\,e^{-\phi}\,\big(\,e^{-4\lambda}\,+\,2e^{2\lambda}
\big)\, \Big]\,\Gamma_{r}\,\hat\Gamma_{123}\,\epsilon = 0 ~.
\label{conitres}
\eeq
From
this last equation, it is clear that the supersymmetric parameter
must satisfy a projection of the sort: 
\beq
\Gamma_{r}\,\hat\Gamma_{123}\,\epsilon\,=\,-\big(\, \beta +
\tilde\beta\,\hat\Gamma_1\,\Gamma_{\theta}\,\big)\,\epsilon ~,
\label{conirproj}
\eeq
where $\beta$ and $\tilde\beta$ are (functions of the radial
coordinate) proportional to the first derivatives of $\phi'$ and
$g'$: 
\beq
\phi' = \Big[\, {1\over
2}\,e^{\phi-2\lambda-2h}\,\big(g^2-1)\,+\, {1\over 8} e^{-\phi}
\big(\,e^{-4\lambda}\,+\,2e^{2\lambda}\,\big)\, \Big]\,\beta ~,
\label{conidilatoneq}
\eeq
\beq
e^{\phi+\lambda-h}\,g' = \Big[\, {1\over 2}
e^{\phi-2\lambda-2h} \big(g^2-1)\,+\, {1\over 8} e^{-\phi} \big(
e^{-4\lambda} +\,2e^{2\lambda}\,\big)\, \Big]\,\tilde\beta ~.
\label{conigeq}
\eeq
This radial projection
 encodes a non-trivial fibering of the two sphere with the
external three sphere as will become clear below. Since $(\Gamma_r\,
\hat\Gamma_{123})^2\epsilon\,=\,\epsilon$ and $\{\Gamma_{r}
\hat\Gamma_{123}, \hat\Gamma_1 \Gamma_{\theta}\}=0$, one must have:
\beq
\beta^2\,+\,\tilde\beta^2 = 1\,\,\,\,.
\label{beta2}
\eeq
 Thus, we can
represent $\beta$ and $\tilde\beta$ as:
\beq
\beta\,=\,\cos\alpha ~, ~~~~~\,\, \tilde\beta\,=\,\sin\alpha ~.
\label{betas}
\eeq
Also, it is clear that both independent projections 
(\ref{coniangproj}) and
(\ref{conirproj}) leave unbroken eight supercharges as expected.
Inserting the radial projection (\ref{conirproj}),
 as well as (\ref{conidilatoneq}), in
(\ref{conidos}), we get an equation determining $\lambda'$: 
\beq
\lambda' =\,ge^{-h}\,\sinh 3\lambda\,\,\tilde\beta\,-\,
\Big[\,{1\over 3}\,e^{\phi-2\lambda-2h}\,\big(g^2-1)\,-\, {1\over
6}\,e^{-\phi}\,\big(\,e^{-4\lambda}\,-\,e^{2\lambda}\,\big)\,
\Big]\,\beta ~,
\label{conilambdaeq}
\eeq
together with an algebraic constraint:
\beq
ge^{-h}\,\sinh 3\lambda\,\,\beta\,+\, \Big[\,{1\over
2}\,e^{\phi-2\lambda-2h}\,\big(g^2-1)\,-\, {1\over
8}\,e^{-\phi}\,\big(\,e^{-4\lambda}\,-\,2e^{2\lambda}\,\big)\,
\Big]\,\tilde\beta\,=\,0 ~.
\label{constr}
\eeq
Let us now consider the equations
obtained from the supersymmetric variation of the gravitino. From
the components along the unwrapped directions one does not get
anything new, while from the angular components we get:
\bear
h'\epsilon &= & -\,ge^{-h}\,\cosh 3\lambda
\hat\Gamma_1 \Gamma_{\theta}\, \Gamma_{r}\,\hat\Gamma_{123}
\epsilon + {2\over 3} e^{\phi+\lambda-h} g' \hat\Gamma_1
\Gamma_{\theta}\,\epsilon \cr && - {1\over 6}\Big[-5
e^{\phi-2\lambda-2h}\,\big(g^2-1)\,+\, {1\over 4}\,e^{-\phi}
\big(\,2e^{2\lambda}\,+\,e^{-4\lambda}\,\big)\, \Big]\,\Gamma_{r}
\hat\Gamma_{123}\,\epsilon ~.
\label{conicuatro}
\eear
By using the projection (\ref{conirproj})
 we obtain an equation for $h'$: 
\beq
h'\,=\,-ge^{-h}\,\cosh
3\lambda\, \tilde\beta +\, {1\over 6}\Big[\,-5
e^{\phi-2\lambda-2h} \big(g^2-1)\,+\, {1\over 4}\,e^{-\phi}
\big(\,2e^{2\lambda} + e^{-4\lambda} \big)\, \Big]\,\beta ~,
\label{coniheq}
\eeq
together with a second algebraic constraint:
\beq
-ge^{-h}\,\cosh 3\lambda\, \beta +\, \Big[\,{1\over
2}\,e^{\phi-2\lambda-2h} \big(g^2-1)\,-\, {1\over
8}\,e^{-\phi}\,\big(\,2e^{2\lambda} + e^{-4\lambda} \big)\,
\Big]\,\tilde\beta\,=\,0 ~.
\label{constr2}
\eeq
Finally, from the radial component of the gravitino we get the
functional dependence of the supersymmetric parameter $\epsilon$: 
\beq
\partial_r\,\epsilon = {5\over
6}\,e^{\phi+\lambda-h}\,g'\,\, \hat\Gamma_1\,\Gamma_{\theta}
\epsilon - {1\over 12}\, \Big[\,e^{\phi-2\lambda-2h} \big(g^2-1) +
{1\over 4}\, \big(\,2e^{2\lambda} + e^{-4\lambda}\,\big)\,\Big]\,
\Gamma_{r}\,\hat\Gamma_{123}\,\epsilon ~.
\label{conirad}
\eeq

The projection (\ref{conirproj}) 
gives the generalized twisting conditions
first studied in \cite{eprii} and applied to this case
in \cite{epriii}.
 Its interpretation goes as 
follows: using the trigonometric
parametrization (\ref{betas}), the
generalized projection can be written as: 
\beq
\Gamma_{r}\,\hat\Gamma_{123} ~\epsilon = - e^{\alpha\hat\Gamma_1
\Gamma_{\theta}} ~\epsilon ~,
\label{conirot}
\eeq
which can be solved as:
\beq
\epsilon\,=\,e^{-{1\over 2}\alpha\hat\Gamma_1 \Gamma_{\theta}}\,
\epsilon_0 ~, ~~~~~~~~ \Gamma_{r} \hat\Gamma_{123}
~\epsilon_0 = -\epsilon_0 ~.
\label{coniKilling}
\eeq
We can determine $\epsilon$ by plugging (\ref{coniKilling})
 into the equation for
the radial component of the gravitino (\ref{conirad}). 
Using (\ref{conirot}), we get
two equations. The first one gives the characteristic radial dependence
of $\epsilon_0$ in terms of the eight dimensional dilaton, namely:
\beq
\partial_r \epsilon_0\,= {\phi'\over 6}\,\epsilon_0 ~
\,\,\,\,\,\,\,\Rightarrow \,\,\,\,\,\,\,
\epsilon_0\,= e^{{\phi\over 6}}\,\eta ~,
\label{conicinco}
\eeq
with $\eta$
being a constant spinor. The other equation determines the radial
dependence of the phase $\alpha$: 
\beq
\alpha' =
-2e^{\phi+\lambda-h} g' ~.
\label{conialpha}
\eeq
 Thus, the spinor $\epsilon$ can be
written as: 
\beq
\epsilon\,=\,e^{{\phi\over 6}}\,
e^{-{1\over 2}\alpha\hat\Gamma_1 \Gamma_{\theta}}\,\eta\,\,,
~~~~~ \,\,\,\, \Gamma_{r}\,\hat\Gamma_{123}\,\eta\,=\,-\eta\,\,,
~~~~~ \,\,\,\,\,\, \Gamma_{\theta\varphi}\,\hat\Gamma_{12}\,\eta =
\eta ~.
\label{conispinor}
\eeq
The meaning of the phase $\alpha$ can be better
understood by using the  identity
(\ref{gammaprod}), so that:
\beq
\Gamma_{x^0\cdots
x^4}\,\big(\cos\alpha ~ \Gamma_{\theta\varphi} - \sin\alpha
~\Gamma_{\theta}\hat\Gamma_{2} \big) \epsilon =\,\epsilon ~,
\label{conirot2}
\eeq
which shows that the D6--brane is wrapping a two--cycle which is
non-trivially embedded in the $K3$ manifold as seen from the
uplifted perspective that is implied in (\ref{conirot2}).
 
Let us turn to explicitly finding the functions $\phi$,
$\la$, $h$ and $g$. We start by solving the two algebraic
constraints (\ref{constr}), (\ref{constr2}).
 By adding and subtracting the two
equations, we get: 
\beq
\tan\alpha \equiv {\tilde \beta\over
\beta}\,=\,-2\,g e^{\phi+\lambda-h}\,=\,{g\,e^{-3\lambda-h}\over
e^{\phi-2\lambda-2h} \big(g^2-1)\,-\,{1\over 4}\,
e^{-\phi-4\lambda}} ~.
\eeq
Whereas the first part of this equation allows us to write
$\alpha$ in terms of the remaining functions, the
last equality provides an algebraic constraint that restricts our
ansatz. It is not hard to arrive at the following simple equation:
\beq
g\,\Big[\,g^2-1\,+\,\,{1\over 4}\,e^{-2\phi-2\lambda+2h} 
\, \Big]\,=\,0 ~.\eeq
There are obviously two solutions:
\bear
g&=&0  \,\,\,, \label{resol}\\
g^2\,&=&\,1\,-\,{1\over 4}\,e^{-2\phi-2\lambda+2h} ~.
\label{deform}
\eear
In the following sections, it will be proved that 
(\ref{resol}) leads to the resolved conifold metric
while (\ref{deform}) leads to the deformed conifold
metric. They can also be imposed simultaneously, and
the regularized conifold is found.

%**********************************************************

\setcounter{equation}{0}
\section{The generalized resolved conifold}
\label{rescon}

Let us first consider the possibility (\ref{resol}), {\it i.e.} the case
$g=0$.
In view of (\ref{constr}), (\ref{constr2}) this implies:
\beq
\tilde{\beta}=0 \Rightarrow 
\cases{\alpha=0 \cr \beta=1}\,\,\,\,,
\label{alphacero}
\eeq
and so, the radial projection on the spinor (\ref{conirproj}) is
unrotated. This is a consistent truncation of the system
of equations (notice that (\ref{conigeq}) and
(\ref{conialpha}) are automatically satisfied).
It leads to the case studied in \cite{en},
whose integral is the generalized resolved conifold (see also
 \cite{epr}). The system of differential equations
(\ref{conidilatoneq}), (\ref{conilambdaeq}),
(\ref{coniheq})
becomes:
\bear
\phi' &=&  -{1\over
2}\,e^{\phi-2\lambda-2h}\,+\, {1\over 8} \,e^{-\phi}\,
\big(\,e^{-4\lambda}\,+\,2e^{2\lambda}\,\big)\,\,\,\,,\rc
\lambda' &=&{1\over 3}\,e^{\phi-2\lambda-2h}\,+\,
{1\over 6}\,e^{-\phi}\,\big(\,e^{-4\lambda}\,
-\,e^{2\lambda}\,\big)\,
\,\,\,\,\,\,\,\,\,,\rc
h'&=& {5\over 6}\,
e^{\phi-2\lambda-2h} \,+\, {1\over 24}\,e^{-\phi}\,
\big(\, e^{-4\lambda}+2e^{2\lambda } \,\big)\,
\,\,\,\,\,.
\label{resolsyst}
\eear
{\it A priori}, this system seems hard to solve. However, there
is a procedure that sometimes works for this kind of problems.
First, we look for a combination of the fields such that some
dependence cancels in the resulting differential equation.
Then, we try to redefine the radial variable in a way that
only the new field appears in the equation. In this case, it
is convenient to define the new field $x$ and the new
radial variable $t$:
\beq
x\equiv 4e^{2\phi-2h+2\lambda}\,\,,\,\,\,\,\,\,\,
\,\,\,\,\,
{dr\over dt}\,=\,e^{\phi\,+\,4\lambda}\,\,,
\label{defs1}
\eeq
and then we can derive from (\ref{resolsyst}):
\beq
{dx\over dt}\,=\,{1\over 2}\,x\,(1\,-\,x)\,\,,
\label{cotreinta}
\eeq
which is solved by:
\beq
x\,=\,{1\over 1\,+\,c\,\,e^{-{t\over 2}}}\,\,,
\label{condos}
\eeq
with $c$ being an integration constant. It follows from the first-order
system (\ref{resolsyst}) that $\lambda$ satisfies the equation:
\beq
{d\lambda\over dt}\,=\,{1\over 6}\,\big(\,1\,-\,e^{6\lambda}\,)\,
+\,{x\over 12}\,\,.
\label{contres}
\eeq
By using the explicit dependence of $x$ on $t$, displayed in eq. 
(\ref{condos}),
the integral of eq. (\ref{contres}) is easy to find. 
In order to express this
integral in a convenient way, let us parametrize $\lambda$ as:
\beq
\lambda\,=\,{1\over 6}\,\Big[\,\log\big(\,{3\over 2}\,\big)
\,-\,\log \kappa\,\Big]\,\,.
\label{concuatro}
\eeq
In general, the function $\kappa(t)$ is
given by:
\beq
\kappa(t)\,=\,{e^{{3\over 2}t}\,+\,{3\over 2}\,c\,e^{t}\,+\,d\over
e^{{3\over 2}t}\,+\,\,c\,e^{t}}\,\,,
\label{concinco}
\eeq
where $d$ is a new integration constant. Knowing $x$ and
$\la$, it is immediate to get the expressions of 
$h$ and $\phi$ by integrating in (\ref{resolsyst}):
\beq
e^{2h}\,=\,e^{{3t \over 4}}\,(1+c\,e^{-\frac{t}{2}})\,
\kappa(t)^{{1\over 6}}\,,\,\,\,\,\,\,\,\,\,\,\,
\,\,\,\,\,\,\,\,\,\,\,\,
e^{\phi}\,=\,96^{-\frac{1}{6}}\,e^{{3t \over 8}}\,
\kappa(t)^{{1\over 4}}\,\,.
\label{hphieq}
\eeq
In order to obtain the metric in a
more familiar way, let us redefine again the radial variable
and the integration constants:
\beq
e^{{t\over 2}}\,=\,{1\over 6\,(96)^{{1\over 9}}}\,\,\,\rho^2\,\,,
\,\,\,\,\,\,\,\,\,\,\,\,\,\,\,\,\,\,\,\,\,\,
c\,=\,{1\over (96)^{{1\over 9}}}\,\,a^2\,\,,
\,\,\,\,\,\,\,\,\,\,\,\,\,\,\,\,\,\,\,\,\,\,
d\,=\,-{1\over 6^3\,(96)^{{1\over 3}}}\,\,b^6\,\,.
\label{coctuno}
\eeq
(we are assuming that $d\le 0$). With these definitions the function
$\kappa$ becomes:
\beq
\kappa(\rho)\,=\,{\rho^6\,+\,9a^2\,\rho^4\,-\,b^6\over
\rho^6\,+\,6a^2\rho^4}\,\,,
\label{coctsiete}
\eeq
while eq. (\ref{hphieq}) turns out to be:
\beq
e^{2h}\,=\,{1 \over 6\,(12)^{\frac{2}{3}}}\,\rho\, (\rho^2+6a^2) \,
\,\kappa(\rho)^\frac{1}{6}\,,\,\,\,\,\,\,\,\,\,\,\,
\,\,\,\,\,\,\,\,\,\,\,\,
e^{\phi}\,=\,\frac{1}{12}\,\rho^\frac{3}{2}\,\kappa(\rho)^\frac{1}{4}\,,
\label{hphieq2}
\eeq
and the eleven dimensional metric (\ref{coniuplift}) is:
\bear
ds^2_{11} &= & dx_{1,4}^2\,+\,
\frac{1}{6}\,(\rho^2+6\,a^2)\,d\Omega_2^2\,+\,
{d\rho^2 \over \kappa(\rho)}
\,\,+\,\frac{1}{6}\,\rho^2\,\left((\tilde{w}^1)^2
+(\tilde{w}^2)^2\right)\cr
 && +\, 
\frac{\rho^2}{9}\,\kappa(\rho)\left(\tilde{w}^3+\cos\theta
d\varphi\right)^2\,\,\,,
\label{coctocho}
\eear
where (\ref{defs1}), (\ref{coctuno}) have been used to
calculate $dr^2$. Finally, by using the expression
of the left invariant one forms (\ref{omegas}),
the metric can be written as $ds_{11}^2=dx_{1,4}^2
+ds_6^2$, where the six-dimensional metric $ds_6^2$
is:
\bear
ds_6^2&=&{d\rho^2 \over \kappa(\rho)}\,+\,
\frac{\rho^2}{9}\,\kappa(\rho)\left(d\tilde{\psi}\,+
\,\cos\tilde{\theta}\,d\tilde{\varphi}\,+\,
\cos\theta\,d\varphi\right)^2\,+\rc
&+&\,\frac{1}{6}\rho^2\,(d\tilde{\theta}^2\,+
\,\sin^2\tilde{\theta}\,d\tilde{\varphi}^2)\,+\,
\frac{1}{6}\,(\rho^2+6\,a^2)(d\theta^2\,+
\,\sin^2\theta\, d\varphi^2)\,\,,
\label{resmetric}
\eear
which is the known metric of the generalized resolved
conifold \cite{pt,pztdos}. The inclusion of the parameter $b$
generalizes the  resolved conifold metric \cite{cdlo,pzt},
that is recovered when $b=0$.

%****************************************************

\subsubsection{First order system from a superpotential}

We are going to rederive here the system of differential
equations (\ref{resolsyst}) using the method presented
in section \ref{superpotential}.

By directly plugging in (\ref{conilagr})
the ansatz for the fields (\ref{d6s2metric}),
(\ref{coniscalars}), (\ref{conifstr}), (\ref{resol}), one
gets the effective
lagrangian\footnote{Integration by parts has been
performed in order to get rid of the second derivatives
that appear in the calculation of the Ricci scalar.}
 in the radial variable $r$:
\bear
L_{eff}&=&e^{5f+2h}\,\bigg[\,5 f'^2\,+\,5f'h'\,+\,
\half h'^2\,-\,\frac{3}{2}\la'^2\,-\,\half\phi'^2\,-\,
\half e^{2\phi-4h-4\la}\,+\rc
&+&\frac{1}{16}\,e^{-2\phi}\,(-\half e^{-8\la}+2e^{-2\la})\,+\,
\half \,e^{-2h}\bigg]\,\,\,.
\eear
As we want a lagrangian of the type (\ref{newlagr}),
we must define:
\beq
A\,=\,f\,+\,\half\, h\,\,\,\,,
\eeq
so  the term in $f'h'$ disappears, and then redefine
the radial coordinate $r\rightarrow \eta$:
\beq
{dr \over d\eta}\,=\,e^{-\half h}\,\,\,\,,
\eeq
in order to have the field $A$ in the exponent of the common
factor. The new lagrangian (now in the variable $\eta$) is:
\bear
\hat L_{eff}&=&e^{5A}\,\bigg[\,5\,(\partial_\eta A)^2\,-\,
\frac{3}{4}\,(\partial_\eta h)^2\,-\,
\frac{3}{2}\,(\partial_\eta \la)^2\,-\,
\half\,(\partial_\eta \phi)^2\,-\,
\half \,e^{2\phi-5h-4\la}\,+\rc
&&+\ \frac{1}{16}\,e^{-2\phi-h}\,(-\half e^{-8\la}\,+\,2e^{-2\la})\,+\,
\half \,e^{-3h}\bigg]\,\,\,.
\label{lhat}
\eear
Notice the extra factor because of
 $\int L \,dr=\int\hat L\, d\eta \,\Rightarrow\,
\hat L =L \,e^{-\half h}$ . So we have an expression
like (\ref{newlagr}) being:
\bear
c_1=5\,\,,\,\,\,\,\,\,\,\,\,\,
c_2=5\,\,,\,\,\,\,\,\,\,\,\,\,
G_{hh}=\frac{3}{2}\,\,,\,\,\,\,\,\,\,\,\,\,
G_{\phi\phi}=1\,\,,\,\,\,\,\,\,\,\,\,\,
G_{\la\la}=3\,\,,\,\,\,\rc
\tilde{V}(h,\phi,\la)=
\half \,e^{2\phi-5h-4\la}\,-
\frac{1}{16}\,e^{-2\phi-h}\,(-\half e^{-8\la}\,+\,2e^{-2\la})\,-\,
\half \,e^{-3h}\,\,.
\eear
According to (\ref{newdefsuperp}), we seek a function
$\tilde{W}(h,\phi,\la)$ such that (take $c_3=1$):
\beq
\tilde{V}(h,\phi,\la)=
\frac{1}{3}\left({\partial \tilde W \over \partial h}\right)^2\,+\,
\frac{1}{6}\left({\partial \tilde W \over \partial \la}\right)^2\,+\,
\frac{1}{2}\left({\partial \tilde W \over \partial \phi}\right)^2\,-\,
\frac{5}{4}\,\tilde W^2\,\,\,.
\eeq
A simple straightforward calculation allows to check
that the condition is fulfilled for:
\beq
\tilde{W}\,=\,-\half \,e^{\phi-\frac{5}{2}h-2\la}\,-\,
\frac{1}{8}\,e^{-\phi-\frac{1}{2}h}\,\,(2\,e^{2\la}\,+\,
e^{-4\la})\,\,.
\eeq
The first order equations
(\ref{neworder1})
obtained from it exactly match the system (\ref{resolsyst})
after changing back to the original radial variable,
and also give the relation $f'=\phi'/3$ .

It is worth pointing out that to use the superpotential 
method, the constraint (\ref{resol}) had to be imposed
from the beginning whereas looking at the susy fermionic
variations, it directly came from the equations. It is not
clear how to start from the general ansatz (\ref{conigauge})
and to get the two possible constraints (\ref{resol}),
(\ref{deform}) following the reasoning of the superpotential.

%***********************************************************

\setcounter{equation}{0}
\section{The generalized deformed conifold}
\label{defcon}

Let us now analyze the second possibility
(\ref{deform}).
It is not difficult to find the values of
$\beta$ and $\tilde\beta$ for this solution of the constraint:
\beq
\beta\,=\,{1\over
2}\,e^{-\phi-\lambda+h} ~, ~~~~~ \,\,\,\,\, \tilde\beta\,=\,-g ~.
\eeq
Notice that they satisfy $\beta^2+\tilde\beta^2=1$ as a consequence of
the relation (\ref{deform}). Moreover, one can verify that 
(\ref{deform})\ is
consistent with the first-order equations. Indeed, by differentiating
eq. (\ref{deform})\ and using the first-order equations for $\phi$,
 $\lambda$
and $h$ (eqs. (\ref{conidilatoneq}),
 (\ref{conilambdaeq})
 and (\ref{coniheq})), we arrive precisely at the
first-order equation for $g$ written in (\ref{conigeq}).
 It can be also checked
that eq. (\ref{conialpha})
 is identically satisfied for this solution of the
constraint. Thus, one can eliminate $g$ from the first-order equations
arriving at the following system of equations for $\phi$, $\lambda$ and
$h$: 
\bear
\phi' &= & {1\over 8}\,e^{-2\phi+\lambda+h}\,\,,\rc
\lambda' &= &
{1\over 24}\,e^{-2\phi+\lambda+h}\,-\, {1\over 2}\,e^{3\lambda-h}
+ {1\over 2}\,e^{-3\lambda-h}\,\,,\rc
 h' &= & -{1\over 12}
e^{-2\phi+\lambda+h} +\,{1\over 2}\,e^{3\lambda-h} + {1\over
2}\,e^{-3\lambda-h} ~.
\label{deformsyst}
\eear
In order to integrate the system (\ref{deformsyst}),  we
will follow again the procedure explained after
(\ref{resolsyst}).
Let us define the function $z\,=\,\phi+\lambda-h$  and a new
radial coordinate $\tau$, $dr\,=\,2\,e^{\phi-2\lambda}\,d\tau$.
Then, 
it follows from (\ref{deformsyst})\ that $z$ satisfies the equation: 
\beq
\partial_\tau z =
{1\over 2} e^{-z} - 2e^{z} ~.
\label{zetaeq}
\eeq
 This equation can be immediately
integrated: 
\beq
e^z\,=\,{1\over 2}\,\,{\cosh
(\tau+\tau_0)\over \sinh(\tau+\tau_0)} ~,
\label{ezeta}
\eeq
where $\tau_0$ is an integration constant, which from now on we will 
absorb in a redefinition of the origin of $\tau$. We can obtain $\phi$ by
noticing that it satisfies the equation: 
\beq
\partial_\tau\phi\,=\,{1\over 4}\,e^{-z} .
\label{phieq}
\eeq
Since we know $z(\tau)$ explicitly, we can obtain immediately
$\phi(\tau)$, namely: 
\beq
e^{\phi}\,=\,\hat\mu
\big(\cosh\tau\big)^{{1\over 2}} ~,
\label{phisolv}
\eeq
where $\hat\mu$ is a constant
of integration. Finally, $h$ satisfies the following differential
equation: 
\beq
\partial_\tau h\,=\,-{1\over 6}\,e^{-z}\,+\,e^{z}\,+
e^{6\phi-5z-6h} ~.
\eeq
If we define, $y=e^{6h}$ and use the
expressions of $z$ and $\phi$ as functions of $\tau$, we get:
\beq
\partial_\tau y\,=\,{\cosh^2\tau+2\over \cosh\tau\sinh\tau}\,y +
192\,\hat\mu^6\,\,{(\sinh\tau)^5\over (\cosh\tau)^2} ~,
\eeq
which is
also easily integrated by the method of variation of constants. In
order to express the corresponding result, let us define the
function: 
\beq
K(\tau)\equiv {\Big(\sinh 2\tau\,-\,2\tau
+\,C\,\Big)^{1\over 3}\over 2^{{1\over 3}}\,\sinh\tau} ~,
\label{Ktau}
\eeq
where $C$ is a new constant of integration. Then, $h$ is given by:
\beq
e^{h}\,=\,3^{{1\over 6}}\,2^{{5\over 6}}\,\hat\mu\,\,
{\sinh\tau\over (\cosh\tau)^{{1\over 3}}}\, \big[ K(\tau)
\big]^{{1\over 2}} ~.
\eeq
As we know $z$, $h$ and $\phi$, we can
obtain $\lambda$. The result is: 
\beq
e^{\lambda} = \Big(
{3\over 2} \Big)^{{1\over 6}}\, (\cosh\tau)^{{1\over 6}}\, \big[
K(\tau) \big]^{{1\over 2}} ~.
\eeq
Finally, we can get $g$ from the
solution of the constraint (\ref{deform}), namely: 
\beq
g\,=\,{1\over
\cosh\tau} ~.
\label{gvalue}
\eeq
It follows immediately from (\ref{gvalue}) that $g\rightarrow 0$ as
$\tau\rightarrow\infty$. Moreover,  by  using the
explicit form of this solution we can find the value of the phase $\alpha$: 
\beq
\cos\alpha\,=\,{\sinh\tau\over\cosh\tau} ~, ~~~~~ \,\,\,\,\,
\sin\alpha\,=\,-{1\over \cosh\tau} ~,
\eeq
Notice that $\alpha\rightarrow -\pi/2$ when $\tau\rightarrow 0$, whereas
$\alpha\rightarrow 0$ for $\tau\rightarrow\infty$. In order to express
neatly the form of the corresponding eleven dimensional metric, let us
define the following set of one-forms:
\bear
g^1 &= & {1\over \sqrt{2}}\,
\big[\,\us^2\,-\, \tilde{w}^2\,\big]\,, \,\,\,\,\,\,\,\,\,\,\,
g^2= {1\over \sqrt{2}}\, \big[\,\us^1\,-\, \tilde{w}^1\,\big]\,,
\,\,\,\,\,\,\,\,\,\,\,
g^3 =  {1\over \sqrt{2}}\, \big[\,\us^2\,+\, \tilde{w}^2\,\big]\,,
\rc
g^4 &=&  {1\over \sqrt{2}}\,
\big[\,\us^1\,+\, \tilde{w}^1\,\big]\,,\,\,\,\,\,\,\,\,\,\,\,
g^5 = 
\big[\,\us^3\,+\, \tilde{w}^3\,\big] \,,
\eear
and a new constant $\mu$, related to $\hat\mu$ as $\mu =
2^{{11\over 4}} 3^{{1\over 4}} \hat\mu$. Then, by using the uplifting
formula (\ref{coniuplift}), the resulting eleven
dimensional metric $ds^2_{11}$ can again be written as $ ds^2_{11} =
dx_{1,4}^2 + ds^2_{6}$, where now the six dimensional metric is:
\bear
ds^2_{6}&=&{1\over 2}\,\mu^{{4\over
3}}\,K(\tau) \Bigg[{1\over 3 K(\tau)^3}\,\big(\,d\tau^2\,+\,
(g^5)^2\,)\,+\, \cosh^2\big({\tau\over 2}\big)\,
\big(\,(g^3)^2\,+\, (g^4)^2\,)\,+\,\rc
& &\,\,\,\,\,\,\,\,\,\,\,\,\,\,\,\,\,\,\,\,\,\,\,\,\,\,\,\,\,\,\,\,
+\,\sinh^2\big({\tau\over
2}\big)\,\big(\,(g^1)^2\,+\, (g^2)^2\,)\, \,\Bigg]\,\,,
\label{gendefcon}
\eear
which, for $C=0$ is nothing but the standard metric of the deformed
conifold, with $\mu$ being the corresponding deformation parameter
\cite{cdlo}. 
The metric (\ref{gendefcon}) for $C\not=0$ was studied in ref.
 \cite{pztdos},
where it was shown to display a curvature singularity when $\mu\not=0$.
The $\mu=0$ case will be addressed in the next section.

This solution may also be obtained from the superpotential
method by imposing the constraint (\ref{deform}) from the
beginning.

%***************************************************************

\setcounter{equation}{0}
\section{The regularized conifold}

Both solutions of the constraint 
(\ref{resol}), (\ref{deform}) can be imposed simultaneously.
Of course, this leads to  particular solutions of the
equations of the previous sections: it amounts to taking
$a=0$ in section \ref{rescon} or to taking $\mu=0$ in
section \ref{defcon}. We now prove this statement by 
making $\mu\rightarrow 0$ in (\ref{gendefcon}). This limit
must be taken with care, since, at first glance, it seems to  
annihilate the metric. The key point is that the constant
$\tau_0$ of eq. (\ref{ezeta}) must be first reintroduced ({\it i.e.} by
changing
$\tau\rightarrow\tau+\tau_0$) and then taken to infinity
in an appropriate fashion. In fact, if we write:
\beq
e^{2\tau_0}\,=\,{32 \over 27 \mu^4}\,\,,
\eeq
the $\mu\rightarrow 0$ (and accordingly $\tau_0\rightarrow  
\infty$) limit of
(\ref{gendefcon}) reads:
\bear
ds_6^2&=&{1\over 9}\,e^{2\tau}\left(e^{2\tau}+{27\mu^4 \over 16}C
\right)^{-\frac{2}{3}}\,\left(d\tau^2+(d\tilde\psi+\cos\tilde\theta
d\tilde\varphi+\cos\theta d\varphi)^2\right)+\rc
&&+\,{1\over 6}\left(e^{2\tau}+{27\mu^4 \over 16}C
\right)^{\frac{1}{3}}
\left(d\tilde\theta^2+\sin^2\tilde\theta d\tilde\varphi^2+
d\theta^2+\sin^2\theta d\varphi^2\right)\,\,.
\eear
The constant $C$ must also be appropriately taken to infinity in
order to keep finite the ${27\mu^4\over 16}C$ term. In fact, let us 
identify \cite{pztdos}:
\beq
b^6\equiv{27\mu^4\over 16}C\,\,,
\eeq
and define a new radial coordinate $\rho$:
\beq
\rho^6\equiv e^{2\tau}\,+\,b^6\,\,.
\eeq
Then, one arrives at:
\bear
ds_6^2&=&{d\rho^2 \over 1-{b^6\over \rho^6}}+
{\rho^2 \over 9}\left(1-{b^6\over \rho^6}\right)
(d\tilde\psi+\cos\tilde\theta
d\tilde\varphi+\cos\theta d\varphi)^2+\rc
&&+{1\over 6}\rho^2
\left(d\tilde\theta^2+\sin^2\tilde\theta d\tilde\varphi^2+
d\theta^2+\sin^2\theta d\varphi^2\right)\,\,,
\label{regucon}
\eear
The metric (\ref{regucon}) coincides with the $a\rightarrow 0$
limit of (\ref{resmetric}). We have proved that both the generalized
deformed and generalized resolved conifolds tend to (\ref{regucon})
when $\mu\rightarrow 0$ or $a\rightarrow 0$ respectively.

This metric was studied in reference \cite{pztdos}, where
it was named regularized conifold. It has no curvature
singularity for $b\ne 0$. It has a conical singularity that
can be removed after a
$\ZZ_2$ identification of the $U(1)$ fiber $\tilde{\psi}$, 
similarly to what happens in the Eguchi-Hanson metric.

It follows from this discussion that the regularized
conifold is a boundary in the moduli space separating the regions that
correspond to the generalized deformed and resolved conifolds
(see section \ref{discconi} for more details).

Moreover, taking $b=0$, the standard singular conifold
metric (\ref{singconi}) is recovered.

%*******************************************************
\setcounter{equation}{0}
\section{Discussion}
\label{discconi}

In this chapter, a unified scenario for conifold singularity
resolutions from the perspective of M-theory has been presented.
A single system of equations encompasses both the (generalized)
deformation and the (generalized) small resolution of the
conifold. Each kind of metric appears as one of the only two
possible solutions of an algebraic constraint. This allows us
to give a unified representation of the moduli space of 
metrics on the conifold. The regularized conifold 
interpolates between the two kinds of singularity resolution,
being the metric obtained when both solutions of the constraint
are fulfilled simultaneously. Notice that we cannot
continuously connect the deformed and resolved conifolds through
a supersymmetric trajectory of non-singular metrics
(as the generalized deformed metric displays a curvature 
singularity).

\begin{figure}[h]
\centerline{\hskip -.8in \epsffile{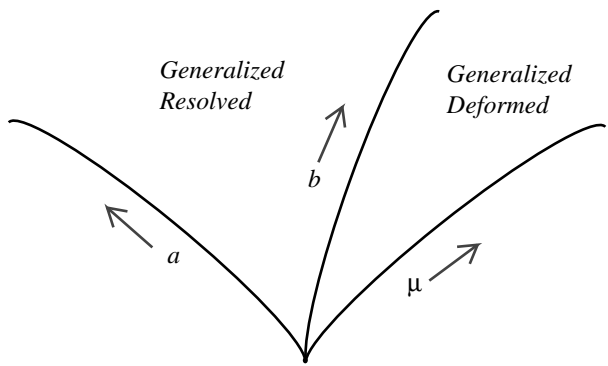}}
\caption{Representation of the moduli space of generalized resolutions of
the conifold singularity. The two regions depicted correspond to the two
solutions of our constraint. The generalized deformed conifold metric is
singular. A point on each of the three lines represents, from left to
right, the resolved, regularized and deformed conifold. They meet at a
single point, the singular conifold.}
\label{buchaca}
\end{figure}

A pictorial scheme is depicted in figure 
\ref{buchaca}. It may be worth to remind that only two
of the three integration constant exist on a solution
($a$, $b$ for the generalized resolved and $\mu$, $b$ 
for the generalized deformed)

As a byproduct of the supersymmetry analysis, the Killing
spinors for these geometries have been obtained\footnote{Although the
eight dimensional spinors have been presented, the eleven dimensional
susy analysis goes much the same way, identifying the hatted gammas with
the three Dirac matrices along the directions of the $S^3$ appearing in
the uplift.}.
In particular, the obtainment of the Killing spinors on the
deformed conifold is remarkable. It is based on a rotated
radial projection (\ref{conirot}). We will see in the 
following chapters that this technical trick is really 
useful, since such kind of rotated projection appears in 
lots of setups.

Moreover, it has been shown that lower-dimensional gauged
supergravities are an appropriate framework to resolve 
singularities in the study of geometries corresponding
to D-branes wrapping supersymmetric cycles, which 
are dual to supersymmetric gauge theories. A conventional
twisting (\ref{conigauge}) with $g=0$ and constant $\lambda$
leads to the singular conifold metric. But as we have seen,
by switching on the appropriate degrees of freedom of
the gauged sugra, the twisting can be generalized and
the singularity smoothened. However, asymptotically
(for large $r$), the solution, and therefore, the 
twisting, remains unaltered.

Finally, it is worth to point out that the mechanism
presented in this chapter for desingularizing the 
conifold is also useful to study other singularity 
resolutions. Some examples will be presented in the
following.

%****************************************

\chapter{$G_2$ holonomy metrics from gauged supergravity}
\label{g2chapter}

\setcounter{equation}{0}
\section{Introduction}

\subsubsection{$G_2$ holonomy manifolds and their physical significance}

On a Riemannian manifold, one can move tangent vectors along
a path by parallel transport. The holonomy group is defined as
the set of linear transformations arising from parallel transport along
closed loops. As parallel transport preserves the length of vectors,
the holonomy group of an orientable $n$-manifold must be contained in 
$SO(n)$. If it is indeed a proper subgroup, then we say that the manifold
has special holonomy. Special holonomy manifolds are important
in physics because they preserve some fraction of the supersymmetry
(we have already seen an example in the previous chapter,
the conifold, which  has $SU(3)$ holonomy  
and, hence, it is 1/4-supersymmetric). In this chapter we will
consider manifolds with $G_2$ holonomy.

The group $G_2$ is a subgroup of $SO(7)$, so we will deal with
seven dimensional manifolds. It leaves
invariant the multiplication table of the imaginary octonions. Therefore,
its action preserves the
 form\footnote{We write here the (possibly) most standard notation for the
octonion algebra, although in the following sections, this
form will be defined with some redefinition of the labels.}:
\beq
\Phi_{(3)}=dx_{123}+dx_{145}+dx_{167}+dx_{246}-dx_{257}-dx_{347}-
dx_{356}\,\,.
\label{g2form}
\eeq
Then in order to have $G_2$ holonomy, there must exist such
a three-form  that does not change under parallel transport
throughout the manifold.
Referring the three-form to some vielbein basis, we need its covariant
derivative to vanish:
$D_\mu \Phi_{\nu\rho\sigma}=0$. It is known that this is
equivalent to having:
\beq
d\Phi\,=\,d{}^*\Phi\,=\,0\,\,.
\label{cococlo}
\eeq
$\Phi$ is a calibrating three-form which calibrates a
minimal three-cycle inside the seven dimensional $G_2$ holonomy geometry.
This cycle is called associative (the four-cycle calibrated by ${}^*\Phi$
is called coassociative).

Let us now turn to the relation of $G_2$ holonomy with supersymmetry.
The spinor representation of $SO(7)$ can be decomposed, in terms of
representations of $G_2$, as:
\beq 
{\bf 8}\,\rightarrow\,{\bf 7}\,+\,{\bf 1}\,\,,
\eeq
{\it i.e.} it splits into a fundamental  and a singlet of $G_2$.
Only the singlet gives a covariantly constant spinor $D_\mu \epsilon=0$.
The rest of the spinors get changed under transport over the manifold.
Hence,
$G_2$ holonomy manifolds are 1/8-supersymmetric. Going now to eleven
dimensional supergravity, one can conclude that, being supersymmetric, it
solves the equations of motion, and, therefore, $G_2$ holonomy
manifolds are Ricci-flat (we set the 3-form of D=11 sugra to zero).
It is also worth to point out that the three-form $\Phi$
can be computed as a bilinear in the covariantly constant spinor.
This property will be explicitly used in sections \ref{calibrating}
and \ref{calibrating2} below. It gives a relation between the
projections on the Killing spinors and the calibrating form.

For a better introduction to all the topics presented above, see
\cite{gubser}.

\medskip

By studying M-theory on $G_2$ holonomy manifolds, one 
finds a dual description of ${\cal N}=1$ four dimensional 
supersymmetric Yang-Mills (SYM). If the manifold
is large enough and smooth, the low energy description is given in
terms of a purely gravitational configuration of eleven
dimensional supergravity. The gravity/gauge theory correspondence
then allows for a geometrical approach to the study of important
aspects of the strong coupling regime of SYM theory such as the existence
of a mass gap, chiral symmetry breaking, confinement, gluino
condensation, domain walls
and chiral fermions \cite{vafa,g2things,amv,GHP}. These facts have led,
in the last  years, to a concrete and important physical
motivation to study compact and non--compact seven-manifolds of
$G_2$ holonomy.

Up to  year 2001, there were only three known examples of
complete metrics with $G_2$ holonomy on Riemannian manifolds
\cite{bs,gpp}. They correspond to $\RR^3$ bundles over $S^4$ or
${\CC\IP}^2$, and to an $\RR^4$ bundle over $S^3$. These manifolds
develop isolated conical singularities corresponding,
respectively, to cones on ${\CC\IP}^3$, $SU(3)/U(1)\times U(1)$,
or $S^3\times S^3$.

 Constructing
additional complete metrics can be an important issue in improving our
understanding of the strongly coupled infrared dynamics of ${\cal
N}=1$ supersymmetric gauge theories. For example, a new $G_2$
holonomy manifold with an asymptotically stabilized $S^1$ 
was recently found \cite{bggg}. This solution is asymptotically
locally conical (ALC) --near infinity it approaches a circle
bundle with fibres of constant length over a six--dimensional
cone--, as opposed to the asymptotically conical (AC) solutions
found in \cite{bs,gpp}.

\subsubsection{The r\^ole of gauged supergravity}

Salam-Sezgin eight dimensional gauged supergravity (see section
\ref{ss8d}) is an 
appropriate tool for obtaining a large family of $G_2$
holonomy metrics. Once again, the reason is that the natural
framework to study D6-branes is  an eight dimensional theory, where the
D6's
are domain walls.

A ten dimensional configuration where D6-branes wrap a  special
Lagrangian three-cycle inside a Calabi-Yau three-fold uplifts
to $G_2$ holonomy manifolds in eleven dimensions. This is
deduced from supersymmetry and holonomy matching conditions \cite{jaume}.
We will consider an eight dimensional ansatz corresponding to
D6-branes wrapping a cycle which is topologically $S^3$.
When uplifting to M-theory, there is another $S^3$, so the
metrics obtained will have $S^3\times S^3$ principal orbits. Furthermore,
they will have cohomogeneity one (corresponding to the radial
direction). At small $r$, one of the spheres shrinks yielding
the topology of an $\RR^4$ bundle over $S^3$ (so the solutions
with the other topologies cannot be found by this procedure).

All known metrics of this kind will be found below from 
gauged supergravity, as solutions of a single system
 of equations \cite{eprii}. 
Actually, we will recover Hitchin's general system  \cite{hit} from
a different approach.
The  calibrating forms and Killing spinors are
also computed.

Notably, the set of solutions comprises the ALC ones, what proves
what a powerful tool gauged sugra can be. That such metrics
are obtained may
seem surprising, as the archetypical ALC metric is Taub-NUT
space, which corresponds to the eleven dimensional description
of the full supergravity solution describing D6-branes in flat space.
It was argued in \cite{boonstra} that D=8 gauged supergravity
only accounts for the near-horizon physics of D6-branes, thus
leading to the Eguchi-Hanson solution, in which there is no stabilized
circle.

Indeed, let us take the eight dimensional ansatz \cite{ed}:
\beq
ds_8^2\,=\,e^{2\phi\over 3}\,dx_{1,6}^2\,+\,dr^2\,\,,
\eeq
and  excite one of the scalars in the coset, as in eq.
(\ref{coniscalars}). There is no gauge field in this case.
By imposing the projection:
\beq
\G_r \hat\G_{123}\,\epsilon\,=\,-\,\epsilon\,\,,
\eeq
the BPS eqs. coming from the vanishing of the fermion
field susy variations (\ref{susy8}) read:
\bear
\phi'&=&{1\over 8}\,e^{-\phi}\,(e^{-4\la}\,+\,2\,e^{2\la})\,\,,\rc
\la'&=&{1\over 6}\,e^{-\phi}\,(e^{-4\la}\,-\,e^{2\la})\,\,,
\eear
whereas if $\eta$ is a constant spinor satisfying $\G_r
\hat\G_{123}\eta=-\eta$, one has $\epsilon=e^{{\phi\over 6}}\eta$. 
Changing the radial variable as $dr=e^{\phi-2\la}dt$, the BPS system can
be easily solved:
\beq
\la={1 \over 6}\left[\log(e^t+a^4)-t\,\right]\,\,,\qquad\qquad
\phi={3\over 4}\left[\la+\half t-\log(16)\,\right]\,\,,
\eeq
where we have appropriately defined the integration constants.
Let us make another redefinition of the radial variable:
$e^t=\rho^4-a^4$.
Then, the eleven dimensional solution (\ref{uplift8d}) reads
$ds_{11}^2=dx_{1,6}^2+ ds_{EH}^2$ and we find the (1/2-supersymmetric)
Eguchi-Hanson metric (see, for example, \cite{eguchi}):
\beq
ds_{EH}^2\,=\,{1 \over 1-{a^4 \over \rho^4}}\,d\rho^2\,+\,
{\rho^2 \over 4}\,\left[ (\tilde{w}^1)^2+(\tilde{w}^2)^2+
\left(1-{a^4 \over \rho^4}\right)(\tilde{w}^3)^2\,\right]\,\,.
\eeq
On the other hand, the ALC Ricci-flat Taub-NUT metric reads \cite{eguchi}:
\beq
ds_{Taub-NUT}^2\,=\,{1 \over 4}{r+m\over r-m}dr^2+
{1\over 4}(r^2-m^2)\left[ (\tilde{w}^1)^2+(\tilde{w}^2)^2\right]+
m^2\,{r-m\over r+m}\,(\tilde{w}^3)^2\,\,.
\eeq
There is also a multi-center version of this solution corresponding
to parallel, but not coincident, D6-branes.

%*************************************************************

\setcounter{equation}{0}
\section{D6-brane wrapped on $S^3$}
\label{roundg2}

For the sake of clarity, let us start by analyzing the case that
 corresponds to a D6-brane wrapping a three--cycle
in such a way that the corresponding eight dimensional metric $ds^2_8$
contains a round three-sphere. The  general, much more involved, ansatz
(allowing anisotropy of the $S^3$) will be addressed in section
\ref{secsqs3}. Accordingly, we consider now the following
 metric:
\beq
ds^2_8\,=\,e^{{2\phi \over 3}}\,dx_{1,3}^2\,+\,
e^{2h}\,d\Omega_3^2\,+\,dr^2\,\,,
\label{uno}
\eeq
where $dx_{1,3}^2=-(dx^0)^2+(dx^1)^2+(dx^2)^2+(dx^3)^2$, $\phi$ and $h$
are functions of the radial coordinate $r$ and $d\Omega_3^2$ is the
metric of the unit $S^3$. The factor on the unwrapped directions has
been chosen as in (\ref{fphi3}) to have  Minkowski space in those
directions after uplifting. It is
convenient to parametrize the three-sphere by means of a new
set of left invariant one-forms $\bo^i$, $i=1,2,3$, of the $SU(2)$ 
group manifold like the ones defined in (\ref{omegas}). (The symbol
$\tilde{w}^i$ will continue to denote the forms of the $S^3$ involved in
the reduction from eleven to eight dimensions.) So we have:
\beq
d\Omega_3^2\,=\,{1\over 4}\,\sum_{i=1}^3\,(\,\bo^i\,)^2\,\,.
\label{cuatro}
\eeq

 In the configurations studied in the
present section, apart from the metric, we will only need to
excite the dilatonic scalar $\phi$ and the $SU(2)$ gauge potential
$A_{\mu}^i$. 
As the scalars in the coset are trivial ($\lambda=0$ in the notation
of (\ref{coniscalars})), we get the following
simple values for the tensors
appearing in the Salam-Sezgin formalism:
\beq
T_{ij}\,=\,\delta_{ij}\,\,,
\,\,\,\,\,\,\,\,\,\,
P_{ij}\,=\,0\,\,,
\,\,\,\,\,\,\,\,\,\,
Q_{ij} =
\pmatrix{0&&-A^3&&A^2\cr
A^3&&0&&-A^1\cr
-A^2&&A^1&&0}\,\,.
\label{roundTPQ}
\eeq

We will assume that the non-abelian gauge potential $A_{\mu}^i$ has only
non-vanishing components along the directions of the $S^3$. Actually, we will
adopt an ansatz in which this field, written as a one-form, is given by:
\beq
A^i\,=\,\Big(\,g\,-\,{1\over 2}\,\Big)\,  \bo^i\,\,,
\label{seis}
\eeq
with $g$ being a function of the radial coordinate $r$.
 The field strength
corresponding to the potential (\ref{seis}) is (the prime denotes
derivative with respect to $r$):
\beq
F^i\,=\,g'\,dr\,\wedge\,\bo^i\,
+\,{1\over 8}\,(\,4g^2-1\,)\,\epsilon^{ijk}
\,\bo^j\,\wedge\,\bo^k\,\,.
\label{siete}
\eeq
The eleven dimensional metric reads (\ref{uplift8d}):
\beq
ds_{11}^2\,=\,dx_{1,3}^2\,+\,e^{-\frac{2\phi}{3}}\,dr^2\,+\,
{1\over 4} \,e^{2h-\frac{2\phi}{3}}\,w_i^2\,+\,4\,e^\frac{4\phi}{3}\,
\left(\tilde{w}^i+(g-\half)\,w^i\right)^2
\label{addon1}
\eeq
This ansatz extends  the one used in \cite{en}
to get a metric of $G_2$ holonomy by the inclusion
of the function $g$. We will see that the key point for having $g\neq 0$
is that a rotated projection similar to (\ref{conirot}) is needed.

%*************************************************************

\subsection{Supersymmetry analysis}
\label{roundg2susy}

In this section we will find a system of 
first-order equations by analyzing the supersymmetry transformations of
the fermionic fields,
along the lines of section \ref{vffv}.
 As we will verify soon, this approach will
give us the hints we need to extend our analysis to metrics more general
than the one written in eq. (\ref{uno}).

In what follows, the Dirac matrices along
the sphere $S^3$  shall be denoted by
$\{\Gamma_1, \Gamma_2,\Gamma_3\}$, while, as before,
$\{\hat\Gamma_1, \hat\Gamma_2,\hat\Gamma_3\}$
will be the matrices along the
$SU(2)$ group manifold.

The first-order BPS equations we are trying to find are obtained by requiring
that $\delta\chi_i=\delta\psi_{\lambda}=0$ (\ref{susy8})
for some Killing spinor $\epsilon$,
which must satisfy some projection conditions. First of all, 
we need to impose
that:
\beq
\Gamma_{12}\,\epsilon\,=\,-\hat\Gamma_{12}\,\epsilon\,\,,
\,\,\,\,\,\,\,\,\,\,\,\,\,\,\,
\Gamma_{23}\,\epsilon\,=\,-\hat\Gamma_{23}\,\epsilon\,\,,
\,\,\,\,\,\,\,\,\,\,\,\,\,\,\,
\Gamma_{13}\,\epsilon\,=\,-\hat\Gamma_{13}\,\epsilon\,\,.
\label{nueve}
\eeq
Notice that in eq. (\ref{nueve}) the projections along the sphere
 $S^3$ and
the $SU(2)$ group manifold are related. Actually,
 only two of these equations
are independent and, for example, the last one can be obtained from the
first two. Moreover, it follows from (\ref{nueve}) that:
\beq
\Gamma_1\hat\Gamma_1\epsilon\,=\,
\Gamma_2\hat\Gamma_2\epsilon\,=\,
\Gamma_3\hat\Gamma_3\epsilon\,\,.
\label{diez}
\eeq
These projections are imposed by the ambient Calabi--Yau three--fold in
which the three--cycle lives, from the conditions $J_{ab}~\epsilon =
\Gamma_{ab}~\epsilon$, where $J$ is the K\"ahler form. By using eqs.
(\ref{nueve})  and (\ref{diez}) to evaluate the right-hand side of
(\ref{susy8}), together with the ansatz for the metric, dilaton and
gauge field, one gets some equations which give the
radial derivative of $\phi$, $h$ and
$\epsilon$. Actually, one arrives at the following equation for the radial
derivative of the dilaton:
\beq
\phi'\epsilon\,=\,{3\over 8}\,\Big[\,4\,(\,1-4g^2\,)\,e^{\phi-2h}\,-\,
e^{-\phi}\,\Big]\,\Gamma_r\,\hat\Gamma_{123}\,\epsilon\,+\,
3\,e^{\phi-h}\,g'\,\Gamma_1\,\hat\Gamma_1\,\epsilon\,\,,
\label{once}
\eeq
while the derivative of the function $h$ is:
\bear
h'\epsilon&=&2ge^{-h}\,
\Gamma_1\,\hat\Gamma_1\,\Gamma_r\,\hat\Gamma_{123}\,\epsilon\,-\,
{1\over 8}\,\Big[\,12\,(\,1-4g^2\,)\,e^{\phi-2h}\,+\,
e^{-\phi}\,\Big]\,\Gamma_r\,\hat\Gamma_{123}\,\epsilon\,-\,\rc\rc
&&-\,e^{\phi-h}\,g'\,\Gamma_1\,\hat\Gamma_1\,\epsilon\,\,.
\label{doce}
\eear
Moreover, the radial dependence of the spinor $\epsilon$ is determined by:
\bear
&&\partial_{r}\,\epsilon\,-\,{1\over 16}\,\Big[\,
4(\,1-4g^2\,)\,e^{\phi-2h}\,\,-\,e^{-\phi}\,\Big]\,
\Gamma_r\,\hat\Gamma_{123}\,\epsilon\,+\,
{5\over 2}\,e^{\phi-h}\,g'\,\Gamma_1\,\hat\Gamma_1\,\epsilon\,=\,0\,\,.
\label{trece}
\eear
In order to proceed further, we need to impose 
some additional condition  to
the spinor $\epsilon$. It is clear from the right-hand side of eqs.
(\ref{once})-(\ref{trece}) that we must specify 
the action on $\epsilon$ of
the radial projector $\Gamma_r\,\hat\Gamma_{123}$. The choice made in ref.
\cite{en} was to take $g=0$ and impose the condition
$\Gamma_r\,\hat\Gamma_{123}\,\epsilon\,=\,-\epsilon$. It is immediate to
verify that in this case the eqs. (\ref{once})--(\ref{trece}) reduce to
those obtained in ref. \cite{en}. Here we will not take any {\sl a priori}
particular value of
$\Gamma_r\,\hat\Gamma_{123}\,\epsilon$. 
Instead we will try to determine it
in general from eqs. (\ref{once})--(\ref{trece}). 
Notice that in our approach
$g$ will not be constant and, therefore,
 we will have to find a differential
equation which determines it.  
It is clear  from  eq. (\ref{once}) that our
spinor $\epsilon$ must satisfy a relation of the sort of
(\ref{conirproj}):
\beq
\Gamma_r\,\hat\Gamma_{123}\,\epsilon\,=\,-
(\beta\,+\,\tilde\beta\,\Gamma_1\,\hat\Gamma_1\,)\,\epsilon\,\,,
\label{catorce}
\eeq
where $\beta$ and $\tilde\beta$ are again
functions of the radial coordinate $r$,
which in this case 
 can be easily extracted from eq. (\ref{once}), namely:
\bear
\beta&=&-{8\over 3}\,\,
{\phi'\over 4(1-4g^2)\,e^{\phi-2h}\,-\,e^{-\phi}}\,\,,\rc\rc
\tilde\beta&=&8\,\,
{e^{\phi-h}\,g'\over 4(1-4g^2)\,e^{\phi-2h}\,-\,e^{-\phi}}\,\,.
\label{quince}
\eear
As in the previous chapter, the consistency condition:
\beq
\beta^2\,+\,\tilde\beta^2\,=\,1\,\,,
\label{beta2g2}
\eeq
holds. This can be proved by
noticing  that
$(\Gamma_r\,\hat\Gamma_{123})^2\epsilon\,=\,\epsilon$
and that $\{\Gamma_r\,\hat\Gamma_{123},\Gamma_1\,\hat\Gamma_1\}=0$.
By using in eq. (\ref{beta2g2}) the explicit values of $\beta$ and
$\tilde\beta$ given in eq. (\ref{quince}), one gets:
\beq
{(\phi')^2\over 9}\,+\,e^{2\phi-2h}\,(g')^2\,=\,
{1\over 64}\,\Big[\,4(1-4g^2)\,e^{\phi-2h}\,-\,e^{-\phi}\,\Big]^2\,\,,
\label{dsiete}
\eeq
which relates the radial derivatives of $\phi$ and $g$. Let us now
consider the equation for $h'$ written in eq. (\ref{doce}). By using the value
of  $\Gamma_r\,\hat\Gamma_{123}\,\epsilon$ given in eq. (\ref{catorce}),
and separating the terms with and without $\Gamma_1\,\hat\Gamma_1\,\epsilon$,
we get two equations:
\bear
&&h'\,=\,2ge^{-h}\,\tilde\beta\,+\,{1\over 8}\,
\Big[\,\,12(1-4g^2)\,e^{\phi-2h}\,+\,e^{-\phi}\,\Big]\,\beta\,\,,\rc\rc
&&2g\,e^{-h}\,\beta\,-\,{1\over 8}\,
\Big[\,\,12(1-4g^2)\,e^{\phi-2h}\,+\,e^{-\phi}\,\Big]\,\tilde\beta\,+\,
e^{\phi-h}\,g'\,=\,0\,\,.
\label{docho}
\eear
Moreover,  by using in the latter
the values of $\beta$ and $\tilde\beta$ given in
eq.  (\ref{quince}), we get the following relation between $g'$ and
$\phi'$:
\beq
g'\,=\,-{8g\over 3}\,
{e^{2h}\,\phi'\over 4(1-4g^2)\,e^{2\phi}\,+\,e^{2h}}\,\,.
\label{dnueve}
\eeq
Plugging back this equation in the consistency condition (\ref{dsiete}),
we can determine $\phi'$, $g'$, $\beta$ and $\tilde\beta$ in terms of
$\phi$, $g$ and $h$. Moreover, by substituting these results on the
first equation in (\ref{docho}), we get a first-order equation for $h$.
In order to write these
equations, let us define the function:
\beq
K\equiv
\sqrt{\Big(\,4\,(1-2g)^2\,e^{2\phi}+\,e^{2h}\,\Big)\,
\Big(\,4\,(1+2g)^2\,e^{2\phi}+\,e^{2h}\,\Big)}\,\,.
\label{veinte}
\eeq
Then, the BPS equations are:
\bear
\phi'&=&\,{3\over 8}\,\,{e^{-2h-\phi}\over K}\,\,\Big[\,
e^{4h}\,-\,16\,(1-4g^2)^2\,e^{4\phi}\,\Big]\,\,,\rc\rc
h'&=&{e^{-2h-\phi}\over 8K}\,\,\Big[\,
e^{4h}\,+\,16\,(1+4g^2)\,e^{2h+2\phi}\,+\,48\,
(1-4g^2)^2\,e^{4\phi}\,\Big]\,\,,\rc\rc
g'&=&{ge^{-\phi}\over K}\,\,\Big[\,4\,(1-4g^2)\,e^{2\phi}\,-\,
e^{2h}\,\Big]\,\,.
\label{vuno}
\eear
Notice that $g'=g=0$ certainly solves the last of these equations
 and, in this
case, the square root disappears from $K$ and
the first two equations in (\ref{vuno}) reduce to the ones
written in ref. \cite{en}. Moreover, the system (\ref{vuno}) 
is identical to
that found in ref. \cite{CGLP1} by means of the superpotential 
method (see
section \ref{g2superp}).
 The solutions of (\ref{vuno}) have been obtained in ref.
\cite{CGLP1}, and they depend on two parameters (see 
section \ref{solveg2}).

The radial projection can be interpreted
as in section \ref{wrappingS2},
where a similar expression was reached studying
the conifold. We can
parametrize $\beta=\cos\alpha$ and $\tilde\beta=\sin\alpha$, as in
(\ref{betas}), and by substituting  the value of $\phi'$ and $g'$ given by
the first-order equations (\ref{vuno}) into the definition of $\beta$ and
$\tilde\beta$ (eq. (\ref{quince})), one arrives at:
\beq
\tan\alpha\,=\,8g\,{e^{\phi+h}\over 4(1-4g^2)\,e^{2\phi}\,+\,e^{2h}}\,\,.
\label{vtres}
\eeq
Then, by using the representation (\ref{betas}), it is immediate to
rewrite eq. (\ref{catorce}) as:
\beq
\Gamma_r\,\hat\Gamma_{123}\,\epsilon\,=\,-
e^{\alpha\Gamma_1\hat\Gamma_1}\,\,\epsilon\,\,.
\label{vcuatro}
\eeq
Since $\{\Gamma_r\,\hat\Gamma_{123},\Gamma_1\,\hat\Gamma_1\}=0$,
eq.  (\ref{vcuatro}) can be solved as:
\beq
\epsilon\,=\,e^{-{1\over 2}\alpha\Gamma_1\hat\Gamma_1}
\,\,\epsilon_0\,\,,
\label{vcinco}
\eeq
where $\epsilon_0$ is a spinor satisfying the standard radial projection
condition with $\alpha=0$, \ie:
\beq
\Gamma_r\,\hat\Gamma_{123}\,\epsilon_0\,=\,-\epsilon_0\,\,.
\label{vseis}
\eeq
To determine completely $\epsilon_0$ we must use eq. (\ref{trece}), which
dictates the radial dependence of the Killing spinor. Actually, by using the
first-order equations (\ref{vuno}) one can compute $\partial_r\,\alpha$ from
eq. (\ref{vtres}). The result is remarkably simple, namely:
\beq
\partial_r\,\alpha\,=\,6\,e^{\phi-h}\,g'\,\,.
\label{vsiete}
\eeq
By using eq. (\ref{vcinco}) in eq. (\ref{trece}), one
consistently obtains (\ref{vsiete}) again. Furthermore,
the typical radial dependence of $\epsilon_0$ in terms
of the dilaton (\ref{conicinco}) is found.
Thus, after collecting all
results, it follows that $\epsilon$ can be written as:
\beq
\epsilon\,=\,e^{{\phi\over 6}}\,
e^{-{1\over 2}\alpha\Gamma_1\hat\Gamma_1}\,\eta\,\,.
\label{treinta}
\eeq
The projections conditions satisfied by  $\eta$ are simply:
\beq
\Gamma_{12}\,\hat\Gamma_{12}\,\eta\,=\,\eta\,\,,
\,\,\,\,\,\,\,\,\,\,\,\,\,\,\,\,\,\,\,\,\,\,\,\,
\Gamma_{23}\,\hat\Gamma_{23}\,\eta\,=\,\eta\,\,,
\,\,\,\,\,\,\,\,\,\,\,\,\,\,\,\,\,\,\,\,\,\,\,\,
\Gamma_{r}\,\hat\Gamma_{123}\,\eta\,=\,-\eta\,\,.
\label{tuno}
\eeq
In order to find out the meaning  of the  phase $\alpha$, let us 
use (\ref{gammaprod}) to
 write the radial projection (\ref{vcuatro}) as:
\beq
\Gamma_{x^0\cdots x^3}\,\big(\,\cos\alpha\Gamma_{123}\,-\,
\sin\alpha\hat\Gamma_{123}\,\big)\,\epsilon\,=\,\epsilon\,\,,
\label{ttres}
\eeq
which is the projection corresponding to a D6--brane wrapped on a
three--cycle,
which is non-trivially embedded in the two three--spheres, with $\alpha$
measuring the contribution of each sphere. This equation must be understood
as seen from the uplifted perspective. The case $\alpha = 0$ corresponds
to the D6--brane wrapping a three--sphere that is fully contained in the
eight--dimensional space-time where supergravity lives, and has been
studied earlier \cite{en}. Notice that $\alpha = \pi/2$ is not a solution
of the system. This is an important consistency check as this would mean
that the D6--brane is not wrapping a three--cycle contained in the
eight--dimensional space-time and the twisting would make no sense.
However, solutions that asymptotically approach $\alpha = \pi/2$ are
possible. In the next subsection we will describe a
quantity for which the rotation by the angle $\alpha$ plays an
important r\^ole.

%*************************************************************

\subsection{The calibrating three-form}
\label{calibrating}

Given a solution of the BPS equations (\ref{vuno}), one can get an eleven
dimensional metric $ds^2_{11}$ by means of the uplifting formula
(\ref{uplift8d}). The 
corresponding eleven dimensional manifold is a direct product of four
dimensional Minkowski space and a seven dimensional manifold, \ie:
\beq
ds^2_{11}\,=\,dx^2_{1,3}\,+\,ds^2_7\,=\,
dx^2_{1,3}\,+\,\sum_{A=1}^{7}\,\big(\,e^{A}\,)^2\,\,,
\label{tcuatro}
\eeq
where we have written $ds^2_7$ in terms of  a basis of one-forms
$e^{A}$ ($A=1,\cdots, 7)$. It follows from (\ref{uplift8d}) that this
basis can be taken as:
\bear
e^{i}&=&{1\over 2}\,e^{h-{\phi\over 3}}\,\,\bo^{i}\,,
\,\,\,\,\,\,\,\,\,\,(i=1,2,3)\,\,,\rc\rc
e^{3+i}&=&2\,e^{{2\phi\over 3}}\,
\big(\,\tilde{w}^i\,+\,(\,g-{1\over 2}\,)\,\bo^{i}\,\,\big)\,,
\,\,\,\,\,\,\,\,\,\,(i=1,2,3)\,\,,\rc\rc
e^{7}&=&e^{-{\phi\over 3}}\,dr\,\,.
\label{tcinco}
\eear
It is a well-known fact that a manifold of $G_2$ holonomy is endowed with a
calibrating three-form $\Phi$, which must be closed and co-closed (as
stated in eq. (\ref{cococlo})) with respect
to the seven dimensional metric $ds^2_7$. We shall denote by
$\phi_{ABC}$  the components of $\Phi$ in the basis $(\ref{tcinco})$, namely:
\beq
\Phi\,=\,{1\over 3!}\,\phi_{ABC}\,e^{A}\wedge e^{B}\wedge e^{C}\,\,.
\label{tseis}
\eeq
The relation between $\Phi$ and the Killing spinors of the metric is also
well-known. Indeed, let $\tilde \epsilon$ be the Killing spinor uplifted to
eleven dimensions, which in terms of $\epsilon$ is simply
$\tilde \epsilon= e^{-{\phi\over 6}}\,\epsilon$. Then, one has:
\beq
\phi_{ABC}\,=\,i\,\tilde \epsilon^{\dagger}\,\Gamma_{ABC}\,\tilde
\epsilon\,\,.
\label{tsiete}
\eeq
By using the relation between $\epsilon$ and the constant spinor
$\eta$, one can rewrite eq. (\ref{tsiete}) as:
\beq
\phi_{ABC}\,=\,i\, \eta^{\dagger}\,
e^{{1\over 2}\alpha\Gamma_1\hat\Gamma_1}
\Gamma_{ABC}\, e^{-{1\over 2}\alpha\Gamma_1\hat\Gamma_1}\eta\,\,.
\label{tocho}
\eeq
Let us now denote by $\phi_{ABC}^{(0)}$  the  above matrix element when
$\alpha=0$, \ie:
\beq
\phi_{ABC}^{(0)}\,=\,i\, \eta^{\dagger}\,\Gamma_{ABC}\,\eta\,\,.
\label{tnueve}
\eeq
It is not difficult to obtain the non-zero matrix elements
of (\ref{tnueve}). Recall that $\eta$ is characterized as an eigenvector
of the set of projection operators written in eq. (\ref{tuno}). Thus, if
${\cal O}$ is an operator which anticommutes with any of
these projectors, ${\cal O}\eta$ and $\eta$ are
eigenvectors of the projectors with different eigenvalues and, therefore,
they are orthogonal (\ie\ $\eta^{\dagger}{\cal O}\eta=0$). Moreover, by
using the projection conditions (\ref{tuno}), one can relate the
non-vanishing matrix elements to $\eta^{\dagger}\,\Gamma_{123}\,\eta$.
If we normalize $\eta$ such that $i\, \eta^{\dagger}\,\Gamma_{123}\,
\eta\,=\,1$ and if $\hat i\,=\,i+3$ for $i=1,2,3$, one can easily prove
that the non-zero $\phi_{ijk}^{(0)}$'s are:
\beq
\phi_{ijk}^{(0)}\,=\,\epsilon_{ijk}\,\,,
\,\,\,\,\,\,\,\,\,\,\,\,\,\,\,\,\,\,
\phi_{i\hat j\hat k}^{(0)}\,=\,-\epsilon_{ijk}\,\,,
\,\,\,\,\,\,\,\,\,\,\,\,\,\,\,\,\,\,
\phi_{7i\hat j}^{(0)}\,=\,\delta_{ij}\,\,,
\label{cuarenta}
\eeq
as in \cite{ads}. This form is like the one written in (\ref{g2form})
after a relabelling of the indices. By expanding the exponential in
(\ref{tocho}) and using (\ref{cuarenta}), it is straightforward to find
the different components of $\Phi$ for arbitrary
$\alpha$. Actually, one can write the result as:
\bear
\Phi&=&e^{7}\,\wedge\,\big(\,e^{1}\,\wedge\,e^{4}\,+\,
e^{2}\,\wedge\,e^{5}\,+\,e^{3}\,\wedge\,e^{6}\,\big)\,+\,\rc\rc
&&+\,\big(\,e^{1}\cos\alpha+e^{4}\sin\alpha\,\big)\,\wedge\,
\big(\,e^{2}\wedge e^{3}-e^{5}\wedge e^{6}\,\big)\,+\rc\rc
&&+\,\big(\,-e^{1}\sin\alpha+e^{4}\cos\alpha\,\big)\,\wedge\,
\big(\,e^{3}\wedge e^{5}-e^{2}\wedge e^{6}\,\big)\,\,,
\label{cuno}
\eear
which shows that the effect on $\Phi$ of introducing the phase $\alpha$ is
just a (radial dependent) rotation in the $(e^{1}, e^{4})$ plane
(alternatively, the same
expression can be written as a rotation in the $(e^{2}, e^{5})$ or
$(e^{3}, e^{6})$ plane). As mentioned above, $\Phi$
should be closed and co-closed:
\beq
d\Phi=0\,\,,
\,\,\,\,\,\,\,\,\,\,\,\,\,\,\,\,\,\,
d*_7\Phi=0\,\,,
\label{cdos}
\eeq
where $*_7$ denotes the Hodge dual in the seven dimensional metric. There
is an immediate consequence of this fact which we shall now exploit. 
Let us
denote by $p$ and $q$ the components of $\Phi$ along the volume forms of
the two three spheres, \ie:
\beq
\Phi\,=\,p\,\,\bo^1\wedge \bo^2\wedge \bo^3\,+
\,q\,\,\tilde{w}^1\wedge \tilde{w}^2\wedge \tilde{w}^3\,+\,\cdots\,\,.
\label{ctres}
\eeq
From the condition $d\Phi=0$, it follows immediately that $p$ and $q$ 
must be
constant. By plugging the explicit expression
of the forms $e^A$,
given in eq. (\ref{tcinco}), on the right-hand side 
of eq. (\ref{cuno}), one
easily gets $p$ and $q$ in terms of $\phi$, $h$ and $g$. Thus,
the two constant of motion are:
\bear
p&=&{1\over 8}\,\Big[\,e^{3h-\phi}\,-\,12\,e^{h+\phi}\,
\,(\,1-2g\,)^2\,\Big]\,\cos\alpha\,-\,
{1\over 4}\,(1-2g)\,
\Big[\,3e^{2h}\,-\,4\,e^{2\phi}\,(\,1-2g\,)^2\,\Big]
\,\sin\alpha\,\,,\rc\rc
q&=&-8e^{2\phi}\,\sin\alpha\,\,.
\label{ccuatro}
\eear
Notice that $\alpha = 0$ implies $q = 0$ which is precisely the case
studied in \cite{en}.
By explicit calculation one can check that $p$ and $q$ are constants as a
consequence of the BPS equations. Actually, by using (\ref{vuno})
one can show that, indeed, $\Phi$ is closed and co-closed as it should.

%*************************************************************

\subsection{Solving the equations}
\label{solveg2}

\subsubsection{The Bryant-Salamon metric}

As argued above, one can take consistently $g=0$,
going back to the analysis of \cite{en}. In this case, the
system of equations (\ref{vuno}) gets drastically simplified:
\bear
\phi'\,&=&\,\frac{3}{8}\,e^{-\phi}\,-\,\frac{3}{2}\,
e^{\phi-2h}\,\,,\rc\rc
h'\,&=&\,\frac{1}{8}\,e^{-\phi}\,+\,\frac{3}{2}\,
e^{\phi-2h}\,\,.
\label{addon5}
\eear
Let us define a new function $x$ and a new radial variable $t$:
\beq
x\equiv 12\,e^{2\phi-2h}\,\,,\,\,\,\,\,\,\,\,\,\,\,\,\,\,\,
\frac{dr}{dt}\,=\,e^\phi\,\,.
\label{addon6}
\eeq
Then, eq. (\ref{cotreinta}) is obtained again, and therefore:
\beq
x\,=\,{1 \over 1\,+\,b\,e^{-\frac{t}{2}}} \,\,\,,
\label{addon7}
\eeq
that immediately leads to:
\beq
e^{2h}\,=\,e^\frac{t}{4}\,(b\,+\,e^\frac{t}{2})^\half
\,\,,\,\,\,\,\,\,\,\,\,\,\,\,\,\,\,
e^{2\phi}\,=\,\frac{1}{12}\,e^\frac{3t}{4}\,
(b\,+\,e^\frac{t}{2})^{-\half}\,\,.
\label{addon2}
\eeq
The solution is better expressed by redefining the integration
constant and the radial coordinate:
\beq
b\,\equiv \frac{a^3}{18}\,\,,\,\,\,\,\,\,\,\,\,\,\,\,\,\,\,
e^\frac{t}{2}\equiv {1 \over 18}\,(\rho^3-a^3)\,\,,
\label{addon3}
\eeq
in terms of which (\ref{addon2}) reads:
\beq
e^{2h}\,=\,\frac{\rho^3}{18}\,\,
\left(1-\frac{a^3}{\rho^3}\right)^\half
\,\,,\,\,\,\,\,\,\,\,\,\,\,\,\,\,\,
e^{2\phi}\,=\,\frac{\rho^3}{216}\,\,
\left(1-\frac{a^3}{\rho^3}\right)^\frac{3}{2}\,\,.
\label{addon8}
\eeq
Then, the eleven dimensional metric (\ref{addon1}) can
be written as $ds_{11}^2=dx_{1,3}^2+ds_7^2$ where:
\beq
ds_7^2\,=\,{d\rho^2 \over \left(1-\frac{a^3}{\rho^3}\right)}\,+\,
{\rho^2 \over 12}\,(w^i)^2\,+\,
{\rho^2 \over 9}\,\left(1-\frac{a^3}{\rho^3}\right)\,
\left(\tilde{w}^i-\frac{w^i}{2}\right)^2\,\,.
\label{brysal}
\eeq
This is the $G_2$ holonomy metric with $\RR^4\times S^3$
topology which
was first obtained by Bryant and
Salamon \cite{bs} (see also \cite{gpp}). When 
$\rho\rightarrow a$, one of the three-spheres shrinks to
a point while the other one remains finite, so $a$ plays
a r\^ole similar to the resolution or deformation parameters
of the conifold. As in that case, it cures the 
curvature singularity at the origin.

\subsubsection{The general solution}

The general solution of the system (\ref{vuno}) was worked out
in  ref. \cite{CGLP1}.  This 
system is harder
to solve than (\ref{resolsyst}) or (\ref{deformsyst}), as no decoupled
first order equation can be found. However, as will be
shown next, a second order  equation can be decoupled
by proceeding cunningly.

First of all, it is convenient to write the equation in
terms of three new functions
 $\{y,\ z,\ \psi\}$:
\beq
y\,=\,4\,e^{{4\phi \over 3}}\,\,,
\,\,\,\,\,\,\,\,\,\,\,\,\,\,\,\,\,\,
z\,=\,{1 \over 4}\,e^{2h-\frac{2\phi}{3}}\,\,,
\,\,\,\,\,\,\,\,\,\,\,\,\,\,\,\,\,\,
\psi\,=\,(2\,g)^{-1/2}\,\,,
\label{yzpsi}
\eeq
and a new radial variable $w$ defined as:
\beq
\frac{dw}{dr}\,=\,2\,e^{\phi}\,K^{-1}\,\,.
\label{dwdr}
\eeq
The system (\ref{vuno}) becomes:
\bear
\frac{dy}{dw}&=&4z\,-\,\frac{y^2}{4z}\,(1-\psi^{-4})^2
\,\,\,,\rc
\frac{dz}{dw}&=&\frac{y}{2}\,(1-\psi^{-4})^2\,+\,
2z\,(1+\psi^{-4})\,\,\,,\rc
\frac{d\psi}{dw}&=&-\psi\,(1-\psi^{-4})+\frac{4\,\psi \,z}{y}\,\,\,.
\label{systw}
\eear
Now, by differentiating the last equation and also making  use of the
first two, the desired decoupled equation is achieved:
\beq
{d^2\psi \over dw^2}\,=\,4\psi\,-\,4\,\psi^{-3}\,\,\,.
\label{psi2}
\eeq
The easiest way to integrate (\ref{psi2}) is to notice that
it can be regarded as an equation describing a classical
system with a potential given by $-\frac{\partial V}
{\partial\psi}=4\psi-4\psi^{-3}$. Then, conservation of
energy reads $\half\left(\frac{d \psi}
{d w}\right)^2+V=const$. This has separate variables
and can be immediately integrated. Calling $const=8/\la$ for
convenience, and denoting by $m$ the other constant of
integration, we get:
\beq
e^{4w}\,=\,8m\,\left[\,\sqrt{1+{2 \over \la}\psi^2+\psi^4}\,+\,
\psi^2+{1 \over \la}\,\right]\,\,.
\label{e4w}
\eeq
As in \cite{CGLP1}, let us further redefine the radial variable:
\beq
\tau\,=\,\frac{\la}{8}\,e^{4w}\,\,.
\eeq
In (\ref{e4w}), it is easy to find the  value of $\psi$, and
therefore, of $g$:
\beq
g\,=\,\la\,\,m\,\tau\,Y(\tau)^{-1}\,\,\,,
\label{sol1}
\eeq
where $Y(\tau)$ has been defined to be:
\beq
Y(\tau)\,\equiv\,\tau^2\,-\,2m\tau\,+\,m^2\,(1-\la^2)\,\,.
\eeq
Knowing $\psi$, one can find $z/y$ from the last equation
of (\ref{systw}) and then it is not difficult to integrate
the other equations:
\beq
y\,=\,F(\tau)^{-\frac{1}{3}}\,Y(\tau)\,\,\,,
\,\,\,\,\,\,\,\,\,
z\,=\,{1 \over 4}\,F(\tau)^{\frac{2}{3}}\,Y(\tau)^{-1}\,\,\,,
\label{sol2}
\eeq
where:
\beq
F(\tau)\,\equiv\,3\tau^4\,-\,8m\tau^3\,+\,
6\,m^2\,(1-\la^2)\,\tau^2-m^4\,(1-\la^2)^2\,\,.
\eeq
The two constants of integration $\la$, $m$ play the same
r\^ole as the constants of motion $p$, $q$ found in section
\ref{calibrating}. In fact, they can be related by directly
plugging the solution (\ref{sol1}), (\ref{sol2}) in the 
expressions (\ref{ccuatro}), getting:
\beq
q\,=\,-\,m\,\la\,\,,\,\,\,\,\,\,\,\,\,\,\,\,\,\,\,\,
\,\,\,\,\,\,\,\,p\,=\,\half\,m\,(1+\la)\,\,.
\eeq
Now the eleven dimensional metric  can be explicitly written.
 We have to insert (\ref{uno}) and (\ref{seis}) in the uplifting
formula (\ref{uplift8d}). The solution
(\ref{sol1}), (\ref{sol2}) and the definitions (\ref{yzpsi}), 
(\ref{dwdr}) are needed. Finally, by dropping the flat
four dimensional Minkowski part of the metric, we arrive at
the sought seven dimensional $G_2$ holonomy metric:
\beq
ds^2_7\,=\,F^{-{1\over 3}}\,d\tau^2\,+\,
{1\over 4}\,F^{{2\over 3}}\,Y^{-1}\,(\,\bo^i\,)^2\,+\,
F^{-{1\over 3}}\,Y\,
\Bigg(\,\tilde{w}^i\,-\,
\big(\,{1\over 2}\,+\,{q\,\tau\over  Y}\big)\,\bo^i\,\Bigg)^2\,\,.
\label{cseis}
\eeq
The analysis of the metrics (\ref{cseis}) has been carried out in ref.
\cite{CGLP1}. It turns out that only in three cases ($p=0$, $q=0$ and
$p=-q$) the metric  (\ref{cseis}) is non-singular. The first two cases
are related by the so-called flop transformation, which is a $\ZZ_2$
action that exchanges $\bo^i$ and $\tilde{w}^i$, while the $p=-q$ case
is flop invariant. It is interesting to point out that, as $g \to 0$
when $\tau \to \infty$, the gauge field (\ref{seis}) asymptotically
approaches that used in \cite{en} to perform the twisting. This agrees
with the fact that the twisting just fixes the value of the gauge field
 at infinity (we have already seen an example in the previous chapter).

Finally, let us make contact with the $g=0$  case. It  
 is recovered from the above results by
taking $q=0$ ($\la=0$), while $p$ plays the r\^ole of the scale 
parameter $a$. In fact, by identifying this parameter as
$a^3\equiv 24\sqrt{3}\,p$ and introducing the radial variable
$\rho\equiv\sqrt{3}(3\tau+2p)^{\frac{1}{3}}$, we obtain again the
 metric (\ref{brysal}).

%*************************************************************

\subsection{First order system from a superpotential}
\label{g2superp}

In this section we are going to derive the first-order equations
(\ref{vuno}) by finding a superpotential for the effective lagrangian
$L_{eff}$ in eight dimensional supergravity. The first step in
this approach is to obtain the form of $L_{eff}$ for the ansatz
given in eqs. (\ref{uno}) and (\ref{seis}). Actually, the expression of
$L_{eff}$ can be obtained by substituting (\ref{uno}) and
(\ref{seis}) into the lagrangian given by eq. (\ref{apatres}). Indeed,
one can check that the
equations of motion of eight dimensional supergravity can be 
derived from the
following effective lagrangian:
\bear
L_{eff}&=&e^{{4\phi \over 3}+3h}\,\Big[\,
(h'\,)^2\,-\,{1\over 9}\,(\phi'\,)^2\,-\,
4e^{2\phi-2h}\,(g'\,)^2\,+\,
{4 \over 3}\,\phi'\,h'\,+\,\rc
&&+\,e^{-2h}\,+\,{1\over 16}\,e^{-2\phi}\,-\,
(4g^2\,-\,1)^2\,e^{2\phi-4h}\,\,\Big]\,\,,
\label{apbuno}
\eear
together with the zero-energy condition.  Next, let us introduce a new
set of functions:
\beq
a\,=\,2\,e^{{2\phi\over 3}}\,\,,
\,\,\,\,\,\,\,\,\,\,\,\,\,\,\,
b\,=\,{1\over 2}\,e^{h\,-\,{\phi\over 3}}\,\,,
\label{apbdos}
\eeq
and a new variable $\eta$, defined as:
\beq
{dr\over d\eta}\,=\,e^{{4\phi\over 3}\,+\,3h}\,\,.
\label{apbtres}
\eeq
The effective lagrangian in these new variables has the kinetic term:
\beq
T\,=\,\Big(\,{\partial_\eta a\over a}\,\Big)^2\,+\,
\Big(\,{\partial_\eta b\over b}\,\Big)^2\,+\,3\,
{(\partial_\eta a)(\partial_\eta b)\over a
b}\,-\, {1\over 4}\,{a^2\over b^2}\,\big(\,
\partial_\eta g\,\big)^2\,\,.
\label{apbcuatro}
\eeq
The potential in $L_{eff}$ is:
\beq
V\,=\,{a^6b^6\over 2}\,\Big[\,(1\,-\,4g^2\,)^2\,
{a^2\over 32 b^4}\,-\,{1\over 2a^2}-\,{1\over 2b^2}\,\Big]\,\,.
\label{apbcinco}
\eeq
The superpotential for $T-V$ in the  variables just introduced has been
obtained in ref. \cite{CGLP1}, starting from eleven dimensional
supergravity. So,
we shall follow here the same steps as in ref. \cite{CGLP1} and
define $\alpha^1=\log a$,
$\alpha^2=\log b$ and $\alpha^3= g$. Then, the kinetic energy $T$ can be
rewritten as:
\beq
T\,=\,{1\over 2}\,g_{ij}\,{d\alpha^i\over d\eta}\,
{d\alpha^j\over d\eta}\,\,,
\label{apbseis}
\eeq
where $g_{ij}$ is the matrix:
\beq
g_{ij}\,=\,\pmatrix{2&&3&&0\cr
                    3&&2&&0\cr
                    0&&0&&-{a^2\over 2b^2}}\,\,.
\label{apbsiete}
\eeq
The superpotential $W$ for this system must satisfy (\ref{defsuperp})
where $g^{ij}$ is the inverse of $g_{ij}$ and $V$ has been written in eq.
(\ref{apbcinco}). By using the values of $g_{ij}$  in (\ref{apbsiete}),
one can write explicitly the relation between $V$ and $W$ as:
\beq
V\,=\,{1\over 5}\,a^2\,\Bigg({\partial W\over \partial a}\Bigg)^2\,+\,
{1\over 5}\,b^2\,\Bigg({\partial W\over \partial b}\Bigg)^2\,-\,
{3\over 5}\,ab\,{\partial W\over \partial a}\,{\partial W\over 
\partial b}\,+\,
{b^2\over a^2}\,\Bigg({\partial W\over \partial g}\Bigg)^2\,\,.
\label{apbnueve}
\eeq
Moreover, it is not difficult to verify, following again ref. \cite{CGLP1},
that
$W$ can be taken as:
\beq
W\,=\,{1\over 8}\,a^2\,b\,
\sqrt{\Big(\,a^2\,(1-2g)^2\,+\,4b^2\,\Big)\,
\Big(\,a^2\,(1+2g)^2\,+\,4b^2\,\Big)}\,\,.
\label{apbdiez}
\eeq
The first-order equations associated to the superpotential $W$ are
(\ref{order1}).

By substituting the expressions of $W$ and $g^{ij}$
 on the right-hand side of
eq. (\ref{order1}), and by writing the result in terms of the
typical eight dimensional variables, one can check that
the system   (\ref{vuno}) is reobtained.

It is worth mentioning that the superpotentials with
a square root are typical of systems whose supersymmetry
is based on a projection of the kind (\ref{vcuatro}).
This is because, as we have seen, a square root typically
appears in the equations of motion after working
with a expression of the kind (\ref{dsiete}), which
directly comes from (\ref{beta2g2}).

%*************************************************************

\setcounter{equation}{0}
\section{D6-brane wrapped on squashed $S^3$}
\label{secsqs3}

In this section we are going to generalize the analysis performed in
section \ref{roundg2}
 to a much more general situation, in which the eight
dimensional metric takes the form:
\beq
ds^2_8\,=\,e^{{2\phi\over 3}}\,dx_{1,3}^2\,+\,{1\over 4}\,
e^{2h_i}\,(\,\bo^i\,)^2\,+\,dr^2\,\,.
\label{csiete}
\eeq
Once again, we have  implemented the
$e^{{2\phi\over 3}}$ factor, which ensures that we are going to have a
direct product of four dimensional Minkowski space and a seven
dimensional manifold in the uplift to eleven dimensions. As in the
previous case, we are going to switch on a $SU(2)$ gauge field potential
with components along the squashed $S^3$. The ansatz we shall adopt for
this potential is:
\beq
A^i\,=\,G_i\,\bo^i\,\,,
\label{cocho}
\eeq
which depends on three functions $G_1$,  $G_2$ and  $G_3$. It should be
understood that there is no sum on the right-hand side of eq.
(\ref{cocho}). Moreover, we shall excite coset scalars in the diagonal and,
therefore, the corresponding $L_{\alpha}^i$ matrix will be taken as:
\beq
L_{\alpha}^i\,=\,{\rm diag}\,
(\,e^{\lambda_1}\,,\,e^{\lambda_2}\,,\,e^{\lambda_3}\,)\,\,,
\,\,\,\,\,\,\,\,\,\,\,\,
\lambda_1\,+\,\lambda_2\,+\,\lambda_3\,=\,0\,\,.
\label{cnueve}
\eeq
The matrices $P_{\mu ij}$ and $Q_{\mu ij}$ defined in eq.
(\ref{apauno}) are easily evaluated from eqs. (\ref{cocho}) and
(\ref{cnueve}). Written as differential forms, they are:
\beq
P_{ij}\,+\,Q_{ij}\,=\,
\pmatrix{d\lambda_1&&-A^{3}\,e^{\lambda_{21}}&&A^{2}\,e^{\lambda_{31}}\cr\cr
         A^{3}\,e^{\lambda_{12}}&&d\lambda_2&&-A^{1}\,e^{\lambda_{32}}\cr\cr
        -A^{2}\,e^{\lambda_{13}}&&A^{1}\,e^{\lambda_{23}}&&d\lambda_3}\,\,,
\label{cincuenta}
\eeq
where $\lambda_{ij}\,\equiv\,\lambda_i\,-\,\lambda_j$ and
$P_{ij}$ ($Q_{ij}$) is the symmetric (antisymmetric) part of the matrix
appearing on the right-hand side of (\ref{cincuenta}). Notice that our present
ansatz depends on nine functions, since there are only two independent
$\lambda_i$'s (see eq. (\ref{cnueve})). On the other hand, it would be
convenient in what follows to define the following combinations of these
functions:
\bear
M_1\,\equiv &\,e^{\phi+\lambda_1-h_2-h_3}\,(\,G_1\,+\,G_2\,G_3)\,\,,\rc\rc
M_2\,\equiv&\,e^{\phi+\lambda_2-h_1-h_3}\,(\,G_2\,+\,G_1\,G_3)\,\,,\rc\rc
M_3\,\equiv&\,e^{\phi+\lambda_3-h_1-h_2}\,(\,G_3\,+\,G_1\,G_2)\,\,.
\label{ciuno}
\eear

%*************************************************************

\subsection{Supersymmetry analysis}
\label{secgeng2}

With the setup just described, and the experience acquired in 
section \ref{roundg2susy},
 we will now face the problem of finding supersymmetric
configurations
for this more general ansatz. Notice that, using this method, it
is quite straightforward (although rather lengthy) to obtain
the system of first order equations and constraints that 
 gives the most general supersymmetric configurations.
On the contrary, it seems very hard to achieve the same
results from the superpotential method.

As before, we must guarantee that
$\delta\chi_i=\delta\psi_{\lambda}=0$ for some spinor $\epsilon$. 
We begin by
imposing again the angular projection condition (\ref{nueve}). 
Then, the
equation $\delta\chi_1=0$ yields:
\bear
\Big(\,{1\over 2}\lambda_1'+{1\over 3}\phi'\,\Big)\,\epsilon\,&=&\,
e^{\phi+\lambda_1-h_1}\,G_1'\,\Gamma_1\hat\Gamma_1\,\epsilon\,-\,
2\Big[\,M_1\,-\,{1\over 16}\,e^{-\phi}\,
(\,e^{2\lambda_1}-e^{2\lambda_2}-e^{2\lambda_3}\,)\,\Big]\,
\Gamma_r\hat\Gamma_{123}\,\epsilon\,-\,\rc\rc
&&-\Big[\,e^{-h_2}\,G_2\,\sinh\lambda_{13}\,+\,
e^{-h_3}\,G_3\,\sinh\lambda_{12}\,\Big]\,
\Gamma_1\hat\Gamma_1\,\Gamma_r\hat\Gamma_{123}\,\epsilon\,\,,
\label{cidos}
\eear
and, obviously, $\delta\chi_2=\delta\chi_3=0$ give rise to
other two similar equations which are obtained by permutation of the indices
$(1,2,3)$ in eq. (\ref{cidos}). Adding these three equations and using eq.
(\ref{diez}) and the fact that  $\lambda_1+\lambda_2+\lambda_3=0$, we get the
following equation for $\phi'$:
\bear
\phi'\,\epsilon\,&=&\,e^{\phi}\,\Big[\,
e^{\lambda_1-h_1}\,G_1'\,+\,e^{\lambda_2-h_2}\,G_2'\,+\,
e^{\lambda_3-h_3}\,G_3'\,\Big]\Gamma_1\hat\Gamma_1\,\epsilon\,-\,\rc\rc
&&-2\Big[\,M_1+M_2+M_3+{1\over 16}\,e^{-\phi}\,
\big(\,e^{2\lambda_1}\,+\,e^{2\lambda_2}\,+\,e^{2\lambda_3}\,\big)\,\Big]\,
\Gamma_r\hat\Gamma_{123}\,\epsilon\,\,.
\label{citres}
\eear
It can be checked that this same equation is obtained from the variation of
the gravitino components along the unwrapped directions. Moreover, it
follows from  eq. (\ref{citres}) that
$\Gamma_r\,\hat\Gamma_{123}\,\epsilon$ has
the same structure as in eq. (\ref{catorce}), where now $\beta$ and
$\tilde\beta$ are given by:
\bear
\beta&=&\,{8\,\phi'\over
16\,(\,M_1+M_2+M_3)\,+\,e^{-\phi}\,
(\,e^{2\lambda_1}\,+\,e^{2\lambda_2}\,+\,e^{2\lambda_3}\,)}\,\,,\rc\rc\rc
\tilde\beta&=&\,-\,{8\,e^{\phi}\,(\,
e^{\lambda_1-h_1}\,G_1'\,+\,e^{\lambda_2-h_2}\,G_2'\,+\,
e^{\lambda_3-h_3}\,G_3'\,)
\over
16\,(\,M_1+M_2+M_3)\,+\,e^{-\phi}\,
(\,e^{2\lambda_1}\,+\,e^{2\lambda_2}\,+\,e^{2\lambda_3}\,)}\,\,.
\label{cicuatro}
\eear
It is also immediate to see that in the present case $\beta$
 and $\tilde\beta$
must also satisfy the constraint (\ref{beta2g2}). Thus, in this case we
are going to have the same type of radial projection as in the round
metric of section \ref{roundg2}.
Actually, we shall obtain a set of first-order differential equations in
terms of $\beta$ and $\tilde\beta$ and then we shall find some consistency
conditions which, in particular, allow to determine  the  values of
$\beta$ and $\tilde\beta$. From this point of view it is straightforward to
write the equation for $\phi'$. Indeed, from the definition of $\beta$
(eq. (\ref{cicuatro})), one has:
\beq
\phi'\,=\,\Big[\,2\,(\,M_1\,+\,M_2\,+\,M_3)\,+\,{1\over 8}\,
e^{-\phi}\,(\,e^{2\lambda_1}\,+\,e^{2\lambda_2}\,+\,e^{2\lambda_3}\,)\,
\Big]\,\beta\,\,.
\label{cicinco}
\eeq
In order to obtain the equation for $\lambda_i'$ and $G_i'$, let us consider
again the equation derived from the condition $\delta\chi_i=0$
(eq. (\ref{cidos})). Plugging the projection condition on the right-hand side
of eq. (\ref{cidos}), using the value of $\phi'$ displayed in eq.
(\ref{cicinco}), and considering the terms without $\Gamma_1\,\hat\Gamma_1$,
one gets the equation for $\lambda_1'$, namely:
\bear
\lambda_1'&=&{4\over 3}\,
\Big[\,2M_1\,-\,M_2\,-\,M_3\,-\,{1\over 8}\,e^{-\phi}\,\big(\,
2e^{2\lambda_1}\,-\,e^{2\lambda_2}\,-\,e^{2\lambda_3}\,\big)\,
\Big]\,\beta\,-\,\rc\rc
&&-\,2\,\Big[\,e^{-h_2}\,G_2\,\sinh\lambda_{13}\,+\,
e^{-h_3}\,G_3\,\sinh\lambda_{12}\,\Big]\,\tilde\beta\,\,.
\label{ciseis}
\eear
while the terms with $\Gamma_1\,\hat\Gamma_1$ of eq. (\ref{cidos}) yield the
equation for $G_1'$:
\bear
e^{\phi\,+\,\lambda_1-h_1}\,G_1'&=&\Big[\,
-2M_1\,+\,{1\over 8}\,e^{-\phi}\,
(\,e^{2\lambda_1}\,-\,e^{2\lambda_2}\,-\,e^{2\lambda_3}\,)
\,\Big]\,\tilde\beta\,-\,\rc\rc
&&-\,\Big[\,
e^{-h_2}\,G_2\,\sinh\lambda_{13}\,+\,
e^{-h_3}\,G_3\,\sinh\lambda_{12}\,\Big]\,\beta\,\,.
\label{cisiete}
\eear
By cyclic permutation of the indices in eqs. (\ref{ciseis}) and
(\ref{cisiete}), one obtains the first-order differential equations of 
$\lambda_i'$ and
$G_i'$ for
$i=2,3$.

It remains to obtain the equation for  $h_i'$. With this purpose let us
consider the supersymmetric variation of the gravitino components along the
sphere. One gets:
\bear
h_1'\,\epsilon&=&-{1\over 3}\,e^{\phi}\,
\Big[\,5\,e^{\lambda_1-h_1}\,G_1'\,-\,e^{\lambda_2-h_2}\,G_2'\,-\,
e^{\lambda_3-h_3}\,G_3'\,\Big]\,\Gamma_1\,\hat\Gamma_1\,\epsilon
\,-\,\rc\rc
&&-{1\over 3}\,\Big[\,2\,(M_1-5M_2-5M_3)\,+\,{1\over 8}\,
e^{-\phi}\,(\,e^{2\lambda_1}\,+\,e^{2\lambda_2}\,+\,e^{2\lambda_3}\,)
\,\Big]\,\Gamma_r\,\hat\Gamma_{123}\,\epsilon\,-\,\rc\rc
&&-\,\Big[\,{e^{2h_1}\,-\,e^{2h_2}\,-\,e^{2h_3}\over
e^{h_1+h_2+h_3}}\,-\,2e^{-h_1}\,G_1\,\cosh\lambda_{23}\,\Big]\,
\Gamma_1\,\hat\Gamma_1\,\Gamma_r\,\hat\Gamma_{123}\,\epsilon\,\,,
\label{ciocho}
\eear
and two other equations obtained by cyclic permutation. By considering
the terms without $\Gamma_1\,\hat\Gamma_1$ in eq. (\ref{ciocho})
we get the desired first-order equation for $h_1'$, namely:
\bear
h_1'&=&
{1\over 3}\,\Big[\,2\,(M_1-5M_2-5M_3)\,+\,{1\over 8}\,
e^{-\phi}\,(\,e^{2\lambda_1}\,+\,e^{2\lambda_2}\,+\,e^{2\lambda_3}\,)
\,\Big]\,\beta\,-\,\rc\rc
&&-\,\Big[\,{e^{2h_1}\,-\,e^{2h_2}\,-\,e^{2h_3}\over
e^{h_1+h_2+h_3}}\,
-\,2e^{-h_1}\,G_1\,\cosh\lambda_{23}\,\Big]\,
\tilde\beta\,\,.
\label{cinueve}
\eear
On the other hand, the terms with $\Gamma_1\,\hat\Gamma_1$ of
eq. (\ref{ciocho}) give rise to new equations for the $G_i'$'s:
\bear
&&e^{\phi}\,
\Big[\,5\,e^{\lambda_1-h_1}\,G_1'\,-\,e^{\lambda_2-h_2}\,G_2'\,-\,
e^{\lambda_3-h_3}\,G_3'\,\Big]\,=\,\rc\rc
&&=\,\Big[\,2\,(M_1-5M_2-5M_3)\,+\,{1\over 8}\,
e^{-\phi}\,(\,e^{2\lambda_1}\,+\,e^{2\lambda_2}\,+\,e^{2\lambda_3}\,)
\,\Big]\,\tilde\beta\,+\,\rc\rc
&&+\,3\,\Big[\,{e^{2h_1}\,-\,e^{2h_2}\,-\,e^{2h_3}\over
e^{h_1+h_2+h_3}}\,-\,2e^{-h_1}\,G_1\,\cosh\lambda_{23}\,\Big]\,\beta\,\,.
\label{sesenta}
\eear
This equation (and the other two obtained by cyclic permutation) must be
compatible with  the equation for $G_i'$ written in eq.
(\ref{cisiete}). Actually, by
substituting in eq. (\ref{sesenta}) the value of $G_i'$ given by eq.
(\ref{cisiete}), and by combining appropriately the equations so obtained,
we arrive at three algebraic relations of the type:
\beq
{\cal A}_i\,\beta\,-\,{\cal B}_i\,\tilde\beta\,=\,0\,\,,
\label{suno}
\eeq
where ${\cal A}_1$ and ${\cal B}_1$ are given by:
\bear
{\cal A}_1&=&e^{h_1-h_2-h_3}\,+\,
e^{\lambda_1-\lambda_3-h_2}\,G_2\,+\,
e^{\lambda_1-\lambda_2-h_3}\,G_3\,\,,\rc\rc
{\cal B}_1&=&-4M_1\,+\,{1\over 4}\,e^{-\phi+2\lambda_1}\,\,,
\label{sdos}
\eear
while the values of ${\cal A}_i$ and ${\cal B}_i$ for $i=2,3$ are obtained
from (\ref{sdos}) by cyclic permutation. Notice that the above relations
do not
involve derivatives of the fields and, in particular, they allow to obtain
the values of $\beta$ and  $\tilde\beta$. Indeed, by using the constraint
$\beta^2\,+\,\tilde\beta^2=1$,  and eq. (\ref{suno})  for $i=1$, we get:
\beq
\beta\,=\,{{\cal B}_1\over \sqrt{{\cal A}_1^2\,+\,{\cal B}_1^2}}\,\,,
\,\,\,\,\,\,\,\,\,\,\,
\tilde\beta\,=\,{{\cal A}_1\over \sqrt{{\cal A}_1^2\,+\,{\cal B}_1^2}}\,\,.
\label{stres}
\eeq
Moreover, it is clear from (\ref{suno}) that the ${\cal A}_i$'s and ${\cal
B}_i$'s must satisfy the following  consistency conditions:
\beq
{\cal A}_i\,{\cal B}_j\,=\,{\cal A}_j\,{\cal B}_i\,\,,
\,\,\,\,\,\,\,\,\,\,\,\,
(i\not= j)\,\,.
\label{scuatro}
\eeq
Eq. (\ref{scuatro}) gives two independent algebraic constraints that the
functions of our generic ansatz must satisfy if we demand it  to be a
supersymmetric
solution. Notice that these constraints are trivially satisfied in the round
case of section \ref{roundg2}. On the other hand, if we adopt the radial
projection of refs. \cite{en, hs1}, \ie\ when  $\beta=1$ and
$\tilde\beta=0$, they imply that ${\cal A}_i=0$ (see eq. (\ref{suno})),
this leading precisely to the values of
the $SU(2)$ gauge potential used in those references. Moreover, by using
eq. (\ref{suno}), the differential equation satisfied by the $G_i$'s
can be simplified. One gets:
\bear
e^{\phi\,+\,\lambda_1-h_1}\,G_1'&=&{1\over 2}\,\,\Big[\,
e^{h_1-h_2-h_3}\,+\,e^{\lambda_3-\lambda_1-h_2}\,G_2\,+\,
e^{\lambda_2-\lambda_1-h_3}\,G_3\,\Big]\,\beta\,-\,\rc\rc 
&&\,-\,{e^{-\phi}\over 8}\,\big(\,e^{2\lambda_2}+e^{2\lambda_3}\,\big)\,
\tilde\beta\,\,,
\label{scinco}
\eear
and similar expressions for $G_2$ and $G_3$.

Let us now parametrize $\beta$ and $\tilde\beta$ 
in the usual way (\ref{betas}),
\ie\ $\beta\,=\,\cos\alpha$, $\tilde\beta\,=\,\sin\alpha$. 
Then, it follows
from
eq. (\ref{stres}) that one has:
\beq
\tan\alpha\,=\,{{\cal A}_1\over {\cal B}_1}\,=\,
{{\cal A}_2\over {\cal B}_2}\,=\,
{{\cal A}_3\over {\cal B}_3}\,\,.
\label{sseis}
\eeq
Notice that by taking $\alpha = 0$, eq. (\ref{scinco}) precisely leads
to the expression for the gauge field in terms of scalar fields used in
\cite{hs1} to perform the twisting. Moreover, the radial projection
condition can be written as in eq. (\ref{vcuatro}) and, thus, the
natural solution to the Killing spinor equations is just the one written
in eq. (\ref{treinta}), where $\eta$ is a constant spinor satisfying the
conditions (\ref{tuno}). To check that this is the case, one can plug the
expression of $\epsilon$  given in eq. (\ref{treinta}) in the equation
arising from  the supersymmetric variation of the radial component of
the gravitino. It turns out that this equation is satisfied provided
$\alpha$ satisfies the equation:
\beq
\partial_r\alpha\,=\,-\,\Big[\,
4\,\big(\,M_1\,+\,M_2\,+\,M_3\,\big)\,+\,
{1\over 4}\,e^{-\phi}\,\big(\,
e^{2\lambda_1}\,+\,e^{2\lambda_2}\,+\,e^{2\lambda_3}\,\big)\,\Big]\,
\sin\alpha\,\,.
\label{ssiete}
\eeq
In general, this equation for $\alpha$ does not 
follow from the first-order
equations and the algebraic constraints we have found.
 Actually, by using the
value of $\alpha$ given in eq. (\ref{sseis}) and the 
first-order equations to
evaluate the left-hand side of eq. (\ref{ssiete}), 
we could derive a third
algebraic constraint. However, this new constraint is rather complicated.
Happily,  we will not need to do this explicitly since eq. (\ref{ssiete})
will serve to our purposes.

It is worth pointing out that a seven dimensional manifold
has holonomy contained in $G_2$ if the spin connection on the
seven manifold is
self-dual: $\omega^{ab}=\half \psi_{abcd} \omega^{cd}$, 
where
$\psi_{abcd}$ is the dual of the structure constants $\psi_{abc}$
of the octonions \cite{ads} (see section \ref{secspin7}). It can be
checked
\cite{rafapriv} that the differential equations and 
constraints just found can also be obtained by imposing
this condition on a rotated frame. The calculation is similar
to the one that will be showed in section \ref{secspin7}
for $Spin(7)$ holonomy manifolds.

%*************************************************************

\subsection{The calibrating three-form}
\label{calibrating2}

In order to find the calibrating three-form $\Phi$ in this case, let us
take the following vielbein basis:
\bear
e^{i}&=&{1\over 2}\,e^{h_i-{\phi\over 3}}\,\,\bo^{i}\,,
\,\,\,\,\,\,\,\,\,\,(i=1,2,3)\,\,,\rc\rc
e^{3+i}&=&2\,e^{{2\phi\over 3}+\lambda_i}\,
\big(\,\tilde{w}^i\,+\,G_i\,\bo^{i}\,\,\big)\,,
\,\,\,\,\,\,\,\,\,\,(i=1,2,3)\,\,,\rc\rc
e^{7}&=&e^{-{\phi\over 3}}\,dr\,\,,
\label{socho}
\eear
which is the natural one for the uplifted metric. The different components of
$\Phi$ can be computed by using eq. (\ref{tseis}) and it is obvious from the
form of the projection that the result is just the one given in eq.
(\ref{cuno}), where now $\alpha$ is given by eq. (\ref{sseis}) and the
one-forms $e^A$ are the ones written in eq. (\ref{socho}). If, as in eq.
(\ref{ctres}), $p$ and $q$ denote the components of
$\Phi$ along the two three spheres, it follows from the closure of $\Phi$ that
$p$ and $q$ should be constants of motion. By plugging the expressions of the
$e^A$'s, taken from eq. (\ref{socho}), on the right-hand side of eq.
(\ref{cuno}), one can find the explicit expressions of $p$ and $q$. The result
is:
\bear
p&=&{1\over 8}\,\Big[\,e^{h_1+h_2+h_3-\phi}\,-\,
16e^{\phi}\,\big(\,e^{h_1-\lambda_1}G_2G_3+\,
e^{h_2-\lambda_2}G_1G_3+\,e^{h_3-\lambda_3}G_1G_2\,\big)\Big]\,
\cos\alpha\,+\,\rc\rc
&&+\,{1\over 2}\,\Big[\,
e^{h_2+h_3+\lambda_1}G_1\,+\,e^{h_1+h_3+\lambda_2}G_2\,+\,
e^{h_1+h_2+\lambda_3}G_3\,-\,16e^{2\phi}\,G_1G_2G_3\,\Big]\,
\sin\alpha\,\,,\rc\rc
q&=&-8e^{2\phi}\sin\alpha \,\,.
\label{snueve}
\eear
It is a simple exercise to verify that, when restricted to the round case
studied above, the expressions of $p$ and $q$ given in eq.
(\ref{snueve}) coincide with those written in eq. (\ref{ccuatro}). Moreover,
the proof of the constancy of $p$ and $q$ can be performed by combining
appropriately the
first-order equations and the constraints. Actually, by using eq.
(\ref{cicinco}) to compute the radial derivative of $q$ in eq.
(\ref{snueve}), it
follows that the condition  $\partial_r q=0$ is equivalent to eq.
(\ref{ssiete}). Although the proof of $\partial_r p=0$ is much more involved,
one can demonstrate that $p$ is indeed constant by using the BPS equations and
the constraints (\ref{scuatro}) and (\ref{ssiete}).

%*************************************************************

\setcounter{equation}{0}
\section{The Hitchin system}
\label{hitchsec}

A simple counting argument can be used to determine the  number of
independent functions left out by the constraints. Indeed, we have already
mentioned that our ansatz depends on nine functions. However, we have found
two constraints in eq. (\ref{scuatro}) and one extra condition which fixes
$\partial_r \alpha$ in eq. (\ref{ssiete}). It is thus natural to think that
the number of independent functions is six and, thus, in principle, one should
be able to express the metric and the BPS equations in terms of them. By
looking at the complicated form of the first-order equations and  constraints
one could be tempted to think that this is a hopeless task. However, we will
show that this is not the case and that there exists a set of variables,
which are precisely those introduced by Hitchin in ref. \cite{hit}, in which 
the BPS equations drastically
simplify. These equations involve the constants $p$ and $q$ just discussed,
together with the components of the calibrating three-form $\Phi$. Actually,
following refs. \cite{hit,bran,Chong}, we shall parametrize $\Phi$ as:
\beq
\Phi\,=\,e^{7}\,\wedge\,\omega_{(2)}\,+\,\rho_{(3)}\,\,,
\label{setenta}
\eeq
where the two-form $\omega_{(2)}$ is given in terms of three functions
$y_i$ as:
\beq
\omega_{(2)}\,=\,\sqrt{{y_2y_3\over y_1}}\,\bo^1\wedge \tilde{w}^1\,+\,
\sqrt{{y_3y_1\over y_2}}\,\bo^2\wedge \tilde{w}^2\,+\,
\sqrt{{y_1y_2\over y_3}}\,\bo^3\wedge \tilde{w}^3\,,
\label{stuno}
\eeq
and $\rho_{(3)}$ is a three-form which depends on another set of three
functions
$x_i$, namely:
\bear
\rho_{(3)}\,&=&\,p\,\bo^1\wedge \bo^2\wedge \bo^3\,+
\,q\, \tilde{w}^1\wedge  \tilde{w}^2\wedge  \tilde{w}^3\,+\rc\rc
&&+\,x_1\,\big(\, \bo^1\wedge  \tilde{w}^2\wedge  \tilde{w}^3\,-\,
\bo^2\wedge  \bo^3\wedge  \tilde{w}^1\,\big)\,+\,{\rm cyclic}\,\,.
\label{stdos}
\eear
Notice that the terms appearing in $\omega_{(2)}$ are precisely those
which follow from our expression (\ref{cuno}) for $\Phi$. Moreover, by
plugging on the right-hand side of eq. (\ref{cuno}) the relation
(\ref{socho}) between the one-forms
$e^A$ and the $SU(2)$ left invariant forms, one can find the explicit
relation between the new and old variables, namely:
\bear
y_1&=&e^{{2\phi\over 3}+h_2+h_3-\lambda_1}\,\,,\rc\rc
x_1&=&-2\,\big[\,e^{\phi+h_1-\lambda_1}\,\cos\alpha\,+\,
4\,e^{2\phi}\,G_1\sin\alpha\,\big]\,\,,
\label{sttres}
\eear
and cyclically in $(1,2,3)$. Notice that the coefficients of
$\bo^1\wedge  \tilde{w}^2\wedge  \tilde{w}^3$ and of
$-\bo^2\wedge  \bo^3\wedge  \tilde{w}^1$ in the expression
(\ref{stdos}) of $\rho_{(3)}$ must be necessarily equal if $\Phi$ is
closed. Actually,
by computing the latter in our formalism, we get an alternative expression for
the $x_i$'s. This other expression  is:
\bear
x_1&=&\,2\,\big[\,e^{h_3-\lambda_3}\,G_2\,+\,e^{h_2-\lambda_2}\,G_3\,\big]\,
e^{\phi}\,\cos\alpha\,+\,\rc\rc
&&+\,\big[\,8\,e^{2\phi}\,G_2\,G_3\,-\,{1\over
2}\,e^{\lambda_1+h_2+h_3}\,\big]\,
\sin\alpha\,\,,
\label{stcuatro}
\eear
and cyclically in $(1,2,3)$. As a matter of fact, these 
two alternative expressions for the $x_i$'s are equal as a consequence of
the constraints  (\ref{suno}). In fact, we can regard eqs. (\ref{suno})
and (\ref{ssiete}) as conditions needed to ensure the
closure of $\Phi$. On the other hand, by using, at our convenience, eqs.
(\ref{sttres}) and (\ref{stcuatro}), one can prove the  following useful
relations:
\bear
&&{x_2x_3-px_1\over y_1}\,=\,{1\over 4}\,e^{2h_1-{2\phi\over 3}}\,+\,
4\,e^{{4\phi\over 3}+2\lambda_1}\,\,G_1^2\,\,,\rc\rc
&&{x_1^2-x_2^2-x_3^2-pq\over y_1}\,=\,
8\,e^{{4\phi\over 3}+2\lambda_1}\,\,G_1\,\,,\rc\rc
&&{x_2x_3+qx_1\over y_1}\,=\,4\,e^{{4\phi\over 3}+2\lambda_1}\,\,,
\label{stcinco}
\eear
and cyclically in $(1,2,3)$. As a first application of eq. (\ref{stcinco}),
let us point out that, making use of this equation, one can easily invert the
relation (\ref{sttres}). The result is:
\bear
e^{2\phi}\,&=&\,{1\over 8}\,\,
{(qx_1+x_2x_3)^{1/2}(qx_2+x_1x_3)^{1/2}(qx_3+x_1x_2)^{1/2}
\over \sqrt{y_1y_2y_3}}\,\,,\rc\rc\rc
e^{2\lambda_1}\,&=&\,{(y_2y_3)^{1/3}\over (y_1)^{2/3}}\,\,
{(qx_1+x_2x_3)^{2/3}
\over (qx_2+x_1x_3)^{1/3}(qx_3+x_1x_2)^{1/3}}\,\,,\rc\rc\rc
e^{2h_1}\,&=&\,2\,\,{(y_2y_3)^{5/6}\over (y_1)^{1/6}}\,\,
{(qx_2+x_1x_3)^{1/6}(qx_3+x_1x_2)^{1/6}
\over (qx_1+x_2x_3)^{5/6}}\,\,,\rc\rc\rc
G_1&=&{1\over 2}\,\,{x_1^2-x_2^2-x_3^2-pq\over
qx_1+x_2x_3}\,\,,
\label{stseis}
\eear
and cyclically in $(1,2,3)$.
Moreover, in order to make  contact with the formalism of refs. \cite{hit,
Chong},  let us define now the following ``potential":
\bear
U&\equiv&p^2q^2\,+\,2pq\,(\,x_1^2\,+\,x_2^2\,+\,x_3^2\,)\,+\,
4(p-q)\,x_1x_2x_3\,+\,\rc\rc
&&+\,x_1^4\,+\,x_2^4\,+\,x_3^4\,-\,2x_1^2x_2^2\,
-\,2x_2^2x_3^2\,-\,2x_3^2x_1^2\,\,.
\label{stsiete}
\eear
A straightforward calculation shows that $U$ can be rewritten as:
\bear
U&=&{1\over 3}\,(\,x_1^2\,-\,x_2^2\,-\,x_3^2\,-\,pq\,)^2\,-\,
{4\over 3}\,(\,x_2x_3\,+\,qx_1\,)\,(\,x_2x_3\,-\,px_1\,)\,+\,\rc\rc
&&+\,{\rm cyclic \,\,\,permutations}\,\,.
\label{stocho}
\eear
By using (\ref{stcinco}) to evaluate the right-hand side of eq.
(\ref{stocho}),
together with the definition of the $y_i$'s written in eq. (\ref{sttres}), one
easily verifies that $U$ is given by:
\beq
U\,=\,-4y_1y_2y_3\,\,.
\label{stnueve}
\eeq
It is important to stress the fact that in the general Hitchin formalism the
relation (\ref{stnueve}) is a constraint, whereas here this equation is just
an identity which follows from the definitions of $p$, $q$, $x_i$ and $y_i$.
Another
important consequence of the identities (\ref{stcinco}) is the form of the
metric in the new variables. Indeed, it is immediate from eqs.
(\ref{socho}) and (\ref{stcinco}) to see that the seven dimensional metric
$ds^2_{7}$ takes the form:
\bear
&&ds^2_{7}=dt^{2}\,+\,\rc\rc
&&+\,{1\over y_1}\,\Big[\,\big(\,x_2x_3-px_1\,\big)\,(\bo^1)^2\,+\,
\big(\,x_1^2-x_2^2-x_3^2-pq\,\big)\,\bo^1 \tilde{w}^1\,+\,
\big(\,x_2x_3+qx_1\,\big)\,( \tilde{w}^1)^2\,\Big]\,+\rc\rc
&&+\,{1\over y_2}\,\Big[\,\big(\,x_3x_1-px_2\,\big)\,(\bo^2)^2\,+\,
\big(\,x_2^2-x_3^2-x_1^2-pq\,\big)\,\bo^2 \tilde{w}^2\,+\,
\big(\,x_3x_1+qx_2\,\big)\,(\tilde{w}^2)^2\,\Big]\,+\rc\rc
&&+\,{1\over y_3}\,\Big[\,\big(\,x_1x_2-px_3\,\big)\,(\bo^3)^2\,+\,
\big(\,x_3^2-x_1^2-x_2^2-pq\,\big)\,\bo^3 \tilde{w}^3\,+\,
\big(\,x_1x_2+qx_3\,\big)\,(\tilde{w}^3)^2\,\Big]\,\,,\rc
\label{ochenta}
\eear
where $dt^2\,=\,e^{-2\phi/3}\,dr^2$.

It remains to determine the first-order system of differential equations
satisfied by the new variables. First of all, recall that, in the old
variables,  the BPS equations depend on the phase $\alpha$. Actually, from the
expression of $q$ (eq. (\ref{snueve})), and the first equation in
(\ref{stseis}), one can easily determine $\sin\alpha$, whereas $\cos\alpha$
can be obtained from the second equation in (\ref{sttres}). The result is:
\bear
\sin\alpha&=&-q\,\,
{\sqrt{y_1y_2y_2}\over
(qx_1+x_2x_3)^{1/2}(qx_2+x_1x_3)^{1/2}(qx_3+x_1x_2)^{1/2}}\,\,\,,\rc\rc\rc
\cos\alpha&=&-\,\,
{2x_1x_2x_3\,+\,q\,(x_1^2+x_2^2+x_3^2\,)\,+\,pq^2\over
2(qx_1+x_2x_3)^{1/2}(qx_2+x_1x_3)^{1/2}(qx_3+x_1x_2)^{1/2}}\,\,.
\label{ouno}
\eear
As a check of eq. (\ref{ouno}) one can easily verify that
$\sin^2\alpha+\cos^2\alpha=1$ as a consequence of the relation
(\ref{stnueve}).
It is now straightforward to compute the derivatives of $x_i$ and $y_i$.
Indeed, one can differentiate eq. (\ref{sttres}) and use eqs.
(\ref{cicinco}), (\ref{ciseis}), (\ref{cinueve}), (\ref{scinco}) and
(\ref{ssiete}) to evaluate the result in the old variables. This result can be
converted back to the new variables by means of eqs. (\ref{stseis}) and
(\ref{ouno}). The final result of these calculations is remarkably simple,
namely:
\bear
\dot x_1&=&-\sqrt{{y_2y_3\over y_1}}\,\,,\rc\rc
\dot y_1&=&{pqx_1\,+\,(p-q)x_2x_3\,+\,x_1(x_1^2-x_2^2-x_3^2)
\over \sqrt{y_1y_2y_3}}\,\,,
\label{odos}
\eear
and cyclically in $(1,2,3)$. In eq. (\ref{odos}) the dot denotes derivative
with respect to the variable $t$ defined after eq. (\ref{ochenta}). The
first-order system (\ref{odos}) is, with our notations, the one derived in
refs. \cite{hit, Chong}. Indeed, one can show that the equations satisfied
by the $x_i$'s are a consequence of the condition $d\Phi=0$, whereas, if
the seven dimensional Hodge
dual is computed with the metric (\ref{ochenta}), then $d*_7\Phi=0$ implies
the
first-order equations for the $y_i$'s. Therefore, we have shown that eight
dimensional gauged supergravity provides an explicit realization of the
Hitchin
formalism for general values of the constants $p$ and $q$. Notice that a
non-zero phase $\alpha$ is needed in order to get a system with $q\not=0$.
Recall (see eq. (\ref{ttres})) that the phase $\alpha$ parametrizes the
tilting of the three-cycle on which the D6-brane is wrapped with respect
to the three sphere of the eight dimensional metric. Notice that the
analysis of \cite{hs1} corresponds to the case $q = \alpha = 0$.

Let us finally point out that the first-order equations  (\ref{odos})  are
invariant if we change the constants $(p,q)$ by $(-q,-p)$. In the metric
(\ref{ochenta}) this change is equivalent to the exchange of $\bo^i$ and
$\tilde{w}^i$, \ie\ of the two $S^3$ of the principal orbits of the
cohomogeneity
one metric (\ref{ochenta}). As mentioned above, this is the so-called flop
transformation. Thus, we have proved that:
\beq
\bo^i\leftrightarrow \tilde{w}^i\,\,\,\,
\Longleftrightarrow\,\,\,\,(p,q)\,\leftrightarrow (-q,-p)\,\,.
\label{otres}
\eeq
Notice that the three-form $\Phi$ given in eqs. 
(\ref{setenta})-(\ref{stdos})
changes its sign when both $(\bo^i, \tilde{w}^i)$ and $(p,q)$ are
transformed as in eq. (\ref{otres}).

%*************************************************************

\setcounter{equation}{0}
\section{Some particular cases}

With the kind of ansatz we are adopting for the eight-dimensional
solutions, the
corresponding  eleven dimensional metrics are of  the type:
\beq
ds^2_{11}\,=\,dx^2_{1,3}\,+\,B_i^2\,\,
(\,\bo^i\,)^2\,+\,D_i^2\,
(\,\tilde w^i\,+\,G_i\,\bo^i\,)^2\,+\,dt^2\,\,,
\label{ocuatro}
\eeq
where the coefficients $B_i$, $D_i$ and the variable $t$ are related to eight
dimensional quantities as follows:
\beq
B_i^2\,=\,{1\over 4}\,e^{2h_i\,-\,{2\phi\over 3}}\,\,,
\,\,\,\,\,\,\,\,\,\,\,\,\,\,\,\,
D_i^2\,=\,4\,e^{{4\phi\over 3}+2\lambda_i}\,\,,
\,\,\,\,\,\,\,\,\,\,\,\,\,\,\,\,
dt^2\,=\,e^{-{2\phi\over 3}}\,dr^2\,\,.
\label{ocinco}
\eeq
Moreover, we have found that, for a supersymmetric solution, the nine
functions appearing in the metric are not independent but rather they are
related by some algebraic constraints which are, in general, quite
complicated. Notice that, in this case, the gauged supergravity approach
forces the six function ansatz, this possibly clarifying the reasons
behind this {\it a priori} requirement in previous cases in the
literature. To illustrate
this point, let us write eq. (\ref{scuatro}) in terms of $B_i$, $D_i$ and
$G_i$. One gets:
\bear
&&\big[\,B_1\,D_2^2\,D_1\,G_2\,-\,(\,1\leftrightarrow 2\,)\,\big]\,D_3^2\,
(\,1\,-\,G_3^2\,)\,=\,\rc\rc
&&=\,B_3\,D_2\,\big[\,B_1\,B_3\,D_1\,D_2\,G_2\,+\,
D_1^2\,D_2\,D_3\,G_1^2\,+\,B_1^2\,D_2\,D_3\,\big]\,-\,
(\,1\leftrightarrow 2\,)\,\,,
\label{oseis}
\eear
and cyclically in $(1,2,3)$. In addition, we must ensure that eq.
(\ref{ssiete}) is also satisfied. Despite the terrifying aspect of eq.
(\ref{oseis}), it is not hard to find expressions for, say, $G_2$ and
$G_3$ in terms of the remaining functions. Moreover, we will be able to
find some particular solutions, which
correspond to the different cohomogeneity one metrics with $S^3\times S^3$
principal orbits and $SU(2)$ $\times$ $SU(2)$ isometry which have been
studied in the literature.

%*************************************************************

\subsection{The $q=0$ solution}

The simplest way of solving the constraints imposed by supersymmetry is by
taking $q=0$, which leads to the case studied in \cite{hs1}. A glance at
the second equation in (\ref{snueve}) reveals that in this case
$\sin\alpha=0$ and, thus,
$\beta=1$,
$\tilde\beta=0$. Notice, first of all, that this is a consistent solution
of eq. (\ref{ssiete}). Moreover, it follows from eq. (\ref{suno}) that
one must have:
\beq
{\cal A}_i\,=\,0\,\,.
\label{osiete}
\eeq
By combining the three conditions (\ref{osiete}) it  is easy to find the
values of the gauge field components $G_i$ in terms of the other
functions $B_i$ and
$D_i$ \cite{hs1}. One gets:
\beq
G_1\,=\,{1\over 2}\,\,{D_2D_3\over B_2B_3}\,\,
\Big[\,\,\Big({B_1\over D_1}\Big)^2\,-\,\Big({B_2\over D_2}\Big)^2\,-\,
\Big({B_3\over D_3}\Big)^2\,\,\Big]\,\,,
\label{oocho}
\eeq
and cyclically in $(1,2,3)$, which is precisely the result of \cite{hs1}.
This is the solution of the constraints we were
looking for. One can check that, assuming that the $G_i$'s are given by  eq.
(\ref{oocho}), then eq.  (\ref{scinco}) for $G_i'$  is satisfied if eqs.
(\ref{cicinco}), (\ref{ciseis}) and (\ref{cinueve})  hold. Thus, eq.
(\ref{oocho}) certainly gives a consistent truncation of the first-order
differential equations. On the other hand, by using the value of the $G_i$'s
given in eq.  (\ref{oocho}), one can eliminate them and obtain a system of
first-order equations for the remaining functions $B_i$ and $D_i$. These
equations are:
\bear
\dot B_1&=&-{D_2\over 2B_3}\,(\,G_2\,+\,G_1G_3\,)\,-\,
{D_3\over 2B_2}\,(\,G_3\,+\,G_1G_2)\,\,,\rc\rc
\dot D_1&=&{D_1^2\over 2B_2B_3}\,(\,G_1\,+\,G_2G_3\,)\,+\,
{1\over 2D_2D_3}\,(\,D_2^2\,+\,D_3^2\,-\,D_1^2\,)\,\,,
\label{onueve}
\eear
together with the other permutations of the indices $(1,2,3)$. In
(\ref{onueve}) the $G_i$'s are  the functions of $B_i$ and $D_i$
displayed in
eq. (\ref{oocho}). The constant $p$ can be immediately obtained from
(\ref{snueve}), namely:
\beq
p=B_1B_2B_3\,-\,B_1D_2D_3G_2G_3\,-\,B_2D_1D_3G_1G_3\,-\,B_3D_1D_2G_1G_2\,\,.
\label{noventa}
\eeq

Let us now give the Hitchin variables in this case. By taking $\alpha=0$ on
the right-hand side of (\ref{sttres}) and using the relation (\ref{ocinco}),
one readily arrives at:
\beq
x_1\,=\,-B_1D_2D_3\,\,,
\,\,\,\,\,\,\,\,\,\,\,\,\,\,\,\,
y_1\,=\,B_2B_3D_2D_3\,\,.
\label{nuno}
\eeq
The values of the other $x_i$ and $y_i$ are obtained by cyclic permutation. As
a verification of these expressions, it is not difficult to demonstrate, by
using eq. (\ref{onueve}), that the functions $x_i$ and $y_i$ of
eq. (\ref{nuno}) satisfy the first-order equations (\ref{odos}) for $q=0$.
Finally, let us point out that, by means of a flop transformation, one
can pass from the $q=0$ metric
described above to a metric with $p=0$.

%*************************************************************

\subsection{The flop invariant solution}

It is also possible  to solve our constraints by requiring that the metric be
invariant under the $\ZZ_2$ flop transformation $\bo^i\leftrightarrow
 \tilde{w}^i$. It follows from eq. (\ref{otres}) that, in this case, we
must necessarily have $p=-q$. Moreover, it is also clear that the forms
$\bo^i$ and $\tilde{w}^i$ must enter the metric in the combinations
$(\,\bo^i-\tilde{w}^i\,)^2$ and $(\,\bo^i+\tilde{w}^i\,)^2$, which are the
only quadratic combinations which
are invariant under the flop transformation. Thus the metric we are seeking
must be of the type:
\beq
ds^2_{11}\,=\,dx^2_{1,3}\,+\,a_i^2\,(\,\bo^i\,-\, \tilde{w}^i\,)^2\,+\,
b_i^2\,(\,\bo^i\,+\, \tilde{w}^i\,)^2\,+\,dt^2\,\,,
\label{ndos}
\eeq
where $a_i$ and $b_i$ are functions which obey some system of first-order
differential equations to be determined. In general \cite{CGLP1}, a metric
of the type written in eq. (\ref{ocuatro}) can be put in the form
(\ref{ndos}) only if
$G_i$, $B_i$ and $D_i$ satisfy the following relation:
\beq
G_i^2\,=\,1\,-\,{B_i^2\over D_i^2}\,\,.
\label{ntres}
\eeq
It is easy to show that our constraints are solved for $G_i$ given as in eq.
(\ref{ntres}). Indeed, after some calculations, one can rewrite the constraint
(\ref{scuatro}) for $i=1$ and $j=2$  as:
\bear
&&\Big(\,1\,-\,{1\over 16}\,e^{-2\phi+2h_1-2\lambda_1}\,-\,G_1^2\,\Big)\,
\,e^{-2\lambda_3}\,-\,
\Big(\,1\,-\,{1\over 16}\,e^{-2\phi+2h_2-2\lambda_2}\,-\,G_2^2\,\Big)\,
\,e^{-2\lambda_3}\,+\,\rc\rc
&&+\,\Big(\,1\,-\,{1\over 16}\,e^{-2\phi+2h_3-2\lambda_3}\,-\,G_3^2\,\Big)\,
\Big[\,G_2\,e^{h_1-h_3+\lambda_2}\,-\,
G_1\,e^{h_2-h_3+\lambda_1}\,\Big]\,\,=0\,\,,
\label{ncuatro}
\eear
which is clearly solved for:
\beq
G_i^2\,=\,1\,-\,{1\over 16}\,e^{-2\phi+2h_i-2\lambda_i}\,\,.
\label{ncinco}
\eeq
Similarly, one can verify that eq. (\ref{ncinco}) also solves eq.
(\ref{scuatro}) for the remaining values of $i$ and $j$. After taking into
account the identifications (\ref{ocinco}), we easily conclude that the
solution (\ref{ncinco}) coincides with the condition  (\ref{ntres}) and,
thus, it corresponds to $\ZZ_2$ invariant metric of the type (\ref{ndos}).
Moreover, it
can be checked that the relation (\ref{ncinco}) gives a consistent
truncation of
the first-order differential equations found in section \ref{secgeng2} and
that eq. (\ref{ssiete}) is also satisfied. On the other hand, the
identification  of the $a_i$ and $b_i$ functions with the ones
corresponding to 8d gauged supergravity is easily established by
comparing the uplifted metric with (\ref{ndos}), namely:
\bear
dr&=&e^{{\phi\over 3}}\,dt\,\,,\rc\rc
e^{2h_i-{2\phi\over 3}}&=&16\,{a_i^2\,b_i^2\over
a_i^2\,+\,b_i^2}\,\,,\rc\rc
e^{{4\phi\over 3}+2\lambda_i}&=&{1\over 4}\,(\,a_i^2\,+\,b_i^2\,)\,\,.
\label{nseis}
\eear
This relation allows to obtain $\phi$, $\lambda_i$ and $h_i$ in terms of
$a_i$ and $b_i$:
\bear
e^{2\phi}\,&=&\,{1\over 8}\,\,\prod_i\,
(\,a_i^2\,+\,b_i^2)^{1\over 2}\,\,,\rc\rc
e^{2\lambda_i}\,&=&\,{a_i^2\,+\,b_i^2\over
\prod_j\,(\,a_j^2\,+\,b_j^2)^{1\over 3}}\,\,,\rc\rc
e^{2h_i}\,&=&\,8\,{a_i^2\,b_i^2\over a_i^2\,+\,b_i^2}\,\,
\prod_j\,(\,a_j^2\,+\,b_j^2)^{1\over 6}\,\,,
\label{nsiete}
\eear
while $G_i$ in terms of the $a_i$ and $b_i$ is given by:
\beq
G_i\,=\,{b_i^2-a_i^2\over  b_i^2+a_i^2}\,\,.
\label{nocho}
\eeq
The inverse relation is also useful:
\beq
a_i^2\,=\,2\,e^{{4\phi\over 3}\,+\,2\lambda_i}\,\,
(\,1\,-\,G_i\,)\,\,,
\,\,\,\,\,\,\,\,\,\,\,\,\,
b_i^2\,=\,2\,e^{{4\phi\over 3}\,+\,2\lambda_i}\,\,
(\,1\,+\,G_i\,)\,\,,
\label{nnueve}
\eeq
where $G_i$ is the function of $\phi$, $h_i$ and $\lambda_i$ written in eq.
(\ref{ncinco}). By using eqs. (\ref{nsiete}) and (\ref{nocho}) one can obtain
the values of $\cos\alpha$ and $\sin\alpha$ for this case. One gets:
\bear
\cos\alpha&=&
{\,b_1a_2a_3\,+\,a_1b_2a_3\,+\,a_1a_2b_3\,-\,b_1b_2b_3\,\over
\sqrt{(\,a_1^2\,+\,b_1^2\,)\,(\,a_2^2\,+\,b_2^2)\,(\,a_3^2\,+\,b_3^2\,)}}
\,\,,\rc\rc
\sin\alpha&=&
{\,a_1b_2b_3\,+\,b_1a_2b_3\,+\,b_1b_2a_3\,-\,a_1a_2a_3\over
\sqrt{(\,a_1^2\,+\,b_1^2\,)\,(\,a_2^2\,+\,b_2^2)\,(\,a_3^2\,+\,b_3^2\,)}}
\,\,.
\label{cien}
\eear
Moreover, by differentiating eq. (\ref{nnueve}) and using the first-order
equations of section \ref{secgeng2}, together with eqs. (\ref{nsiete})
and (\ref{cien}), one can find the BPS equations in the $a_i$ and $b_i$
variables. They are:
\bear
\dot a_1&=&
-\,{a_1^2\over 4a_2b_3}\,-\,{a_1^2\over 4a_3b_2}\,+\,
{a_2\over 4b_3}\,+\,{b_2\over 4a_3}\,+\,
{a_3\over 4b_2}\,+\,{b_3\over 4a_2}\,\,,\rc\rc
\dot b_1&=&
-\,{b_1^2\over 4a_2a_3}\,+\,{b_1^2\over 4b_2b_3}\,-\,
{b_2\over 4b_3}\,+\,{a_2\over 4a_3}\,-\,
{b_3\over 4b_2}\,+\,{a_3\over 4a_2}\,\,,
\label{ctuno}
\eear
and cyclically for the other $a_i$'s and $b_i$'s. These are precisely the
equations found in ref. \cite{bggg}  for this type of metrics, where it
was proved that some solutions of this system yield ALC metrics. Moreover,
it is now straightforward to compute the constants $p$ and $q$ in this
case. Indeed, by substituting eqs. (\ref{nsiete}), (\ref{nocho}) and
(\ref{cien}) on the right-hand side of eq. (\ref{snueve}), one easily
proves that:
\beq
p\,=\,-q\,=\,\,a_1b_2b_3\,+\,b_1a_2b_3\,+\,b_1b_2a_3\,-\,a_1a_2a_3\,\,.
\label{ctdos}
\eeq
Similarly, from eq. (\ref{sttres}) one can find
the Hitchin variables in terms of the $a_i$'s and $b_i$'s. The result for
$x_1$ and $y_1$ is:
\bear
x_1&=&a_1b_2b_3\,-\,b_1a_2b_3\,-\,b_1b_2a_3\,-\,a_1a_2a_3\,\,,\rc\rc
y_1&=&4a_2a_3b_2b_3\,\,,
\label{cttres}
\eear
while the expressions of $x_2$, $x_3$, $y_2$ and $y_3$ are obtained from
(\ref{cttres}) by cyclic permutations.

%*************************************************************

\subsection{The conifold unification metrics}

There exists a class of $G_2$ metrics with $S^3\times S^3$ principal
orbits which have an extra $U(1)$ isometry and generic values of $p$ and
$q$. They are the
so-called conifold--unification metrics and they were introduced in
ref. \cite{CGLP2} as a
unification, via M--theory, of the deformed and resolved 
conifolds \footnote{Notice the difference between the
kind of conifold unification described in
chapter \ref{conichapter} and the one presented here. In the former, the
approach leads to a system of equations and several solutions of it are
the distinct conifold metrics in eleven dimensions. In the latter, the 
non-trivial part of the eleven dimensional
metric describes a $G_2$ holonomy manifold and different $U(1)$
compactifications of it lead to resolved and deformed 
conifold-like metrics in ten dimensions.}. 
Following
ref. \cite{CGLP2}, let us parametrize them as:
\bear
ds^2_7&=&a^2\,\big[\,\big(\,\tilde{w}^1\,+\,
{\cal G}\,\bo^1\,\big)^2\,+\,
\big(\,\tilde{w}^2\,+\,
{\cal G}\,\bo^2\,\big)^2\,\big]\,+\,
b^2\,\big[\,\big(\,\tilde{w}^1\,-\,
{\cal G}\,\bo^1\,\big)^2\,+\,
\big(\,\tilde{w}^2\,-\,
{\cal G}\,\bo^2\,\big)^2\,\big]\,+\rc\rc
&&+\,c^2\,\big(\,\tilde{w}^3\,-\,\bo^3\,\big)^2\,+\,
f^2\,\big(\,\tilde{w}^3\,+\,{\cal G}_3\,\bo^3\,\big)^2\,+\,dt^2\,\,.
\label{ctcuatro}
\eear
It is clear that, in order to obtain in our eight-dimensional supergravity
approach a metric such as the one written in eq. (\ref{ctcuatro}), one must
take  $h_1=h_2$, $\lambda_1=\lambda_2\,=\,-\lambda_3/2\,=\,\lambda$ and
$G_1=G_2$ in our general formalism. Then, it is an easy exercise to find the
gauged supergravity variables in terms of the functions appearing in the
ansatz (\ref{ctcuatro}). One has:
\bear
&&e^{\phi}\,=\,{1\over 2\sqrt{2}}\,\big(\,a^2\,+\,b^2\,\big)^{{1\over 2}}\,\,
\big(\,f^2\,+\,c^2\,\big)^{{1\over 4}}\,\,,\rc\rc
&&e^{\lambda}\,=\,\big(\,a^2\,+\,b^2\,\big)^{{1\over 6}}\,\,
\big(\,f^2\,+\,c^2\,\big)^{-{1\over 6}}\,\,,\rc\rc
&&e^{h_1}\,=\,2\sqrt{2}\,a\,b\,{\cal G}\,
\big(\,a^2\,+\,b^2\,\big)^{-{1\over 3}}\,\,
\big(\,f^2\,+\,c^2\,\big)^{{1\over 12}}\,\,,\rc\rc
&&e^{h_3}\,=\,\sqrt{2}\,f\,c\,(1+{\cal G}_3)\,
\big(\,a^2\,+\,b^2\,\big)^{{1\over 6}}\,\,
\big(\,f^2\,+\,c^2\,\big)^{-{5\over 12}}\,\,,\rc\rc
&&G_1\,=\,{\cal G}\,\,{a^2-b^2\over a^2+b^2}\,\,,\rc\rc
&&G_3\,=\,\,{{\cal G}_3\,\,f^2-c^2\over f^2+c^2}\,\,.
\label{ctcinco}
\eear
With the parametrization given above, it is not difficult to solve the
constraints (\ref{scuatro}). Actually, one of these constraints is trivial,
while the other allows to obtain ${\cal G}_3$ in terms of the other variables,
namely:
\beq
{\cal G}_3\,=\,{\cal G}^2\,+\,{c\,(\,a^2-b^2\,)(\,1-{\cal G}^2\,)
\over 2abf}\,\,.
\label{ctseis}
\eeq
The relation (\ref{ctseis}), with $a\rightarrow -a$, is precisely the one
obtained in ref. \cite{CGLP2}. One can also prove that eq. (\ref{ctseis})
solves
eq.  (\ref{ssiete}). Actually, the phase $\alpha$ in this case is:
\beq
\cos\alpha\,=\,{2abc\,+\,(b^2-a^2)\,f\over
(\,a^2+b^2\,)\sqrt{c^2+f^2}}\,\,,
\,\,\,\,\,\,\,\,\,\,\,\,\,\,\,\,\,\,
\sin\alpha\,=\,{2abf\,+\,(a^2-b^2)\,c\over
(\,a^2+b^2\,)\sqrt{c^2+f^2}}\,\,.
\label{ctsiete}
\eeq

With all these ingredients  it is now straightforward, although tedious, to
find the first-order equations for the five independent functions of the
ansatz
(\ref{ctcuatro}). The result coincides again with the one written in
ref.\cite{CGLP2}, after changing $a\rightarrow -a$, and is given by:
\bear
\dot a&=&{c^2\,(\,b^2-a^2\,)\,+\,[\,4a^2\,(\,b^2-a^2\,)\,+\,
c^2\,(\,5a^2-b^2\,)\,-\,4abcf\,]\,{\cal G}^2\over
16a^2\,bc\,{\cal G}^2}\,\,,\rc\rc
\dot b&=&{c^2\,(\,a^2-b^2\,)\,+\,[\,4b^2\,(\,a^2-b^2\,)\,+\,
c^2\,(\,5b^2-a^2\,)\,+\,4abcf\,]\,{\cal G}^2\over
16ab^2\,c\,{\cal G}^2}\,\,,\rc\rc
\dot c&=&-{c^2\,+\,(\,c^2\,-\,2a^2\,-\,2b^2\,)\,{\cal G}^2\over
4ab{\cal G}^2}\,\,,\rc\rc
\dot f&=&-{(\,a^2-b^2\,)\,\big[\,4abf^2\,{\cal G}^2\,+\,
c\,(\,a^2\,f\,-\,b^2\,f\,-\,4abc\,)\,(\,1-{\cal G}^2\,)\,\big]\over
16a^3\,b^3\,{\cal G}^2}\,\,,\rc\rc
\dot {\cal G}&=&{c\,(\,1-{\cal G}^2\,)\over 4ab{\cal G}}\,\,.
\label{ctocho}
\eear
Furthermore, the constants $p$ and $q$ are also easily obtained, with the
result:
\bear
p&=&(\,a^2\,-\,b^2\,)\,c\,{\cal G}^2\,+\,
2abf{\cal G}_3\,{\cal G}^2\,\,,\rc\rc
q&=&(\,b^2\,-\,a^2\,)\,c\,-\,2abf\,\,,
\label{ctnueve}
\eear
while the Hitchin variables are:
\bear
&&x_1\,=\,x_2\,=\,-(\,a^2+b^2\,)\,c\,{\cal G}\,\,,
\,\,\,\,\,\,\,\,\,\,\,\,\,\,\,\,\,\,
x_3\,=\,(\,a^2-b^2\,)\,c\,-2abf{\cal G}_3\,\,,\rc\rc
&&y_1\,=\,y_2\,=\,2abcf{\cal G}(\,1\,+\,{\cal G}_3\,)\,\,,
\,\,\,\,\,\,\,\,\,\,\,\,\,\,\,\,\,\,
y_3\,=\,4a^2b^2{\cal G}^2\,\,.
\label{ctdiez}
\eear
Eqs. (\ref{ctnueve}) and (\ref{ctdiez}) are again in agreement with
those given in ref. \cite{CGLP2}, after changing $a$ by $-a$ as before.

%*************************************************************
\setcounter{equation}{0}
\section{Discussion}

In this chapter, a large set of Ricci-flat seven dimensional metrics
with $G_2$ holonomy have been studied, using
 eight dimensional gauged supergravity to obtain eleven dimensional
solutions.
Concretely, all the cohomogeneity one metrics with $S^3\times S^3$
principal orbits are obtained. This is proved by making contact
with the Hitchin formalism (see section \ref{hitchsec}), which
was originally formulated from a completely different perspective.

Indeed, eight dimensional gauged supergravity proves itself
very useful in the computation of such metrics. There is a 
unique, although quite involved, system of BPS equations
(those written is section \ref{secgeng2}), which include first order
differential equations along with algebraic constraints, yielding all the
solutions. Moreover, the Killing spinors and the calibrating three-form
have been computed. As in chapter
\ref{conichapter}, the main technical trick to obtain the most general
set of  solutions is a rotation on the Killing spinor (eqs.
(\ref{catorce}), (\ref{vcinco})). It amounts to the introduction of a
phase
$\alpha$ in the radial projection of the Killing spinor and,
correspondingly, the
twist is implemented by a non-abelian gauge field which is not  fixed
{\it a priori}, but determined
by a first-order differential equation. This gauge field encodes the 
non-trivial fibering of the two three-spheres in the special holonomy
manifold, while the corresponding radial projection determines the
wrapping of the D6-brane in the supersymmetric three-cycle. Actually we
have seen that, for non-zero $\alpha$, the three-cycle on which the
D6-brane is wrapped has components along the two
$S^3$'s (see eq. (\ref{ttres})).

Some particular $G_2$ holonomy metrics have a particular physical
interest. The asymptotically locally conical (ALC) stand among them.
As explained above, they have an $S^1$ that does not grow at large
$r$, so a compactification can be made there without getting a 
growing string coupling constant. We have seen how gauged sugra
can describe such solutions. This  seems to be at odds with the
fact that gauged supergravity is unable to describe the Taub-NUT
metric. This metric is the uplift of the full D6-brane solution, while,
in principle, 8d gauged sugra can only account for the near horizon
physics of the D6-branes. It would be interesting to study whether
there is some way out of this limitation and the full solution can
be obtained in lower dimensional sugra.

%*************************************************************

\chapter{Adding fluxes in 8d gauged supergravity}
\label{fluxes}

%*************************************************************
\medskip

The low energy dynamics of a collection of D--branes  wrapping
supersymmetric cycles is governed, when the size of the cycle is taken to
zero, by a lower dimensional supersymmetric gauge theory with less than
sixteen supercharges. The gravitational description of  these gauge
theories allows for a geometrical approach to the study of  important
aspects of their dynamics. We have seen in chapters \ref{conichapter}
and \ref{g2chapter} that eight dimensional gauged supergravity is
useful in finding the metrics of non-trivial geometries of
reduced supersymmetry and holonomy. The goal of this chapter is to
show how it is possible to generalize some of those supergravity
solutions by the addition of fluxes, and how the same gauged supergravity
is an appropriate framework for such a task \cite{epr}.
We will also try to give an interpretation from the point of view of the
associated brane configurations and dual gauge theories.

Gauged supergravities have several forms coming from the dimensional
reduction of the highest dimensional supergravities \cite{ssch}. Turning
them on amounts to the introduction of other branes into the system in the
form of either localized or smeared intersections and overlappings. Many
of these configurations correspond to extremely interesting supersymmetric
gauge theories. In particular, these configurations give rise to a
world--volume dynamics whose description, at different energy scales, is
given by increasingly richer phases connected by RG flows. See, for 
example, \cite{gns,ps}.

Concretely, we will study the effect of turning on 4-form
fluxes in the non-compact directions of the solutions of the previous
chapters. When uplifted, this amounts to the
addition of a 4-form field strength of eleven dimensional supergravity, 
\ie \ to
the presence of a M2-brane charge, which halves the number of
supercharges. Then, by reducing to ten dimensions and performing
T-dualities, it is possible to find several associated solutions with
different kinds of D-branes which are useful for the construction of
gravity duals.

The structure of the chapter is the following:
In section \ref{general4f}, we show a general procedure to find the
new supergravity solutions \cite{epr,hs2}, that works under some general
assumptions that will always be fulfilled in the cases considered below.
The deformation of the background produced by the inclusion of a 4-form
amounts to the appearance of warp factors. Then, 
in section \ref{efflagr4} we explain how
to find an effective lagrangian with a given ansatz for the eight
dimensional fields when the 4-form $G$ is turned on. This involves a
subtle sign flip when constructing the Routhian after integrating out
the corresponding potential. Sections \ref{fluxS2} and \ref{fluxS3}
are devoted to the study of particular cases, corresponding to 
the addition of flux to solutions analyzed in chapters
\ref{conichapter} (D6-branes wrapping two-cycles) and \ref{g2chapter}
(D6-branes wrapping three-cycles) respectively. In both cases,  we derive
the BPS equations both through the vanishing condition for supersymmetry
transformations of the fermions as well as from the domain wall equations
resulting from the effective Routhian obtained by inserting the ansatz into
the 8d gauged supergravity Lagrangian. We obtain the general solution and
uplift it to eleven dimensions. Then, we find some solutions related
by duality and elaborate on the corresponding field theories.
Some of the solutions display the effect of supersymmetry 
without supersymmetry \cite{bko,dlp}.

Let us finally point out that it would be desirable to address
the problem of adding more general kinds of fluxes in gauged supergravity.
For instance, it would be interesting to look for solutions with
non-vanishing components of the 4-form along the compact directions of
the special holonomy background.

%*************************************************************
\setcounter{equation}{0}
\section{General procedure}

In this section, some  methods for the calculation
of this kind of solutions will be presented. First, in
\ref{general4f}, from a general formalism, it is
 shown how the effect of the
introduction of the 4-form  is the
introduction of  warp factors in the metric, which
distinguish between directions parallel and orthogonal to the
2-brane source of the 4-form. For the sake of clarity,
a simple example where the flux is added to flat space
in eight dimensions is developed. 

Then, in \ref{efflagr4}, the problem of finding an
effective lagrangian in order to use the superpotential
method for finding the first order systems is addressed.
A simple but important subtlety that must be taken
into account is explained.

%*************************************************************

\subsection{General dependence of the metric
on the 4-form}
\label{general4f}

Let us suppose that we adopt the following ansatz for the 
 eight-dimensional
metric:
\beq
ds_8^2\,=\,e^{2f}\,dx_{1,2}^2\,+\,\sum_{i=1}^{4}\,
e^{2h_i}\,\big(\,E^i\,)^2\,+\,dr^2\,\,,
\label{apcuno}
\eeq
where $E^i$ are some vierbiens, which we will assume to be independent of
the radial coordinate $r$. We want to add a 3-form potential
depending on $r$ and spanning the $x_0$, $x_1$, $x_2$ 
space-time directions. Therefore, the 4-form will be:
\beq
G_{\underline{x^0x^1x^2r}}\,=\,\Lambda\,
e^{-\sum h_i\,-\,2\phi}\,\equiv\,\Lambda\,e^{-\phi}\,
\xi(\,\phi, h_i\,)\,\,,
\label{gen4form}
\eeq
where the underlining is to remark the fact that the indices are
flat. $\Lambda$ is a constant and we have defined the
function $\xi(\,\phi, h_i\,)$. This ansatz for $G$ ensures
that the equation of motion for the 3-potential (which comes from
the lagrangian (\ref{apatres})):
\beq
D_{\mu}\,\Big(\, \sqrt{-g_{(8)}}\,\,e^{2\phi}\, 
G^{\mu\nu\tau\sigma}\,\big)\,=0\,\,,
\eeq
is satisfied. Notice that also the consistency condition
(\ref{consist}) is fulfilled, as long as there is no
gauge field strength in the $x$ directions.

All the dependence on $r$ is
included in the functions $f$, $h_i$ and $\phi$. We will assume that we
have also some scalar fields $\lambda_i$. In all the cases studied
below, these functions satisfy certain first-order BPS equations of the 
type:
\bear
{d\over dr}\,f&=&\Upsilon_{f}\,(\,\phi,h_i, \lambda_i)\,+\,
{\Lambda\over 2}\,\,\xi(\,\phi, h_i\,)\,\,,\rc\rc
{d\over dr}\,h_i&=&\Upsilon_{h_i}\,(\,\phi,h_i, \lambda_i)\,-\,
{\Lambda\over 2}\,\,\xi(\,\phi, h_i\,)\,\,,\rc\rc
{d\over dr}\,\phi&=&\Upsilon_{\phi}\,(\,\phi,h_i, \lambda_i)\,-\,
{\Lambda\over 2}\,\,\xi(\,\phi, h_i\,)\,\,,\rc\rc
{d\over dr}\,\lambda_i&=&\Upsilon_{\lambda_i}\,(\,\phi,h_i, \lambda_i)\,\,,
\label{apcdos}
\eear
where the functions $\Upsilon$ of the right-hand side depend on the particular
case we are considering. The only property we will need of these functions is
that they satisfy the following homogeneity condition:
\beq
\Upsilon(\,\phi\,+\,\gamma\,,\,h_i+\,\gamma\,, \lambda_i\,)\,=\,
e^{-\gamma}\,\Upsilon(\,\phi\,,\,h_i\,, \lambda_i\,)\,\,,
\label{apctres}
\eeq
where $\gamma$ is an arbitrary function. In all the cases studied here and in
refs. \cite{gns, hs2} the $\Upsilon$'s satisfy (\ref{apctres}). On the other
hand, from the definition of $\xi(\,\phi\,,\,h_i\,)$ one has:
\beq
\xi(\,\phi\,+\,\gamma\,,\,h_i+\,\gamma\,)\,=\,
e^{-5\gamma}\,\xi(\,\phi\,,\,h_i\,)\,\,.
\label{apccuatro}
\eeq
Let us now consider a function $\chi$ such that solves the following
differential equation:
\beq
{d\chi\over dr}\,=\,-{\Lambda\over 2}\,\,\xi(\,\phi, h_i\,)\,\,,
\label{apccinco}
\eeq
and let us define the functions:
\beq
\tilde f\,=\,f\,+\,\chi\,\,,
\,\,\,\,\,\,\,\,\,\,\,\,\,\,
\tilde h_i\,=\,h_i\,-\,\chi\,\,,
\,\,\,\,\,\,\,\,\,\,\,\,\,\,
\tilde \phi\,=\,\phi\,-\,\chi\,\,.
\label{apcseis}
\eeq
If we now introduce a new radial variable $\tilde r$ such that:
\beq
{d  r\over d\tilde r}\,=\,e^{\chi}\,\,,
\label{apcsiete}
\eeq
then,  it is straightforward to prove that $\tilde f\,$, $\tilde h_i\,$ and
$\tilde \phi$ and $\lambda$ satisfy the following differential equations:
\bear
{d\over d\tilde r}\,\tilde f&=&
\Upsilon_f\,(\,\tilde\phi,\tilde h_i, \lambda_i)\,\,,\rc\rc
{d\over d\tilde r}\,\tilde h_i&=&
\Upsilon_{h_i}\,(\,\tilde \phi,\tilde h_i,\lambda_i)\,\,,\rc\rc
{d\over d\tilde r}\,\tilde\phi&=&
\Upsilon_{\phi}\,(\,\tilde\phi,\tilde h_i, \lambda_i)\,\,,\rc\rc
{d\over d\tilde r}\,\lambda_i&=&
\Upsilon_{\lambda_i}\,(\,\tilde \phi,\tilde h_i, \lambda_i)\,\,,
\label{apcocho}
\eear
which are   the same  as  those for the same system without the 4-form.
Moreover, if we define the function $H$ as:
\beq
H\,\equiv\,e^{4\chi}\,\,,
\label{apcnueve}
\eeq
then, the uplifted metric is:
\bear
ds^2_{11}\,&=&\,H^{-{2\over 3}}\,
e^{2\tilde f\,-\,{2\over  3}\,\tilde\phi}\,dx^2_{1,2}\,+\,\rc\rc
&&+\,H^{{1\over 3}}\,\Big[\,\sum_i\,e^{2\tilde h_i\,-\,{2\over  3}\tilde\phi}
\big(\,E^i\,)^2\,+\,e^{-{2\over  3}\tilde\phi}\,d\tilde r^2
\,+\,4\,e^{{4\over  3}\tilde\phi}
\,\Big(\,A_i\,+\,{1\over 2}\,L_i\,\Big)^2\,\Big]\,\,.
\label{apcdiez}
\eear

It is clear from eq. (\ref{apcdiez}) that
the effect of the 4-form on the metric is the introduction of some powers
of $H$ which distinguish among  the directions parallel and 
orthogonal to the
form. Moreover, it is easy to verify from the equation satisfied 
by $\chi$ that
the harmonic function $H$ satisfies:
\beq
{dH\over d\tilde r}\,=\,-2\Lambda\,\,
\xi(\,\tilde \phi\,, \,\tilde h^i\,)\,=\,-2\Lambda\,
e^{-\sum \tilde h_i\,-\,\tilde\phi}\
\,\,,
\label{apconce}
\eeq
and, thus, if we know the solution without form, we can integrate the
right-hand side of the last equation and find the expression of $H$. 
Notice
that when $\Lambda=0$ we can take $H={\rm constant}$. In this case the
components of the metric parallel to the 4-form are constant provided
that $\tilde\phi=3\tilde f$ solves eq. (\ref{apcocho}), which 
can only happen
if $\Upsilon_\phi=3\Upsilon_f$ (this is just the (\ref{fphi3}) 
condition needed to have a flat Minkowski part of the
eleven dimensional metric without form). This condition holds for
all  the systems studied
here and in refs. \cite{gns, hs2}. Moreover, if $\tilde\phi=3\tilde f$
one can verify that the uplifted 4-form is such that:
\beq
F_{x^0x^1x^2\tilde r}\,=\,\partial_{\tilde r}\,(\,H^{-1}\,)\,\,,
\label{uplifted4form}
\eeq
where the indices are curved ({\it i.e.} they refer to the 
coordinate basis of (\ref{apcdiez})).

As an illustration of the general formalism we have developed above, let
us consider the case of a flat D6-brane with flux. 
In this situation there are no
scalar fields $\lambda$ excited and the ansatz for the metric is \cite{en}:
\beq
ds^2_8\,=\,e^{2 f}\,dx_{1,2}^2\,+\,e^{2 h}\,dy_4^2\,+\,d r^2\,\,.
\label{apcdoce}
\eeq
The functions $\Upsilon$ appearing in the first-order system 
(\ref{apcdos}) are
$\Upsilon_f=\Upsilon_h={\Upsilon_{\phi}\over 3}={1\over 8}\,
e^{-\phi}$. If we change
to a new variable $t$ such that  $d\tilde r\,=\,e^{-\tilde\phi}\,dt$,
 we can
write the solution of the system (\ref{apcocho}) as
$\tilde f=\tilde h={\tilde \phi\over 3}={1\over 8}\,t$. Moreover, 
for the case
at hand 
$\xi(\,\tilde \phi\,, \,\tilde h\,)\,=\,e^{-4\tilde h-\tilde\phi}$ and,
by plugging this result in eq. (\ref{apconce}), we get
that the
harmonic function is:
\beq
H\,=\,-2\Lambda\,\int e^{-4\tilde h-\tilde\phi} d\tilde r\,=\,
-2\Lambda\,\int e^{-4\tilde h} dt\,=\,1\,+\,4\Lambda\,e^{-{t\over 2}}\,\,,
\label{apctrece}
\eeq
where we have fixed the integration constant to recover the solution with
$\Lambda=0$ at $t\rightarrow\infty$.  The eleven dimensional 
metric is readily
obtained from the uplifting formula (\ref{uplift8d}). Since
 there are no $SU(2)$
gauge fields excited in this flat case \cite{en}, we get:
\beq
ds^2_{11}\,=\,H^{-{2\over 3}}\,dx^2_{1,2}\,+\,H^{{1\over 3}}          
\,\,\Big(\,dy_4^2\,+\,e^{{t\over 2}}\,(dt^2\,+\,16\,d\Omega_3^2)
\,\Big)\,\,.
\label{apccatorce}
\eeq
Introducing a new variable
$\rho$ as $\rho\,=\,{4\over {\sqrt N}}\,e^{{t\over 4}}$, the metric
(\ref{apccatorce}) can be put in the form:
\beq
ds^2_{11}\,=\,\Big[\,H(\rho)\,\Big]^{-{2\over 3}}\,dx^2_{1,2}\,+\,
\Big[\,H(\rho)\,\Big]^{{1\over 3}}
\,\,\Big(\,dy_4^2\,+\,
N(d\rho^2\,+\,\rho^2\,d\Omega_3^2\,)\,\Big)\,\,,
\label{apcquince}
\eeq
where $H(\rho)$ is given by:
\beq
H(\rho)\,=\,1\,+\,{64\Lambda\over N}\,{1\over \rho^2}\,\,.
\label{apcdseis}
\eeq
Notice that  the harmonic 
function of the
D2-brane $H(\rho)$ appearing  in the metric (\ref{apcquince}) is 
not in its
near horizon limit. Actually, if one drops the $1$ on the
 right-hand side of
eq. (\ref{apcdseis}), one can check that (\ref{apccatorce}) 
coincides with the
metric of the standard near horizon D2-D6 intersection.

%*************************************************************

\subsection{Effective lagrangian with 4-form}
\label{efflagr4}

In this section, we explain how to find effective lagrangians for a given
ansatz for the eight dimensional fields when the four-form $G$ 
is non-zero. Let
us imagine that we substitute our ansatz for the metric and gauge field
$A_{\mu}^i$ in the Salam-Sezgin lagrangian (\ref{apatres}) and let us
denote by $f_i$ the different functions $f$, $h$, $\dots$ of the
ansatz (including the dilaton and other scalar fields).
As the four-form field has a radial component, we can represent it as $B'$,
where $B$ is a potential and the prime denotes radial derivative. After
integrating by parts to eliminate the second derivatives, the resulting
lagrangian will be of the type:
\beq
L\,=\,\tilde L(\,f_i, f_i'\,)\,+\,a(\,f_i\,)\,
(\,B'\,)^2\,\,,
\label{elagr1}
\eeq
where $a(\,f_i\,)$ does not depend on the derivatives of the $f_i$'s.
The equations of motion for $L$  are:
\bear
{d\over dr}\,{\partial \tilde L\over \partial f_i'}&=&
{\partial \tilde L\over \partial f_i}\,+\,(\,B'\,)^2\,
{\partial a\over \partial f_i}\,\,,\rc\rc
{d\over dr}\,\Big[\,a\,B'\,\Big]&=&0\,\,.
\label{elagr2}
\eear
 Integrating the equation for $B$ we get:
\beq
B'\,=\,{\Lambda\over a(\,f_i\,)}\,\,,
\label{elagr3}
\eeq
where $\Lambda$ is a constant. This is precisely our typical
ansatz for $G$ (\ref{gen4form}), which will be explicitly used in
eqs. (\ref{fstnueve}) and (\ref{fdos}). Substituting the value of $B'$
given in eq. (\ref{elagr3}) in the equation for the $f_i$'s, one gets:
\beq
{d\over dr}\,{\partial \tilde L\over \partial f_i'}=
{\partial \tilde L\over \partial f_i}\,+\,
{\Lambda^2\over a^2}\,{\partial a\over \partial f_i}\,=\,
{\partial \over \partial f_i}\,\Big(\,
\tilde L\,-\,{\Lambda^2\over a}\,\Big)\,\,,
\label{elagr4}
\eeq
and, therefore,  the effective lagrangian for the $f_i$'s is:
\beq
L_{eff}\,=\,
\tilde L(\,f_i, f_i'\,)\,-\,
{\Lambda^2\over a(\,f_i\,)}\,\,.
\label{elagr5}
\eeq
Indeed, the Euler-Lagrange equations for $L_{eff}$ are precisely
(\ref{elagr4}). Notice the change of sign in the last term of
$L_{eff}$ as compared with the corresponding one in 
$L$ (so one cannot naively introduce the ansatz for
the 4-form in (\ref{apatres})).
This sign flip will be taken into account is eqs. (\ref{fodos}) and
(\ref{fquince}) and is crucial to find the superpotentials.
Equivalently,  one can
obtain $L_{eff}$ by eliminating the cyclic coordinate $B$ by
constructing
the Routhian ${\cal R}$ as:
\beq
{\cal R}\,=\,L\,-\,B'\,{\partial L\over \partial B'}\,\,.
\label{elagr6}
\eeq
Clearly ${\cal R}=L_{eff}$.

%*************************************************************

\setcounter{equation}{0}
\section{D6-branes wrapped on a 2-sphere with 4-form}
\label{fluxS2}

In this section we are going to extend the results of 
chapter \ref{conichapter} by the inclusion of the
4-form. More concretely, only the generalized resolved
conifold case (section \ref{rescon}) will be treated.
The extension of the solution of section \ref{defcon} would
go similarly.

Therefore, the ansatz for the scalars is given in
(\ref{coniscalars}) while the ansatz for the gauge
field is (\ref{conigauge}). However, as the constraint
(\ref{resol}) is imposed, the gauge field and its field 
strength are just:
\beq
A^3\,=\,\cos\theta \,d\varphi\,\,,\,\,\,\,\,\,\,\,\,\,
\,\,\,\,\,\,
F^3\,=\,-\,\sin\theta\,d\theta\,\wedge\,d\varphi\,\,.
\eeq
The natural ansatz for the metric is:
\beq
ds^2_8\,=\,e^{2f}\,dx_{1,2}^2\,+\,e^{2\zeta}\,dy_2^2\,+\,
e^{2h}\,d\Omega_2^2\,+\,dr^2\,\,,
\label{fstocho}
\eeq
where, $f$, $\zeta$ and $h$ are functions of $r$
and
$dy_2^2=(dy^1)^2+(dy^2)^2$.
Notice that the $x$ and $y$ coordinates must be 
distinguished in the metric, because  the 3-form potential
is directed along the $x$ ones. For the metric (\ref{fstocho}), the
equation of motion of the four-form is satisfied if one 
writes (see (\ref{gen4form})):
\beq
G_{\underline{x^{0}x^{1}x^{2}r}}\,=\,\Lambda\,e^{-2\zeta-2h-2\phi}\,\,,
\label{fstnueve}
\eeq
where $\Lambda$ is a constant.

%*************************************************************

\subsubsection{Supersymmetry analysis}

 Let us look for the BPS configurations  by requiring the vanishing of
the supersymmetry variations of the fermionic fields. 
The angular projection related to the K\"ahler structure of the
compact space (\ref{coniangproj}) remains unchanged. For the resolved
conifold, the function $\alpha$  parametrizing the rotation of the spinor
is zero (\ref{alphacero}) and, hence, the radial
projection (\ref{conirot}) gets reduced to:
\beq
\G_r\hat\G_{123}\,\epsilon\,=\,-\epsilon\,\,,
\eeq
 Furthermore, the presence of the
flux (or equivalently of a D2-brane in the
type IIA theory in ten
dimensions) makes necessary an extra projection which
further halves the preserved supersymmetry. It reads:
\beq
\G_{x^0x^1x^2}\epsilon\,=\,-\epsilon\,\,.
\eeq
The number of unbroken supercharges is then four. 
It is now straightforward to find the first-order equations 
which follow from
the conditions $\delta\psi_{\lambda}=\delta\chi_i=0$. One gets:
\bear
f'&=&-{1\over 6}\,e^{\phi-2h-2\lambda}\,
+\,{1\over 24}\,e^{-\phi}\,(\,2e^{2\lambda}\,+\,e^{-4\lambda}\,)
\,+\,{\Lambda\over 2}\, e^{-\phi-2h-2\zeta}\,\,,\rc
\zeta'&=&-{1\over 6}\,e^{\phi-2h-2\lambda}\,
+\,{1\over 24}\,e^{-\phi}\,(\,2e^{2\lambda}\,+\,e^{-4\lambda}\,)
\,-\,{\Lambda\over 2}\, e^{-\phi-2h-2\zeta}\,\,,\rc
h'&=&{5\over 6}\,e^{\phi-2h-2\lambda}\,
+\,{1\over 24}\,e^{-\phi}\,(\,2e^{2\lambda}\,+\,e^{-4\lambda}\,)
\,-\,{\Lambda\over 2}\, e^{-\phi-2h-2\zeta}\,\,,\rc
\phi'&=&-{1\over 2}\,e^{\phi-2h-2\lambda}\,
+\,{1\over 8}\,e^{-\phi}\,(\,2e^{2\lambda}\,+\,e^{-4\lambda}\,)
\,-\,{\Lambda\over 2}\, e^{-\phi-2h-2\zeta}\,\,,\rc
\lambda'&=&{1\over 3}\,e^{\phi-2h-2\lambda}\,
-\,{1\over 6}\,e^{-\phi}\,(\,e^{2\lambda}\,-\,e^{-4\lambda}\,)\,\,.
\label{founo}
\eear
As a check, it is interesting to verify in eq. (\ref{founo}) that, when
$\Lambda=0$, we have $f'=\zeta'=\phi'/3$, and the resulting equations
coincide with those of (\ref{resolsyst}).

%*************************************************************

\subsubsection{First order equations from a superpotential}

 Let us briefly present  here how to obtain the
first order system by finding a superpotential. By 
inserting the ansatz in the lagrangian (\ref{apatres}),
and taking into account the sign change explained in
section \ref{efflagr4}, one finds the effective lagrangian:
\bear
L_{eff}&=&e^{3f+2\zeta+2h}\,\Big[\,{3\over 2}\,\big(f\,'\,)^2\,+\,
{1\over 2}\,\big(\zeta\,'\,)^2\,+\,{1\over 2}\,\big(h\,'\,)^2\,-\,
{3\over 2}\,\big(\lambda\,'\,)^2\,-\,{1\over 2}\,\big(\phi\,'\,)^2\,\rc\rc
&&+\,3f\,'\,\zeta\,'\,+\,3f\,'\,h\,'\,+\,2h\,'\zeta\,'\,+\,
{1\over 2}\,e^{-2h}\,
+\,{1\over 16}\,e^{-2\phi}\,\big(\,2e^{-2\lambda}\,-
\,{1\over 2}\,e^{-8\lambda}\,\big)\,\rc\rc
&&-\,{1\over 2}\,e^{2\phi-4h-4\lambda}\,-\,
{\Lambda ^2\over 2}\,e^{-4\zeta-2\phi-4h}\,\,\Big]\,\,,
\label{fodos}
\eear
where some integration by parts has been made.
Let us define a new
radial coordinate $\eta$:
\beq
{dr\over d\eta}\,=\,e^{-h\,-\,\zeta}\,\,.
\label{fotres}
\eeq
After taking into account the jacobian for the change of variable
(\ref{fotres}), one concludes that the effective lagrangian in the new
variable is
$\hat L_{eff}\,=\,e^{-h\,-\,\zeta}\,L_{eff}$. Moreover, it
is easy to check that
$\hat L_{eff}$ can be put in the form (\ref{newlagr}), with
$A\,\equiv\, f\,+\,h\,+\,\zeta$. The
constants in eq. (\ref{newlagr}) become $c_1=3$, $c_2=3/2$, and now
$\varphi^a$ has  four components, namely, $\varphi^a\,=\,(\zeta, h, \phi,
\lambda)$. The non-vanishing elements of the metric $G_{ab}$ are
$G_{\zeta\,\zeta}
\,=\,G_{h\,h}\,=\,2$, $G_{\zeta\,h}\,=\,G_{\phi\,\phi}\,=\,1$ and
$G_{\lambda\,\lambda}\,=\,3$, and the potential $\tilde{V}$ is given by:
\beq
\tilde{V}={1\over 2}\,e^{2\phi-6h-2\zeta-4\lambda}\,+\,
{1\over 32}\,e^{-2\phi-2h-2\zeta}\,
\big(\,e^{-8\lambda}\,-\,4e^{-2\lambda}\,\big)\,-\,\rc\rc
{1\over 2}\,e^{-4h-2\zeta}\,+\,
{\Lambda ^2\over 2}\,e^{-6\zeta-2\phi-6h}\,\,.
\label{foseis}
\eeq
The corresponding superpotential $\tilde{W}$ must satisfy eq.
(\ref{newdefsuperp}), which in this case becomes (fixing $c_3=1$):
\beq
\tilde{V} = {1\over 3}\,\Bigg(\,{\partial \tilde{W}\over \partial
\zeta}\,\Bigg)^2\,+\, {1\over 3}\,\Bigg(\,{\partial \tilde{W}\over
\partial h}\,\Bigg)^2\,+\, {1\over 2}\,\Bigg(\,{\partial \tilde{W}\over
\partial
\phi}\,\Bigg)^2\,+\, {1\over 6}\,\Bigg(\,{\partial \tilde{W}\over \partial
\lambda}\,\Bigg)^2\,-\,\rc\rc {1\over 3}\,{\partial \tilde{W}\over
\partial h}\, {\partial \tilde{W}\over \partial \zeta}\,-\,{3\over
2}\,\,\tilde{W}^2\,\,.
\label{fosiete}
\eeq
After some elementary calculation, one can prove that $W$ can be taken as:
\beq
\tilde{W}=-{1\over 2}\,\,e^{\phi\,-\,3h\,-\,\zeta\,-\,2\lambda}\,-\,
{1\over 8}\,\,e^{-\phi\,-\,h\,-\,\zeta}\,\,
\big(\,e^{-4\lambda}\,+\,2e^{2\lambda}\,\big)\,
+\,{\Lambda\over 2}\,\,e^{-3\zeta-3h-\phi}\,\,.
\label{foocho}
\eeq
The first-order equations for this superpotential can be obtained by
substituting (\ref{foocho}) on the right-hand side of
eq. (\ref{neworder1}). It is not difficult to check that, in terms of the
original variable $r$, one gets exactly the first-order system
(\ref{founo}).

%*************************************************************

\subsection{Getting the eleven dimensional solution}
\label{Felf}

In order to find the solution of eleven dimensional
supergravity arising from the uplifting of
the eight dimensional configuration
described by the system (\ref{founo}),
 we can use the reasoning of
section \ref{general4f}. Indeed, (\ref{founo}) is
of the type (\ref{apcdos}). The system without 4-form
is written in eq. (\ref{resolsyst}), and its solutions in
the subsequent equations. In the following, tilded
functions will refer to the solution of that system
(\ref{concuatro}), (\ref{hphieq2}).
Using (\ref{fstnueve}), (\ref{gen4form}) in (\ref{apconce}),
a differential equation for the warp factor is obtained:
\beq
{dH \over d\tilde{r}}\,=\,-2\,\Lambda\,e^{-2\tilde{h}-
\frac{5}{3}\tilde{\phi}}\,\,,
\eeq
where $\tilde{\zeta}=\tilde{\phi}/3$ has been used 
(which is simply the condition (\ref{fphi3}) for the
system without flux). It is convenient to express the
warp factor in terms of the radial variable $\rho$ which
was used in the result (\ref{resmetric}). Taking into 
account (\ref{defs1}) and (\ref{coctuno}), it is
immediate to get:
\beq
{d\tilde{r} \over d\rho}\,=\,\frac{4}{\rho}\,\,
e^{\tilde{\phi}+4\la}\,\,\,,
\eeq
and therefore, substituting (\ref{concuatro}), (\ref{hphieq2}),
one arrives at:
\beq
{dH \over d\rho}\,=\,-\,4\,{432\, \Lambda \over
\rho^3\,(\rho^2+6a^2)\,\kappa(\rho)}\,\,.
\label{difeqH}
\eeq

%*************************************************************

\subsubsection{Smeared M2--branes at the tip of the conifold}

Let us consider first the case of the singular conifold
with $a=b=0$ and $\kappa(\rho)=1$.
Then,   the integration of  (\ref{difeqH}) is trivial:
\beq
H(\rho)\,=\,1\,+\,{k\over \rho^4}\,\,,
\label{fcttres}
\eeq
where the 1 has been fixed in order to recover the solution
without flux at $\rho\rightarrow\infty$,
and the constant $k$ is related to $\Lambda$ by means of the 
expression\footnote{The different factor with respect 
to \cite{epr} is due to  a different constant of
integration in the eq. for $f$ which
in \cite{epr} made necessary a
rescaling of the $x$ and $y$ coordinates.}:
\beq
k\,=\,432\,\,\Lambda\,\,.
\label{fctcuatro}
\eeq
We get   the corresponding eleven dimensional metric, which takes the
form:
\beq
ds^2_{11}\,=\,
\big[\,H(\rho)\,\big]^{-{2\over 3}}\,dx_{1,2}^2\,+\,
\big[\,H(\rho)\,\big]^{{1\over 3}}\,\Big[\,
dy_2^2\,+\,ds^2_6\,\Big]\,\,,
\label{fctdos}
\eeq
where $ds^2_6$ is the singular conifold metric written in
(\ref{singconi}).
 Finally, the 4-form $F$ can be obtained from (\ref{uplifted4form}):
\beq
F_{x^0x^1x^2\rho}\,=\,\partial_{\rho}\,
\big[\,H(\rho)\,\big]^{-1}\,\,.
\label{fcsiete}
\eeq

It follows from
these results that this solution can be
interpreted as the geometry created by a smeared distribution of M2-branes
located at the tip of the singular conifold. Notice that we are now
 smearing
the M2- brane along two coordinates, which agrees with the 
power of $\rho$ in the harmonic function (\ref{fcttres}).

%*************************************************************

%%%%%%%%%%   Subsubsection 3.4.2
\subsubsection{Smeared M2--branes and generalized resolved conifold}

Integrating (\ref{difeqH}) for arbitrary values of $a$, $b$ is
somewhat involved. By fixing again $H(\infty)=1$, the 
explicit expression of the integral is:
\beq
H(\rho)\,=\,1\,+\,4k\,\int_{\rho}^{\infty}\,
{\tau d\tau\over \tau^6\,+\,9a^2\tau^4\,-\,b^6}\,\,,
\label{fctonce}
\eeq
with $k$ again given by eq. (\ref{fctcuatro}). 
The metric can be written as (\ref{fctdos}) where now
$ds_6^2$ corresponds to
the small resolution of the generalized conifold 
(\ref{resmetric}).
On the other hand, the
4-form
$F$ for this solution can be put in the form (\ref{fcsiete}) with
$H(\rho)$ given by eq. (\ref{fctonce}).

It is immediate to
conclude  that $H(\rho)$ behaves for
$\rho\rightarrow\infty$ exactly as the right-hand side of
eq. (\ref{fcttres}). In order to find out the behavior of $H$ at small
$\rho$, let us perform explicitly the integral  (\ref{fctonce}) in some
particular cases. First of all, we consider the case $b=0$, for which
$H(\rho)$ is given by:
\beq
H(\rho)\,=\,1\,+\,{2k\over 9a^2}\,{1\over \rho^2}\,-\,
{2k\over 81a^4}\,\log\big(\,1\,+\,{9a^2\over \rho^2}\,\big)\,\,,
\,\,\,\,\,\,\,\,\,\,\,\,\,\,\,
(b=0)\,\,.
\label{fctdoce}
\eeq
This  expression for $H(\rho)$ coincides exactly with the one found in
\cite{pzt} for the case of a D3-brane at the tip of the small resolution
of the conifold, which can be obtained from our solution by dimensional
reduction and T-duality (see below). For $\rho\approx 0$ the harmonic
function behaves as:
\beq
H(\rho)\,\approx\,{2k\over 9a^2}\,{1\over \rho^2}\,\,,
\,\,\,\,\,\,\,\,\,\,\,\,\,\,\, (b=0)\,\,.
\label{fcttrece}
\eeq
When $a=0$ the integral (\ref{fctonce}) can also explicitly performed ,
with the result:
\beq
H(\rho)\,=\,1\,-\,{2k\over b^4}\,\,
\Big[\,{1\over 6}\,\log{(\rho^2-b^2)^3\over \rho^6-b^6}\,+\,{1\over \sqrt{3}}\,
{\rm arccot}{2\rho^2+b^2\over \sqrt{3}\,b^2}\,\Big]\,\,,
\,\,\,\,\,\,\,\,\,\,\,\,\,\,\, (a=0)\,\,,
\label{fctcatorce}
\eeq
and, again, this result coincides with that of ref. \cite{pztdos}. For
$\rho\approx b$ the function in (\ref{fctcatorce}) has a logarithmic
behavior of the form:
\beq
H(\rho)\,\approx\,-{2k\over 3b^4}\,\log{\rho-b\over b}\,\,,
\,\,\,\,\,\,\,\,\,\,\,\,\,\,\, (a=0)\,\,.
\label{fctquince}
\eeq
For general values of $a$ and $b$ the
integral (\ref{fctonce}) can be  performed by factorizing the polynomial
in the denominator. The result depends on the sign of the ``discriminant"
$\Delta\,=\,b^6\,-\,108\,a^6$. The analysis of the different cases has
been carried out in ref. \cite{pztdos}.

%*************************************************************

\subsection{Reduction to D=10 and T-duality}

 Some ten dimensional solutions 
 associated to the eleven dimensional ones of section \ref{Felf}
will be obtained here. This can
be achieved by means of Kaluza-Klein reduction to type IIA
theory and by afterwards implementing T-duality. Finding
this kind of solutions is interesting because they can
be related to D-branes, and therefore, used as gravity
duals of gauge theories.

%%%%%%%%%%   Subsubsection 3.5.1
\subsubsection{D3--branes at the tip of the generalized resolved conifold}

Let us consider first a reduction along a direction orthogonal to the six
dimensional conifold metric. Notice that $\partial/\partial y^1$
and
$\partial/\partial y^2$ are Killing vectors of (\ref{fctdos}). Let us
reduce along $y^2$ followed by a T-duality transformation along $y^1$. The
resulting metric in the IIB theory is:
\beq
ds^2_{10}\,=\,\big[\,H(\rho)\,\big]^{-{1\over 2}}\,
\Big[\,dx^2_{1,2}\,+\,(dy^1)^2\,\Big]\,+\,
\big[\,H(\rho)\,\big]^{{1\over 2}}\,ds^2_6\,\,,
\label{fctdseis}
\eeq
where $H(\rho)$ is written in (\ref{fctonce}) and $ds_6^2$
in (\ref{resmetric}).
The dilaton is constant and there is an RR 5-form:
\beq
F^{(5)}=\partial_{\rho}\,\Big[\,H(\rho)\,\Big]^{-1}\,\,
dx^0\wedge dx^1\wedge dx^2\wedge  dy\wedge d\rho\,\,+\,\,
{\rm Hodge\,\, dual}\,\,.
\label{fctdsiete}
\eeq
This solution is precisely the one studied in ref. \cite{pztdos} and
corresponds to a D3--brane located at the tip of the generalized
resolved conifold.

%%%%%%%%%%   Subsubsection 3.5.2
\subsubsection{Smeared D2--D6 wrapped on a 2-cycle}

Another possibility is to reduce along the fiber $\tilde{\psi}$ of the
$T^{1,1}$ space. The expression (\ref{resmetric}),
when included in (\ref{fctdos}), automatically
gives a reduction ansatz of the type (\ref{KKred}). However,
notice that the vielbein where susy was analyzed 
 and
the spinor was $\tilde{\psi}$-independent is the one
related to (\ref{coctocho}). To go to the vielbein naturally
associated to (\ref{resmetric}), a  $\tilde{\psi}$-dependent
local lorentzian rotation is necessary.
This  introduces a functional
dependence on the eleven dimensional Killing spinor  and
so it  renders a non--supersymmetric supergravity
solution \cite{bgm}. This is nothing but the phenomenon of supersymmetry
without supersymmetry first discussed in \cite{bko}. In order
to write the result of the reduction along $\tilde{\psi}$,
let us define the function:
\beq
\Gamma(\rho)\,\equiv\,{\rho^2\over 9}\,\kappa(\rho)\,\,.
\label{fctdocho}
\eeq
Then, the solution of the type IIA theory that one obtains by reducing along
$\tilde{\psi}$ is:
\bear
ds^2_{10}&=&\Bigg[\,{\Gamma(\rho)\over H(\rho)}\,\Bigg]^{{1\over 2}}\,\,
\Big[\,dx^2_{1,2}\,+\,H(\rho)\,\big(\,dy^2_2\,+\,
{d\rho^2\over \kappa(\rho)}\,+\,{\rho^2+6a^2\over 6}\,d\Omega^2\,+\,
{\rho^2\over 6}\,d\tilde\Omega^2
\,\,\big)\,\Big]\,\,,\rc\rc
e^{\phi}&=&\Big[\,\Gamma(\rho)\,\Big]^{{3\over 4}}\,\,
\Big[\,H(\rho)\,\Big]^{{1\over 4}}\,\,,\rc\rc
F^{(2)}&=&\epsilon_{(2)}\,+\,\tilde\epsilon_{(2)}\,\,,\rc\rc
F^{(4)}&=&\partial_{\rho}\,\Big[\,H(\rho)\,\Big]^{-1}\,\,
dx^0\wedge dx^1\wedge dx^2\wedge  d\rho\,\,,
\label{fctdnueve}
\eear
where $d\tilde\Omega^2\,=\,d\tilde\theta^2\,+\,\sin^2\tilde\theta
d\tilde\varphi^2$, $\epsilon_{(2)}\,=\,\sin\theta d\varphi\wedge
d\theta$ and
$\tilde\epsilon_{(2)}\,=\,\sin\tilde\theta d\tilde\varphi\wedge
d\tilde\theta$.
 This
(non--supersymmetric) solution  (of IIA supergravity) corresponds to a
system of (D2-D6)-branes, with the D2-brane extended along $(x^1, x^2)$
and smeared in $(y^1,y^2)$ and the D6-brane wrapped on a two-cycle.
Notice that the KK reduction somehow disentangled the bundle and the
resulting ten dimensional metric exhibits a product of the two-spheres
instead of a fibration.  This is characteristic of what has been called
supersymmetry without supersymmetry: the supergravity solution does not
display supersymmetry even when it may be present at the level of full
string theory
\cite{dlp,bgm,mt}.

%%%%%%%%%%   Subsubsection 3.5.3
\subsubsection{D4--branes}

If we now perform  T-duality transformations along the coordinates $(y^1,
y^2)$, we arrive at a system composed by D4-branes, for which the metric
and dilaton are:
\bear
ds^2_{10}&=&\Bigg[\,{\Gamma(\rho)\over H(\rho)}\,\Bigg]^{{1\over 2}}\,\,
\Big[\,dx^2_{1,2}\,+\,{dy^2_2\over \Gamma(\rho)}\,
+\,H(\rho)\,\big(\,
{d\rho^2\over \kappa(\rho)}\,+\,{\rho^2+6a^2\over 6}\,d\Omega^2\,+\,
{\rho^2\over 6}\,d\tilde\Omega^2
\,\,\big)\,\Big]\,\,,\rc\rc
e^{\phi}&=&
\Bigg[\,{\Gamma(\rho)\over H(\rho)}\,\Bigg]^{{1\over 4}}\,\,.
\label{fctveinte}
\eear
Moreover, the direct application of the T-duality rules gives the following RR
potentials:
\bear
C^{(3)}&=&\cos\theta\,d\varphi\wedge dy^1\wedge dy^2\,+\,
\cos\tilde\theta\,d\tilde\varphi\wedge dy^1\wedge dy^2\,\,,\rc\rc
C^{(5)}&=&\big[\,H(\rho)\,\big]^{-1}\,
dx^0\wedge dx^1\wedge dx^2\wedge dy^1\wedge dy^2\,\,.
\label{fctvuno}
\eear
However, since  $C^{(5)}$ is really the potential of
$F^{(6)}\,=\,{}^{*}\,F^{(4)}$, we will only have a four-form  RR field
strength, given by:
\beq
F^{(4)}\,=\,(\,\epsilon_{(2)}\,+\,
\tilde\epsilon_{(2)}\,)\,\wedge dy^1\wedge dy^2\,+\,{k\over 27}\,
\epsilon_{(2)}\wedge \tilde\epsilon_{(2)}\,\,,
\label{fctvdos}
\eeq
where $k$ is the constant appearing in the harmonic function $H(\rho)$.
Again, this solution displays the supersymmetry without supersymmetry
behavior.

%*****************************************************************

\setcounter{equation}{0}
\section{D6-branes wrapped on a 3-sphere with 4-form}
\label{fluxS3}

In this section, the 4-form $G$ flux will be added to the
geometry with $G_2$ holonomy studied in chapter
\ref{g2chapter}. We will closely follow the steps of
the previous section for this case.

For concreteness, we will only deal with the
simpler case of the Bryant-Salamon metric (see the 
beginning of section \ref{solveg2}). Therefore, the
gauge field and its field strength are just:
\beq
A^i\,=\,-{1\over 2}\,  w^i\,\,,\,\,\,\,\,\,\,\,\,
\,\,\,
F^i\,=\,-\frac{1}{8}\,\epsilon_{ijk}\,w^j\wedge w^k\,\,.
\label{focho}
\eeq
The metric must take the form:
\beq
ds^2_8\,=\,e^{2f}\,dx_{1,2}^2\,+\,e^{2\zeta}\,dy^2\,+\,
e^{2h}\,d\Omega_3^2\,+\,dr^2\,\,.
\label{funo}
\eeq
The corresponding ansatz
for the 4-form $G$ in flat coordinates is (\ref{gen4form}):
\beq
G_{\underline{x^{0}x^{1}x^{2}r}}\,=\,\Lambda\,e^{-\zeta-3h-2\phi}\,\,,
\label{fdos}
\eeq
with $\Lambda$ being a constant and $\phi$ the eight-dimensional dilaton.

%*************************************************************

\subsubsection{Supersymmetry analysis}

As in section \ref{fluxS2}, the supersymmetric projections for
the configuration without flux have to be kept. In this case,
there are three of them which can be read from eqs.
(\ref{nueve}) and (\ref{catorce}) (where $\tilde{\beta}=0$,
$\beta=1$ as we are looking at the $g=0$ case). They read:
\beq
\G_{12}\hat\G_{12}\,\epsilon\,=\,
\G_{23}\hat\G_{23}\,\epsilon\,=\,
-\G_{r}\hat\G_{123}\,\epsilon\,=\,
\epsilon\,\,.
\eeq
Additionally, the presence of the flux makes necessary
a new projection, reducing to two the total number of
supercharges preserved by the solution:
\beq
\G_{x^0x^1x^2}\epsilon\,=\,-\epsilon\,\,.
\eeq
With these conditions, it is straightforward to obtain the
BPS equations, by demanding, as usual, the vanishing of the
supersymmetric variation of  the gravitino and dilatino fields:
\bear
f'&=&-{1\over 2}\,e^{\phi-2h}\,+\,{1\over 8}\,e^{-\phi}\,+\,{\Lambda\over
2}\, e^{-\phi-3h-\zeta}\,\,,\rc
\zeta'&=&-{1\over 2}\,e^{\phi-2h}\,+\,{1\over
8}\,e^{-\phi}\,-\,{\Lambda\over 2}\, e^{-\phi-3h-\zeta}\,\,,\rc
h'&=&{3\over 2}\,e^{\phi-2h}\,+\,{1\over 8}\,e^{-\phi}\,-\,{\Lambda\over
2}\, e^{-\phi-3h-\zeta}\,\,,\rc
\phi'&=&-{3\over 2}\,e^{\phi-2h}\,+\,{3\over
8}\,e^{-\phi}\,-\,{\Lambda\over 2}\, e^{-\phi-3h-\zeta}\,\,.
\label{fcatorce}
\eear
Notice that, as it should, eqs. (\ref{fcatorce}) reduce to 
(\ref{addon5}) when $\Lambda=0$ (and $f=\zeta=\phi/3$).

%*************************************************************

\subsubsection{First order equations from a superpotential}

Before finding the general integral of the BPS equations
(\ref{fcatorce}), let us derive them again by means of the alternative
superpotential method . Actually, the
equations of motion of eight dimensional supergravity for our ansatz can
be derived from the effective Lagrangian:
\bear
L_{eff}&=&e^{3f+\zeta+3h}\,\Big[\,(f'\,)^2\,+\,
(h'\,)^2\,-\,{1\over 3}\,(\phi'\,)^2\,+\,
3\,f'\,h'\,+\,f'\,\zeta'\,+\,\zeta'\,h'\,\rc
&&+\,e^{-2h}\,+\,{1\over 16}\,e^{-2\phi}\,-\,
e^{2\phi-4h}\,-\,{\Lambda^2\over 3}\,
e^{-2\zeta\,-\,6h\,-\,2\phi}\,\Big]\,\,.
\label{fquince}
\eear
Let us now introduce a new radial variable $\eta$, whose relation to our
original coordinate $r$ is given by:
\beq
{dr\over d\eta}\,=\,e^{-{3\over 2}\,h\,-\,{1\over 2}\zeta}\,\,.
\label{fdseis}
\eeq
The lagrangian in the new variable is $\hat L_{eff}\,=\,e^{-{3\over
2}\,h\,-\,{1\over 2}\zeta}\,L_{eff}\,$, where we have taken into
account the corresponding jacobian. If we define a new scalar field
$A \equiv f\,+\,{1\over 2}\,\zeta\,+{3\over 2}\,h$, so the 
lagrangian is of the sort (\ref{newlagr}) with parameters
$c_1=3$, $c_2=1$ and where the non-vanishing elements of the metric
$G_{ab}$ are $G_{\zeta\,\zeta}\,=\,G_{\zeta\,h}\,=\,{1\over 2}$,
$G_{h\,h}\,=\,{5\over 2}$ and $G_{\phi\,\phi}\,=\,{2\over 3}$.
The potential $\tilde V(\varphi)$ appearing in $\hat L_{eff}$ is:
\beq
\tilde{V}(\varphi)\,=\,
e^{2\phi\,-\,7h\,-\,\zeta}\,-\,{1\over 16}\,
e^{-2\phi\,-\,3h\,-\,\zeta}\,-\,e^{-\,5h\,-\,\zeta}\,+\,
{\Lambda^2\over 3}\,e^{-2\phi\,-\,9h\,-\,3\zeta}\,\,.
\label{fvuno}
\eeq
In view of (\ref{newdefsuperp}), we have to look for
a function $\tilde{W}(\phi,h,\zeta)$ such that:
\beq
\tilde V\,=\,{5\over 4}\,\Bigg(\,{\partial\tilde  W\over \partial
\zeta}\,\Bigg)^2\,+\, {1\over 4}\,\Bigg(\,{\partial\tilde  W\over \partial
h}\,\Bigg)^2\,+\, {3\over 4}\,\Bigg(\,{\partial\tilde  W\over \partial
\phi}\,\Bigg)^2\,-\, {1\over 2}\,{\partial\tilde  W\over \partial
\zeta}\,{\partial\tilde  W\over \partial h}
\,-\,{9\over 4}\,W^2\,\,.
\label{fvcuatro}
\eeq
For the value of $\tilde V$ given above (eq. (\ref{fvuno})) one can check
that eq. (\ref{fvcuatro}) is satisfied by:
\beq
W\,=\,-\,e^{\phi\,-\,{7\over 2}\,h\,-\,{\zeta\over 2}}\,-\,
{1 \over 4}\,e^{-\phi\,-\,{3\over 2}\,h\,-\,{\zeta\over 2}}\,+\,
{\Lambda\over 3}\,e^{-\phi\,-\,{9\over 2}\,h\,-\,{3\over 2}\,\zeta}\,\,.
\label{fvcinco}
\eeq
It is now easy to verify that the first-order  domain wall equations
for this superpotential (\ref{neworder1})
are exactly the same (when expressed in terms of the old variable $r$) as
those obtained from the supersymmetric variation of the fermionic fields
(eqs. (\ref{fcatorce})).

%*************************************************************

\subsection{Eleven dimensional solution}

Let us now write the solution that comes from integrating
eqs. (\ref{fcatorce}). Once again, the general procedure of section
\ref{general4f} simplifies the task drastically. It is easy to
check that all the conditions needed for (\ref{apcdiez}),
(\ref{uplifted4form}) to be the solution are fulfilled. 
The system of equations for $\Lambda=0$ and its solution are
written in (\ref{addon5})-(\ref{brysal}). In the following, as in section
\ref{Felf}, the tilded functions will refer to quantities of
the system without four-form. The equation for the warp factor
(\ref{apconce})  is just, using $\tilde\zeta=\tilde\phi/3$:
\beq
{dH \over d\tilde r}\,=\,-2\,\Lambda\,
e^{-3\tilde h-\frac{4\tilde\phi}{3}}\,\,.
\eeq
In order to express the equation in
terms of the radial variable $\rho$ used in (\ref{brysal}), we
need: 
\beq
{d\tilde r\over d\rho}\,=\,\frac{6}{\rho}\,\left(1-\frac{a^3}{\rho^3}
\right)^{-1}\,e^{\tilde\phi}\,\,,
\eeq
which can be easily derived from (\ref{addon6}), (\ref{addon3}).
Finally, reading from (\ref{addon8})
the values of $\tilde h$, $\tilde\phi$,
 we arrive at the simple equation:
\beq
{dH \over d\rho}\,=\,-\,\frac{1296\,\sqrt{3}\,\Lambda}
{(\rho^3-a^3)^2}\,\,.
\label{warpg2}
\eeq
One gets the
following metric in D=11:
\beq
ds^2_{11}\,=\,
\big[\,H(\rho)\,\big]^{-{2\over 3}}\,dx_{1,2}^2\,+\,
\big[\,H(\rho)\,\big]^{{1\over 3}}\,\Big[\,
dy^2\,+\,ds^2_7\,\Big]\,\,,
\label{fcdos}
\eeq
where $ds^2_7$ is the Bryant-Salamon metric (\ref{brysal}).
The 4-form $F$ in the solution has the usual form
(\ref{uplifted4form}).

\subsubsection{Smeared M2--branes on the tip of a $G_2$ cone}
Consider first the particular solution $a=0$, where the
warp factor is really simple:
\beq
H(\rho)\,=\,1\,+\,{k\over \rho^5}\,\,.
\label{fctres}
\eeq
with $k$ being:
\beq
k\,=\,{1296\over 5}\,\sqrt{3}\,\,
\Lambda\,\,.
\label{fccuatro}
\eeq
Notice that
$H(\rho)$ is an harmonic function in the transverse seven
dimensional space.
It is clear from the result of the uplifting that our 
solution corresponds to
a smeared distribution of M2--branes in the tip 
of the singular cone over
$S^3 \times S^3$ with a $G_2$ holonomy metric found in \cite{bs,gpp}.
Notice that the power of $\rho$ in the harmonic 
function (\ref{fctres}) is
the one expected within this interpretation.

%%%%%%%%%%   Subsubsection 2.4.2
\subsubsection{Smeared M2--branes on the resolved $G_2$ cone}

For a general value of $a$, fixing
$H(\infty)=1$, the warp factor is (\ref{warpg2}):
\beq
H(\rho)\,=\,1\,+\,k\,\int_\rho^\infty
{5 \over (\tau^3-a^3)^2}\,d\tau\,\,,
\eeq
where the constant $k$ is the same as in eq. (\ref{fccuatro}). 
After some calculation, an explicit expression can be 
obtained, namely:
\beq
H(\rho)\,=\,1+\,k\,\Bigg[\,{5\over 3 a^3\rho^2}\,
{1\over 1\,-\,{a^3\over \rho^3}}+{10\over  3\sqrt{3}\,a^5}\,
{\rm arccot}\,\big[\,{2\rho +a\over a\sqrt{3}}\big]\,
-\,{5\over 9a^5}\,\log\big(\,1\,+\,{3 a\rho\over
(\rho-a)^2}\big)\,\Bigg]\,\,.
\label{fcitres}
\eeq
 This solution represents a smeared
distribution of M2--branes on the resolved manifold of $G_2$ holonomy
$X_7$ (the Bryant-Salamon solution), whose singular limit is the cone 
obtained above. It is an $\RR^4$ bundle over $S^3$.  Actually, the
function
$H(\rho)$ can also be determined by solving the Laplace equation on the
seven dimensional
$G_2$ manifold
\cite{cghp}. It is also interesting to analyze the large and small
distance behavior of this harmonic function. When
$\rho\rightarrow\infty$, $H(\rho)$ can be approximated as:
\beq
H(\rho)\approx 1\,+\,{k\over \rho^5}\,+\,{5a^3k\over 4\rho^8}
\,+\,\cdots\,\,,
\label{fciseis}
\eeq
\ie\ it has the same  leading asymptotic behavior as the function
(\ref{fctres}). On the other hand, for $\rho\approx a$, $H(\rho)$
diverges as:
\beq
H(\rho)\approx {5k\over 9a^4}\,{1\over \rho-a}\,+\,
{10k\over 9a^5}\,\log{\rho-a\over a}\,+\,\cdots\,\,.
\label{fcisiete}
\eeq
It is tempting to argue at this point that this supergravity smeared
solution might be the dual of some gauge theory at a given low energy
range. The resolution of the conical singularity must render the theory
non-conformal in the IR. In order to better understand our solutions,
it is important to go to ten dimensions. There are different reductions
to type IIA string theory: we can reduce on the smeared direction, or
we can embed the M--theory circle in the $\RR^4$ fiber or the $S^3$
base in $X_7$. We will study them in the following subsection.

%*************************************************************

\subsection{Reduction to D=10 and T-duality}

%%%%%%%%%%   Subsubsection 2.5.1
\subsubsection{D2--branes on the tip of a (resolved) $G_2$ cone}

There are several possible choices for a coordinate to reduce.
The simplest election --and the most meaningful from the point of view of
gauge/string duality, as long as the smearing is removed-- is 
to reduce along $y$, for
which the metric and dilaton of the IIA theory are:
\bear
ds^2_{10}&=&
\big[\,H(\rho)\,\big]^{-{1\over 2}}\,dx_{1,2}^2\,+\,
\big[\,H(\rho)\,\big]^{{1\over 2}}\,ds^2_7\,\,,\rc\rc
e^{\phi}&=&\big[\,H(\rho)\,\big]^{{1\over 4}}\,\,,
\label{fcinueve}
\eear
while the 4-form field strength of D=11 becomes the RR 4-form $F^{(4)}$ of
the type IIA theory and $C^{(1)}$ vanishes. It is clear that this D=10
solution represents a D2 sitting at the tip of the $G_2$ holonomy
manifold $X_7$, whose principal orbits are topologically trivial $\tilde
S^3$ bundles over $S^3$. In the singular limit, when the base $S^3$ has
vanishing volume, we end with D2--branes at the tip of the $G_2$ cone
over the Einstein manifold $Y_6$. This configuration is reminiscent of
the Klebanov--Witten's $D3$--branes placed at the tip of the conifold
\cite{kw}. Indeed, it is a sort of lower supersymmetric version of it.
Notice, however, that the solution resulting from gauged supergravity is
the complete D2--brane solution and not its near horizon limit. This
might look strange since gauged supergravity usually gives directly the
near horizon metric. The reason is that the near horizon limit of the
D6--branes (that we would obtain through a different reduction, see
below), which are the {\em host} branes of D=8 gauged supergravity,
do not imply, in general, the near horizon limit of the D2--branes that
are intersecting them.  In summary,
in order to get the supergravity dual of the system of D2--branes on the
tip of the $G_2$ cone, we must consider the near horizon limit. We should
reintroduce $l_p$ units everywhere and take $\rho$, $a$ and $l_p$ to zero
such that
\beq
U \equiv {a\rho\over l_p^3} ~~~~~~~ \mbox{and}
~~~~~~~ L \equiv {a^2\over l_p^3}
\label{nearh}
\eeq
are kept fixed. The resulting expression for the harmonic function
(\ref{fcitres}), for large $U$, admits the following asymptotic expansion
\beq
H(U)\,=\,{5\, g_{YM}^3\, N\over 3 \, l_s^4\, L^3\, U^2} \sum_{n=1}^\infty
{3n\over 3n+2} \Big( {L\over U} \Big)^{3n} \,,
\label{harexp}
\eeq
where $g_{YM}^2 \approx L$ is the three dimensional coupling constant,
$a l_s^2 = l_p^3$, and $N$ is the number of $D2$--branes. The asymptotic
background gives the near horizon limit of $N$ $D2$--branes transverse
to the $G_2$ holonomy manifold:
\bear
ds^2_{10} & = & l_s^2 \left( {U^{5 \over 2}\over \sqrt{g_{YM}^2 N}}
\,dx_{1,2}^2\,+\, {\sqrt{g_{YM}^2 N} \over U^{{5\over 2}}} \,ds^2_7\,
\right) \,,\rc\rc
e^{\phi} & = & \left( {g_{YM}^{10} N \over U^5} \right)^{{1\over 4}}\,,
\label{cinuevenh}
\eear
and the 4-form field strength $F$ is still given by (\ref{fcsiete}).
It is analogous to the flat D2--brane \cite{imsy} except for the fact
that the transverse $\RR^7$ has been replaced by the $G_2$ cone over $S^3
\times S^3$. This is the valid description for intermediate high energies,
$g_{YM}^2 N > U > g_{YM}^2 N^{1\over 5}$, where the string coupling and
the curvature are small, and the radius of the eleventh circle vanishes.

In the UV we can trust the super Yang--Mills theory description. It is an
${\cal N}=1$ theory in $2+1$ dimensions. We can obtain its field content
following the arguments in \cite{kw}. In the case of a single $D2$--brane,
it is a $U(1) \times U(1)$ gauge theory with four complex scalars $Q_i$,
$\tilde Q_i$, $i=1,2$, and a vector multiplet whose gauge field can be
dualized to a compact scalar that would parametrize the position of the
$D2$--branes along the M--theory circle. The vacuum moduli space is given
by
\beq
|q_1|^2 + |q_2|^2 - |\tilde q_1|^2 - |\tilde q_2|^2 = L^2 ~,
\label{vmoduli}
\eeq
where $q_i$, $\tilde q_i$ are the scalar components of the superfields
$Q_i$, $\tilde Q_i$, which precisely provides an algebraic--geometric
description of the manifold $X_7$ \cite{amv}.

%%%%%%%%%%   Subsubsection 2.5.2
\subsubsection{D2--D6 system wrapping a special Lagrangian $S^3$}

The second possibility we shall explore is 
the reduction along some compact
direction of the $G_2$ manifold. 
Let us consider first the three-sphere $\tilde
S^3$, parametrized by the $SU(2)$
 left-invariant 1-forms $\tilde w^i$. Notice
that $\tilde S^3$ is external to the 
D6-brane worldvolume in the D=8 gauged
supergravity approach. We shall 
regard the $\tilde S^3$ sphere as a Hopf
bundle over a two-sphere, and we 
will reduce along the fiber of this bundle.
Denoting the $\tilde w^i$'s as in eq. (\ref{omegas}), we shall
choose $z=\tilde
\varphi$ as the coordinate along which the dimensional reduction will take
place. Accordingly
\cite{clp}, let us  define the vector $\tilde \mu^i$ and the 1-forms
$\tilde e^i$ by means of the following decomposition of the $\tilde w^i$'s:
\beq
\tilde w^i\,=\,\tilde e^i\,+\,\tilde \mu^i\,d\tilde \varphi\,\,.
\label{fsesenta}
\eeq
The components of $\tilde \mu^i$ and $\tilde e^i$ are:
\bear
&&\tilde \mu^1\,=\,\sin\tilde\theta\sin\tilde\psi\,\,,
\,\,\,\,\,\,\,\,\,\,
\tilde \mu^2\,=\,-\sin\tilde\theta\cos\tilde\psi\,\,,
\,\,\,\,\,\,\,\,\,\,
\tilde \mu^3\,=\,\cos\tilde\theta\,\,,\rc\rc
&&\tilde e^1\,=\,\cos\tilde\psi\, d\tilde\theta\,\,,
\,\,\,\,\,\,\,\,\,\,\,\,\,\,\,\,\,
\tilde e^2\,=\,\sin\tilde\psi\, d\tilde\theta\,\,,
\,\,\,\,\,\,\,\,\,\,\,\,\,\,\,\,\,\,\,\,\,\,\,
\tilde e^3\,=\,d\tilde\psi\,\,.
\label{fsuno}
\eear
Notice that $\tilde\mu^i\tilde\mu^i\,=\,1$. One can also check the following
relation:
\beq
\tilde e^i\,=\,\epsilon_{ijk}\,\tilde \mu^j\,d\tilde \mu^k\,+\,
\cos\tilde\theta\,d\tilde\psi\,\tilde \mu^i\,\,,
\label{fsdos}
\eeq
{}from which it follows that
$\tilde e^i\tilde\mu^i\,=\,\cos\tilde\theta\,d\tilde\psi$. Next, let us
define the one-forms $D\tilde \mu^i$ as:
\beq
D\tilde \mu^i\,\equiv\,d\tilde \mu^i\,-\,{1\over 2}\,\epsilon_{ijk}\,
w^j\,\tilde \mu^k\,\,.
\label{fscuatro}
\eeq
It is important to point out  that the $D\tilde \mu^i$ one-forms are not
independent since $\tilde\mu^i\,D\tilde \mu^i\,=\,0$. Moreover, after some
calculation one verifies \cite{clp} that:
\beq
\,\sum_{i=1}^3\,\big(\,\tilde w^i\,-\,{1\over 2}\,w^i\,\big)^2\,=\,
\,\sum_{i=1}^3\,\big(\,D\tilde \mu^i\,\big)^2\,+\,\sigma^2\,\,,
\label{fscinco}
\eeq
where $\sigma$ is given by:
\beq
\sigma\,=\,d\tilde \varphi\,+\,\cos\tilde\theta\,d\tilde\psi\,-\,
{1\over 2}\,\tilde \mu^i\,w^i\,\,.
\label{fsseis}
\eeq
Using eq. (\ref{fscinco}) to rewrite the right-hand side of 
(\ref{brysal}), one is able to put the metric (\ref{fcdos}) in the
form (\ref{KKred}) with
$z=\tilde\varphi$. Before giving the form of the resulting D=10
supergravity background, let us write a more explicit expression for
$\big(\,D\tilde \mu\,\big)^2$,
\bear
\,\sum_{i=1}^3\,\big(\,D\tilde \mu^i\,\big)^2&=&\Bigg(\,d\tilde\theta\,-\,
\cos\tilde\psi\,{w^1\over 2}\,-\,\sin\tilde\psi\,{w^2\over 2}
\,\Bigg)^2\,\rc\rc
&&+\,\sin^2\tilde\theta\,\Bigg(\,d\tilde\psi\,+\,
\cot\tilde\theta\,\sin\tilde\psi\,{w^1\over 2}\,-\,
\cot\tilde\theta\,\cos\tilde\psi\,{w^2\over 2}\,-\,
{w^3\over 2}\Bigg)^2\,.
\label{fssiete}
\eear
If we  define $\gamma(\rho)$ as:
\beq
\gamma(\rho)\,\equiv\,{\rho^2\over 9}\,\,\big(\,1\,-\,{a^3\over
\rho^3}\,\big)\,\,,
\label{fsocho}
\eeq
then, the D=10 metric and dilaton obtained by reducing along
$\tilde\varphi$ are:
\bear
ds^2_{10}&=&\Bigg[\,{\gamma(\rho)\over H(\rho)}\,\Bigg]^{{1\over 2}}\,
\Big[\,dx^2_{1,2} + H(\rho)\,\big(\,dy^2\,+\,
{d\rho^2\over 1\,-\,{a^3\over \rho^3}}\,+\,{\rho^2\over 12}\,\sum_{i=1}^3\,
(\,w^i\,)^2\,+\,\gamma(\rho)\,\sum_{i=1}^3\,\big(\,D\tilde \mu^i\,\big)^2
\,\big)\,\Big]\,,\rc\rc
e^{\phi}&=&\Big[\,\gamma(\rho)\,\Big]^{{3\over 4}}\,
\Big[\,H(\rho)\,\Big]^{{1\over 4}}\,.
\label{fsnueve}
\eear
As the dilaton $\phi$ diverges at $\rho\rightarrow\infty$, it follows
that this solution has infinite string coupling constant. Moreover, the
RR potentials $C^{(1)}$ and $C^{(3)}$ of the type IIA theory are:
\bear
C^{(1)}&=&\cos\tilde\theta\,d\tilde\psi\,
-\,{1\over 2}\,\tilde\mu^i\,w^i\,\,,\rc\rc
C^{(3)}&=&-\Big[\,H(\rho)\,\Big]^{-1}\,\,
dx^0\wedge dx^1\wedge dx^2\,\,,
\label{fsetenta}
\eear
whose field strengths are:
\bear
F^{(2)}&=&-{1\over 2}\,\epsilon_{ijk}\,\tilde\mu^k\,
\Big[\,D\tilde\mu^i\wedge D\tilde\mu^j\,+\,{1\over 4}\,
w^i\wedge w^j\,\Big]\,\,,\rc\rc
F^{(4)}&=&\partial_{\rho}\,\Big[\,H(\rho)\,\Big]^{-1}\,\,
dx^0\wedge dx^1\wedge dx^2\wedge d\rho\,\,,
\label{fstuno}
\eear
which clearly correspond to a (D2-D6)-brane system with the D2--brane
smeared in one of the directions of the D6--brane worldvolume (\ie\ along
the $y$ direction). Three of the directions of the D6--brane are
wrapping a supersymmetric 3-cycle in a complex deformed Calabi--Yau.
Yet, the smearing in D=10 makes this solution a bit awkward from the
point of view of the AdS/CFT correspondence. Instead, we can perform a
T--duality transformation along that direction.

%%%%%%%%%%   Subsubsection 2.5.3
\subsubsection{Curved D3--branes and deformed conifold}

Notice that $\partial/\partial y$ is still a Killing vector of the D=10
metric (\ref{fsnueve}). Therefore, we can perform a T-duality
transformation along the direction of the coordinate $y$ and, in this
way, we get the following solution of the type IIB theory:
\bear
ds^2_{10}&=&\Bigg[\,{\gamma(\rho)\over H(\rho)}\,\Bigg]^{{1\over 2}}\,
\Big[\,dx^2_{1,2} + {dy^2\over\gamma(\rho)} +
H(\rho)\,\left( {d\rho^2\over 1\,-\,{a^3\over \rho^3}}\,+\,{\rho^2\over 12}
\sum_{i=1}^3\,(w^i)^2 + \gamma(\rho) \sum_{i=1}^3\,\big(\,D\tilde
\mu^i\,\big)^2 \,\right)\,\Big]\,,\rc\rc
e^{\phi}&=&\Big[\,\gamma(\rho)\,\Big]^{{1\over 2}}\,,\rc\rc
F^{(3)}&=&{1\over 2}\,\epsilon_{ijk}\,\tilde\mu^k\,
\Big[\,D\tilde\mu^i\wedge D\tilde\mu^j\,+\,{1\over 4}\,
w^i\wedge w^j\,\Big]\wedge dy\,,\rc\rc
F^{(5)}&=&\partial_{\rho}\,\Big[\,H(\rho)\,\Big]^{-1}\,
dx^0\wedge dx^1\wedge dx^2\wedge  dy\wedge d\rho\,+\,
{\rm Hodge\,\, dual}\,\,.
\label{fstdos}
\eear
The solution (\ref{fstdos}) contains a D3-brane extended along
$(x^1,x^2,y)$, with the $y$-direction distinguished from the other two.
For large $\rho$ the space transverse to the D3-brane is topologically a
cone over $S^3\times S^2$. Moreover, since $\gamma(\rho)\rightarrow 0$ as
$\rho\rightarrow a$, the $S^2$ part of the transverse space shrinks to
zero near $\rho=a$ and, thus, the transverse space has the same topology
as the deformed conifold.

%%%%%%%%%%   Subsubsection 2.5.4
\subsubsection{Type IIA background with RR fluxes}

Another possible reduction to the type IIA theory is obtained by choosing
the M-theory circle as the Hopf fiber of the three sphere $S^3$ (the one
parametrized by the one-forms $w^i\,$). In order to proceed in this way,
let us first rewrite  the seven dimensional metric (\ref{brysal}) as:
\beq
ds^2_7\,=\,{d\rho^2\over 1\,-\,{a^3\over \rho^3}}\,+\,
{\rho^2\over 12}\,\xi(\rho)\,\,\sum_{i=1}^3\,(\,\tilde w^i\,)^2\,
+\,\beta(\rho)\,\,\sum_{i=1}^3\,\big(\,
w^i\,-\,{\xi(\rho)\over 2}\,\tilde w^i\,\big)^2\,\,.
\label{fsttres}
\eeq
with $\xi(\rho)$ and $\beta(\rho)$ being:
\beq
\xi(\rho)\,\equiv\,{1-{a^3\over \rho^3}\over 1-{a^3\over 4\rho^3}}\,\,,
\,\,\,\,\,\,\,\,\,\,\,\,\,\,\,\,\,\,
\beta(\rho)\,\equiv\,{\rho^2\over 9}\,
\big(\,1\,-\,{a^3\over 4 \rho^3}\,\big)\,.
\label{fstcuatro}
\eeq
As in eq. (\ref{fsesenta}), we decompose $w^i$ as
$w^i\,=\,e^i\,+\,\mu^i\,d\varphi$. The components of $e^i$ and $\mu^i$ are
similar to the ones written in eq. (\ref{fsuno}). Moreover, if we define
the 1-forms
$D\mu^i$ as:
\beq
D\mu^i\,\equiv\,d\mu^i\,-\,{\xi(\rho)\over 2}\,\epsilon_{ijk}\,
\tilde w^j\,\mu^k\,\,,
\label{fstcinco}
\eeq
then, one can easily find expressions of the type of
eqs. (\ref{fscinco})--(\ref{fsseis}) and the D=10 solution is readily
obtained. For the metric, dilaton and RR 1-form potential one gets:
\bear
ds^2_{10}&=&\Bigg[\,{\beta(\rho)\over H(\rho)}\,\Bigg]^{{1\over 2}}\,\,
\Big[\,dx^2_{1,2}\,+\,H(\rho)\,\big(\,dy^2\,+\,
{d\rho^2\over 1\,-\,{a^3\over \rho^3}}\,+\,{\rho^2\over 12}\,\xi(\rho)\,
(\,\tilde w^i\,)^2\,+\,\beta(\rho)\,\big(\,D\mu^i\,\big)^2
\,\,\big)\,\Big]\,\,,\rc\rc
e^{\phi}&=&\Big[\,\beta(\rho)\,\Big]^{{3\over 4}}\,\,
\Big[\,H(\rho)\,\Big]^{{1\over 4}}\,\,,\rc\rc
C^{(1)}&=&\cos\theta\,d\phi\,
-\,{\xi(\rho)\over 2}\,\mu^i\,\tilde w^i\,\,,
\label{fstseis}
\eear
while the RR potential $C^{(3)}$ is the same as in eq. (\ref{fsetenta}).

%%%%%%%%%%   Subsubsection 2.5.5
\subsubsection{Curved D3--branes and resolved conifold}

We can make a T-duality transformation to the background
(\ref{fstseis}) in the direction of the coordinate $y$. The resulting
metric and dilaton are:
\bear
ds^2_{10}&=&\Bigg[\,{\beta(\rho)\over H(\rho)}\,\Bigg]^{{1\over 2}}\,\,
\Big[\,dx^2_{1,2}\,+\,{dy^2\over\beta(\rho)} \,+\,
H(\rho)\,\big(\,
{d\rho^2\over 1\,-\,{a^3\over \rho^3}}\,+\,{\rho^2\over 12}\,\xi(\rho)\,
(\,\tilde w^i\,)^2\,+\,\beta(\rho)\,\big(\,D \mu^i\,\big)^2
\,\,\big)\,\Big]\,\,,\rc\rc
e^{\phi}&=&\Big[\,\beta(\rho)\,\Big]^{{1\over 2}}\,\,,
\label{extrastseis}
\eear
which for large $\rho$ corresponds, again,  to a D3-brane with a transverse
space with the topology of a cone over $S^3\times S^2$.
However, since $\xi(\rho)$ vanishes at
$\rho=a$, in this case the $S^3$  part of the cone shrinks to zero as
$\rho\rightarrow a$ and, therefore, the transverse space has a structure
similar to the resolved conifold.

%*************************************************************

\chapter{The Maldacena-N\'u\~nez model}
\label{MNchapter}

\medskip
%****************************************************

In ref. \cite{MN}, Maldacena and N\' u\~nez proposed a duality between a
supergravity solution (previously found by  Chamseddine and Volkov
\cite{CV}) and
${\cal N}=1$ super Yang-Mills with $SU(N)$ gauge group\footnote{Other
duals of similar gauge theories have
 been proposed \cite{KS,civ,PS}.}. The setup consists in a stack of
$N$ D5-branes wrapping a (compact) supersymmetric two-cycle inside a 
(non-compact) Calabi-Yau three-fold. The Calabi-Yau is 
1/4 supersymmetric and the presence of D5-branes further halves
the number of susys (and also spoils conformal symmetry) leaving
a total of 4 supercharges.

Then, if one looks at the low energy dynamics of open strings
on the D5-branes (discarding Kaluza-Klein modes and stringy excitations),
one finds a Yang-Mills theory living in the 1+3 unwrapped dimensions.
On the other hand, the closed string dynamics shows up in the generated
supergravity background. Hence, one can think of an
open/closed string duality which becomes a gauge/gravity duality, 
in the same spirit as AdS/CFT (although the dualities with reduced
supersymmetry are never as clean as the original AdS/CFT). As in that
case, the relation is  holographic, and the non-compact direction of the
Calabi-Yau  plays the r\^ole of the energy scale
of the gauge theory.

More concretely, we start with D5-branes wrapped in the finite,
topologically non-trivial two-cycle of a resolved conifold.
The backreaction of the branes deforms the geometry, and one
has a geometric transition as those studied in
\cite{gopavafa,vafa}. The final geometry, which is described by the
solution of next section, is topologically like a deformed conifold,
where the
$S^2$ is contractible but the
$S^3$ is not. Moreover, the branes disappear and are replaced by fluxes.
The total RR-charge associated to these fluxes is that of the initial
number of branes.

The Lorentz symmetry of the configuration is
$SO(1,3)\times SO(2)\times SO(4)$. In order to keep the
desired amount of supersymmetry, one must perform the  twisting, {\it
i.e.}, appropriately embed the spin connection on the $SO(2)$ inside the
$SO(4)=SU(2)\times SU(2)$. This will be explicitly done in the
next section.

The degrees of freedom of $D=4$,  ${\cal N}=1$ SYM can be arranged
into a vector multiplet composed by a gauge vector field $A_\mu$ (two
on-shell bosonic degrees of freedom) and a Majorana spinor $\la$
(two on-shell fermionic degrees of freedom), both of
them transforming in the adjoint representation of the gauge group.
An important difference with other gauge theories with more supersymmetry
is that there are no scalars, so there is no moduli space.
This theory has some similarities with QCD and, therefore, it can be used
to address, in a simpler context, some phenomena like confinement.

In section \ref{secMNsol}, the derivation of the sugra solution from a
supersymmetry analysis will be explained thoroughly. The explicit
expression for the Killing spinors will be found. In section 
\ref{MNreview}, we  briefly review how some aspects of the
field theory can be read from the supergravity solution.
In chapter \ref{flavoring}, supersymmetric probes in the background
are studied in detail, and it is argued that their presence
is dual to the addition of flavor to the dual gauge theory.

%****************************************************

\setcounter{equation}{0}
\section{Supersymmetry and the gravity solution}
\label{secMNsol} 

The supergravity solution which is the topic of this
chapter was first found by Chamseddine and Volkov \cite{CV} by
studying  non-abelian supersymmetric monopoles in $D=4$ gauged
supergravity. In ten dimensions, it represents  D5-branes
wrapping a two-cycle inside a resolved conifold.
Therefore, in the spirit of previous chapters, the natural
supergravity where one should try to find the solution is
$D=7$, where a 5-brane is a domain wall.
 The BPS equations are obtained in section \ref{FMN7dim}
from the seven dimensional point of view
and a neat expression for the Killing spinors is given.
Then, in \ref{FMN10D}, the process is repeated in $D=10$
type IIB supergravity. The interest of this repetition is 
two-fold: the relation between 7d and 10d supersymmetry is 
clearly seen and the ten dimensional Killing spinors 
(which are useful for the gauge-gravity correspondence)
are found. For completeness, in \ref{FMNsolve} the integration
of the equations is  explicitly carried out,  following the
steps of \cite{CV}.

%****************************************************

\subsection{$D=7$ supersymmetry analysis}
\label{FMN7dim}

The aim is to describe the solution using the seven
dimensional supergravity of section \ref{TVN}. The natural
ansatz for the metric of a 5-brane wrapping an $S^2$ reads
(string frame):
\beq
ds_7^2\,=\,dx_{1,3}^2\,+\,e^{2h}\,(d\theta^2\,+\,
\sin^2\theta d\varphi^2)\,+\,dr^2\,\,.
\label{FMNmetric7}
\eeq
The other excited degrees of freedom are the dilaton
$\phi_7$ (the subindex is to remind that this is the
seven dimensional dilaton, which will be different from
the IIB dilaton that will appear in section \ref{FMN10D})
and the $SU(2)$ gauge field. Its ansatz is:
\beq
A^1\,=\,-a(r) d\theta\,,
\,\,\,\,\,\,\,\,\,
A^2\,=\,a(r) \sin\theta d\varphi\,,
\,\,\,\,\,\,\,\,\,
A^3\,=\,- \cos\theta d\varphi\,.
\label{Foneform}
\eeq
Notice the similarity \footnote{Different signs are 
related to the different signs in conventions for the
uplifting formulae.}
 with (\ref{conigauge}). Like
there, the $A^3$ component is uniquely determined by the twisting
condition, and the $A^1$ and $A^2$
play the r\^ole of smoothing the singularity in the 
infrared, in complete analogy to what happens with the 
resolution of the Dirac string by the 't Hooft-Polyakov
monopole. The field strength, calculated with the
expression (\ref{gaugefstr}) is:
\beq
F^1\,=\,-a'\,dr\wedge d\theta\,\,,
\,\,\,\,\,\,\,\,\,\,
F^2\,=\,a'\sin\theta dr\wedge d\varphi\,\,,
\,\,\,\,\,\,\,\,\,\,
F^3\,=\,(\,1-a^2\,)\,\sin\theta d\theta\wedge d\varphi\,\,.
\label{FMNfstr}
\eeq
As $F\wedge F=0$, the 3-form potential $B$ can be consistently
taken to vanish. We want to impose $\delta\la=
\delta\psi_\mu=0$ in (\ref{susyTVN}). First of all, the
angular projection is needed:
\beq
\G_{\theta\varphi}\,\epsilon\,=\,\sigma^1\sigma^2\,\epsilon\,\,.
\label{FMNangproj7}
\eeq
By using eq. (\ref{FMNangproj7}), the dilatino equation reduces to:
\bear
\phi_7'\,\epsilon\,+\,\big(\,1\,+\,{e^{-2h}\over 4}\,(a^2-1)\,\big)\,
\Gamma_r\,\epsilon\,-\,{1\over 2}\,a'\,e^{-h}\,
\Gamma_\theta\,i\,\sigma^1\,\epsilon\,=\,0\,\,,
\label{Fdilatino7}
\eear
while $\delta\psi_\theta=\delta\psi_\varphi=0$ yield:
\beq
h'\epsilon\,-\,{1\over 2}\,a'\,e^{-h}\,\Gamma_\theta
\,i\,\sigma^1\,\epsilon\,+\,
a\,e^{-h}\,\Gamma_r\G_\theta\,i\,\sigma^1\,\epsilon\,+\,
{1\over
2}\,(a^2-1)\,e^{-2h}\,\Gamma_r\,\epsilon\,=\,0\,\,.
\label{Fgravitino7}
\eeq
From the transformation of the radial component of the
gravitino $\delta\psi_r=0$, one just gets:
\beq
\partial_r\epsilon\,=\,{1\over 2}\,a'\,e^{-h}\,\Gamma_\theta
\,i\,\sigma^1\,\epsilon\,\,.
\label{Fpsir7}
\eeq
From (\ref{Fdilatino7}), we find a rotated projection for
the Killing spinor, in full analogy with (\ref{conirproj}) or
(\ref{catorce}). 
\beq
\Gamma_{r}\,\epsilon\,=\,
\big(\,\beta+\tilde\beta\,\Gamma_\theta\,i\,\sigma^1
\,\big)\,\epsilon\,\,,
\label{Fradial7}
\eeq
where $\beta$ and $\tilde\beta$ can be read from eq. (\ref{Fdilatino7}),
namely:
\beq
\beta\,=\,{-\phi_7'\over 1+{e^{-2h}\over 4}\,(a^2-1)}\,\,,
\,\,\,\,\,\,\,\,\,\,\,
\tilde\beta\,=\,{1\over 2}\,
{e^{-h}a'\over 1+{e^{-2h}\over 4}\,(a^2-1)}\,\,.
\label{Fbeta}
\eeq
On the other hand, it is easy to check that the consistency
condition is the same of previous cases
$\beta^2+\tilde\beta^2=1$,
and, therefore, we can represent again $\beta$ and $\tilde\beta$ as:
\beq
\beta=\cos\alpha\,,\,\,\,\,\,\,\,\,\,\,\,
\tilde\beta=\sin\alpha\,\,.
\label{FFalfa}
\eeq
Substituting the radial projection (eq. (\ref{Fradial7})) into eq.
(\ref{Fgravitino7}), and considering the terms with and without 
$\Gamma_\theta i\sigma^1$, we get the following two equations:
\beq
h'\,=\,-{1\over 2}\,e^{-2h}\,(a^2-1)\,\beta-ae^{-h}\tilde\beta\,\,,
\,\,\,\,\,\,\,\,\,\,\,
a'\,=\,-2a\beta\,+\,e^{-h}\,(a^2-1)\,\tilde\beta\,\,.
\label{Fgravitinodos}
\eeq
By using the definition of $\beta$  and $\tilde\beta$ (eq. (\ref{Fbeta}))
into the second equation  in (\ref{Fgravitinodos}), we get the following
relation between
$\phi_7'$  and $a'$:
\beq
\phi_7'\,=\,{a'\over 2a}\,\Big[\,1\,-\,{1\over 4}\,e^{-2h}\,
(a^2-1)\,\Big]\,\,.
\label{Fphia}
\eeq
Furthermore, 
from the condition $\beta^2+\tilde\beta^2=1$ , one obtains a new relation
between
$\phi_7'$ and $a'$, namely:
\beq
\phi_7'^2\,+\,{1\over 4}\,e^{-2h}\,a'^2\,=\,
[\,1\,+\,{1\over 4}\,e^{-2h}\,(\,a^2-1\,)\,]^2\,\,.
\label{Fsquare}
\eeq
By combining eqs. (\ref{Fphia}) and  (\ref{Fsquare}) one can get 
the expression of $\phi_7'$ and $a'$ in terms of $a$ and $h$. 
Moreover, by using these results in eq. (\ref{Fbeta}),
one can get
$\beta$ and $\tilde\beta$ as functions of $a$ and $h$ and, by plugging the
corresponding expressions on the first eq. in (\ref{Fgravitinodos}), 
one can
obtain the differential equation for $h$. In order to write
 these expressions
in a compact form, let us define:
\beq
Q\,\equiv\,\sqrt{e^{4h}\,+\,{1\over 2}\,e^{2h}\,(a^2+1)\,+\,{1\over 16}\,
(a^2-1)^2}\,\,.
\label{FdefQ}
\eeq
Then, one has the following system of first-order differential 
equations for
$\phi_7$, $h$ and $a$:
\bear
\phi_7'&=&-{1\over Q}\,
\Big[\,e^{2h}\,-\,{e^{- 2h}\over 16}\,(a^2-1)^2\,\Big]\,\,,\rc\rc
h'&=&{1\over 2Q}\,
\Big[\,a^2+1+{e^{- 2h}\over 4}\,(a^2-1)^2\,\Big]\,\,,\rc\rc
a'&=&-{2a\over Q}\,
\Big[\,e^{ 2h}+{1\over 4}\,(a^2-1)\,\Big]\,\,,
\label{Fdifferentialeqs}
\eear 
and the values of $\beta=\cos\alpha$ and $\tilde\beta=\sin\alpha$,
 which are given by:
\beq
\sin\alpha\,=\,-{ae^h\over Q}\,\,,
\,\,\,\,\,\,\,\,\,\,\,\,
\cos\alpha\,=\,{e^{2h}\,-\,{1\over 4}\,(\,a^2-1\,)\over Q}\,\,.
\label{Falpha}
\eeq
It is interesting to notice that, when solving the quadratic eq. 
(\ref{Fsquare})
to obtain (\ref{Fdifferentialeqs}) and (\ref{Falpha}), we have a 
sign ambiguity.
We have fixed this sign by requiring that $h'$ is always positive.  
It remains to verify the fulfillment of equation (\ref{Fpsir7}). Notice,
first of all, that the radial projection (\ref{Fradial7}) can be written
as:
\beq
\Gamma_r\,\,\epsilon\,=\,e^{\alpha\,\Gamma_\theta\,i\,\sigma^1}\,
\epsilon\,\,,
\eeq
which, after taking into account that 
$\{\Gamma_{r},\Gamma_\theta\,i\,\sigma^1\}=0$,  can be solved as:
\beq
\epsilon\,=\,e^{-{\alpha\over
2}\,\Gamma_\theta\,i\,\sigma^1}\,\epsilon_0\,\,,
\,\,\,\,\,\,\,\,\,\,\,\,\,\,\,\,\,\,\,\,
\Gamma_r\,\epsilon_0=\epsilon_0\,\,.
\label{Fkilling7}
\eeq
Finally, inserting this parametrization of $\epsilon$ into eq.
(\ref{Fpsir7}), we get:
\beq
\alpha'\,=\,-e^{-h}\,a'\,\,.
\label{Fphase}
\eeq
 But, by differentiating eq.
(\ref{Falpha}) and using eq. (\ref{Fdifferentialeqs}), one can verify
that  (\ref{Fphase}) is automatically satisfied. 

In summary, Eq. (\ref{Fdifferentialeqs}) is  a system of first-order 
differential equations whose solution determines the metric, dilaton and
RR three-form of the background. The explicit expression of the 
Killing spinor can be read from (\ref{Fkilling7}), where
$\epsilon_0$  must also fulfil the projection
(\ref{FMNangproj7}). Clearly, this solution preserves four
supersymmetries.

%****************************************************

\subsection{$D=10$ supersymmetry and solution}
\label{FMN10D}

The type IIB ten dimensional metric corresponding to
the analysis of the previous section, after performing an 
S-duality transformation in
(\ref{upliftTVN}), is (Einstein frame):
\beq
ds^2_{10}\,=\,e^{{\phi\over 2}}\,\,\Big[\,
dx^2_{1,3}\,+\,e^{2h}\,\big(\,d\theta^2+\sin^2\theta d\varphi^2\,\big)\,+\,
dr^2\,+\,{1\over 4}\,(\om^i-A^i)^2\,\Big]\,\,,
\label{Fmetric}
\eeq
where the $\om^i$ were defined in
(\ref{otherom}) and $\phi$ is the dilaton:
\beq
\phi\,=\,-\,\phi_7\,\,.
\eeq
 The unwrapped coordinates $x^{\mu}$ 
 have been
rescaled and all distances are measured in units of $N g_s\alpha'$.
The solution of the
type IIB supergravity includes a
Ramond-Ramond three-form $F_{(3)}$ given by:
\beq
F_{(3)}\,=\,-{1\over 4}\,\big(\,\om^1-A^1\,\big)\wedge 
\big(\,\om^2-A^2\,\big)\wedge \big(\,\om^3-A^3\,\big)\,+\,{1\over 4}\,\,
\sum_a\,F^a\wedge \big(\,\om^a-A^a\,\big)\,\,,
\label{FRRthreeform}
\eeq
where $A$ and $F$ are the ones in (\ref{Foneform}), 
(\ref{FMNfstr}). As usual, we want to plug this ansatz
in the corresponding susy transformations (\ref{SUSYIIB})
and enforce $\delta\lambda=\delta\psi_{\mu}=0$.
For the metric ansatz of eq. (\ref{Fmetric}), 
let us consider the frame:
\bear
e^{x^i}&=&e^{{\phi\over 4}}\,d x^i\,\,,
\,\,\,\,\,\,\,(i=0,1,2,3)\,\,,\rc\rc
e^{1}&=&e^{{\phi\over 4}+h}\,d\theta\,\,,
\,\,\,\,\,\,\,\,\,\,\,\,\,\,
e^{2}=e^{{\phi\over 4}+h}\,\sin\theta d\varphi\,\,,\rc\rc
e^{r}&=&e^{{\phi\over 4}}\,dr\,\,,
\,\,\,\,\,\,\,\,\,\,\,\,\,\,
e^{\hat i}={e^{{\phi\over 4}}\over 2}\,\,
(\,\om^i\,-\,A^i\,)\,\,,\,\,\,\,\,\,\,(i=1,2,3)\,\,.
\label{Fframe}
\eear
The projection condition corresponding to the SUSY two-cycle reads:
\beq
\Gamma_{12}\,\epsilon\,=\,\hat\Gamma_{12}\,\epsilon\,\,,
\label{Fprojectionone}
\eeq
This is the ten dimensional equivalent of (\ref{FMNangproj7}).
Furthermore, the following projection is also needed:
\beq
\epsilon\,=\,i\epsilon^*\,\,.
\label{Fprojectiontwo}
\eeq
From the seven dimensional point of view, this was imposed
from the beginning (remember that the gauged supergravity we
are using only has half of the maximal susy). Then, the
supersymmetry calculation runs in complete analogy to section
\ref{FMN7dim}. The analogous to (\ref{Fradial7}) is:
\beq
\Gamma_{r}\hat\Gamma_{123}\,\epsilon\,=\,
\big(\,\beta+\tilde\beta\,\Gamma_2\hat\Gamma_2
\,\big)\,\epsilon\,\,,
\label{Fradial}
\eeq
where $\beta$, $\tilde\beta$ are the same as in eq. (\ref{Fbeta}).
Parametrizing them as in eq. (\ref{FFalfa}), we can rewrite
(\ref{Fradial}) as:
\beq
\Gamma_r\,\hat\Gamma_{123}\,\epsilon\,=\,e^{\alpha\Gamma_2\hat\Gamma_2}\,
\epsilon\,\,,
\eeq
which, after taking into account that 
$\{\Gamma_{r}\hat\Gamma_{123},\Gamma_2\hat\Gamma_2\}=0$,  can be solved as:
\beq
\epsilon\,=\,e^{-{\alpha\over 2}\,\Gamma_2\hat\Gamma_2}\,\,\epsilon_0\,\,,
\,\,\,\,\,\,\,\,\,\,\,\,\,\,\,\,\,\,\,\,
\Gamma_r\,\hat\Gamma_{123}\,\epsilon_0=\epsilon_0\,\,.
\eeq
Moreover, from the transformation of the radial component of the
dilatino, an additional equation appears, governing the radial
dependence of the spinor:
\beq
\partial_r\epsilon_0-{1\over 8}\,\phi'\,\epsilon_0\,=\,0\,\,,
\,\,\,\,\,\,\,\,\,\,\,\,\,\,\,\,\,\,\,\,
\eeq
Thus, the explicit form of the ten dimensional Killing spinor is:
\beq
\epsilon\,=\,e^{{\alpha\over 2}\,\Gamma_1\hat\Gamma_1}\,\,
e^{{\phi\over 8}}\,\,\eta\,\,,
\eeq
where $\eta$ is a constant spinor satisfying:
\beq
\Gamma_{x^0\cdots x^3}\,\Gamma_{12}\,\eta\,=\,\eta\,\,,
\,\,\,\,\,\,\,\,\,\,\,\,
\Gamma_{12}\,\eta\,=\,\hat\Gamma_{12}\,\eta\,\,,
\,\,\,\,\,\,\,\,\,\,\,\,
\eta\,=\,i\eta^*\,\,.
\eeq
We have made use of the fact that $\epsilon$ is a
spinor of definite chirality of type IIB supergravity, 
so it satisfies  $\Gamma_{x^0\cdots
x^3}\Gamma_{12}\Gamma_{r}\hat\Gamma_{123}\epsilon=\epsilon$. 
If we multiply the
radial projection condition (\ref{Fradial}) by  
$\Gamma_{x^0\cdots x^3}\Gamma_{12}$, we
obtain a expression that will be useful for the kappa
symmetry analysis that will be carried out in the next chapter:
\beq
\Gamma_{x^0\cdots x^3}\,\big(\,
\cos\alpha\,\Gamma_{12}\,+\,\sin\alpha\,\Gamma_1\hat\Gamma_2\,)\,
\epsilon\,=\,\epsilon\,\,.
\label{Falphaproj}
\eeq

\subsubsection{The explicit solution}

By solving the system (\ref{Fdifferentialeqs})
(and taking into account $\phi_7=-\phi$),
one gets (see next section):
\bear
a(r)&=&{2r\over \sinh 2r}\,\,,\rc\rc
e^{2h}&=&r\coth 2r\,-\,{r^2\over \sinh^2 2r}\,-\,
{1\over 4}\,\,,\rc
e^{-2\phi}&=&e^{-2\phi_0}{2e^h\over \sinh 2r}\,\,,
\label{FMNsol}
\eear
where $\phi_0$ is the value of the dilaton at $r=0$. Near the origin $r=0$ the function 
$e^{2h}$ behaves as $e^{2h}\sim r^2$ and the metric is non-singular.
 By plugging in eq. (\ref{FdefQ}) the values of $h$ and
$a$ given in eq. (\ref{FMNsol}),  one verifies that
\beq
Q=r\,\,.
\eeq
Then, one gets the following simple expression for $\cos\alpha$:
\beq
\cos\alpha\,=\,{\rm \coth} 2r\,-\,{2r\over \sinh^22r}\,\,.
\label{Falphaexplicit}
\eeq
It is interesting to write here the UV and
IR limits of $\alpha$, namely
\beq
\lim_{r\rightarrow \infty}\,\alpha\,=\,0\,\,,
\,\,\,\,\,\,\,\,\,\,\,\,\,\,\,\,\,\,\,\,
\lim_{r\rightarrow 0}\,\alpha\,=\,-{\pi\over 2}\,\,.
\eeq

The BPS equations (\ref{Fdifferentialeqs}) also admit a  solution in which
the function
$a(r)$ vanishes, \ie\ in which the one-form $A^i$ has  only one
non-vanishing component, namely $A^{3}$. We will refer to this solution
as the abelian ${\cal N}=1$ background, and it is important as it
corresponds to the UV limit of the associated gauge theory. Its explicit
form can be easily obtained by taking the
$r\rightarrow\infty$ limit of the functions given  in eq. (\ref{FMNsol}).
Notice that, indeed $a(r)\rightarrow 0$ as $r\rightarrow\infty$ in  eq.
(\ref{FMNsol}).  Neglecting exponentially suppressed terms, one gets:
\beq
e^{2h}\,=\,r\,-\,{1\over 4}\,\,,
\,\,\,\,\,\,\,\,\,\,\,\,\,\,\,\,\,\,(a=0)\,\,,
\eeq
while $\phi$ can be obtained from the last equation  in (\ref{FMNsol}).
The metric of the abelian background is singular at $r=1/4$ (the position
of the singularity can be moved to $r=0$ by a redefinition of the radial
coordinate). This IR singularity of the abelian background is removed in
the non-abelian metric by switching on the $A^1, A^2$ components of the
one-form (\ref{Foneform}). Moreover, when $a=0$, the angle
$\alpha$ appearing in the expression of the Killing  spinors is zero, as 
follows from eq. (\ref{Falpha}). 

\medskip
\medskip

To finish this section, the potentials associated to the 
RR 3-form field strength and its Hodge dual will be given.
Since $dF_{(3)}=0$, one can represent $F_{(3)}$ in terms of a two-form potential 
$C_{(2)}$ as $F_{(3)}\,=\,dC_{(2)}$. Actually, it is not difficult to verify that 
$C_{(2)}$ can be taken as:
\bear
C_{(2)}&=&{1\over 4}\,\Big[\,\tilde\psi\,(\,\sin\theta d\theta\wedge
d\varphi\,-\,
\sin\tilde\theta d\tilde\theta\wedge d\tilde\varphi\,)
\,-\,\cos\theta\cos\tilde\theta d\varphi\wedge d\tilde\varphi\,-\rc\rc
&&-a\,(\,d\theta\wedge \om^1\,-\,\sin\theta d\varphi\wedge
\om^2\,)\,\Big]\,\,.
\eear
Moreover, the equation of motion of $F_{(3)}$ in the Einstein frame is
$d\Big(\,e^{\phi}\,{}^*F_{(3)}\,\Big)=0$, where $*$ denotes Hodge duality. Therefore
it follows that, at least locally, one must have:
\beq
e^{\phi}\,{}^*F_{(3)}\,=\,d C_{(6)}\,\,,
\eeq
with $C_{(6)}$ being a six-form potential. It is readily checked that 
$C_{(6)}$ can be taken as:
\beq
C_{(6)}\,=\,dx^0\wedge dx^1\wedge dx^2\wedge dx^3\wedge
{\cal C}\,\,,
\eeq
where ${\cal C}$ is the following two-form:
\bear
{\cal C}&=&-{e^{2\phi}\over 8}\,\,\Big[\,
\Big(\,(\,a^2-1\,)a^2\,e^{-2h}\,-\,16\,e^{2h}\,\Big)\,\cos\theta
d\varphi\wedge dr
\,-\,(\,a^2-1\,)\,e^{-2h}\,\om^3\wedge dr\,+\rc\rc
&&+\,a'\,\Big(\,\sin\theta d\varphi\wedge \om^1\,+\,d\theta\wedge
\om^2\,\Big)\,\Big]\,\,.
\eear

%****************************************************

\subsection{Integrating the equations}
\label{FMNsolve}

For the sake of completeness, the procedure for integrating
the system (\ref{Fdifferentialeqs})  is briefly described
in this section \cite{CV}.

The idea is to divide the equation for $h'$ by that for
$a'$ in order to have a first order equation for $\frac{dh}{da}$
in which the annoying factor $Q$ has disappeared. To use the same
notation as \cite{CV}, let us define:
\beq
x\equiv a^2\,\,,
\,\,\,\,\,\,\,\,\,\,\,\,\,\,
R^2\equiv 4\,e^{2h}\,\,.
\label{Fchamsdef}
\eeq
Then, eqs. (\ref{Fdifferentialeqs}) yield:
\beq
x\,(R^2+x-1)\,\frac{d(R^2)}{dx}\,+\,R^2\,(x+1)\,+\,
(x-1)^2\,=\,0\,\,.
\label{Fchams}
\eeq
Eq. (\ref{Fchams}) can be drastically simplified by using
the parametrization:
\beq
x\,=\,\rho^2\,e^{\xi(\rho)}\,\,,
\,\,\,\,\,\,\,\,\,\,\,\,\,\,
R^2\,=\,-\rho\frac{d\xi(\rho)}{d\rho}\,-\,
\rho^2\,e^{\xi(\rho)}\,-\,1\,\,,
\label{Fchamssol}
\eeq
so (\ref{Fchams}) reduces to:
\beq
{d^2\xi(\rho) \over d\rho^2}\,=\,2e^{\xi(\rho)}\,\,.
\eeq
Up to a meaningless constant, which cancels out in the
final expressions, the physical 
solution\footnote{$\xi(\rho)=-2\log(\sin(\rho-\rho_0)) $
also solves this equation, but then, $R^2$ becomes negative.} is:
\beq
\xi(\rho)=-2\log(\sinh (\rho-\rho_0))\,\,.
\label{Fchamsxi}
\eeq
It is not difficult to find the relation between $\rho$ and
the original radial variable $r$: $\rho=2r +c$, where $c$ is
a new integration constant that  is only a 
redefinition of the origin of $r$. Taking $c=0$, and 
substituting (\ref{Fchamsxi}) into (\ref{Fchamssol}), and
back into (\ref{Fchamsdef}), one can easily get:
\bear
a(r)&=&{2r\over \sinh 2(r-r_0)}\,\,,\rc\rc
e^{2h}&=&r\coth 2(r-r_0)\,-\,{r^2\over \sinh^2 2(r-r_0)}\,-\,
{1\over 4}\,\,,
\eear
which clearly is (\ref{FMNsol}) when $r_0$ is taken to zero.
On the other hand, this is the only way of having a 
smooth metric at the origin.
Once these functions are known, the dilaton can be
immediately found by direct integration (see (\ref{FMNsol})).

%****************************************************

\setcounter{equation}{0}
\section{Achievements of the 
Maldacena-N\'u\~nez model}
\label{MNreview}

In order to verify that the Maldacena-N\'u\~nez  solution is dual to
${\cal N}=1$ SYM, we should be able to find the gauge theory
information encoded by this background.
This section will be devoted to making a brief overview of the
literature on the subject and it will be shown how  
many field theory features have been successfully addressed
from the gravitational perspective. Reviews on this topic
can be found in \cite{MerReview,n1n2,Merlatti,imeroni,DV04}.
Only the ideas and results will be discussed, without getting
into deep details, the goal being just to give some insight on 
the duality. Problems like decoupling limits, comparison of
scales and validity
regimes will not be covered at all. 

It is also worth pointing out that generalizations of
this duality have been explored: the dual of non-commutative
${\cal N}=1$ SYM was constructed in \cite{mpt} while
scenarios where supersymmetry is softly broken were
considered in \cite{ass}. Moreover, some aspects regarding the
complex geometry of the solution were studied in 
\cite{pt}.

%****************************************************
\subsubsection{Confinement and magnetic monopoles}

The quark-antiquark potential is basically  the energy
of a fundamental string extended along one of the $x$ directions
where the gauge theory lives. The action for such a string is
given by the Nambu-Goto action\footnote{see however the
subsection below on string tensions for a more detailed analysis.},
 $S=(2\pi\alpha')^{-1}
\int d\tau d\sigma \sqrt{-{\mathrm det}\, g_{ab}}$, where
$g_{ab}$ is the pull-back of the string frame metric on
the worldvolume of the string.
Looking at the metric (\ref{Fmetric}), one can immediately
see that the string will prefer to stretch out sitting at $r=0$, 
as the value of $e^{\phi}$ is minimum there (see eq. (\ref{FMNsol})). Its
action reads  (note that (\ref{Fmetric}) is in Einstein frame, and to go 
to string frame, it must be multiplied overall by $e^{\phi/2}$):
\beq
S={e^{\phi_0} \over 2\pi\alpha'}\int dx \,dt\qquad
\Rightarrow \qquad
T_s={e^{\phi_0} \over 2\pi\alpha'}\,\,.
\label{confine}
\eeq
Therefore, the string tension does not vanish and the theory is
confining.

As stated in \cite{MN}, magnetic monopole sources correspond
to D3-branes wrapping an $S^2$ and extending in the radial 
direction. The monopole-antimonopole potential is similar to
the quark-antiquark one, but in the action one must now further
multiply by the volume of the $S^2$. But this sphere shrinks
in the $r=0$ limit, so the tension of the monopole-antimonopole
string vanishes. Therefore, they are screened, not confined.

\subsubsection{$U(1)_R$ symmetry breaking, instantons and the gluino
condensate}

The action of $SU(N)$ ${\cal N}=1$ SYM has an $U(1)_R$ symmetry at the 
classical level, which amounts to giving a phase to the gluino
field:
\beq
\la\rightarrow e^{-i \varepsilon}\la\,\,,
\eeq
where we define the parameter $\varepsilon \in [0,2\pi)$.
However, at the quantum level, this symmetry is anomalous. From
instanton calculation, it can be proved  that the Yang-Mills
angle gets modified:
\beq
\theta_{YM}\rightarrow \theta_{YM}+2 N \varepsilon\,\,.
\eeq
The transformation is a symmetry of the theory only if 
the $\theta_{YM}$ does not get modified. Being defined with
periodicity $2\pi$, one needs $\theta_{YM}\rightarrow 
\theta_{YM}+2n\pi$, where $n$ is some integer. So the
parameter $\varepsilon$ can take the values
$\varepsilon=n\pi/N$ with $n=0,\dots,2N-1$, and $U(1)_R$ gets
broken down to $\ZZ_{2N}$. Furthermore, it is known that 
this symmetry group is spontaneously broken 
to $\ZZ_2$ in the IR because of the
formation of a gluino condensate $<\la^2>$, whose transformation
reads $<\la^2>\rightarrow e^{-2i\pi n/N}<\la^2>$, so only two
values of $n$ leave it unchanged. Hence, the gauge theory has $N$
inequivalent vacua.

All this symmetry breaking can be nicely found in the gravity
solution \cite{MN}. In the UV (\ie\ when $a=0$), there is an isometry 
of the metric (\ref{Fmetric}):
\beq
\tilde\psi\rightarrow\tilde\psi+2\varepsilon\,\,.
\label{psitrans}
\eeq
We have written $2\varepsilon$ in (\ref{psitrans}) so taking $\varepsilon$
from 0 to $2\pi$ corresponds to a period in $\tilde\psi$.
This shift in $\tilde\psi$ is the gravitational counterpart of the
$U(1)_R$. However, (\ref{psitrans}) is not a symmetry of the full
solution since it changes $C_{(2)}$. Changing $\tilde\psi$
amounts to adding a closed, but not exact, form to the
potential $C_{(2)}$. This is a large gauge transformation,
which is generically quantized.

This effect can be seen quantitatively by obtaining
 the explicit expression
of the $\theta_{YM}$ angle in the gravity approach. With this purpose, let
us consider a D5-brane probe wrapping the $S^2$ sphere of the geometry
with a worldvolume gauge field strength $F$ excited \cite{DV}. The 
quadratic action for this
$F$ will be the bosonic action of the gauge field of
${\cal N}=1$ SYM, after integrating over the $S^2$. The probe
action is a sum of a Born-Infeld and a Wess-Zumino term:
\beq
S=-T_5\,\int d^6 \sigma\,e^{-\phi}\sqrt{g_{\mathrm str}+2\pi\alpha'F}
\,+\,T_5\int C\wedge e^{2\pi\alpha' F}\,\,.
\eeq
By inserting the solution and comparing with the bosonic action:
\beq
S_{YM}\,=\,-\,{1 \over 4\, g_{YM}^2}\int d^4x\,F_{\alpha\beta}^A\,
F^{\alpha\beta}_A\,+\,{\theta_{YM} \over 32\pi^2}\int d^4x\,
F_{\alpha\beta}^A\,
{}^*F^{\alpha\beta}_A\,\,,
\eeq
we obtain the following expressions (the last term comes from
the $C_{(2)}\wedge F\wedge F$ coupling):
\bear
{1 \over g_{YM}^2}&=&{1\over 2(2\pi)^3\alpha'g_s}
\int_{S^2} e^{-\phi}\sqrt{{\mathrm det}\,G}\rc\rc
\theta_{YM}&=&{1 \over 2\pi\,\alpha'g_s}\int_{S^2}C_{(2)}\,\,,
\label{gym}
\eear
where $G$ is the induced metric on the $S^2$. We have made use
of the expression for the tension of a D5 brane: 
$T_5=\left((2\pi)^5g_s\alpha'^3\right)^{-1}$. 
(\ref{gym}) can also be obtained by considering  an instanton in the
gravitational setup, which is an euclidean D1-brane wrapping the same
$S^2$ \cite{MN}, and
comparing its action to the field theory instanton action
\cite{loewy,MerReview}. 

In order to perform the explicit integration of (\ref{gym}), we
need the parametrization of the two-sphere. There are two
equivalent choices for this cycle \cite{BertMer}\footnote{A first choice
for this cycle \cite{DV} was to take constant $\tilde\theta$,
$\tilde\varphi$ and $\tilde\psi$. It was corrected in \cite{BertMer} after
some problems with the beta function were pointed out in
\cite{sannino}.}. Notice that
the $S^2$ shrinks as $r\rightarrow 0$:
\bear
&&\tilde\theta=\pi-\theta\,\,,\qquad\qquad
\tilde\varphi=\varphi\,\,,\qquad\qquad\,\,\,\,\,\,\,\,\,\,\,\,
\tilde\psi=\tilde\psi_0\,\,\rc
&&\tilde\theta=\theta\,\,,\qquad\qquad\,\,\,\,\,\,\,\,\,\,
\tilde\varphi=2\pi-\varphi\,\,,\qquad\qquad
\tilde\psi=\tilde\psi_0\,\,.
\label{bmcycles}
\eear
Inserting it in (\ref{gym}), one gets the gauge coupling (we
now insert in the solution the factor $Ng_s\alpha'$ that
 had not been considered up to now and  which comes
from the quantization of the RR charge corresponding to $N$ D5-branes. It
multiplies the RR forms and all the components of the metric except those
on the 1+3 unwrapped $x$ directions):
\beq
{1\over g_{YM}^2}\,=\,{N \over 4\,\pi^2} \,r\,\tanh r\,\,,
\label{gymder}
\eeq
and, at large r, the value of the Yang-Mills angle:
\beq
\theta_{YM}\,=\,N\,\tilde\psi_0\,\,.
\eeq
Now, imposing that $\theta_{YM}\rightarrow 
\theta_{YM}+2n\pi$ under (\ref{psitrans}) transformations,
the parameter can only be $\varepsilon=n\pi/N$ with $n=0,\dots,2N-1\,\,$,
in perfect agreement with the field theory analysis. 
Therefore, we have found the UV quantum anomaly $U(1)_R\rightarrow
\ZZ_{2N}$ from the gravity approach. 

By considering the full solution up to the IR, which amounts to taking
$a\neq 0$, the only surviving symmetry from the initial $U(1)_R$
is $\tilde\psi\rightarrow \tilde\psi+2\pi$ and
$\tilde\psi$ has only two possible values for each case 
in (\ref{bmcycles}). The function
$a$ plays the same r\^ole as the gluino condensate in the
spontaneous breaking to $\ZZ_2$, so it is natural to think
that it is its gravitational counterpart (the same conclusion
can be reached by looking for the fields to which $a$ 
couples \cite{apreda}):
\beq
<\la^2>\,\leftrightarrow a(r)\,\,.
\label{gaugid}
\eeq
We see that the function $a(r)$, which is needed for the sugra
solution to be smooth when $r\rightarrow 0$, is also responsible of
the gravity description of non-trivial gauge theory effects
in the IR, \ie\ of the $\ZZ_{2N}\rightarrow\ZZ_2$ spontaneous R-symmetry
breaking.

\subsubsection{The $\beta$-function}

Eq. (\ref{gymder}) shows the evolution of the coupling constant
in terms of the radial variable $r$. From general grounds in
holography, we know that it has to be related to the energy
scale of the gauge theory. Large values of $r$ correspond to the
UV (where one finds asymptotic freedom, as expected) while 
$r\rightarrow 0$ is the IR. In order to obtain the precise
radius-energy relation, the statement (\ref{gaugid}) has been
used \cite{DV}. The operator $<\la^2>$ has non-anomalous 
dimension 3, and so $<\la^2>= c\,\Lambda^3$, where $\Lambda$
is the dynamically generated scale of the gauge theory. Thus, we
can write:
\beq
a(r)\,=\,{\Lambda^3 \over \mu^3}\,\,,
\label{enradrel}
\eeq
$\mu$ being an arbitrary mass scale introduced to regulate the
theory. By using (\ref{gymder}) and (\ref{enradrel}), one can
now easily compute the beta function:
\beq
\beta(g_{YM})={\partial g_{YM} \over \partial \log (\mu/\Lambda)}=
{\partial g_{YM} \over \partial r}
{\partial r \over \partial \log (\mu/\Lambda)}\approx
-3{N\,g_{YM}^3 \over 16\,\pi^2}\left(1-{N\,g_{YM}^2 \over 8\,\pi^2}
\right)^{-1}\,\,.
\eeq
For the last step, terms exponentially suppressed in $r$ have
been neglected. This is exactly the NSVZ $\beta$-function, which
was calculated in \cite{NSVZ}, using a Pauli-Villars 
regularization scheme.

This result seems surprising since in the AdS/CFT duality, the
validity of the gravity approach is limited to the strong
coupling regime of the gauge theory. 
Therefore, it is puzzling that we
are finding the correct $\beta$-function in the perturbative regime. The
answer to this question may go along the lines of \cite{vlmp}. There, by
computing an  annulus diagram, it was proved that the open/closed string
duality allows the perturbative regime of a non-conformal gauge theory
to be encoded in a supergravity solution. The calculation
is made for  particular cases where the sugra dual is  constructed
with fractional branes in  orbifolds. The Maldacena-N\'u\~nez
case might have the same property.
However, an analogous 
calculation in the background we are dealing with is much more
difficult and has not been done. 

For a deeper discussion on the $\beta$-function
just obtained, see \cite{MerReview}.

\subsubsection{String tensions}

A $q$-string is a tube connecting a set of $q$ quarks
with a set of $q$ antiquarks in a $SU(N)$ gauge theory.
If $T_{q+q'}<T_q+T_{q'}$, it will not decay in  $q$ separate 1-strings.
In ref. \cite{klebherz}, these objects were studied from
the gravitational point of view. They are represented by a bunch
of fundamental strings placed at $r=0$ and extended in the 
$x$ direction. Because of the presence of the RR potential,
the Myers polarization effect blows the F-strings up into
a D3-brane, extending in the $x$ direction and wrapping an
$S^2$ inside the finite $S^3$. The size of the $S^2$ depends
on the number $q$. The tension of the $q$-string is the energy
density of the D3-brane after integration in the $S^2$ directions.

The calculation goes like that
of \cite{bachas} where a D2-brane in an NS background was considered.
The generalization to RR background was performed in 
\cite{pawelczyk}, and the relation to Myers effect was explicitly
shown in \cite{swb}.
One finds:
\beq
T_q = c \,\sin {\pi \,  q \over N},
\eeq
where $c$ is a constant, related to an IR scale of the gauge
theory. This result agrees with the ones obtained from
other approaches (see \cite{klebherz} for references).
The constant $c$ can be determined by direct calculation or by noting
that for
$q=1$ (and large $N$), the result of (\ref{confine}) should be recovered.
Then we have $c={e^{\phi_0} N \over 2\pi^2\alpha'}$.
 Notice that when
$q=N$, the set of quarks and the set of antiquarks form separate colorless
states (a baryon and an antibaryon) and therefore the tension
vanishes.

\subsubsection{BPS domain walls}

As we have seen, ${\cal N}=1$ SYM is characterized by a set
of $N$ different vacua. There exist domain wall configurations
that interpolate between them. They are BPS states and preserve
half of the supersymmetries. Their tension is related to the
different vevs for the gluino condensate at both sides of the
domain wall.

The corresponding object in the supergravity setup is a 
stack of $n$ D5-branes
wrapped on the $S^3$ of the geometry, and extended in three
of the unwrapped space-time directions, say, $x_0$, $x_1$, $x_2$.
Then, going from $x_3=-\infty$ to $x_3=\infty$ amounts to crossing
the domain wall and therefore moving from one vacuum into another.
It can be shown \cite{loewy} that crossing the domain wall 
implies  a shift in the angle $\tilde\psi$ such that
$\Delta\tilde\psi_0=2\pi n/N$. Hence, choosing $n$ among $n=1,\dots N-1$
one can have a domain wall between any pair of vacua, in perfect
agreement with what is expected from field theory.

The D5-branes will wrap the $S^3$ at $r=0$, where its volume is
minimum. This agrees with the fact that the domain walls (even
the existence of $N$ vacua) is an infrared effect. Furthermore,
the fact that QCD-strings can end in domain walls is perfectly
reproduced as their gravity counterpart are $F$-strings
(or their blow-up to D3 {\it a la Myers}), which can end in
the D5-brane domain walls.

It is worth pointing out that the field theory interpretation of
different possible brane probes is summarized in ref.
\cite{loewy}.

\subsubsection{The Veneziano-Yankielowicz potential}

In ref. \cite{VY}, Veneziano and Yankielowicz constructed an effective
action for ${\cal N}=1$ SYM. They showed that non-perturbative (IR)
effects give rise to an effective superpotential of the form:
\beq
W_{VY}\,=\,-N\,S\,\left(1-\log {S\over \Lambda^3}\right)\,\,,
\eeq
where $S$ is a gauge invariant superfield that contains the
composite operator $\la^2$. Indeed, the fact that this potential
has a minimum leads to the existence of a non-trivial vacuum
expectation value for $<\la^2>$, the gluino condensate.

It was argued by Vafa \cite{vafa} that such kind of potentials
may be found in string theory duals by integrating over certain cycles the
fluxes present in the solution (that appear after geometric transitions).
The Veneziano-Yankielowicz potential was found following this
approach in different supergravity duals of ${\cal N}=1$ SYM
\cite{vafa,civ,IL}. However, it is difficult to use such a  
procedure in the MN model because of the varying dilaton.

An alternative approach was presented in \cite{muck}.  The  proposal
is to relate the VY potential to the potential that feels a brane
probe. Partial success was achieved, but further research
may be required.

\subsubsection{Glueballs}

The glueball spectrum of the theory was analyzed in ref. \cite{Pons}.
In the spirit of AdS/CFT correspondence, the idea is to look for
the supergravity mode that couples to the gauge invariant operator
of the field theory. In this case, the dilaton field couples
to ${\mathrm Tr}\,F^2$, which corresponds to a glueball with
quantum numbers $J^{PC}=0^{++}$. Therefore, the fluctuation 
spectrum of the dilaton should yield the mass spectrum
of this kind of glueballs (the fluctuations of the RR potential
$C_{(2)}$, dual to $1^{--}$ glueballs, were also studied).
By considering an ansatz for the dilaton fluctuation:
\beq
\Phi (x,r)\,=\,\tilde\Phi (r)e^{ik\cdot x}\,\,,
\eeq
one reaches the equation of motion:
\beq
\partial_r (e^{2\phi+2h}\partial_r \tilde\Phi)\,+\,
M^2 e^{2\phi+2h}\,\tilde\Phi\,=\,0\,\,,
\label{glueeq}
\eeq
where $M^2=k_0^2-|\vec{k}|^2\,$ is the mass in the four dimensional
theory. Dependence of the fluctuation field on the angular coordinates of
the compact space would lead to  solutions where the Kaluza-Klein modes
(not present in the field theory)  contribute. 

Unfortunately, eq. (\ref{glueeq}) does not lead to a discrete
spectrum. As argued in \cite{Pons}, this is due to the fact
that one cannot trust the solution (\ref{Fmetric}), (\ref{FRRthreeform}),
(\ref{FMNsol}) all the way to the UV because the dilaton grows
unbounded. One should perform an S-duality at some point. This
can be simulated by taking a cutoff $\Lambda$ and imposing by hand
that the fluctuations for $r>\Lambda$ vanish. This may seem
awkward at first sight, but glueballs are an IR effect, and
there is a scale really present in the theory: the scale at
which gluinos condense and the IR regime is reached. The 
precise value of $\Lambda$ is not clear but it must be in the
region $2 <\Lambda < 5$. The form of the spectrum obtained
depends on the concrete value of $\Lambda$. A
 pattern similar to what is obtained from other supergravity 
models is found by taking $\Lambda\approx 3.5$

In section \ref{secmesons}, the same problem will appear when calculating
the meson spectrum of the theory with flavor.

\subsubsection{The theory with flavor}

This topic will be  developed in the
next chapter. However, for the sake of completeness of this section, some
ideas and results are summarized here.

When there is a gauge theory living on some brane worldvolume, matter
transforming in the fundamental representation is described by 
fundamental strings with one end on the gauge theory brane. The
other end of the string must be attached to some other brane, which
will be called flavor brane. 
From the analysis of the  ${\cal N}=1$ SYM  theory, it has been known
for a long time that one can add flavor (massive quarks) without
reducing the number of supercharges.
Therefore, if one wants to describe this effect from the
supergravity dual, one must find a locus where the flavor
brane can be placed without further breaking the supersymmetry
of the background.

Moreover, as discussed in \cite{KK}, the flavor brane must be 
space-time filling in the space-time dimensions where the
field theory lives and also in the holographic (radial)
direction. That the brane extends to infinity in the radial
direction is pleasant from a holographic point of view, 
because to introduce something new in the field theory, 
something should be modified on the boundary of the gravity setup.
In \cite{KK}, it was also argued that if the quarks are to have
a finite mass, the flavor brane should extend only up to a minimum
value of $r$ ({\it vanishing in thin air}),  
 because the quark mass is related to the minimum energy
that a string stretched between both branes can have.

In \cite{WangHu}, by analyzing the spectrum of massless modes
of strings going from one brane to another, it was concluded
that this addition of supersymmetric flavor can be done with
D9-branes or with D5-branes extending {\it orthogonally} to
the gauge theory branes in two directions of the Calabi-Yau
space.

In \cite{flavoring} (see next chapter), the possible
positions for these D5-branes
were found explicitly by $\kappa$-symmetry analysis. The solutions
fulfil all the conditions stated above. Moreover, there is one
parameter that can be naturally related to the mass of the quark.
With the explicit expression of the solutions, one can see
the gravity counterpart of a number of known phenomena of the
gauge theory: $U(1)_R$ symmetry breaking by the formation of an
squark condensate, non-smoothness of the limit $m_q\rightarrow 0$ and
$U(1)_B$ baryonic symmetry preservation. Moreover, a formula for the mass
spectrum of the mesons is given. All the analysis is made in the
probe approximation, that corresponds to the limit of the gauge theory
with
$N_f
\ll N_c$.

%*******************************************

\chapter{
Supersymmetric probes in the MN model:
flavor}
\label{flavoring}

\setcounter{equation}{0}
\section{Introduction}

Two related problems in the context of the supergravity dual to  ${\cal
N}=1$ SYM will be studied in this chapter \cite{flavoring}. One of the
problems  is finding kappa symmetric D5-brane probes in this  particular
background. The other is the use of these probes to add flavors to the
gauge theory. We will find a rich and mathematically appealing  structure
of the supersymmetric  embeddings of a D5-brane probe in this background.
Besides, we compute the mass spectrum of the  low energy excitations of
${\cal N}=1$ SQCD (mesons) and match our results with some field  theory
aspects known from the study of supersymmetric gauge theories with a
small number of flavors.

Most of the analysis carried out with the background of \cite{MN} 
(see the previous chapter) do
not incorporate quarks in the fundamental representation  which, in a
string theory setup, correspond to open strings. In order to introduce an
open string sector in a supergravity dual it is quite natural to add
D-brane probes and see whether one can extract some information about the
quark dynamics. As usual, if the number of brane probes is much smaller
than those of the background, one can assume that there is no backreaction
of the probe in the bulk geometry. In this chapter, we follow this
approach and we will probe with D5-branes the supergravity dual of 
${\cal N}=1$ SYM. Since we will interpret the  brane probes as
introducing flavor, the results for the dual gauge theory can only be
valid for the so-called quenched approximation, where the number of
flavors is much less than the number of colors ($N_f\ll N$). 
Obviously, one cannot go beyond this limit without finding the
backreacted supergravity background.

The
main technique to determine the supersymmetric brane probe configurations is
kappa symmetry
\cite{swedes}, which
tells us that, if $\epsilon$ is a Killing spinor of the  background, only
those embeddings for which a certain matrix $\Gamma_{\kappa}$ satisfies:
\beq
\Gamma_{\kappa}\,\epsilon\,=\,\epsilon\,\,
\label{Fkappaprojection}
\eeq
preserve the supersymmetry of the background \cite{bbs}. The matrix 
$\Gamma_{\kappa}$  depends on the metric induced on the worldvolume of the
brane. Therefore, if the Killing spinors $\epsilon$ are known, we can
regard  (\ref{Fkappaprojection}) as an equation for the embedding of the
brane. We will be able to find embeddings where the brane probe preserves
exactly the same susy as the background and no additional projections are
needed.

The starting point in this program will be the simple expression for the
Killing spinor of the background found in section \ref{FMN10D}.

The probes we are going to  consider are D5-branes wrapped on a
two-dimensional submanifold. By inserting in
(\ref{Fkappaprojection}) the projections (\ref{Fprojectionone}),
(\ref{Fprojectiontwo}) and (\ref{Fradial}), we will be able to find some
differential equations for the embedding. They are, in general, quite
complicated to solve. The first obvious configuration one should look at is
that of a fivebrane wrapped at a fixed distance from the origin. In this
case the equations simplify drastically and we will be able to prove a
no-go theorem which states that, unless we place the brane  at an infinite
distance from the origin, the probe breaks supersymmetry. This result is
consistent with the fact that these ${\cal N}=1$ theories do not have a
moduli space. In this analysis we will make contact with the two-cycle
considered in ref.
\cite{BertMer} and show that it preserves supersymmetry at an
asymptotically large distance from the origin.

Guided by the negative result obtained when   trying to wrap the D5-brane
at constant distance, we will allow this distance to vary within the
two-submanifold of the embedding. To simplify the  equations that
determine the embeddings, we first consider the singular version of the
background, in which the vector field of the seven dimensional gauged
supergravity is abelian. This geometry coincides with the non-singular
one, in which the vector field is non-abelian, at large distances from
the origin. By choosing an appropriate set of variables we will be able
to write the differential equations for the embedding as two pairs of
Cauchy-Riemann equations which are straightforward to integrate in
general. Among all possible solutions, we will concentrate on some of
them characterized by integers, which can be interpreted as winding
numbers. Generically these solutions have spikes, in which the probe is
at infinite distance from the origin and, thus, they correspond to
fivebranes wrapping a non-compact submanifold. Moreover, these
configurations are worldvolume solitons and we will verify that they
saturate an energy bound
\cite{GGT}.

With the insight gained by the analysis of the   worldvolume solitons in
the abelian background we will consider the equations for the embeddings
in the non-abelian background. In principle, any solution for the smooth
geometry must coincide in the UV with one of the configurations found for
the singular metric. This observation will allow us to formulate an
ansatz to solve the complicated equations arising from kappa symmetry.
Actually, in some cases,  we will be able to find analytical solutions
for the embeddings, which behave as those found for the singular metric
at large distance from the origin and also saturate an energy bound,
which ensures their stability. 

One of our motivations to study brane  probes is to use these results to
explore the quark sector of the gauge/gravity duality. Actually, it was
proposed in refs. \cite{KK, KKW} that one can add flavor to this
correspondence by considering space-time filling branes. Open strings
coming into the gauge theory brane from the flavor brane represent the
quarks, and the fluctuations of the branes introducing flavor will
be low energy excitations of the gauge theory, which are mesons.
In ref.
\cite{KMMW} this program has been made explicit for the $AdS_5\times S^5$
geometry of a stack of D3-branes and a D7-brane probe. When the D3-branes
of the background and the D7-brane of the probe are separated, the
fundamental matter arising from the strings stretched between them becomes
massive and a  discrete spectrum  of mesons for an
${\cal N}=2$ SYM with a matter hypermultiplet can be obtained analytically
from the fluctuations of a D7-brane probe. In ref. \cite{Sonnen} a similar
analysis was performed for the ${\cal N}=1$ Klebanov-Strassler background
\cite{KS}, while in refs. \cite{Johana,KMMW-two} the meson spectrum for some
non-supersymmetric backgrounds was found (for recent related work see refs.
\cite{IL,Ouyang}).

It was suggested in ref. \cite{WangHu}  that one possible way to add flavor
to the  
${\cal N}=1$ SYM background is  by considering supersymmetric embeddings of
D5-branes which wrap a two-dimensional  submanifold and are space-time
filling. Some of the configurations we will find in our kappa symmetry
analysis have the right ingredients to be used as flavor branes. They are
supersymmetric by construction, extend infinitely and have some parameter
which determines the minimal distance between the brane probe and the
origin. This distance should be interpreted as the mass scale of the
quarks. It corresponds to what in \cite{KK} is called {\it branes
vanishing in thin air}. They vanish from a five dimensional point of
view (the four space-time dimensions plus the radial, holographic
dimension) since, as explained above, there is a minimal radial distance.
However, from a ten dimensional point of view, the configuration is 
perfectly smooth.

 Moreover, these brane probes capture geometrically the pattern of
R-symmetry breaking of SQCD with few flavors
\cite{Affleck:1983mk}. Consequently,  we will study the quadratic
fluctuations around the static probe configurations found by integrating
the kappa symmetry equations. We will verify that these fluctuations decay
exponentially at large distances. However, we will not be able to define a
normalizability condition which could give rise to a discrete spectrum. The
reason for this is the exponential blow up of the dilaton at large
distances. Actually, the same difficulty was found in ref.
\cite{Pons} in the study of the glueball   spectrum for this background.
As proposed in ref. \cite{Pons}, we shall introduce a cut-off and impose
boundary conditions which ensure that the fluctuation takes place in a
region in which the supergravity approximation remains valid. The resulting
spectrum is discrete and, by using numerical methods, we will be able to
determine its form. 

In section \ref{seckappa} we obtain the  kappa symmetry equations which
determine the supersymmetric embeddings. In section \ref{secrconst} we
obtain the no-go theorem for branes wrapped at fixed distance. In section
\ref{secabel} the kappa symmetry equations for the abelian background are
integrated in general and some of the particular solutions are studied in
detail. Section \ref{secnonabel}  deals with the integration of the
equations for the supersymmetric embeddings in the full non-abelian
background. 
Readers more
interested in the gravity version of the addition of  flavors to ${\cal
N}=1$ SYM may take for granted all these results and look at the solutions
exhibited in eqs. (\ref{Fabspikes}), (\ref{FBM}) and (\ref{Fnoabflavor}),
which are what we called ``abelian and non-abelian unit-winding solutions".
Then, they should go straight to section \ref{secmesons}, where  the
spectrum of the quadratic fluctuations is analyzed and the gauge theory
interpretation is explained. Moreover, an appendix is devoted to
the asymptotic form of the fluctuations.

\setcounter{equation}{0}
\section{Kappa symmetry}
\label{seckappa}

As mentioned above, 
the kappa symmetry condition for a supersymmetric embedding  of a D5-brane
probe is
$\Gamma_{\kappa}\,\epsilon\,=\,\epsilon$ (see eq.
(\ref{Fkappaprojection})),  where $\epsilon$ is a Killing spinor of the
background. For  $\epsilon$ such that
$\epsilon=i\epsilon^*$  and when there is no worldvolume  gauge field, one
has:
\beq
\Gamma_{\kappa}\,=\,{1\over 6!}\,\,{1\over \sqrt{-g}}\,\,
\epsilon^{m_1\cdots m_6}\,\,\gamma_{m_1\cdots m_6}\,\,,
\eeq
where $g$ is the determinant of the induced metric $g_{mn}$  on the
worldvolume 
\beq
g_{mn}\,=\,\partial_m X^{\mu}\,\partial_n X^{\nu}\,G_{\mu\nu}\,\,,
\eeq
with $G_{\mu\nu}$ being the ten-dimensional metric
and
$\gamma_{m_1\cdots m_6}$ are antisymmetrized products of  worldvolume Dirac
matrices 
$\gamma_{m}$, defined as:
\beq
\gamma_{m}\,=\,\partial_{m}X^{\mu}\,e_{\mu}^{a}\,
\Gamma_{a}\,\,.
\eeq
The vierbeins $e_{\mu}^{a}$ are the coefficients  which
relate the one-forms $e^a$ of the frame and the
differentials of the coordinates, \ie\
$e^{a}=e_{\mu}^{a}dX^{\mu}$.  The ten dimensional vierbein that we will
use for the Maldacena-N\'u\~nez background is written in (\ref{Fframe}).
Let us take as worldvolume coordinates
$(x^0,
\cdots, x^3,
\theta,\varphi)$. Then, for an embedding with
$\tilde\theta=\tilde\theta(\theta,\varphi)$, 
$\tilde\varphi=\tilde\varphi(\theta,\varphi)$,  
$\tilde{\psi}=\tilde{\psi}(\theta,\varphi)$ and $r=r(\theta, \varphi)$, 
the kappa symmetry matrix
$\Gamma_{\kappa}$ takes the form:
\beq
\Gamma_{\kappa}\,=\,{e^{\phi}\over \sqrt{-g}}\,\Gamma_{x^0\cdots x^3}\,
\gamma_{\theta\varphi}\,\,,
\eeq	
with $\gamma_{\theta\varphi}$ being the antisymmetrized  product of the two
induced matrices
$\gamma_{\theta}$ and $\gamma_{\varphi}$, which can be written as:
\bear
e^{-{\phi\over 4}}\,\gamma_{\theta}&=&
e^h\,\Gamma_1\,+\,(V_{1\theta}+{a\over 2}\,)\,\hat\Gamma_1\,+\,
V_{2\theta}\,\hat\Gamma_2\,+\,V_{3\theta}\,\hat\Gamma_3\,+\,
\partial_{\theta} r\Gamma_{r}\,\,,\rc\rc
{e^{-{\phi\over 4}}\over \sin\theta}\,\gamma_{\varphi}&=&
e^h\,\Gamma_2\,+\,
V_{1\varphi}\,\hat\Gamma_1\,+\,(V_{2\varphi}\,-{a\over 2}\,)\,\hat\Gamma_2
\,+\,V_{3\varphi}\,\hat\Gamma_3\,+\,
{\partial_{\varphi} r\over \sin\theta}\,
\Gamma_r
\,\,,
\label{Fgthetaphi}
\eear
where the $V$'s can be obtained by computing the pullback  on the
worldvolume of the left invariant one-forms
$\om^i$ (see (\ref{otherom})), and are given by:
\bear
V_{1\theta}&=&{1\over
2}\,\cos\tilde{\psi}\,\partial_{\theta}\,\tilde\theta\,+\, {1\over
2}\sin\tilde{\psi}\sin\tilde\theta\,\partial_{\theta}\,
\tilde\varphi\,\,,\rc\rc
\sin\theta\,V_{1\varphi}&=&
{1\over 2}\,\cos\tilde{\psi}\,\partial_{\varphi}\,\tilde\theta\,+\,
{1\over
2}\sin\tilde{\psi}\sin\tilde\theta\,\partial_{\varphi}\,\tilde\varphi
\,\,,\rc\rc V_{2\theta}&=&-{1\over
2}\,\sin\tilde{\psi}\,\partial_{\theta}\,\tilde\theta\,+\, {1\over
2}\cos\tilde{\psi}\sin\tilde\theta\,\partial_{\theta}\,
\tilde\varphi\,\,,\rc\rc
\sin\theta\,V_{2\varphi}&=&-
{1\over 2}\,\sin\tilde{\psi}\,\partial_{\varphi}\,\tilde\theta\,+\,
{1\over
2}\cos\tilde{\psi}\sin\tilde\theta\,\partial_{\varphi}\,\tilde\varphi
\,\,,\rc\rc V_{3\theta}&=&{1\over
2}\,\partial_{\theta}\tilde{\psi}\,+\,{1\over 2}\cos\tilde\theta\,
\partial_{\theta}\,\tilde\varphi\,\,,\rc\rc
\sin\theta\,V_{3\varphi}&=&
{1\over 2}\,\partial_{\varphi}\tilde{\psi}\,+\,{1\over 2}\cos\tilde\theta\,
\partial_{\varphi}\,\tilde\varphi\,+\,{1\over 2}\,\cos\theta\,\,.
\label{FVs}
\eear
By using the projections (\ref{Fprojectionone}) and (\ref{Fradial})  one can
compute the action of $\gamma_{\theta\varphi}$ on the Killing spinor
$\epsilon$. It is clear that one arrives at an expression of the type:
\bear
{e^{-{\phi\over 2}}\over \sin\theta}\,\,
\gamma_{\theta\varphi}\,\epsilon\,&=&\,\big[\,
c_{12}\,\Gamma_{12}\,+\,c_{1\hat 2}\,\Gamma_{1}\hat\Gamma_{2}\,+\,
c_{1\hat 1}\,\Gamma_{1}\hat\Gamma_{1}\,+\,
c_{1\hat 3}\,\Gamma_{1}\hat\Gamma_{3}\,+\,
\rc\rc
&&+\,c_{\hat 1\hat 3}\,\hat\Gamma_{13}\,
+\,c_{\hat 2\hat 3}\,\hat\Gamma_{23}\,
+\, c_{2\hat 3}\,\Gamma_{2}\hat\Gamma_{3}\,\big]\,\epsilon\,\,,
\label{Fces}
\eear
where the $c$'s are coefficients that can be explicitly computed.  By using
eq. (\ref{Fces}) we can obtain the action of $\Gamma_{\kappa}$ on $\epsilon$
and we can use this result to write the kappa symmetry projection
$\Gamma_{\kappa}\epsilon=\epsilon$. Actually, eq. (\ref{Fkappaprojection})
is automatically satisfied if it can be reduced to 
eq. (\ref{Falphaproj}). If we want this to happen, all terms except the
ones containing $\Gamma_{12}\,\epsilon$ and 
$\Gamma_{1}\,\hat\Gamma_{2}\,\epsilon$ on the right-hand side of  eq.
(\ref{Fces}) should vanish. Then, we should require:
\beq
c_{1\hat 1}\,=\,c_{1\hat 3}\,=\,c_{\hat 1\hat 3}\,=\,c_{\hat 2\hat 3}\,=\,
c_{2\hat 3}\,=\,0\,\,.
\label{Fcesnull}
\eeq
By using the explicit expressions of the $c$'s one can obtain  from eq.
(\ref{Fcesnull}) five conditions that our supersymmetric embeddings must
necessarily satisfy. These conditions are:
\bear
&&e^{h}\,(V_{1\varphi}\,+\,V_{2\theta}\,)=0\,\,,\label{Fkappauno}\\\rc
&&e^h\,(\,V_{3\varphi}\,+\,\cos\alpha\partial_{\theta}r\,)\,
+\,(\,V_{2\varphi}\,-\,{a\over 2}\,)\,\sin\alpha\,\partial_{\theta}r
\,-\,V_{2\theta}\,\sin\alpha\,\,{\partial_{\varphi} r\over \sin\theta}
\,=\, 0\,\,,\label{Fkappados}\\\rc
&&(\,V_{1\theta}\,+\,{a\over 2}\,)\,V_{3\varphi}\,-\,
V_{3\theta}\,V_{1\varphi}\,-\,e^{h}\sin\alpha\,\partial_{\theta}r\,
+\,\rc\rc &&\,\,\,\,\,\,\,\,\,\,\,\,\,\,\,\,\,\,\,\,\,\,\,
+\,(\,V_{2\varphi}\,-\,{a\over 2}\,)\,\cos\alpha\,\partial_{\theta}r
\,-\,V_{2\theta}\cos\alpha\,{\partial_{\varphi} r\over \sin\theta}
\,=0\,\,,\label{Fkappatres}\\\rc
&&V_{3\varphi}\,V_{2\theta}\,-\,
V_{3\theta}\,(\,V_{2\varphi}\,-\,{a\over 2}\,)\,-\,V_{1\varphi}\cos\alpha
\,\partial_{\theta}r\,+\,\rc\rc
&&\,\,\,\,\,\,\,\,\,\,\,\,\,\,\,\,\,\,\,\,\,\,\,
+\,\bigg(\,e^h\sin\alpha\,+\,(V_{1\theta}\,+\,{a\over 2}\,)
\cos\alpha\bigg)\, {\partial_{\varphi} r\over
\sin\theta}\,=0\,\,,\label{Fkappacuatro}\\\rc &&\sin\alpha
\,V_{1\varphi}\partial_{\theta}r
\,-\,e^h\,V_{3\theta}
+\,\bigg(\,e^h\cos\alpha\,-\,(V_{1\theta}\,+\,{a\over 2}\,)\sin\alpha\,
\bigg)\,{\partial_{\varphi} r\over \sin\theta}
=0\,\,.
\label{Fkappacinco}
\eear
Moreover, if we want the kappa symmetry projection 
$\Gamma_{\kappa}\epsilon=\epsilon$ to  coincide with the SUGRA projection,
the ratio of the coefficients of the terms with
$\Gamma_{1}\,\hat\Gamma_{2}\,\epsilon$  and $\Gamma_{12}\,\epsilon$ must be
$\tan\alpha$, \ie\ one must have:

\beq
\tan\alpha\,=\,{c_{1\hat 2}\over c_{1 2}}\,\,.
\label{Ftanalpha}
\eeq
The explicit form of $c_{1 2}$ and $c_{1\hat 2}$ is:
\bear
c_{1 2}&=&
e^{2h}+V_{1\theta}\,V_{2\varphi}-
V_{2\theta}\,V_{1\varphi}-
{a\over 2}\,(\,V_{1\theta}\,-\,V_{2\varphi}\,)-{a^2\over 4}-
\cos\alpha V_{3\varphi}\,\partial_{\theta} r+
\cos\alpha V_{3\theta}{\partial_{\varphi}r\over\sin\theta}\,\,,\rc\rc
c_{1\hat 2}&=&
e^h\,(\,V_{2\varphi}\,-\,V_{1\theta}\,-a\,)\,-\,\sin\alpha
V_{3\varphi}\,\partial_{\theta} r\,+\,\sin\alpha
V_{3\theta}\,{\partial_{\varphi}r\over\sin\theta}\,\,.
\eear
Amazingly, except when $r$ is constant and takes some value  in the
interval
$0<r<\infty$ (see section \ref{secrconst}), eq. (\ref{Ftanalpha}) is a
consequence of eqs.  (\ref{Fkappauno})-(\ref{Fkappacinco}). Actually, by 
eliminating
$V_{3\theta}$ of eqs. (\ref{Fkappacuatro}) and (\ref{Fkappacinco}), and
making use of eqs.  (\ref{Fkappauno}) and (\ref{Fkappados}), one arrives at 
the following expression of $\tan\alpha$:
\beq
\tan\alpha\,=\,{
e^h\,(\,V_{2\varphi}\,-\,V_{1\theta}\,-a\,)\over
e^{2h}\,+\,V_{1\theta}\,V_{2\varphi}\,-\,
V_{2\theta}\,V_{1\varphi}\,-\,
{a\over 2}\,(\,V_{1\theta}\,-\,V_{2\varphi}\,)\,-\,{a^2\over 4}}
\,\,.
\eeq
Notice that the terms of $c_{1\hat 2}$ ($c_{1 2}$) which do not contain
$\sin\alpha$ ($\cos\alpha$) are just the ones in the  numerator
(denominator) of the right-hand side of this equation. It follows from this
fact that eq.  (\ref{Ftanalpha})  is satisfied if eqs.
(\ref{Fkappauno})-(\ref{Fkappacinco}) hold. Moreover, by using the values of
$\cos\alpha$ and $\sin\alpha$ given in eq. (\ref{Falpha}), one obtains the
interesting relation:
\beq
\Big(\,1\,+\,a^2\,+\,4e^{2h}\,)\,(\,V_{1\theta}\,-\,V_{2\varphi}\,)=
4a\,\Big(\,V_{2\theta}^2\,+\,V_{1\theta}\,V_{2\varphi}\,-\,{1\over 4}
\,\Big)\,\,.
\label{Fcondition}
\eeq
The system of eqs. (\ref{Fkappauno})-(\ref{Fkappacinco})  is rather involved
and, although it could seem at first sight very difficult and even hopeless
to solve, we will  be able to do it in some particular cases. Moreover, it
is interesting to notice that, by simple manipulations, one can obtain the
following expressions of the partial derivatives of $r$:
\bear
\partial_{\theta} r&=&-\cos\alpha V_{3\varphi}\,+\,\sin\alpha\,e^{-h}\,\,
\big[\,(V_{1\theta}\,+\,{a\over 2}\,)\,V_{3\varphi}\,-\,
V_{3\theta}\,V_{1\varphi}\,\big]\,\,,\rc\rc
\partial_{\varphi} r&=&\cos\alpha\sin\theta V_{3\theta}\,+\,
\sin\alpha\,\sin\theta e^{-h}\,\,
\big[\,(V_{2\varphi}\,-\,{a\over 2}\,)\,V_{3\theta}\,-\,
V_{3\varphi}\,V_{2\theta}\,\big]\,\,,
\label{Frderivatives}
\eear
which will be very useful in our analysis.

\setcounter{equation}{0}
\section{Branes wrapped at fixed distance}
\label{secrconst}

In this section, we will consider the possibility of  wrapping the
D5-branes at a fixed distance $r>0$ from the origin. It is clear that, in
this case, we have 
$\partial_{\theta}r\,=\,\partial_{\varphi}r=0$ and many  of the terms on
the left-hand side of eqs. (\ref{Fkappauno})-(\ref{Fkappacinco}) cancel. 
Moreover $e^{h}$ is non-vanishing when $r>0$ and it can be
factored out in these equations. Thus, 
the equations (\ref{Fkappauno})-(\ref{Fkappacinco}) of  kappa symmetry when
the radial coordinate $r$ is constant and non-zero reduce to:
\beq
V_{1\varphi}\,+\,V_{2\theta}\,=\,V_{3\varphi}\,=\,V_{3\theta}\,=\,0\,\,.
\eeq
From the equations $V_{3\varphi}\,=\,V_{3\theta}\,=\,0$  we obtain the
following differential equations for $\tilde{\psi}$ :
\beq
\partial_{\theta}\,\tilde{\psi}\,=\,-\cos\tilde\theta\,
\partial_{\theta}\tilde\varphi\,\,,
\,\,\,\,\,\,\,\,\,\,\,\,\,\,\,\, 
\partial_{\varphi}\,\tilde{\psi}\,=\,-\cos\tilde\theta\,
\partial_{\varphi}\tilde\varphi\,-\,\cos\theta\,\,.
\label{Feqpsi}
\eeq
The integrability condition for this system gives:
\beq
\partial_{\varphi}\tilde\theta\,
\partial_{\theta}\tilde\varphi\,-\,
\partial_{\theta}\tilde\theta\,
\partial_{\varphi}\tilde\varphi\,=\,
{\sin\theta\over\sin\tilde\theta}\,\,.
\label{Fintegrability}
\eeq
By using this condition and the definition of the  $V$'s (eq. (\ref{FVs}))
one can prove that:
\beq
V_{1\theta}\,V_{2\varphi}\,-\,
V_{1\varphi}\,V_{2\theta}\,=\,-{1\over 4}\,\,.
\label{FVintegrability}
\eeq
Let us now define $\Delta$ as follows:
\beq
V_{2\varphi}\,-\,V_{1\theta}\,\equiv\,\,\Delta\,\,.
\label{FDelta}
\eeq
By using the expression of the V's in terms of  the angles, one can 
combine eq. (\ref{FDelta})  and the condition 
$V_{1\varphi}\,+\,V_{2\theta}\,=\,0$ in the following matrix equation
\beq
\pmatrix{\cos\tilde{\psi}&\sin\tilde{\psi}\cr\cr
          -\sin\tilde{\psi}&\cos\tilde{\psi}}\,\,
\pmatrix{\sin\theta\partial_{\theta}\tilde\theta\,-\,
\sin\tilde\theta\partial_{\varphi}\tilde\varphi\cr\cr
\sin\theta\sin\tilde\theta\partial_{\theta}\tilde\varphi\,+\,
\partial_{\varphi}\tilde\theta}\,\,=\,\,
\pmatrix{-2\Delta\sin\theta\cr\cr 0}\,\,.
\eeq
Since the matrix appearing on the left-hand side is  non-singular, we can
multiply by its inverse. By doing this one arrives at the following
equations:
\bear
&&\partial_{\theta}\tilde\theta\,-\,{\sin\tilde\theta\over\sin\theta}\,
\partial_{\varphi}\tilde\varphi\,=\,-2\Delta\,\cos\tilde{\psi}\,\,,\rc\rc
&&{\partial_{\varphi}\tilde\theta\over \sin\theta}\,+\,
\sin\tilde\theta\partial_{\theta}\tilde\varphi\,=\,-2\Delta\,
\sin\tilde{\psi}\,\,.
\label{FnonCR}
\eear
Substituting the derivatives of $\tilde\theta$  obtained from the above
equations into the integrability condition (\ref{Fintegrability}) we obtain
after some calculation
\beq
\sin^2\tilde\theta\,
\Bigg(\,\partial_{\theta}\tilde\varphi\,+\,\Delta\,
{\sin\tilde{\psi}\over\sin\tilde\theta}\,\Bigg)^2\,+\,
{\sin^2\tilde\theta\over \sin^2\theta}\,\,
\Bigg(\,\partial_{\varphi}\tilde\varphi\,-\,\Delta\,
\cos\tilde{\psi} \,\,{\sin\theta\over\sin\tilde\theta}\,\Bigg)^2\,=\,
\Delta^2\,-\,1\,\,.
\label{Fsquares}
\eeq
The left-hand side of eq. (\ref{Fsquares}) is  non-negative. Then, one
obtains a bound for
$\Delta$:
\beq
\Delta^2\,\ge\,1\,\,.
\label{FDeltabound}
\eeq
Notice that we have not imposed all the  requirements of kappa symmetry.
Indeed, it still remains to check that the ratios between the coefficients
$c_{12}$ and 
$c_{1\hat 2}$ is the one corresponding to the projection  of the
background. Using eq.  (\ref{FVintegrability}) and the definition of
$\Delta$ (eq. (\ref{FDelta})), one obtains:
\beq
c_{12}\,=\,e^{2h}\,+\,a\,{\Delta\over 2}\,-\,{a^2+1\over 4}\,\,,
\,\,\,\,\,\,\,\,\,\,\,\,\,\,\,\, \,\,\,\,\,\,\,\,\,\,\,\,\,\,\,\,
c_{1\hat 2}\,=\,e^h\,(\Delta-a)\,\,.
\eeq
Then, one must have:
\beq
\tan\alpha\,=\,{e^h\,(\Delta-a)\over 
e^{2h}\,+\,a\,{\Delta\over 2}\,-\,{a^2+1\over 4}}\,=\,
-{ae^{h}\over e^{2h}+{1-a^2\over 4}}\,\,,
\label{FalphaDelta}
\eeq
where we have used the values of $\sin\alpha$ and $\cos\alpha$  given in
the eq. (\ref{Falpha}). If $e^h$ is non-zero (and finite), we
can factor it out  in eq.  (\ref{FalphaDelta}) and obtain the following
expression of
$\Delta$:
\beq
\Delta= {2a\over 1+a^2+4e^{2h}}\,\,.
\eeq
Notice that $\Delta$ depends only on the coordinate $r$ and  is a
monotonically decreasing function such that $0<\Delta<1$ for $0<r<\infty$
and
\beq
\lim_{r\rightarrow0}\,\,\Delta=1\,\,,
\,\,\,\,\,\,\,\,\,\,\,\,\,\,\,\, 
\lim_{r\rightarrow\infty}\,\,\Delta\,=\,0\,\,.
\eeq
As $\Delta<1$, the bound  (\ref{FDeltabound}) is not  satisfied and, thus,
there is no solution to our equations for $0<r<\infty$. Notice that this
was to be expected from the lack of moduli space of the ${\cal N}=1$
theories.

Let us now consider the possibility of placing the brane  probe at
$r\rightarrow\infty$. Notice that in this case eq. (\ref{FalphaDelta}) is
satisfied for any finite value of
$\Delta$. However, the value $\Delta=1$ is special since, in this case, the 
right-hand side of eq. (\ref{Fsquares}) vanishes and we obtain two 
equations that determine
the derivatives of $\varphi$, namely:
\beq
\partial_{\theta}\tilde\varphi\,=\,-
{\sin\tilde{\psi}\over\sin\tilde\theta}\,\,,
\,\,\,\,\,\,\,\,\,\,\,\,\,\,\,\, \,\,\,\,\,\,\,\,\,\,\,\,\,\,\,\, 
\partial_{\varphi}\tilde\varphi\,=\,
\cos\tilde{\psi} \,\,{\sin\theta\over\sin\tilde\theta}
\label{Fpartialvarphi}
\eeq
Using these equations into the system (\ref{FnonCR}) for  $\Delta=1$ one
gets the following equations for the derivatives of $\tilde\theta$:
\beq
\partial_{\theta}\tilde\theta\,=\,-\cos\tilde{\psi}\,\,,
\,\,\,\,\,\,\,\,\,\,\,\,\,\,\,\, \,\,\,\,\,\,\,\,\,\,\,\,\,\,\,\, 
\partial_{\varphi}\tilde\theta\,=\,-\sin\theta\sin\tilde{\psi}\,\,,
\label{Fpartialtheta}
\eeq
and, similarly,  the equations  (\ref{Feqpsi}) for $\tilde{\psi}$ become:
\beq
\partial_{\theta}\tilde{\psi}\,=\,\sin\tilde{\psi}\cot\tilde\theta\,\,,
\,\,\,\,\,\,\,\,\,\,\,\,\,\,\,\, \,\,\,\,\,\,\,\,\,\,\,\,\,\,\,\, 
\partial_{\varphi}\tilde{\psi}\,=\,-\sin\theta\cot\tilde\theta\,
\cos\tilde{\psi}\,-\,
\cos\theta
\label{Fpartialpsi}
\eeq
The equations (\ref{Fpartialvarphi}) and (\ref{Fpartialtheta})  can be
regarded  as coming from the following identifications of the frame forms
in the 
$(\theta,\varphi)$ and $(\tilde\theta,\tilde\varphi)$ spheres:
\beq
\pmatrix{d\tilde\theta\cr\cr \sin\tilde\theta d\tilde\varphi} \,=\,
\pmatrix{\cos\tilde{\psi}&-\sin\tilde{\psi}\cr\cr\sin\tilde{\psi}&
\cos\tilde{\psi}}\,
\pmatrix{-d\theta\cr\cr \sin\theta d\varphi}
\label{Feqnsforms}
\eeq
The differential equations (\ref{Fpartialpsi}) are just the  integrability
conditions of the system (\ref{Feqnsforms}). Another interesting observation
is that one can prove by using the differential eqs.
(\ref{Fpartialvarphi})-(\ref{Fpartialpsi}) that the pullbacks of the
$SU(2)$ left-invariant one-forms are
\beq
P[w^1]\,=\,-d\theta\,\,,
\,\,\,\,\,\,\,\,\,\,\,\,\,\,\,\, \,\,\,\,\,\,\,\,\,\,\,\,\,\,\,\, 
P[w^2]\,=\,\sin\theta d\varphi\,\,,
\,\,\,\,\,\,\,\,\,\,\,\,\,\,\,\, \,\,\,\,\,\,\,\,\,\,\,\,\,\,\,\, 
P[w^3]\,=\,-\cos\theta d\varphi\,\,.
\label{Fpullbacksw}
\eeq

Let us try to find a solution of the differential equations
(\ref{Fpartialvarphi})-(\ref{Fpartialpsi}) in which 
$\tilde\theta=\tilde\theta(\theta)$ and 
$\tilde\varphi=\tilde\varphi(\varphi)$. The vanishing of
$\partial_{\varphi}\tilde\theta$ and $\partial_{\theta}\tilde\varphi$ 
immediately leads to $\sin\tilde{\psi}=0$ or $\tilde{\psi}=0,\pi\,\,\,( 
\mod
\,\,2\pi)$. Thus $\tilde{\psi}$ is constant in this case. Let us put
$\cos\tilde{\psi}=\eta\,=\,\pm 1$. The vanishing of 
$\partial_{\theta}\tilde{\psi}$  is automatic, whereas the condition 
$\partial_{\varphi}\tilde{\psi}\,=\,0$ leads to a relation between 
$\tilde\theta$ and $\theta$:
\beq
\cot\tilde\theta\,=\,-\eta\cot\theta
\eeq
In the case $\tilde{\psi}=0$, one has $\eta=1$ and the previous relation
yields 
$\tilde\theta=\pi-\theta$. Notice that this relation is in  agreement with
the first equation in eq. (\ref{Fpartialtheta}). Moreover,  the second
equation in (\ref{Fpartialvarphi}) gives $\tilde\varphi=\varphi$.
Similarly, one can solve the equations for  $\tilde{\psi}=\pi$. 
The solutions in these two cases are
just the ones used in ref. \cite{BertMer} in the calculation  of the beta
function  (with some correction in the $\tilde{\psi}=0$ case to have the
correct range of $\theta$ and
$\tilde\theta$), namely:
\bear
&&\tilde\theta\,=\,\pi-\theta\,\,,
\,\,\,\,\,\,\,\,\,\,\,\,\,
\tilde\varphi\,=\,\varphi\,\,,
\,\,\,\,\,\,\,\,\,\,\,\,\,
\tilde{\psi}=0 \,\,( \mod \,\,2\pi)\,\,,\rc\rc
&&\tilde\theta\,=\,\theta\,\,,
\,\,\,\,\,\,\,\,\,\,\,\,\,
\tilde\varphi\,=\,2\pi-\varphi\,\,,
\,\,\,\,\,\,\,\,\,\,\,\,\,
\tilde{\psi}=\pi \,\,( \mod \,\,2\pi)\,\,.
\label{FBMembeddings}
\eear

It follows from our results that the  embedding of ref. \cite{BertMer}  is
only supersymmetric asymptotically when $r\rightarrow\infty$. In this
sense, although it is somehow distinguished,   it is not unique since for
any embedding such that the $V$'s are finite when
$r\rightarrow\infty$, the determinant of the induced metric diverges  as
$\sqrt{-g}\sim e^{{3\phi\over 2}+2h}$ and the only term which survives in
the equation
$\Gamma_{\kappa}\epsilon=\epsilon$ is the one with the matrix $\Gamma_{12}$, giving
rise to the same projection as the background for $r\rightarrow\infty$.

\setcounter{equation}{0}
\section{Worldvolume solitons (abelian case)}
\label{secabel}

Let us consider the case $a=\alpha=0$ in the general equations of section
\ref{seckappa}.  From  equations (\ref{Fkappauno})  and  (\ref{Fcondition})
we get the following (Cauchy-Riemann like) equations:
\beq
V_{1\theta}\,=\,V_{2\varphi}\,\,,
\,\,\,\,\,\,\,\,\,\,\,
V_{1\varphi}\,=\,-V_{2\theta}\,\,,
\label{FVCR}
\eeq
whereas, from eq. (\ref{Frderivatives}) we obtain that  the derivatives of
$r$ are given by:
\beq
r_{\theta}\,=\,-V_{3\varphi}\,\,,
\,\,\,\,\,\,\,\,\,\,\,\,\,\,\,\,\,\,\,\,\,\,\,
r_{\varphi}\,=\,\sin\theta\,V_{3\theta}\,\,,
\label{Fabrderiv}
\eeq
where $r_{\theta}\equiv \partial_{\theta} r$ and 
$r_{\varphi}\equiv \partial_{\varphi} r$. 
It can be easily demonstrated that, in this abelian case, the full set of
equations (\ref{Fkappauno})-(\ref{Fkappacinco})  collapses to  the two pairs
of equations  (\ref{FVCR}) and (\ref{Fabrderiv}). Notice that $c_{1\hat
2}=0$ when 
$a=\alpha=0$ and eq. (\ref{FVCR}) holds and, thus, eq.  (\ref{Ftanalpha}) is
satisfied identically. 

Let us study first the two equations (\ref{FVCR}). By using  the same
technique as the one employed in section \ref{secrconst} to derive eq.
(\ref{FnonCR}),  it can be shown easily that they can be written as:
\beq
\sin\theta\partial_{\theta}\tilde\theta\,=\,
\sin\tilde\theta\partial_{\varphi}\tilde\varphi\,\,,
\,\,\,\,\,\,\,\,\,\,\,\,\,\,\,\,\,\,\,\,\,\,
\partial_{\varphi}\tilde\theta\,=\,-
\sin\theta\sin\tilde\theta\partial_{\theta}\tilde\varphi\,\,.
\label{FCR}
\eeq
In order to find the general solution of eq. (\ref{FCR}),  let us introduce
a new set of variables $u$ and $\tilde u$ as follows:
\beq
u\,=\,\log\tan{\theta\over 2}\,\,,
\,\,\,\,\,\,\,\,\,\,\,\,\,\,\,
\tilde u\,=\,\log\tan{\tilde\theta\over 2}\,\,.
\label{Fuvariables}
\eeq
Then,  eq. (\ref{FCR}) can be written as the Cauchy-Riemann equations in
the 
$(u,\varphi)$ and $(\tilde u,\tilde\varphi)$ variables, namely:
\beq
{\partial \tilde u\over \partial u}\,=\,
{\partial \tilde \varphi\over \partial \varphi}\,\,,
\,\,\,\,\,\,\,\,\,\,\,\,\,\,\,
{\partial \tilde u\over \partial \varphi}\,=\,-
{\partial \tilde \varphi\over \partial u}\,\,.
\label{FCRu}
\eeq
Since $u, \tilde u\in (-\infty, +\infty)$ and 
 $\varphi, \tilde \varphi\in (0, 2\pi)$, the above equations are  the
Cauchy-Riemann equations in a band. The general solution of  these
equations is of the form:
\beq
\tilde u+i\tilde\varphi\,=\,f(u+i\varphi)\,\,,
\eeq
where $f$ is an arbitrary function. Given any function $f$,  it is clear
that the above equation provides the general solution $\tilde\theta
(\theta, \varphi)$ and 
$\tilde\varphi (\theta, \varphi)$ of the system (\ref{FCR}).

Let us turn now to the analysis of the system of equations 
(\ref{Fabrderiv}), which determines the radial coordinate $r$. By using the
explicit values of 
$V_{3\varphi}$ and $V_{3\theta}$, these equations can be written as:
\bear
r_{\theta}&=&-{1\over 2\sin\theta}\,\partial_{\varphi}\,\tilde{\psi}\,-\,
{1\over 2}\,{\cos\tilde\theta\over \sin\theta}\,\partial_{\varphi}
\tilde\varphi -\,{1\over 2}\cot\theta
\,\,,\rc\rc
r_{\varphi}&=&{\sin\theta\over 2}\,\partial_{\theta}\,\tilde{\psi}\,+\,
{\sin\theta\over 2}\,\cos\tilde\theta\,\partial_{\theta}\tilde\varphi\,\,,
\eear
where $\tilde\theta (\theta, \varphi)$ and $\tilde\varphi  (\theta,
\varphi)$ are solutions of eq. (\ref{FCR}). In terms of the derivatives
 with respect to variable $u$ defined
above ($\sin\theta\partial_{\theta}=\partial_u$), these equations become:
\bear
r_u&=&-{1\over 2}\,\,\partial_{\varphi}\tilde{\psi}\,-\,{1\over 2}
\,\cos\tilde\theta\,
\partial_{\varphi}\tilde\varphi\,-\,{1\over 2}\,\cos\theta\,\,,\rc\rc
r_{\varphi}&=&{1\over 2}\,\,\partial_{u}\tilde{\psi}\,+\,{1\over 2}\,
\cos\tilde\theta\,\partial_{u}\tilde\varphi\,\,.
\label{Feqr}
\eear
The integrability condition of these equations is just 
$\partial_{\varphi}r_u=\partial_{u}r_{\varphi}$. As any solution 
$(\tilde\theta, \tilde\varphi)$ of the  Cauchy-Riemann  equations
(\ref{FCR}) satisfies:
\beq
\partial_{\varphi}\tilde\theta\partial_{\varphi}\tilde\varphi\,=\,-
\partial_{u}\tilde\theta\partial_{u}\tilde\varphi\,\,,
\eeq
and, since $\tilde\varphi$, being a solution of the Cauchy-Riemann 
equations, is harmonic in $(u,\varphi)$, it follows from (\ref{Feqr})
that 
$\partial_{\varphi}r_u=\partial_{u}r_{\varphi}$ if and only if
$\tilde{\psi}$ is also harmonic in $(u,\varphi)$, \ie\ the differential
equation  for $\tilde{\psi}$ is just the Laplace equation in the
$(u,\varphi)$ plane, namely:
\beq
\partial_{\varphi}^2\tilde{\psi}\,+\,\partial_u^2\tilde{\psi}\,=\,0\,\,.
\label{Flaplacepsi}
\eeq
Remarkably, the form of
$r(\theta,\varphi)$ can be obtained in general. Let us define:
\beq
\Lambda(\theta,\varphi)\,=\,\int_{0}^{\varphi} d\varphi \sin\theta
\partial_{\theta}\tilde{\psi}(\theta,\varphi)\,-\,
\int {d\theta\over \sin\theta}\,\,
\partial_{\varphi}\tilde{\psi}(\theta,0)\,\,,
\eeq
It follows from this definition and the fact that $\tilde{\psi}$ is
harmonic  in $(u,\varphi)$ that $\tilde{\psi}$ and $\Lambda$ also satisfy
the  Cauchy-Riemann equations:
\beq
{\partial \Lambda\over \partial \varphi}\,=\,
{\partial \tilde{\psi}\over \partial u}\,\,,
\,\,\,\,\,\,\,\,\,\,\,\,\,\,\,\,\,\,\,\,\,
{\partial \Lambda\over \partial u}\,=\,-
{\partial \tilde{\psi}\over \partial \varphi}\,\,.
\label{FCRpsi}
\eeq
Thus $\tilde{\psi}$ and $\Lambda$ are conjugate harmonic  functions, \ie\
$\tilde{\psi}+i\Lambda$ is an analytic function of $u+i\varphi$. Notice
that given
$\Lambda$ one can obtain $\tilde{\psi}$ by integrating the previous
differential equations. It can be checked by using the Cauchy-Riemann
equations that the derivatives of $r$, as given by the right hand side of
eq. (\ref{Feqr}), can be written as
$r_{\theta}=\partial_{\theta} F$, $r_{\varphi}=\partial_{\varphi} F$, 
where:
\beq
F(\theta,\varphi)\,=\,{1\over 2}\,\,\bigg[\,\Lambda(\theta,\varphi)\,-\,
\log\big(\sin\theta\sin\tilde\theta(\theta,\varphi)\big)\,\bigg]\,\,.
\eeq
Therefore, it follows that:
\beq
e^{2r}\,=\,C\,\,\,{e^{\Lambda(\theta,\varphi)}\over 
\sin\theta\sin\tilde\theta(\theta,\varphi)}\,\,,
\label{Fsolution}
\eeq
with $C$ being a constant. We will make use of this amazingly  simple
expression  to derive the equation of some particularly interesting
embeddings.

\subsection{$n$-Winding solitons}

First of all, let us consider the particular class of  solutions of the
Cauchy-Riemann eqs. (\ref{FCRu}):
\beq
\tilde u+i\tilde\varphi\,=\,n(u+i\varphi)\,+\,{\rm constant}\,\,,
\label{arrolamento}
\eeq
where $n$ is an integer and the constant is complex.  In terms of the
original variables:
\beq
\tan{\tilde\theta\over 2}\,=\,\tilde c\,
\Bigg(\,\tan{\theta\over 2}\,\Bigg)^{n}\,\,,
\,\,\,\,\,\,\,\,\,\,\,\,\,\,\,\,\,
\tilde\varphi\,=\,n\,\varphi\,+\,\varphi_0\,\,,
\label{FVCRsol}
\eeq
with $\tilde c$ and $\varphi_0$ being constants. 
It is clear that in this solution the $\tilde\varphi$  coordinate of the
probe wraps $n$ times the $[0,2\pi]$ interval as $\varphi$ varies between
$0$ and $2\pi$. Let us now assume that the coordinate $\tilde{\psi}$ is
constant,
\ie\ $\tilde{\psi}=\tilde{\psi}_0$. It is clear from its definition that the
function
$\Lambda(\theta,\varphi)$ is zero in this case. Moreover,  by using the
identities:
\beq
\sin x\,=\,{2\tan{x\over 2}\over
1+\tan^2{x\over 2}}\,\,,
\,\,\,\,\,\,\,\,\,\,\,\,\,\,\,\,\,\,\,\,\,
\tan{x\over 2}\,=\,\sqrt{{1-\cos x\over 1+\cos x}}\,\,,
\eeq
one can prove that:
\beq
\sin\tilde\theta\,=\,2\sqrt c\,\,\,{(\sin\theta\,)^{n}\over
(\,1\,+\,\cos\theta\,)^n\,+\,c\,(\,1\,-\,\cos\theta\,)^n}\,\,,
\label{Ftildetheta}
\eeq
where $c=\tilde c^2$. 
After plugging this result  in eq. (\ref{Fsolution}), one obtains   the
explicit form of the function  $r(\theta)$, namely:
\beq
e^{2r}\,=\,{e^{2r_{*}}\over 1+c}\,\,
{(\,1\,+\,\cos\theta\,)^n\,+\,c\,(\,1\,-\,\cos\theta\,)^n\over
(\sin\theta\,)^{n+1}}\,\,,
\label{Fabspikes}
\eeq
where $r_{*}=r(\theta=\pi/2)$. We will call $n$-winding  embedding to the
brane configuration corresponding to eqs. (\ref{FVCRsol}) and
(\ref{Fabspikes}) for a constant value of the angle $\tilde{\psi}$.

\begin{figure}
\centerline{\hskip -.8in \epsffile{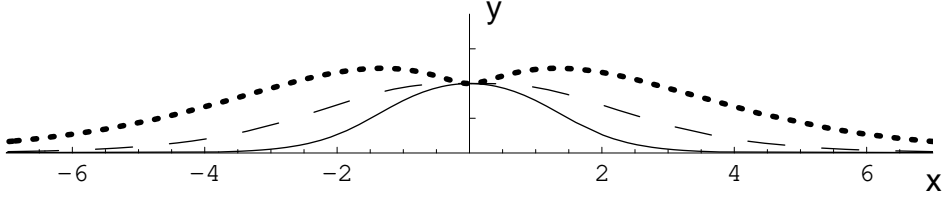}}
\caption{Curves $y=y(x)$ for three values of the winding  number $n$: $n=0$
(solid line), 
$n=1$ (dashed curve) and $n=2$ (dotted line). These three  curves
correspond to $r_*=1$. }
\label{Ffig1}
\end{figure}

Let us pause for a moment to study the function (\ref{Fabspikes}).  First of
all it is easy to verify that this function is invariant if we change
$n\rightarrow -n$ and
$c\rightarrow 1/c$ (or equivalently  changing $\theta\rightarrow
\pi-\theta$ for the same constant $c$). Actually, in what follows we shall
take the integration constant
$c=1$ and thus we can restrict ourselves to the case in which $n$  is
non-negative.  In this $c=1$ case, $r_*$ is the minimal separation between
the brane probe and the origin.
Another observation is that $r$ diverges for $\theta=0,\pi$,
which corresponds to the location of the spikes of the  worldvolume
solitons. Therefore the supersymmetric embedding we have found is
non-compact. Actually, it has the topology of a cylinder whose compact
direction is parametrized by $\varphi$. This cylinder connects the two
poles at $\theta=0,\pi$ of the $(\theta,\varphi)$ sphere at $r=\infty$ and
passes at a distance $r_*$ from the origin.

It is also interesting to discuss the symmetries of our
solutions. Recall that the angle $\tilde{\psi}$ is constant for our
embeddings.  Thus, it is clear that one can shift it by an arbitrary
constant $\epsilon$ as $\tilde{\psi}\to\tilde{\psi}+\epsilon$. This $U(1)$
symmetry corresponds to an isometry of the abelian background which
is broken quantum-mechanically  to $\ZZ_{2N}$ as a consequence of the flux
quantization of the RR two-form potential
\cite{MN, KOW, GHP}. In the gauge theory side this isometry has  been
identified 
\cite{MN, KOW, GHP} with the $U(1)$ R-symmetry of the ${\cal N}=1$  SYM
theory, which is broken down to $\ZZ_{2N}$ by a field theory anomaly
\cite{Affleck:1983mk}. On the other hand, it is also clear that we have an
additional $U(1)$ associated to constant shifts in
$\tilde\varphi$,  which are equivalent to a redefinition of  $\varphi_0$ in
eq. (\ref{FVCRsol}).

To visualize the shape of the
brane in these solutions it is rather convenient to introduce  the
following cartesian coordinates $x$ and $y$ :
\beq
x\,=\,r\cos\theta\,\,,
\,\,\,\,\,\,\,\,\,\,\,\,\,\,\,
y\,=\,r\sin\theta\,\,.
\label{Fcartesian}
\eeq
In terms of $(x,y)$ the D5-brane embedding will be described  by means of a
curve
$y=y(x)$. Notice that $y\ge 0$, whereas $-\infty<x<+\infty$.  The value of
the coordinate
$y$ at $x=0$ is just $r_*$, \ie\ $y(x=0)=r_*$. Moreover, for  large values
of the coordinate $x$, the function $y(x)\rightarrow 0$ exponentially as
\beq
y(x)\,\approx C\,|x|\,e^{-{2\over |n|+1}|x|}\,\,,
\,\,\,\,\,\,\,\,\,\,\,\,\,\,\,
(|x|\rightarrow \infty)\,\,,
\eeq
where $C$ is a constant. To illustrate this behavior we have  plotted in
figure \ref{Ffig1} the curves $y(x)$ for three different values of the
winding
$n$ and the same value of 
$r_*$.

A particularly interesting
case is obtained when $n=\pm 1 $. By adjusting properly the  constant
$\varphi_0$ in eq. (\ref{FVCRsol}), the angular embedding reduces to:
\bear
&&\tilde\theta\,=\,\theta\,\,,
\,\,\,\,\,\,\,\,\,\,\,\,\,\,
\tilde\varphi\,=\,\varphi\,+\,{\rm constant}\,\,,
\,\,\,\,\,\,\,\,\,\,\,\,\,\,\,\,\,\,\,\,\,\,\,\,\,\,\,\,\,\,
\,\,\,\,\,\,\,\,\,\,\,\, (n=1)\,\,,\rc\rc
&&\tilde\theta\,=\,\pi-\theta\,\,,
\,\,\,\,\,\,\,\,\,\,\,\,\,\,
\tilde\varphi\,=\,2\pi-\varphi\,+\,{\rm constant}\,\,,
\,\,\,\,\,\,\,\,\,\,\,\,\,\,(n=-1)\,\,,
\label{FBM}
\eear 
with $\tilde{\psi}$ being constant. 
These types of angular embeddings are similar to the ones   considered in
ref.
\cite{BertMer} (although they are not the same, see eq.
(\ref{FBMembeddings}) )  and we will refer to them as
unit-winding embeddings. Notice that the two cases displayed  in eq. 
(\ref{FBM}) represent the two possible identifications of the
$(\theta,\varphi)$ and
$(\tilde\theta,\tilde\varphi)$ two-spheres. 

When $n=0$ the brane is wrapping the $(\theta,\varphi)$ sphere  at constant
values of
$\tilde\theta$ and $\tilde\varphi$, \ie\ one has:
\beq
\tilde\theta\,=\,{\rm constant}\,=\,\tilde\theta_0\,\,,
\,\,\,\,\,\,\,\,\,\,\,\,\,\,\,\,\,\,\,
\tilde\varphi\,=\,{\rm constant}\,=\,\tilde\varphi_0\,\,,
\,\,\,\,\,\,\,\,\,\,\,\,\,\,\,\,\,\,\,(n=0)
\eeq
We will refer to this case as zero-winding embedding \cite{DV}. 

One can verify that
the brane embeddings we have found are solutions of the probe  equations of
motion. Actually, they are supersymmetric worldvolume solitons of the D5
brane probe. To illustrate this fact let us show that these  configurations
saturate a BPS energy bound. To simplify matters, let us assume that the
angular embedding is the one displayed in eq. (\ref{FVCRsol}) and let
$r(\theta)$ be an arbitrary function.  The Dirac-Born-Infeld (DBI)
lagrangian density for a D5-brane with unit tension is:
\beq
{\cal L}\,=\,-e^{-\phi}\,\sqrt{-g_{st}}\,+\,P\,\Big[\,C_{(6)}\,\Big]\,\,,
\label{FDBI}
\eeq
where $g_{st}$ is the determinant of the induced metric in the string frame
($g_{st}=e^{3\phi}\,g$) and  $P\,\Big[\,C_{(6)}\,\Big]$ is  the pullback on
the worldvolume of the RR six-form of the background. The elements of the
induced metric for the $n$-winding solution along the angular coordinates
are:
\bear
g_{\theta\theta}&=&e^{{\phi\over 2}}\,
\bigg(\,e^{2h}\,+\,{n^2\over 4}\,
{\sin^2\tilde\theta\over \sin^2\theta}\,+\,r_{\theta}^2\,\bigg)\,\,,\cr\cr
g_{\varphi\varphi}&=&e^{{\phi\over 2}}\,
\bigg(\,e^{2h}\,+\,{n^2\over 4}\,
{\sin^2\tilde\theta\over \sin^2\theta}\,+\,V_{3\varphi}^2\,\bigg)
\sin^2\theta
\,\,.
\eear
From this expression one immediately obtains the determinant  of the
induced metric, namely:

\beq 
\sqrt{-g}\,=\,e^{{3\phi\over 2}}\,\sin\theta\,\,
\sqrt{\bigg(\,e^{2h}\,+\,{n^2\over 4}\,
{\sin^2\tilde\theta\over \sin^2\theta}\,+\,V_{3\varphi}^2\,\bigg)\,
\bigg(\,e^{2h}\,+\,{n^2\over 4}\,
{\sin^2\tilde\theta\over \sin^2\theta}\,+\,r_{\theta}^2\,\bigg)}\,\,.
\eeq
Moreover, the pullback on the worldvolume of the two-form ${\cal C}$ is
\footnote{It is worth mentioning that the pullback of the  RR two-form to
the worldvolume is
$$
P[C_{(2)}]\,=\,{\tilde{\psi}\over 4}\,d\varphi\wedge 
\big(\,n\sin\tilde\theta d\tilde\theta\,-\,\sin\theta d\theta\,\big)\,\,,
$$
where $\tilde\theta(\theta)$ is the function displayed in  eq.
(\ref{Ftildetheta}) . From this expression it is straightforward to verify
that the RR two-form flux through the two-submanifold where we are wrapping
our brane is
$$
\int\,P[C_{(2)}]\,=\,\pi\tilde{\psi}\,(|n|-1)\,\,,
$$
and thus it vanishes  iff $n=\pm 1$.
 }:
\beq
P\,\Big[\,{\cal C}\,\Big]\,=\,{e^{2\phi}\over 8}\,
(\,16e^{2h}\,\cos\theta\,-\,ne^{-2h}\,\cos\tilde\theta\,)\,r_{\theta}\,
d\varphi\wedge d\theta
\eeq
The hamiltonian density ${\cal H}$ for a static configuration is just 
${\cal H}\,=\,-{\cal L}$ or:
\bear
&&{\cal H}\,=\,e^{2\phi}\,\Bigg[\,\sin\theta\,
\sqrt{\bigg(\,e^{2h}\,+\,{n^2\over 4}\,
{\sin^2\tilde\theta\over \sin^2\theta}\,+\,V_{3\varphi}^2\,\bigg)\,
\bigg(\,e^{2h}\,+\,{n^2\over 4}\,
{\sin^2\tilde\theta\over \sin^2\theta}\,+\,r_{\theta}^2\,\bigg)}\,
-\,\rc\rc\rc &&\,\,\,\,\,\,\,\,\,\,\,\,\,\,\,\,\,\,\,
\,-\,{1\over 8}\,
(\,16e^{2h}\,\cos\theta\,-\,ne^{-2h}\,\cos\tilde\theta\,)\,r_{\theta}\,
\Bigg]\,\,,
\eear
It can be checked that, for an arbitrary function $r(\theta)$,  one can
write
${\cal H}$ as:
\beq
{\cal H}\,=\,{\cal Z}\,+\,{\cal S}\,\,,
\eeq
where ${\cal Z}$ is a total derivative:
\beq
{\cal Z}\,=\,-\partial_{\theta}\,\Bigg[\,
e^{2\phi}\,\bigg(e^{2h}\,\cos\theta\,+\,{n\over 4}\, \cos\tilde\theta\,
\bigg)\,\Bigg]\,\,,
\eeq
and ${\cal S}$ is non-negative:
\beq
{\cal S}\ge 0\,\,,
\eeq
with ${\cal S}= 0$ precisely when the BPS equations for the embedding are
satisfied. The expression of ${\cal S}$ is:
\bear
{\cal S}&=&\sin\theta\,e^{2\phi}\,\Bigg[\,
\sqrt{\bigg(\,e^{2h}\,+\,{n^2\over 4}\,
{\sin^2\tilde\theta\over \sin^2\theta}\,+\,V_{3\varphi}^2\,\bigg)\,
\bigg(\,e^{2h}\,+\,{n^2\over 4}\,
{\sin^2\tilde\theta\over \sin^2\theta}\,+\,r_{\theta}^2\,\bigg)}\,\,-\rc\rc
&&\,\,\,\,\,\,\,\,\,\,\,\,\,\,\,\,\,\,\,\,\,\,\,\,\,\,\,\,
-\bigg(\,e^{2h}\,+\,{n^2\over 4}\,
{\sin^2\tilde\theta\over \sin^2\theta}\,-\,V_{3\varphi}\,
r_{\theta}\,\bigg)\,
\Bigg]\,\,.
\label{FcalS}
\eear
The BPS equation for $r$ in this case is $r_{\theta}=-V_{3\varphi}$  (see
eq. (\ref{Fabrderiv})). If this equation is satisfied,  the first term on
the right-hand side of eq.  (\ref{FcalS}) is a square root of a perfect
square which cancels against the second term of this equation. Moreover, it
is easy to check that the condition ${\cal S}\ge 0$ is equivalent to:
\beq
\big(r_{\theta}\,+\,V_{3\varphi}\,\big)^2\ge 0\,\,,
\eeq
which is obviously satisfied and reduces to an equality if and  only if the
BPS equation for the embedding is satisfied. 

\subsection{(n,m)-Winding solitons}

The solutions found in the previous section are easily  generalized if we
allow the angle $\tilde{\psi}$ to wind a certain number of times as the
coordinate
$\varphi$ varies from 
$\varphi=0$ to $\varphi=2\pi$. Recalling that $\tilde{\psi}$ ranges  from
$0$ to
$4\pi$, let us write the following ansatz for $\tilde{\psi}(\varphi)$:
\beq
\tilde{\psi}\,=\,\tilde{\psi}_0\,+\,2m\varphi\,\,,
\eeq
where $m$ is an integer. It is obvious that the above function  satisfies
the Laplace equation (\ref{Flaplacepsi}). Moreover, its harmonic conjugate
$\Lambda$ is immediately obtained by solving the Cauchy-Riemann
differential equations (\ref{FCRpsi}), namely:
\beq
\Lambda\,=\,-2mu\,\,.
\eeq
In terms of the angle $\theta$, the above equation becomes:
\beq
e^{\Lambda}\,=\,{1\over \Big(\,\tan{\theta\over 2}\,\Big)^{2m}}\,\,.
\eeq
By plugging this result in eq. (\ref{Fsolution}), and using the value of
$\sin\tilde\theta$ given in eq.  (\ref{Ftildetheta}), it is  straightforward
to obtain the function $r(\theta)$ of the embedding. One gets:
\beq
e^{2r}\,=\,{e^{2r_{*}}\over 1+c}\,\,
{(\,1\,+\,\cos\theta\,)^n\,+\,c\,(\,1\,-\,\cos\theta\,)^n\over
\big[\tan{\theta\over 2}\big]^{2m}
(\sin\theta\,)^{n+1}}\,\,,
\label{Fnmsol}
\eeq
where, as in the $n$-winding case,   $r_{*}=r(\theta=\pi/2)$.

An interesting observation concerning the solution we have  just found is
that, by choosing appropriately the winding number $m$, one of the spikes
of the $m=0$ solutions at $\theta=0$ or $\theta=\pi$ disappears. Indeed if,
for example,  $n$ is non-negative and we take $2m=n+1$, the function
$r(\theta)$ is regular at $\theta=0$. Similarly, also when $n\ge 0$, one
can eliminate the spike at $\theta=\pi$ by choosing
$2m=-n-1$.

\subsection{Spiral solitons}

By considering more general solutions of the Cauchy-Riemann  equations
(\ref{FCRu}) and  (\ref{FCRpsi}) we can obtain many more classes of
supersymmetric configurations of the brane probe. One of the questions one
can address is whether or not one can have embeddings in which $r$ is
finite for all values of the angles. We will now see that the answer to
this question is yes, although the corresponding embeddings seem not to be
very interesting. To illustrate this point, let us see how we can find
functions
$\tilde{\psi}$ and $\Lambda$ such that they make the radial  coordinate of
the
$n$-winding embedding finite at $\theta=0, \pi$. First of all, notice that,
in terms of the Cauchy-Riemann variables $u$ and $\tilde u$ defined in eq.
(\ref{Fuvariables}), we have to explore the behavior of the embedding at
$u, \tilde u\rightarrow \pm\infty$. Since:
\beq
\sin\theta\,=\,{2e^{u}\over 1+e^{2u}}\,\,,
\,\,\,\,\,\,\,\,\,\,\,\,\,\,\,\,\,\,\,\,\,
\sin\tilde\theta\,=\,{2e^{\tilde u}\over 1+e^{2\tilde u}}\,\,,
\eeq
one has that $\sin\theta\rightarrow e^{-|u|}$,  
$\sin\tilde\theta\rightarrow e^{-|\tilde u|}$ as $u, \tilde u\rightarrow
\pm\infty$. Then, the factors multiplying 
$e^{\Lambda}$ in eq. (\ref{Fsolution}) diverge as $e^{|u|+|\tilde u|}$.  In
the
$n$-winding solution $|\tilde u|=|n|| u|$ and, therefore this  divergence
is of the type
$e^{(|n|+1)|u|}$. We can cancel this divergence by adding a
$\Lambda$ such that $e^{\Lambda}\rightarrow 0$ as  $u\rightarrow\pm\infty$
in  such a way that, for example,   $\Lambda+(|n|+1)|u|\rightarrow
-\infty$. This is clearly achieved by taking a function $\Lambda$ such that
$\Lambda\rightarrow-u^2$. It is straightforward to find an
analytic function in the $(u,\varphi)$ plane such that  its imaginary part
behaves as
$-u^2$ for $u\rightarrow\pm\infty$.  One can take
\beq
\tilde{\psi}+i\Lambda\,=\,-i(u+i\varphi)^2\,=\,2u\varphi\,-\,
i(u^2-\varphi^2)\,\,.
\eeq
From this equation we can read the functions $\tilde{\psi}$ and $\Lambda$. 
In terms of 
$\theta$ and $\varphi$ they are:
\beq
\tilde{\psi}\,=\,2u\varphi\,=\,2\varphi\log\tan{\theta\over 2}\,\,,
\,\,\,\,\,\,\,\,\,\,\,\,\,\,\,\,\,\,\,\,\,
\Lambda\,=\,-u^2+\varphi^2\,=\,-\big(\,\log\tan{\theta\over 2}
\,\big)^2\,+\,
\varphi^2\,\,.
\eeq
In this case $r\rightarrow 0$, $\tilde{\psi}\rightarrow\pm\infty$ as 
$\theta\rightarrow 0,\pi$, which means that we describe an  infinite spiral
which winds infinitely in the $\tilde{\psi}$ direction. Notice that,
although $r$ is always finite, the volume of the two-submanifold is
infinite due to this infinite winding. One can try other alternatives to
make the radial coordinate finite. In all the ones we have analyzed, one
obtains the infinite spiral behavior described above.

\setcounter{equation}{0}
\section{Worldvolume solitons (non-abelian case)}
\label{secnonabel}

Let us consider the full non-abelian background and let us  try to obtain
solutions to the kappa symmetry equations
(\ref{Fkappauno})-(\ref{Fkappacinco}). Actually we will restrict ourselves
to the situations in which $r$ only depends on the angle $\theta$.  It
can be easily checked that, in this case, only four of the five equations 
(\ref{Fkappauno})-(\ref{Fkappacinco}) are independent. As an  independent
set of equations we will choose eqs. (\ref{Fkappauno}), (\ref{Fcondition})
and:
\bear
&&\partial_{\theta}r=-{e^{h}\,V_{3\varphi}\over
e^{h}\,\cos\alpha\,+\,(\,V_{2\varphi}\,-\,{a \over 2}\,)\,\sin\alpha}\,\,,
\label{Fnabuno}\\\rc
&&\sin\alpha V_{1\varphi}\,\partial_{\theta}r\,-\,e^{h}\,V_{3\theta}=0\,\,,
\label{Fnabdos}
\eear
which can be obtained from eqs. (\ref{Fkappados})  and (\ref{Fkappacinco})
after taking
$\partial_{\varphi} r=0$.

We will now try to find the non-abelian version of  the solutions found in
the abelian theory for arbitrary winding $n$. With this purpose, let us
consider the following ansatz for $\tilde\varphi$:
\beq
\tilde\varphi(\theta,\varphi)\,=\,n\varphi\,+\,f(\theta)\,\,,
\eeq
while we shall assume that $\tilde\theta$, $\tilde{\psi}$ and $r$ are
functions of
$\theta$ only. We will require that, in the asymptotic UV,
$\tilde\varphi\rightarrow n\varphi$. It is clear that in this ansatz
$\partial_{\varphi}\tilde\varphi\,=\,n$ and that 
$\partial_{\theta}\tilde\varphi\,=\,\partial_{\theta}f$. Moreover, from 
eq. (\ref{Fkappauno}) we can obtain the relation between
$\partial_{\theta}\tilde\varphi$ and 
$\partial_{\theta}\tilde\theta$, namely:
\beq
\partial_{\theta}\tilde\varphi\,=\,\tan\tilde{\psi}\,\,\bigg[\,
{\partial_{\theta}\tilde\theta\over \sin\tilde\theta}\,-\,
{n\over \sin\theta}\,\bigg]\,\,.
\label{Frel}
\eeq
Using this value of $\partial_{\theta}\tilde\varphi$, we get the following
values of the $V$ functions:
\bear
&&V_{1\theta}\,=\,{1\over 2}\,\bigg[\,
{\partial_{\theta}\tilde\theta\over \cos\tilde{\psi}}\,-\,n\,
{\sin\tilde\theta\over \sin\theta}\,\,{\sin^2\tilde{\psi}\over
\cos\tilde{\psi}}\,\bigg]\,\,,\rc\rc
&&V_{1\varphi}\,=\,{n\over 2}\,{\sin\tilde\theta\over
\sin\theta}\,\sin\tilde{\psi}
\,=\,-V_{2\theta}\,\,,\rc\rc
&&V_{2\varphi}\,=\,{n\over 2}\,{\sin\tilde\theta\over
\sin\theta}\,\cos\tilde{\psi}
\,\,,\rc\rc
&&V_{3\theta}\,=\,{1\over 2}\,\partial_{\theta}\tilde{\psi}\,+\,
{1\over
2}\,\cot\tilde\theta\,\tan\tilde{\psi}\,\partial_{\theta}\tilde\theta\,-\,
{n\over 2}\,\tan\tilde{\psi}\,{\cos\tilde\theta\over\sin\theta}\,\,,\rc\rc
&&V_{3\varphi}\,=\,{n\over 2}\,{\cos\tilde\theta\over \sin\theta}\,+\,
{1\over 2}\,\cot\theta\,\,.
\label{FnabVs}
\eear
By using these values in eq. (\ref{Fcondition}), one gets the value of 
$\partial_{\theta}\tilde\theta$ in terms of the other variables:
\beq
\partial_{\theta}\tilde\theta\,=\,
{n\sin\tilde\theta\cosh 2r-\sin\theta\cos\tilde{\psi}\over 
\sin\theta\cosh 2r - n\sin\tilde\theta\cos\tilde{\psi}}\,\,.
\label{Feqtildetheta}
\eeq
On the other hand, by combining the two equations 
(\ref{Frel}) and (\ref{Feqtildetheta}), we obtain:
\beq
\partial_{\theta}\,\tilde\varphi\,=\,
{n^2\sin^2\tilde\theta\,-\,\sin^2\theta\over
\sin\theta\sin\tilde\theta \,
(\,\sin\theta\cosh
2r\,-\,n\sin\tilde\theta\cos\tilde{\psi}\,)}\,\sin\tilde{\psi}\,\,.
\label{Feqtildevarphi}
\eeq
Moreover, plugging the values of
$V_{1\varphi}$, $V_{2\varphi}$, $V_{3\theta}$ and $V_{3\varphi}$ in eqs. 
(\ref{Fnabuno}) and (\ref{Fnabdos}) we can obtain the values of the
derivatives of $\tilde{\psi}$ and $r$. The result is:
\bear
\partial_{\theta}\tilde{\psi}&=&
{n\cot\theta\sin\tilde\theta+\cot\tilde\theta\sin\theta\over
\sin\theta\cosh2r-n\sin\tilde\theta\cos\tilde{\psi}}\,\,\sin\tilde{\psi}
\,\,,\rc\rc
\partial_{\theta}r&=&-{1\over 2}\,\,
{n\cos\tilde\theta+\cos\theta\over 
\sin\theta\cosh2r- n\sin\tilde\theta\cos\tilde{\psi}}\,\,\sinh 2r\,\,.
\label{Feqpsiyr}
\eear

It follows from eq. (\ref{Feqtildetheta}) that, asymptotically in the UV, 
$\sin\theta\,\partial_{\theta}\tilde\theta\rightarrow n\sin\tilde\theta$. 
If one wants to fulfil this relation for arbitrary $r$ it is easy to see
from  eq. (\ref{Feqtildetheta}) that one must have 
$(n\sin\tilde\theta)^2=\sin^2\theta$, which only happens for $n=\pm 1$ and
$\sin\theta=\sin\tilde\theta$. Noticing that for  these values one has
$\partial_{\theta}\tilde\theta\,=\,\pm 1$, one is finally led to the two 
possibilities of eq. (\ref{FBM}): $\tilde\theta=\theta$ for $n=1$ and
$\tilde\theta=\pi-\theta$ for $n=-1$. Notice that  in the two cases of eq. 
(\ref{FBM}) this equation implies that 
$\partial_{\theta}\,\tilde\varphi=0$ and thus when  $n=\pm 1$ the angular
identifications of the abelian unit-winding embeddings (eq. (\ref{FBM}))
also solve the non-abelian equations (\ref{Feqtildetheta}) and 
(\ref{Feqtildevarphi}) for all $r$.

For a general value of $n$, one has  that asymptotically in the UV
$\partial_{\theta}\tilde\varphi\rightarrow 0$ and 
$\partial_{\theta}\tilde{\psi}\rightarrow 0$, as in the abelian solutions.
Moreover  it follows from eqs. (\ref{Feqtildevarphi}) and (\ref{Feqpsiyr})
that
$\tilde\varphi$ and $\tilde{\psi}$ can be kept constant for all $r$ if
$\sin\tilde{\psi}=0$,
\ie\ when $\tilde{\psi}=0,\pi\,\,\,\mod\,\,2\pi$. For these  values of
$\tilde{\psi}$ the equations simplify and, although we will not attempt to
do it here,  one could try to integrate numerically the equations of
$\tilde\theta$ and $r$. It is however interesting to  point out that,
contrary to what happens in the abelian $n$-winding solution, the angle
$\tilde{\psi}$ cannot be an arbitrary constant for the non-abelian probes.
As we will argue below, this is a geometrical realization of the  breaking
of the R-symmetry of the corresponding ${\cal N}=1$ SYM theory in the IR.
On the contrary, the angle
$\tilde\varphi$ can take an arbitrary constant  value, as in the abelian
solution.

\subsection{Non-abelian unit-winding solutions}
\label{secunit}

Let us now obtain the non-abelian generalization  of the unit-winding
solutions. First of all we define:
\beq
\eta\,=\,n=\pm 1\,\,.
\eeq
We have already noticed that for unit-winding  embeddings the values of
$\tilde\theta$ and $\tilde\varphi$ displayed in eq. (\ref{FBM}) solve the
non-abelian differential equations (\ref{Feqtildetheta}) and
(\ref{Feqtildevarphi}). Therefore,   let us try to find a solution in the
non-abelian theory in which the  embedding of the $(\tilde\theta,
\tilde\varphi)$ coordinates is the same as in the abelian theory, \ie\ as
in eq. (\ref{FBM}).  For this type of embeddings 
$\sin\tilde\theta=\sin\theta$, $\partial_{\theta}\tilde\theta\,=\,\eta$ and
eq. (\ref{FnabVs}) reduces to:
\bear
&&V_{1\theta}\,=\,{\eta\cos\tilde{\psi}\over 2}\,\,,
\,\,\,\,\,\,\,\,\,\,\,\,\,\,\,\,\,\,\,\,\,\,\,\,\,\,
V_{1\varphi}\,=\,{\eta\sin\tilde{\psi}\over 2}\,\,,\rc\rc
&&V_{2\theta}\,=\,-{\eta\sin\tilde{\psi}\over 2}\,\,,
\,\,\,\,\,\,\,\,\,\,\,\,\,\,\,\,\,\,\,\,\,\,
V_{2\varphi}\,=\,{\eta\cos\tilde{\psi}\over 2}\,\,,\rc\rc
&&V_{3\theta}\,=\,{\tilde{\psi}_{\theta}\over 2}\,\,,
\,\,\,\,\,\,\,\,\,\,\,\,\,\,\,\,\,\,\,\,\,\,\,\,\,\,\,\,
\,\,\,\,\,\,\,\,\,\, V_{3\varphi}\,=\,\cot\theta\,\,,
\eear
where we have denoted
$\tilde{\psi}_\theta\equiv\partial_{\theta}\,\tilde{\psi}$.  As a check,
notice that
$V_{1\theta}$,  $V_{1\varphi}$, $V_{2\theta}$ and $V_{2\varphi}$  satisfy
eqs. (\ref{FVCR}). It follows from eq.  (\ref{Fcondition}) that they must
also satisfy:
\beq
V_{1\theta}^2\,+\,V_{1\varphi}^2\,=\,{1\over 4}\,\,,
\,\,\,\,\,\,\,\,\,\,\,\,\,\,\,\,\,\,\,\,\,\,
V_{2\theta}^2\,+\,V_{2\varphi}^2\,=\,{1\over 4}\,\,,
\eeq
which indeed they verify. Moreover, by substituting 
$\sin\tilde\theta=\sin\theta$ and
$\cos\tilde\theta=\eta\cos\theta$ in eq. (\ref{Feqpsiyr}), we  obtain the
following differential equations for $\tilde{\psi}(\theta)$ and $r(\theta)$:
\bear
\tilde{\psi}_\theta&=&-{2\eta\sin\tilde{\psi}\over \sinh
2r}\,\,r_\theta\,\,,\rc\rc r_\theta&=&-{\cot\theta\over 
\cosh 2r\,-\,\eta\cos\tilde{\psi}}\,\,
\sinh 2r\,\,.
\label{FnabBPS}
\eear
These equations can be integrated with the result:
\bear
&&\Bigg(\,\tan\,{\tilde{\psi}\over 2}\,\Bigg)^{\eta}\,=\,A\,\coth
r\,\,,\rc\rc && {\sinh r\over \sqrt{A^2+\tanh^2r}}\,=\,{C\over
\sin\theta}\,\,, 
\label{FBMnoab}
\eear
where $A$ and $C$ are constants of integration.  Eq. (\ref{FBMnoab}),
together with eq. (\ref{FBM}), determines the unit-winding embeddings of the
probe in the non-abelian background. 
Notice that, as in the corresponding
abelian solution, $r$ diverges when $\theta=0,\pi$, \ie\  the brane probe
extends infinitely in the radial direction.  On the other hand,
it is also instructive to explore the $r\rightarrow\infty$ limit of the
solution (\ref{FBMnoab}). First of all, it is clear that when 
$r\rightarrow\infty$ the angle
$\tilde{\psi}$ reaches asymptotically  a constant value $\tilde{\psi}_0$,
given by:
\begin{figure}
\centerline{\hskip -.8in \epsffile{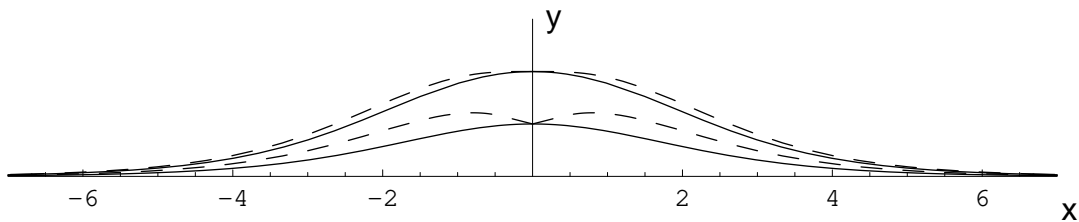}}
\caption{Comparison between the non-abelian (solid line) and  abelian
(dashed line) unit-winding embeddings for the same value of $r_*$ . The
non-abelian embedding is the one corresponding to eq. (\ref{Fnoabflavor})
and the abelian one is that given in  eq. (\ref{Fabspikes}) 
for $n=1$ and $c=1$.  The curves for two different values of $r_*$
($r_*=0.5$ and $r_*=1$) are shown. The variables $(x,y)$ are  the ones
defined in eq. (\ref{Fcartesian}). }
\label{Ffig2}
\end{figure}

\beq
\cos\tilde{\psi}_0\,=\,{1-A^2\over 1+A^2}\,\,\eta\,\,.
\eeq
Moreover, when $r\rightarrow\infty$, the function $r(\theta)$  displayed
in eq. (\ref{FBMnoab}) becomes, after a proper identification of  the
integration constants, exactly the one written in eq. (\ref{Fabspikes}) for
$n=\pm 1$ and $c=1$. Notice that  the angle $\tilde{\psi}$ in the  embedding
(\ref{FBMnoab}) is not constant in general. Actually, only when $A=0$ or
$A=\infty$ the coordinate
$\tilde{\psi}$ remains constant and equal to $0,\pi\,\,\mod\,2\pi$ 
($\cos\tilde{\psi}=\eta$ for $A=0$ and
$\cos\tilde{\psi}=-\eta$ in the case
$A=\infty$). It is interesting to write the dependence of  $r$ on $\theta$
in these two particular cases. When $A=\infty$ the solution is:

\bear
\tilde\theta&=&\cases{\theta, &if $\,\,\eta=+1\,,$\cr
                        \pi-\theta, &if $\,\,\eta=-1\,,$ }\,\,,
\,\,\,\,\,\,\,\,\,\,\,\,
\tilde\varphi\,=\,\cases{\varphi+{\rm constant}, &if $\,\,\eta=+1\,,$\cr
                        2\pi-\varphi+{\rm constant},   &if $\,\,\eta=-1$
}\,\,,\rc\rc
\tilde{\psi}&=&\cases{\pi,3\pi, &if $\,\,\eta=+1\,,$\cr
               0,2\pi, &if $\,\,\eta=-1\,,$ }\,\,,
\,\,\,\,\,\,\,\,\,\,\,\,
\sinh r\,=\,{\sinh r_*\over \sin\theta}\,\,\,,
\label{Fnoabflavor}
\eear
where $r_*$ is the minimal value of $r$ (\ie\ $r_*=r(\theta=\pi/2)$)  and
we have also displayed the angular part of the embedding. Notice that, for
a given sign of the winding number $\eta$, only two  values of
$\tilde{\psi}$ are possible. Thus, in this solution, the $U(1)$ symmetry
of shifts in $\tilde{\psi}$ is broken to a $\ZZ_2$ symmetry. This will be
interpreted in section \ref{secmesons} as the realization, at the level of
the brane probe, of the R-symmetry breaking of the gauge theory.

To have a better understanding of the solution  (\ref{Fnoabflavor}) we have
plotted it in figure \ref{Ffig2}
in terms of the variables $(x,y)$ defined in eq.  (\ref{Fcartesian}). For
comparison we have also plotted the abelian solution corresponding to the
same value of 
$r_*$. In this figure the embeddings for two  different values of the
minimal radial distance $r_*$ are shown. When $r_*$ is large enough
($r_*\ge 2$) the two curves become practically identical. 

\begin{figure}[h]
\centerline{\hskip -.8in \epsffile{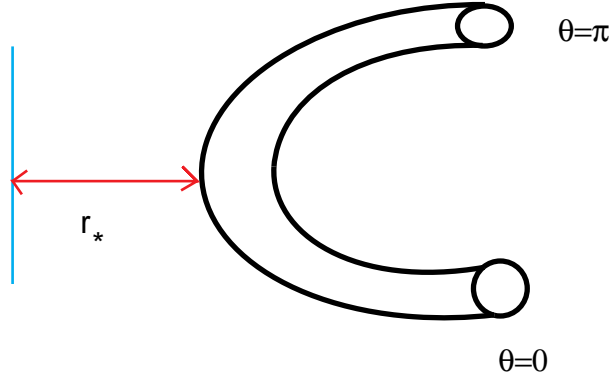}}
\caption{A pictorial representation of the embedding, which
has the topology of a cylinder. The compact direction is the
$\varphi$ angle, while the non-compact one extends in $r(\theta)$.
It goes to infinity for $\theta=0,\pi$ and $r$ is minimum
for $\theta=\pi/2$. Besides, the probe brane is also extended
in the four flat space-time dimensions where the gauge theory
lives.}
\label{cylinderfig}
\end{figure}

In figure \ref{cylinderfig}, a pictorial representation of
the non-abelian unit-winding embedding of the brane has been plotted.
$r_*$ is the
minimal distance to the origin it reaches. 
From the five-dimensional point of view, we see how the 
brane {\sl vanishes in thin air} at $r=r_*$, while it is
space-time filling as it goes to $r\to\infty$.
For $r_*=0$, we
have the cylinder solutions described below. This limit is
somehow pathological as the brane gets disconnected into
two pieces, one at $\theta=\tilde\theta=0$ (suppose $\eta=+1$)         
and the other at $\theta=\tilde\theta=\pi$ without going 
through intermediate values of $\theta$.

\begin{figure}
\centerline{\hskip -.8in \epsffile{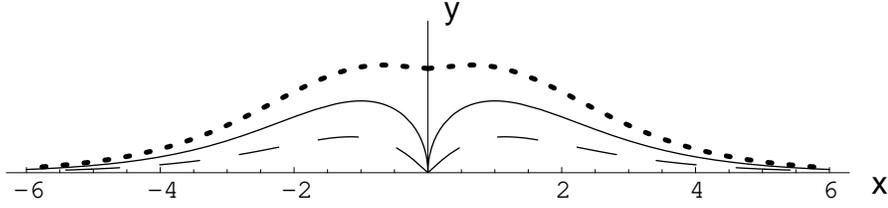}}
\caption{ Graphic representation of the unit winding embedding 
of eq. (\ref{Fnabcosh}) for three values of
the constant $C$: $C=0.5$ (dashed line), $C=1$ (solid line) and $C=1.5$ 
(dotted line). The variables $(x,y)$ are the ones defined in eq.
(\ref{Fcartesian}). }
\label{Ffig3}
\end{figure}

Let us now have a look at the case of the $A=0$   embeddings. The function
$r(\theta)$ in this case can be read from eq. (\ref{FBMnoab}), namely:
\beq
\cosh r\,=\,{C\over \sin\theta}\,\,\,.
\label{Fnabcosh}
\eeq
We have plotted in figure \ref{Ffig3} the profile for these  embeddings in
terms of the variables
$(x,y)$ of eq. (\ref{Fcartesian}). Notice that, when $C$  is in the
interval
$(1,\infty)$ it can be parametrized as $C=\cosh r_*$, with $r_*>0$ being
the minimal radial distance between the probe and the origin. On the
contrary, when $C$ lies in the interval $[0,1]$ the brane reaches the
origin when $\sin\theta=C$. We have thus, in this case, a one-parameter
family of configurations which pass through the origin.

As in their abelian counterparts, all of
these worldvolume solitons for the non-abelian background  saturate an
energy bound. In order to prove this fact,  let us define:
\beq
D\,\equiv\,\coth 2r\,-\,\eta\,{\cos\tilde{\psi}\over \sinh 2r}\,\,.
\eeq
Notice that $D\ge 0$ for any real $\tilde{\psi}$ and $r$. Moreover the 
equation for $r(\theta)$ can be written as
$r_{\theta}=-\cot\theta/D$. For arbitrary functions $r(\theta)$ and 
$\tilde{\psi}(\theta)$ the hamiltonian density takes the form:
\bear
{\cal H}\,=\,e^{2\phi}\sin\theta &\Bigg[&
\sqrt{\Big(r-r_{\theta}\cot\theta\Big)^2\,+\,
rD\,\Big(r_{\theta}+{\cot\theta\over D}\,\Big)^2\,+\,
{rD\over 4}\Big(\tilde{\psi}_{\theta}\,-\,{2\eta\sin\tilde{\psi}\over
D\sinh 2r}\,
\cot\theta
\,\Big)^2}\,-\rc\rc
&&-\,\Big(\,2e^{2h}\,-\,{1\over 8}\,(a^2-1)^2\,e^{-2h}\,
\,\Big)\,\cot\theta \,\,r_{\theta}\Bigg]\,\,.
\eear

It can be verified that ${\cal H}$ can be written as 
${\cal H}={\cal Z}+{\cal S}$, where:
\beq
{\cal Z}=-{d\over d\theta}\,\Big[\,e^{2\phi}\,r\cos\theta\,\Big]\,\,.
\eeq
(this is the same value as in the abelian soliton for $n=1$).  The
expression of 
${\cal S}$ is:
\bear
{\cal S}\,=\,e^{2\phi}\sin\theta &\Bigg[&
\sqrt{\Big(r-r_{\theta}\cot\theta\Big)^2\,+\,
rD\,\Big(r_{\theta}+{\cot\theta\over D}\,\Big)^2\,+\,
{rD\over 4}\Big(\tilde{\psi}_{\theta}\,-\,{2\eta\sin\tilde{\psi}\over
D\sinh 2r}\,
\cot\theta
\,\Big)^2}\,-\rc\rc
&&\,-\Big(\,r\,-\,r_{\theta}\cot\theta\,\big)\,\Bigg]\,\,.
\label{FnabcalS}
\eear
As in the abelian case, if the BPS equations (\ref{FnabBPS}) are  satisfied,
the square root on eq. (\ref{FnabcalS}) can be exactly evaluated and ${\cal
S}$ vanishes. Furthermore, one can easily check that ${\cal S}\ge 0$ is
equivalent to the condition:
\beq
rD\,\Big(r_{\theta}+{\cot\theta\over D}\,\Big)^2\,+\,
{rD\over 4}\Big(\tilde{\psi}_{\theta}\,-\,{2\eta\sin\tilde{\psi}\over
D\sinh 2r}\,
\cot\theta
\,\Big)^2\,\ge\,0\,\,,
\eeq
which, since $D\ge 0$, is trivially satisfied for any functions 
$r(\theta)$ and 
$\tilde{\psi}(\theta)$. Moreover, as $r-r_{\theta}\cot \theta\ge 0$ for  the
solution of the BPS equations, it follows that our BPS embedding saturates
the bound.

\subsection{Non-abelian zero-winding solutions}
The differential equations for the non-abelian version of the  zero-winding
solution can be obtained by putting $n=0$ in our general equations.
Actually, by taking $n=0$ in the second equation in   (\ref{Feqpsiyr}) one
obtains the differential equation which determines the dependence of
$r$ on the angle $\theta$, namely:
\beq
r_{\theta}\,=\,-{\cot\theta\over 2\coth2r}\,\,,
\eeq
which can be easily integrated, namely:
\beq
\sinh 2r\,=\,{C\over \sin\theta}\,\,.
\label{Fzerowindingr}
\eeq
Notice that, as in the abelian case, this solution has two spikes  at
$\theta=0,\pi$, where
$r$ diverges and, thus, the brane probe also extends infinitely in  the
radial direction. Moreover, the minimal value of the radial coordinate,
which we will denote by $r_*$, is reached at $\theta=\pi/2$. This minimal
value is related to the constant $C$ in eq.  (\ref{Fzerowindingr}), namely
$\sinh 2r_*=C$.  It is readily verified  that for large $r$ this solution
behaves exactly as the zero-winding solution  in the abelian theory.
Moreover, it follows from eq. (\ref{Frel}) that, in this $n=0$ case, the
angle $\tilde\varphi$ only depends on $\theta$. Actually, the differential
equations for the angles $\tilde\theta$, $\tilde\varphi$ and $\tilde{\psi}$
as functions of $\theta$ are easily obtained from eqs.
(\ref{Feqtildetheta}),  (\ref{Feqtildevarphi}) and (\ref{Frel}):
\bear
\partial_{\theta} \,\tilde\theta&=&-{\cos\tilde{\psi}\over \cosh
2r}\,\,,\rc\rc
\partial_{\theta} \,\tilde\varphi&=&-{1\over \cosh 2r}\,\,
{\sin\tilde{\psi}\over \sin\tilde\theta}\,\,,\rc\rc
\partial_{\theta}
\,\tilde{\psi}&=&{\cot\tilde\theta\sin\tilde{\psi}\over\cosh 2r}\,\,.
\label{Fzerowindingangles}
\eear
By combining the equations of $\tilde{\psi}$ and $\tilde\theta$ one  can
easily get the relation between these two angles, namely:
\beq
\sin\tilde{\psi}\,=\,{B\over \sin\tilde\theta}\,\,.
\label{Fpsitildetheta}
\eeq
Notice that, for consistency, $B\le 1$ and $\sin\tilde\theta\ge B$.  We can
also obtain $\tilde\varphi=\tilde\varphi(\tilde\theta)$ and 
$\tilde\theta=\tilde\theta(\theta)$:
\bear
&&\tilde\varphi=-\arctan\Bigg[\,
{B\cos\tilde\theta\over \sqrt{\sin^2\tilde\theta-B^2}}\Bigg]\,+\,
{\rm constant}\,\,,\rc\rc
&&-\arcsin\Bigg[\,{\cos\tilde\theta\over \sqrt{1-B^2}}\Bigg]=
\arcsin\Bigg[\,{\cos\theta\over \sqrt{1+C^2}}\Bigg]\,+\,
{\rm constant}\,\,.
\eear
Actually, much simpler equations for the embedding are obtained  if one
considers the particular case in which the angle $\tilde{\psi}$ is
constant. Notice that, as was pointed out after eq. (\ref{Feqpsiyr}), this
only can happen if
$\tilde{\psi}=0,\pi$ ($\mod \,\,2\pi$) (see also the last equation in
(\ref{Fzerowindingangles})). These solutions correspond to taking the
constant $B$ equal to zero in eq. (\ref{Fpsitildetheta}). Moreover, it
follows from the eq.  (\ref{Fzerowindingangles}) that $\tilde\varphi$ is an
arbitrary constant in this case, while the dependence of $\tilde\theta$ on
$\theta$ can be obtained by combining eqs.  (\ref{Fzerowindingr}) and
(\ref{Fzerowindingangles}). If we denote $\cos\tilde{\psi}=\epsilon$, with
$\epsilon=\pm 1$, one has:
\beq
\sinh 2r\,=\,{\sinh 2r_*\over \sin\theta}\,\,,
\,\,\,\,\,\,\,\,\,\,\,\,\,\,\,\,\,\,\,
\sin (\,\tilde\theta\,-\,\tilde\theta_*\,)\,=\,\epsilon\,
{\cos\theta\over \cosh 2r_*}\,\,,
\label{Fzwprofile}
\eeq
where $\tilde\theta_*=\tilde\theta(\theta=\pi/2)$. Notice  that there are
four possible values of $\tilde{\psi}$ in this zero-winding solution and,
thus, the $U(1)$ R-symmetry is broken to $\ZZ_4$ in this case.

Let us finally point out that, also in this case, 
the  hamiltonian density ${\cal H}$ can be put as   ${\cal H}={\cal
Z}+{\cal S}$, where
${\cal S}$ is non-negative (${\cal S}=0$ for the BPS solution) and 
${\cal Z}$ is given by:
\beq
{\cal Z}\,=\,-\partial_{\theta}\,\Bigg[\,
e^{2\phi}\,\cos\theta\bigg(\,r\,-\,{1\over 4}\,\coth 2r\,+\,
{r\over 2\sinh^22r}\,
\bigg)\,\Bigg]\,\,.
\eeq

\subsection{Cylinder solutions}

We shall now show that there exists a general class of supersymmetric
embeddings for the non-abelian background. For convenience, let us
consider $r$ as worldvolume coordinate and let us assume that the
D5-brane is sitting at the north poles of the $(\theta,\varphi)$ and 
$(\tilde\theta,\tilde\varphi)$ two-spheres, \ie\ at 
$\theta=\tilde\theta=0$. In the remaining angular
coordinates  $\varphi$,  $\tilde\varphi$ and $\tilde{\psi}$,   the
embedding  is characterized by the equation:
\beq
{\varphi-\varphi_0\over p}\,=\,
{\tilde\varphi-\tilde\varphi_0\over q}\,=\,
{\tilde{\psi}-\tilde{\psi}_0\over s}\,\,,
\label{Fcylcurve}
\eeq
where $(\varphi_0,\tilde\varphi_0,  \tilde{\psi}_0)$ and $(p,q,s)$ are
constants. Notice that, if one of the constants of the denominator in
(\ref{Fcylcurve}) is zero, then the corresponding angle must be a
constant. Let us parametrize these embeddings by means of two worldvolume
coordinates $\sigma_1$ and $\sigma_2$, defined as follows:
\beq
\sigma_1\,=\,{\varphi-\varphi_0\over p}\,=\,
{\tilde\varphi-\tilde\varphi_0\over q}\,=\,
{\tilde{\psi}-\tilde{\psi}_0\over s}\,\,,
\,\,\,\,\,\,\,\,\,\,\,\,\,\,\,\,\,
\sigma_2\,=\,r\,\,.
\eeq
It is straightforward to demonstrate that the pullback to the
worldvolume of the forms $w^i$ and $A^i$ is given by:
\bear
&&P[\,w^1\,]\,=\,P[\,w^2\,]\,=\,0\,\,,
\,\,\,\,\,\,\,\,\,\,\,\,\,\,\,\,\,
P[\,w^3\,]\,=\,(q+s)\,d\sigma_1\,\,,\rc\rc
&&P[\,A^1\,]\,=\,P[\,A^2\,]\,=\,0\,\,,
\,\,\,\,\,\,\,\,\,\,\,\,\,\,\,\,\,
P[\,A^3\,]\,=\,-p\,d\sigma_1\,\,.
\eear
It follows from these results that the pullback of the frame
one-forms 
$e^i$ and $e^{\hat j}$  is zero for $i,j=1,2$, whereas $P[e^{\hat 3}]$
is non-vanishing. As a consequence, the induced Dirac matrices are:
\beq
\gamma_{\sigma_1}\,=\,{1\over 2}\,(p+q+s)\,e^{{\phi\over 4}}\,
\hat\Gamma_3\,\,,
\,\,\,\,\,\,\,\,\,\,\,\,\,\,\,\,\,\,\,\,\,\,\,\,\,\,\,\,\,\,\,\,\,\,
\gamma_{\sigma_2}\,=\,e^{{\phi\over 4}}\,\,\Gamma_r\,\,.
\eeq
The kappa symmetry matrix $\Gamma_{\kappa}$ for the
embedding at hand is:
\beq
\Gamma_{\kappa}\,=\,{e^{\phi}\over \sqrt{-g}}\,
\Gamma_{x^0\cdots x^3}\,\gamma_{\sigma_1\sigma_2}\,\,.
\eeq
Moreover, by using the projection conditions satisfied by the Killing
spinors $\epsilon$, one can prove that:
\beq
\gamma_{\sigma_1\sigma_2}\,\epsilon\,=\,-
{p+q+s\over 2}\,\,e^{{\phi\over 2}}\,\,\Gamma_r\,\hat\Gamma_3\,
\epsilon\,=\,{p+q+s\over 2}\,\,e^{{\phi\over 2}}\,
\Big(\,\cos\alpha\,\Gamma_{12}\,+\,\sin\alpha\Gamma_1\hat\Gamma_2
\,\Big)\,\epsilon\,\,,
\eeq
and, since the determinant of the induced metric is
$\sqrt{-g}\,=\,e^{{3\phi\over 2}}\,{p+q+s\over 2}$,
it is immediate to
verify that the kappa symmetry projection 
$\Gamma_{\kappa}\epsilon=\epsilon$
coincides with the
projection  satisfied by the Killing spinors of the
background. Therefore, our brane probe preserves all the
supersymmetries of the background. Notice that the induced metric on
the worldvolume along the $\sigma_1,\sigma_2$ directions has the form
\beq
e^{{\phi\over 2}}\big[\,{(p+q+s)^2\over 4}\,
d\sigma_1^2\,+\,d\sigma_2^2
\,\big]\,\,,
\eeq
which is conformally equivalent to the metric of a cylinder. After a
simple calculation one can prove that the energy density of these
solutions is
\beq
{\cal H}\,=\,\partial_r\,\Bigg[\,e^{2\phi}\,
\bigg(\,pr\,+\,(q+s-p)\,\bigg(
{\coth2r\over 4}\,-\,
{r\over 2\sinh^22r}\,\bigg)\,
\bigg)\,\Bigg]\,\,.
\eeq
 One can also have cylinders  located at the south pole of the
$(\theta, \varphi)$ and $(\tilde\theta, \tilde\varphi)$ two-spheres.
Indeed, the above equations remain valid if $\theta=\pi$
($\tilde\theta=\pi$) if one changes $p\rightarrow -p$ 
($q\rightarrow -q$, respectively). On the other hand, if $p=0$ , the
angle $\varphi$ is constant and, as the pullback of the $e^{i}$ frame
one-forms also vanishes when $\theta$ is also constant, it follows
that $\theta$ can have any constant value when $p=0$. Similarly, if
$q=0$ one necessarily has $\tilde\varphi=\tilde\varphi_0$ and 
$\tilde\theta$ can be an arbitrary constant in this case.  

When $p=1$, $q=n$ and $s=2m$, the angular part of the embedding is the
same as in the $(n,m)$-winding solitons. Actually, these cylinder
solutions correspond to formally taking $r_*\rightarrow-\infty$ is the
abelian solution of eq. (\ref{Fnmsol}). This forces one to take
$\theta=0,\pi$ and, thus one can regard the cylinder as a zoom which
magnifies   the region in which the probe goes to infinity. One can
also get cylinder embeddings by consider the limit of the non-abelian
solutions in which the probe reaches the point $r=0$.  For example, by
taking $r_*=0$ in eq. (\ref{Fnoabflavor}) one gets the $p=q=1$, $s=0$
cylinder solutions while the 
$r_*\to 0$ limit of the embedding (\ref{Fzwprofile})  corresponds to a
cylinder with $p=1$ and  $q=s=0$. Actually, when one takes the $r_*\to 0$
limit of these non-abelian embeddings one obtains two cylinder solutions,
one with $\theta=0$ and the other with
$\theta=\pi$. This suggests that, in order to obtain a  consistent
solution, one must combine in general two cylinders located at each of the
two poles of the
$(\theta,\varphi)$ two-sphere. Notice that this is  also required if one
imposes the condition of RR charge neutrality of the two-sphere at
infinity.

\setcounter{equation}{0}
\section[Quadratic fluctuations]{Quadratic fluctuations around the
unit-winding embedding}
\label{secmesons}

As mentioned in the introduction, we are now going to consider  some of the
brane probe configurations previously found as flavor branes, which will
allow us to introduce  dynamical quarks in the ${\cal N}=1$ SYM theory.
Following refs. \cite{KK, KKW, KMMW}, the spectrum of quadratic
fluctuations of the brane probe will be interpreted as the meson spectrum
of ${\cal N}=1$ SQCD. So, let us try to elaborate on the reasons to
consider  these probes as the addition of flavors to the field theory dual.
In fact, when considering the  't Hooft expansion for large number of
colors,  the r\^ole of flavors is played by the boundaries of the Feynman
graph. From  a gravity perspective, these boundaries correspond to the
addition of D-branes and open strings in the game.

In our case, we have a system of $N$ D5-branes wrapping a two-cycle inside 
the 
resolved conifold and $N_f$ D5-branes that wrap another  two-submanifold,
thus  introducing $N_f$ flavors in the $SU(N) $ gauge theory. Taking the 
decoupling limit with $g_s \alpha' N$ fixed and large is  equivalent to
replacing  the $N$ D5-branes by the geometry they generate (the one studied
in  section \ref{FMN10D}), while the $N_f$-D5 branes that do not backreact
(because we  take $N_f$ much smaller than $N$) are treated as 
probes. From a gauge theory perspective, this is equivalent to consider 
the dynamics of gluons and gluinos coupled to  fundamentals, but 
neglecting the backreaction of the latter. Of course it would be of 
great interest to find the backreacted solution.

The way of adding fundamental fields in this gauge theory from a string 
theory perspective was discussed in \cite{WangHu}, where two possible 
ways, adding $D9$ branes or adding $D5$
branes as probes, were proposed. Here, we are  considering the 
cleaner case of 
$D5$ probes. A careful analysis 
of the open string spectrum shows the existence of a four dimensional 
gauge ${\cal N}=1$ vector multiplet and a complex scalar multiplet. This 
is 
the spectrum of SQCD. In the case analyzed below we will consider abelian 
DBI actions for the probes, so that we will be dealing with the 
 $N_f=1$ case.

 We have found several brane configurations
in the non-abelian background which, in principle,  could be  suitable to
generate  the meson spectrum. 
One of the requirements we should demand to these configurations  is that
they must incorporate some scale parameter which could be used  to
generate the mass scale of the quarks. Within our framework such a mass
scale is nothing but the minimal distance between the flavor brane and
the origin, \ie\ what we have  denoted by $r_*$. This requirement allows
to discard the cylinder solutions we have found since they reach the
origin and have no such a mass scale. We are thus left with the
unit-winding solutions  and the zero-winding solutions of
section  \ref{secnonabel} as the only analytical solutions we have found for
the non-abelian background.

In this section we shall analyze the fluctuations around the 
unit-winding solutions. We have several reasons for this
election. First of all, the unperturbed unit-winding embedding is
simpler. Secondly, we will show in appendix \ref{appendixB} that the UV
behavior of the  fluctuations is better in the unit-winding configuration
than in the zero-winding embedding. Thirdly,  the unit-winding embeddings
of constant
$\tilde{\psi}$ incorporate the correct pattern $U(1)\to \ZZ_2$ of
R-symmetry breaking, whereas for the zero-winding embeddings of eq.
(\ref{Fzwprofile}) the $U(1)$ symmetry is broken to a
$\ZZ_4$ subgroup.

Recall from section \ref{secunit} that we have two possible solutions with
$\tilde{\psi}=0,\pi\,\,(\mod\,2\pi)$, which are the ones displayed in eqs. 
(\ref{Fnoabflavor}) and (\ref{Fnabcosh}). As discussed in section
\ref{secunit}, the solution of eq. (\ref{Fnabcosh}) contains a one-parameter
subfamily of embeddings which reach the origin and, thus, they should
correspond to massless dynamical quarks. On the contrary,  the embeddings
of eq. (\ref{Fnoabflavor}) pass through the origin only in one case, \ie\
when
$r_*=0$ and, somehow, the limit in which the quarks are massless is
uniquely defined. Recall that for
$r_*=0$ the solution (\ref{Fnoabflavor})  is identical to the 
unit-winding cylinder. For these reasons we consider the configuration
displayed in eq.  (\ref{Fnoabflavor}) more adequate for our purposes and
we will use it as the unperturbed flavor brane.

 We will consider first in section \ref{secescmes}  the
fluctuations of the scalar transverse to the brane probe,   while in
section  \ref{secvecmes} we will study the fluctuations of the
worldvolume gauge field.  The gauge theory interpretation of the results
will be discussed in section \ref{secgauge}

\subsection{Scalar mesons}
\label{secescmes}

Let us consider a non-abelian unit-winding  embedding with  
$\tilde\theta=\theta$, $\tilde\varphi=\varphi+{\rm constant}$  and
$\tilde{\psi}=\pi \,(\mod\, 2\pi)$. For convenience we take first $r$ as
worldvolume coordinate and consider
$\theta$ as a function of $r$, $\varphi$  and of the unwrapped 
coordinates $x$, \ie\
$\theta=\theta(r,x,\varphi)$.  The lagrangian density for  such embedding
can be easily obtained by computing the induced metric. One gets:
\bear
&&{\cal L}=-e^{2\phi}\,\sin\theta\,\times\rc\rc
&&\times\Bigg[\,
\sqrt{\bigg[\,1+r\tanh r\Big(\,(\partial_r\theta)^2+
(\partial_x\theta)^2\,+\, {1\over \cos^2\theta+r\coth r\sin^2\theta}\,
(\partial_{\varphi}\theta)^2\,\Big)
\bigg]\,
\big(r\coth r+\cot^2\theta\big)}\,+\,\rc\rc
&&\,\,\,\,\,\,\,\,\,\,\,\,\,\,\,\,\,\,\,\,\,\,\,\,\,\,\,\,\,\,\,\,
+\,r\partial_r\theta-\cot\theta\,\Bigg]\,\,,
\eear
where we have neglected the term $\partial_r (r\cos\theta  e^{2\phi})$
which, being a total radial derivative, does not contribute to the
equations of motion. 

We are going to expand this lagrangian around the corresponding 
non-abelian unit-winding configuration obtained in section  \ref{secunit}.
Actually, by taking in eq. (\ref{Fnoabflavor})
$\eta=+1$  one  obtains a  configuration with $\tilde{\psi}=\pi \,\,(\mod
\,2\pi)$,  which corresponds to a function $\theta=\theta_0(r)$, given
by:
\beq
\sin\theta_0(r)={\sinh r_{*}\over \sinh r}\,\,,
\label{Fthetazero}
\eeq
where $r_{*}$ is the minimum value of $r$ and $r_{*}\le r<\infty$.  It is
clear from this equation that with the coordinate $r$ we can only
describe one-half of the brane probe: the one that is wrapped, say, on
the north hemisphere of the two-sphere, in which
$\theta\in (0,\pi/2)$. Outside this interval $\theta_0(r)$ is a 
double-valued function of
$r$. Let us put:
\beq
\theta(r,x,\varphi)\,=\,\theta_0(r)\,+\,\chi(r,x,\varphi)\,\,,
\eeq
and expand ${\cal L}$ up to quadratic order in $\chi$. Using  the
first-order equation satisfied by $\theta_0(r)$, namely:
\beq
\partial_r\theta_0\,=\,-\coth r\tan\theta_0\,\,,
\eeq
we get:
\bear
{\cal L}&=&-{1\over 2}\,\,
{e^{2\phi}\over 1+r\coth r\tan^2\theta_0}\,\,\Bigg[\,\,
r\tanh r\cos\theta_0\,(\partial_r\chi)^2\,+\,
{2r\over \cos\theta_0}\,\chi\,\partial_r\chi\,+\,
{r\coth r\over \cos^3\theta_0}\,\,\chi^2\,\,\Bigg]\,-\rc\rc
&&-\,{1\over 2}\,\,e^{2\phi}\,r\tanh r\,\Big[\,
\cos\theta_0\, (\partial_x\chi)^2\,+\,
{1\over \cos\theta_0\,(\,1+r\coth r\tan^2\theta_0\,)}\,\,
(\partial_{\varphi}\chi)^2
\,\,\Big]\,\,.
\label{Flagrafluct}
\eear

In the equations of motion derived from this lagrangian we will  perform
the ansatz:
\beq
\chi(r,x,\varphi)=e^{ikx}\,e^{il\varphi}\,\,\xi(r)\,\,,
\label{Fchiansatz}
\eeq
where, as $\varphi$ is a periodically identified coordinate,   $l$ must
be an integer and
$k$ is a four-vector whose square determines the four-dimensional  mass
$M$ of the fluctuation mode:
\beq
M^2=-k^2\,\,.
\label{Fmasssquare}
\eeq
By substituting the functions (\ref{Fchiansatz}) in the equation  of
motion that follows from the lagrangian (\ref{Flagrafluct}), one gets a
second order differential equation which is rather complicated and that
can only be solved numerically. However, it is not difficult to obtain
analytically the asymptotic behavior of $\xi(r)$. This has been done in
appendix \ref{appendixB} and we will now use these results to explore the
nature of the fluctuations. For large $r$, \ie\ in the UV, one gets (see
eq. (\ref{FUVasymp})) that
$\xi(r)$ vanishes exponentially in the form:
\beq
\xi(r)\,\sim\,{e^{-r}\over r^{{1\over 4}}}
\cos \big[\sqrt{M^2-l^2}\,\,r\,+\,\delta\,\big]\,\,,
\,\,\,\,\,\,\,\,\,\,\,\,\,\,\,\,\,
(r\rightarrow\infty)\,\,,
\label{FUVasympmain}
\eeq
where $\delta$ is a phase and we are assuming that $M^2\ge l^2$. 
For $M^2< l^2$ the fluctuations do not oscillate in the UV  and we will
not be able to impose the appropriate boundary conditions (see below).
Notice that our
unperturbed solution $\theta_0(r)$ also decreases in the UV as:
\beq
\theta_0(r)\sim e^{-r}\,\,
\,\,\,\,\,\,\,\,\,\,\,\,\,\,\,\,\,
(r\rightarrow\infty)\,\,.
\eeq
Thus $\xi(r)/\theta_0(r)\rightarrow 0$ as $r\rightarrow\infty$  and the
first-order expansion we are performing continues\footnote{For the
zero-winding solution, on the contrary, the ratio $\xi(r)/\theta_0(r)$
diverges in the UV (see appendix \ref{appendixB}).} to be  valid in the
UV. On the other hand, for
$r$ close to $r_*$ there are two independent solutions, one of  them is
finite at 
 $r=r_*$ while the other diverges as:
\beq
\xi(r)\,\sim\,{1\over \sqrt{r-r_*}}\,\,.
\label{FxiUV}
\eeq

Let us now see how one can use the information on the asymptotic 
behavior of the fluctuation modes to extract their value for the full
range of the radial coordinate. First of all, it is clear  that, in
principle,  by consistency with the type of expansion we are  adopting,
one should require that $\xi<<\theta_0$. Thus, one should discard the
solutions which diverge in the infrared (see, however, the discussion
below). Moreover, the behavior of
the fluctuations $\xi$ for large $r$ should be determined  by some
normalizability conditions. The corresponding norm would be an expression
of the form:
\beq
\int_{0}^{\infty}\,dr\sqrt{\gamma}\,\,\,\xi^2\,\,,
\eeq
where $\sqrt{\gamma}$ is some measure, which can be determined  by
looking at the lagrangian (\ref{Flagrafluct}). If we regard $\chi$ as a
scalar field with the standard normalization in a curved space, then
$\sqrt{\gamma}$ is just the coefficient of the kinetic term ${1\over
2}\,(\partial_r\chi)^2$ in ${\cal L}$, namely:
\beq
\sqrt{\gamma}\,=\,e^{2\phi}\,\,
{r\tanh r \cos\theta_0\over 1+r\coth r \tan^2\theta_0}\,\,.
\label{Fmeasure}
\eeq
For large $r$, $\sqrt{\gamma}$  behaves as:
\beq
\sqrt{\gamma} \,\,(\,\,r\rightarrow\infty\,\,)\,\approx\,r^{{1\over 2}}\,
e^{2r}\,\,.
\label{FUVmeasure}
\eeq
Notice that the factors on the right-hand side of eq.  (\ref{FUVmeasure})
cancel against the exponentials and power factors of $\xi^2$ in the UV 
(see eq. (\ref{FUVasympmain})).
As a consequence, all solutions have infinite norm.

The reason for the bad behavior we have just discovered is  the
exponential blow up of the dilaton in the UV which invalidates the
supergravity approximation. Actually, if one wishes to push the theory to
the UV one has to perform an S-duality, which basically changes
$e^{2\phi}$ by 
$e^{-2\phi}$. The S-dual theory corresponds to wrapped  Neveu-Schwarz
fivebranes and is the supergravity dual of a little string theory. Notice
that, by changing  $e^{2\phi}\rightarrow e^{-2\phi} $ in the measure
(\ref{Fmeasure}),  all solutions
become normalizable, which is  as bad as having no normalizable 
solutions at all. Moreover, it is unclear how to perform an S-duality in
our D5-brane probe and convert it in a Neveu-Schwarz fivebrane for large
values of the radial coordinate. 

A problem similar to the one we are facing here appeared in  ref.
\cite{Pons} in the calculation of the glueball spectrum for this
background. It was argued in this reference that, in order to have a
discrete spectrum, one has to introduce a cut-off to discriminate between
the two regimes of the theory. Notice that, since they extend infinitely
in the radial direction, we cannot avoid that our D5-brane probe explores
the deep UV region. However, what we can do is to consider fluctuations
that are significantly non-zero only on scales in which one can safely
trust the supergravity approximation.  In ref. \cite{Pons} it was
proposed to implement this condition by requiring the fluctuation to
vanish at some conveniently chosen UV cut-off $\Lambda$. Translated to
our situation, this proposal amounts to requiring:
\beq
\xi(r)\big|_{r=\Lambda}\,=\,0\,\,.
\label{Fcutoff}
\eeq
This condition, together with the regularity of $\xi(r)$  at $r=r_{*}$,
produces a discrete spectrum which we shall explore below. Notice that,
for consistency with the general picture described above, in addition to
having a node at
$r=\Lambda$ as in eq. (\ref{Fcutoff}), the function $\xi(r)$  should be
small for
$r$ close to the UV cut-off. This condition can be fulfilled by adjusting
appropriately the mass scale of our solution, \ie\ the minimal  distance
$r_*$ between the probe and the origin, in such a way that 
$r_*$ is not too close to $\Lambda$. 

Notice that, by imposing the boundary condition (\ref{Fcutoff}) on the
fluctuations, we are effectively introducing an infinite wall  located at
the UV cut-off. The introduction of this wall allows to have a discrete
spectrum and should be regarded as a physical condition which implements
the correct range of validity of the background geometry as a
supergravity dual of 
${\cal N}=1$  Yang-Mills. Even if this regularization could  appear too
rude and unnatural, the results obtained by using it for the first
glueball masses are not too bad \cite{Pons}.

The cut-off scale $\Lambda$ should not be a new scale but  instead it
should  be obtainable from the background geometry itself. The proposal
of ref. \cite{Pons} is to take $\Lambda$ as the scale of gluino
condensation, which is believed to correspond to the point at which the
function $a(r)$ approaches its asymptotic value $a=0$. A more pragmatic
point of view, to which we will adhere here, is just taking the value of
$\Lambda$ which gives reasonable values for the glueball masses. In ref.
\cite{Pons} the value $\Lambda=2$ was needed to fit the glueball masses
obtained from lattice calculations, whereas with 
$\Lambda=3.5$ one gets a glueball spectrum which resembles  that
predicted by other supergravity models. Notice that from $r=0$ to $r=3$
the effective string coupling constant $e^{\phi}$ increases in an order
of magnitude. From our point of view it is also natural to look at the
effect of the background on our brane probe. In this sense it is
interesting to point out that for $r_*=2-3$ onwards the abelian and
non-abelian embeddings are indistinguishable (see sect. \ref{secunit}). 
\begin{figure}
\centerline{\hskip -.8in \epsffile{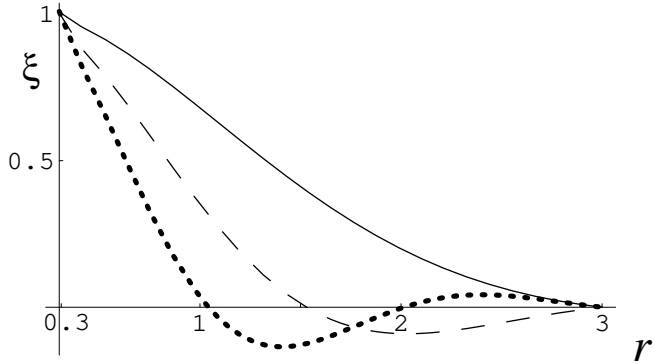}}
\caption{Graphic representation of the first three fluctuation  modes for
$r_*=0.3$, 
$\Lambda=3$ and $l=0$. The three curves have been normalized to  have
$\xi(r_*)=1$. }
\label{Ffig4}
\end{figure}

We have performed the numerical integration of the equation  of motion of
$\xi(r)$ subject to the boundary condition (\ref{Fcutoff}) by means of the
shooting technique. For a generic value of the mass $M$ the solution
diverges at $r=r_*$. Only for some discrete set of masses the
fluctuations are regular in the IR. In figure \ref{Ffig4}, we show the
first three modes obtained by this procedure for $r_*=0.3$ and
$\Lambda=3$. From this figure,  we notice that the number of zeroes of
$\xi(r)$ grows with the mass. In general one observes that the $n^{th}$
mode has $n-1$ nodes in the region $r_*<r<\Lambda$, in agreement with the
general expectation for this type of boundary value problems. Moreover,
for
$l=0$, the mass $M$ grows linearly with the number of nodes (see below).

At this point it is interesting to pause a while and discuss  the
suitability of our election of $r$ as worldvolume coordinate. Although
this coordinate is certainly very useful to extract the asymptotic
behavior of the fluctuations (specially in the UV), we should keep in
mind that we are only describing one half of the brane, \ie\ the one
corresponding to one of the two hemispheres of the two-sphere. On the
other hand, the election of the angle as excited scalar has some
subtleties which we now discuss. Actually, to describe the displacement
of  the brane probe with respect to its unperturbed configuration it is
physically more sensible to use the coordinate $y$, defined in eq.
(\ref{Fcartesian}). Accordingly, let us define the function
$y(r,x,\varphi)$ as
\beq
y(r,x,\varphi)\,=\,r\sin\theta(r,x,\varphi)\,\,.
\eeq
Let us put in this equation $\theta(r,x,\varphi)=\theta_0(r)
\,+\,\chi(r,x,\varphi)$. At the  linear order in 
$\chi$ we are working, $y$ can be written as
\beq
y(r,x,\varphi)\,=\,y_0(r)\,+\,r\cos\theta_0(r)\,\chi(r,x,\varphi)\,\,,
\label{Fyexpanded}
\eeq
where $y_0(r)\equiv r\sin\theta_0(r)$ corresponds to the  unperturbed
brane. Notice, first of all, that for those modes in which
$\chi(r_*,x,\varphi)$ is finite, the  fluctuation term in
$y(r_*,x,\varphi)$  vanishes since $\cos\theta_0(r)\rightarrow 0$ when 
$r\rightarrow r_*$. Then
\beq
y(r_*,x,\varphi)\,=\,y_0(r_*)\,=\,r_*\,\,\,\,
{\rm if}\,\,\, \chi(r_*,x,\varphi)\,\,\,\,{\rm is\,\,\,finite}\,\,.
\eeq
Thus, by considering those modes $\chi$ that are regular at  $r= r_*$ we
are effectively restricting ourselves to the modes which have a node in
the $y$ coordinate at 
$r= r_*$. If, on the contrary $\chi$ diverges for $r\approx r_*$,  we
know from eq. (\ref{FxiUV}) that it behaves as $\chi\,\approx\,1/
\sqrt{r-r_*}$.  But we also know that $\cos\theta_0(r)\rightarrow 0$
when $r\rightarrow r_*$ and, in fact, the second term on the  right-hand
side of eq.  (\ref{Fyexpanded}) remains undetermined. The precise form in
which 
$\cos\theta_0(r_*)$ vanishes can be read from eq. (\ref{Fsincos}), namely
$\cos\theta_0(r)\,\approx\,\sqrt{r-r_*}$ for $r\approx r_*$. 
Therefore, even if $\chi$ diverges at $r=r_*$, we could have,  in the
linearized approximation  we are adopting, a finite value for
$y(r_*,x,\varphi)$. Thus, modes with
$\chi(r_*,x,\varphi)\rightarrow\infty$ should not be discarded. 
Actually, $y(r_*,x,\varphi)-y_0(r_*)$, although finite, is  undetermined 
in eq. (\ref{Fyexpanded}) for these modes and, in order to obtain its
allowed values we should impose a boundary condition at the other half of
the brane. As previously mentioned, this cannot be done when $r$ is taken
as worldvolume coordinate. Therefore, it is convenient to come back to
the formalism of sects. \ref{seckappa}-\ref{secnonabel}, in which
$\theta$ had been chosen as one of the worldvolume coordinates.  In this
approach, the unperturbed brane configuration is described by a function
$r_0(\theta)$, given by:
\beq
\sinh r_0(\theta)\,=\,{\sinh r_*\over \sin\theta}\,\,,
\eeq
and the brane embedding is characterized by a function  $r=r(\theta,
x,\varphi)$, which we expand around $r_0(\theta)$ as follows:
\beq
r(\theta, x,\varphi)\,=\,r_0(\theta)\,+\,\rho(\theta, x, \varphi)\,\,.
\eeq
Plugging this expansion into the DBI lagrangian of  eq. (\ref{FDBI}) and
keeping up to second order terms in $\rho$, one gets the following
lagrangian density:
\bear
{\cal L}&=&-{1\over 2}\,
{e^{2\phi}\sin\theta\, r_0\over r_0+\cot^2\theta\tanh r_0}
\Bigg[\coth r_0\,\big(\partial_{\theta}\rho\big)^2+
{2\cot\theta\over \sinh r_0\cosh r_0}\,\,\rho \partial_{\theta}\rho+
{\cot^2\theta\over \sinh r_0 \cosh^3 r_0}\,\rho^2\,\Bigg]-\rc\rc
&&\,\,-\,\,
{e^{2\phi}\sin\theta\,\, r_0\over 2} \,\Bigg[\,
\big(\partial_{x}\,\rho\,\big)^2
\,+\,{1\over \cos^2\theta\,\big(1+r_0\coth r_0\tan^2\theta\,\big)}\,\,
\big(\partial_{\varphi}\,\rho\,\big)^2\,\Bigg]\,\,.
\label{FLfluct}
\eear
Similarly to what we have done with the lagrangian  (\ref{Flagrafluct}),
we will look for solutions of the equations of motion of ${\cal L}$ which
have the form:
\beq
\rho(\theta, x,\varphi)\,=\,e^{ikx}\,e^{il\varphi}\,\,\zeta (\theta)\,\,,
\eeq
where $l$ is an integer and $k^2=-M^2$. As before,  in order to get a
discrete spectrum one must impose some boundary conditions. In the
present approach we should cutoff the regions close to the two poles of
the two-sphere. Accordingly, let us define the following angle:
\beq
\sin\theta_{\Lambda}\equiv {\sinh r_*\over \sinh \Lambda}\,\,,
\,\,\,\,\,\,\,\,\,\,\,\,\,\,\,
\theta_{\Lambda}\in (0, {\pi\over 2})\,\,,
\eeq
where, as indicated, we are taking $\theta_{\Lambda}$ in the range 
$0<\theta_{\Lambda}<\pi/2$. Notice that $\theta_{\Lambda}\rightarrow 0$ if 
$\Lambda\rightarrow\infty$ as it should. Clearly $\theta_{\Lambda}$ and
$\pi-\theta_{\Lambda}$ are the two angles that  correspond to the radial
scale $\Lambda$. Therefore, we impose the following boundary conditions
to our fluctuation:
\beq
\zeta(\theta_{\Lambda})\,=\,\zeta(\pi-\theta_{\Lambda})\,=\,0\,\,.
\label{Fbdytheta}
\eeq
\begin{figure}
\centerline{\hskip -.8in \epsffile{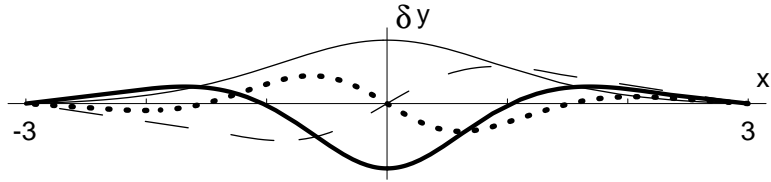}}
\caption{Plot of $\delta y\equiv\zeta\sin\theta$  versus $x= r\cos\theta$
for the first four modes for $r_*=0.3$, $\Lambda=3$ and $l=0$. The dashed
and dotted curves pass through the origin and correspond to the first two
modes of figure \ref{Ffig4}.}
\label{Ffig5}
\end{figure}

The equations of motion derived from (\ref{FLfluct}),  subjected to the
boundary conditions (\ref{Fbdytheta}) can be integrated numerically by
means of the shooting technique. One first enforces the condition at
$\theta=\theta_{\Lambda}$ and then varies the mass $M$ until
$\zeta(\pi-\theta_{\Lambda})$ vanishes. This only happens for a discrete
set of values of the mass $M$. For a given value of $l$, let us order the
solutions in increasing value of the mass. In general one notices that
the $n^{th}$ mode has $n-1$ nodes in the interval
$\theta_{\Lambda}<\theta<\pi-\theta_{\Lambda}$ and for $n$ even (odd) the
function $\zeta(\theta)$ is odd (even) under 
$\theta\rightarrow \pi-\theta$. In figure \ref{Ffig5}, we have  plotted
the first four modes corresponding to $r_*=0.3$, $\Lambda=3$ and $l=0$.
The modes odd under 
$\theta\rightarrow \pi-\theta$ vanish at $\theta=\pi/2$  and their masses
and shapes match those found with the lagrangian (\ref{Flagrafluct}) and
the boundary condition (\ref{Fcutoff}). On the contrary, the modes with an
even number of nodes in 
$\theta_{\Lambda}<\theta<\pi-\theta_{\Lambda}$ are the  ones we were
missing in the formulation in which $r$ is taken as worldvolume
coordinate.

\begin{figure}
\centerline{\hskip -.8in \epsffile{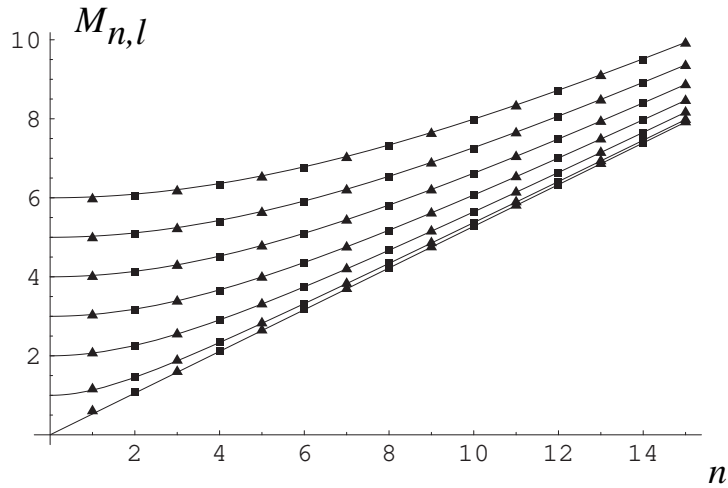}}
\caption{Mass spectrum for $r_*=0.3$ and $\Lambda=3$  for $l=0,\cdots,
6$. The solid lines correspond to the right-hand side of eq.
(\ref{Fspectrum}). The triangles (squares) are the masses of the modes
$\zeta(\theta)$ which are even (odd) under
$\theta\to\pi-\theta$.}
\label{Ffig6}
\end{figure}

Let $M_{n,l}(r_*,\Lambda)$ be the mass corresponding to  the $n^{th}$
mode for a given value $l$ and the mass scales $r_*$ and $\Lambda$. Our
numerical results are compatible with an expression of
$M_{n,l}(r_*,\Lambda)$ of the form:
\beq
M_{n,l}(r_*,\Lambda)\,=\,\sqrt{m^2(r_*,\Lambda)\,\,n^2+l^2}
\label{Fspectrum}
\eeq
To illustrate this fact we have plotted in figure \ref{Ffig6} the  values
of
$M_{n,l}(r_*,\Lambda)$ for $r_*=0.3$ and $\Lambda=3$, together with the
curves corresponding to the right-hand side of (\ref{Fspectrum}).

\begin{figure}
\centerline{\hskip -.8in \epsffile{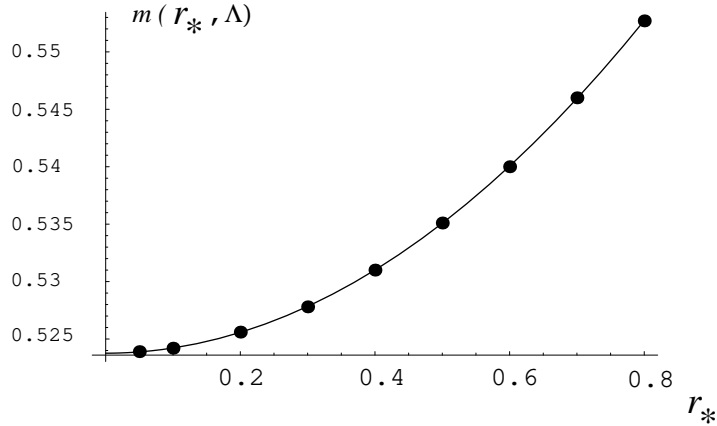}}
\caption{Dependence of $m(r_*,\Lambda)$ on $r_*$ for $\Lambda=3$. The
solid line is a fit to the quadratic function (\ref{Fparabola}).}
\label{Ffig7}
\end{figure}

We have also studied the dependence of the coefficient  $m(r_*,\Lambda)$
on the two scales
$(r_*,\Lambda)$. Recall that $r_*$ is the minimal  separation of the
brane probe from the origin and, thus, can be naturally identified with
the mass of the quarks.  We obtained that   $m(r_*,\Lambda)$ can be
represented by an expression of the type:
\beq
m(r_*,\Lambda)\,=\,{\pi\over 2\Lambda}\,+\,b(\Lambda)\,r^{2}_*\,\,.
\label{Fparabola}
\eeq
The coefficients on the right-hand side of eq. (\ref{Fparabola})  have
been obtained by a fit of $m(r_*,\Lambda)$ to a quadratic expression in
$r_*$. An example of this fit is presented in figure \ref{Ffig7}. The
first term in eq. (\ref{Fparabola}) is a universal term (independent of 
$r_*$) which can be regarded as a finite size effect  induced by our
regularization procedure. We also have determined the dependence of the
coefficient $b$ on $\Lambda$ and it turns out that one can fit
$b(\Lambda)$ to the expression
$b(\Lambda)=0.23\,\Lambda^{-2}\,+\,0.53\,\Lambda^{-3}$\ .

We have obtained a very regular mass spectrum of 
particles, classified by two quantum numbers $n,l$ (see eqs. 
(\ref{Fspectrum}) and  (\ref{Fparabola})). We can offer an interpretation
of these formulas. Indeed, recall that the two-submanifold on which we
are wrapping the brane is topologically like a cylinder, with the compact
direction parametrized by the angle $\varphi$. The quantum number $l$ is
precisely the eigenvalue of the operator
$-i\partial_{\varphi}$, which generates the shifts  of $\varphi$ and,
indeed, the dependence on $l$ of the mass displayed in eq.
(\ref{Fspectrum}) is the typical one for a Kaluza-Klein reduction along a
compact coordinate. Therefore, we should interpret the mesons with
$l\not=0$ as composed by Kaluza-Klein modes.
 Moreover,
the term in (\ref{Fparabola}) proportional to 
$r_*^2$ can be understood as the contribution coming from the mass of the 
constituent quarks, while the term $\frac{\pi}{2 \Lambda}$ can be 
interpreted as a contribution coming from the `finite size` of the meson. 
Indeed, 	it looks like a  Casimir energy and it is originated in the 
presence of the cut-off $\Lambda$.

It is perhaps convenient to emphasize again that the cut-off 
procedure implemented here is a very natural procedure for this 
type of computations in this 
supergravity dual. In fact, given that the mesons are an IR effect in 
SQCD, we expect  the contributions of high energy effects  to be 
irrelevant or negligible in the physical properties of the meson itself.
This is indeed what we are doing when cutting off the integration range. 
We are just taking into account the `non-abelian` part of the probes, 
while neglecting the `abelian` part or, equivalently,  discarding high
energy  contributions.

\subsection{Vector mesons}
\label{secvecmes}

Let us now excite a gauge field that is living in the  worldvolume of the
brane. The linearized equations of motion are:
\beq
\partial_{m}\,\Big[\,e^{-\phi}\,\sqrt{-g_{st}}\, {\cal
F}^{nm}\,\Big]\,=\,0\,\,,
\label{Fvectoreom}
\eeq
where $g_{st}$ is the determinant of the induced  metric in the string
frame and ${\cal F}^{nm}$ is the field strength of the worldvolume gauge
field ${\cal A}_m$,
\ie\ ${\cal F}_{nm}=\partial_n{\cal A}_m-\partial_m{\cal A}_n$. 
Let us assume that the only non-vanishing components of  the gauge field
${\cal A}$ are those along the unwrapped directions $x^{\mu}$ of the
worldvolume of the brane.  In what follows we are going to use $\theta$
as  worldvolume coordinate. Let us put
\beq
{\cal A}_{\mu} (\theta,x,\varphi)\,=\,\epsilon_{\mu}\,\varsigma
(\theta)\,\, e^{ikx}\,\,e^{il\varphi}\,\,,
\eeq
where $\epsilon_{\mu}$ is a constant polarization  four-vector and $l$
must be an integer. It follows from the equations of motion
(\ref{Fvectoreom}) that $\epsilon_{\mu}$ must be transverse,
\ie:
\beq
k^{\mu}\,\epsilon_{\mu}\,=\,0\,\,,
\eeq
and that $\varsigma (\theta)$ must satisfy the following  second-order
differential equation:
\beq
\partial_{\theta}\,\Big[\,{\sin\theta\over \tanh r_0}\,
\partial_{\theta}\,\varsigma\,\Big]\,+\,
{\tanh r_0\over \sin\theta}\,\,\Bigg[\,M^2
\,\,\bigg(\cos^2\theta\,+\,r_0\,\coth r_0\,\sin^2\theta\bigg)\,-
\,l^2\,\Bigg]
\varsigma\,=\,0\,\,,
\label{Fvectorfluct}
\eeq
where, as in eq. (\ref{Fmasssquare}), $k^2=-M^2$.
We have solved numerically this equation by means of the  shooting
technique with the boundary conditions
\beq
\varsigma(\theta_{\Lambda})\,=\,\varsigma(\pi-\theta_{\Lambda})\,=\,0\,\,.
\eeq
Surprisingly, the set of possible values of $M$ is given  by the same
expression as in  the scalar meson case, \ie\ eq. (\ref{Fspectrum}), with
a coefficient function
$m(r_*,\Lambda)$ which is, within the accuracy of our  approximate
calculation, equal numerically to that of the scalar mesons. This is
quite remarkable since the differential equations we are solving in both
cases are quite different (the equation satisfied by
$\zeta(\theta)$, which arises from the lagrangian (\ref{FLfluct}), is much
more complicated than eq.  (\ref{Fvectorfluct})). Thus, to summarize, we
predict a degeneracy between the scalar and vector mesons in the
corresponding
${\cal N}=1$ SYM theory.

\subsection{Gauge theory interpretation}
\label{secgauge}

Let us comment on some gauge theory aspects that can be read from the 
results of this chapter.
For this purpose, we shall concentrate on the main solutions  that have
been  used, namely, the abelian  solution with unit winding 
($n=1$) in eq. (\ref{FBM}) and its non-abelian extension of eq. 
(\ref{Fnoabflavor}).

First of all, let us analyze the R-symmetry of the gauge theory  from the
probe viewpoint. It is clear that the abelian solution (\ref{FBM}) is
invariant under  shifts of $\tilde{\psi}$ by a constant since the value of
this angle is an arbitrary constant in this solution. This symmetry has been
identified as the geometric dual of the R-symmetry
\cite{MN,KOW,GHP}. 
Actually, this is  not a $ U(1)$ invariance of the background because
\cite{MN,KOW,GHP} of the presence of the RR  form that selects,  by
consistency, only some particular values of the angle
$\tilde{\psi}$, \ie\ $\tilde{\psi} = \frac{2 \pi n}{N}$ with $1\le n\le
2N$. So,  the abelian probes can see a 
$\ZZ_{2N}$ symmetry. In contrast, when we consider the non-abelian  probe,
the solution  (\ref{Fnoabflavor}) selects two particular values of  the
angle $\tilde{\psi}$, thus breaking
$\ZZ_{2N}$ down  to  $\ZZ_2$. Notice that the $U(1)\to \ZZ_{2N}$  breaking
is an UV effect (it takes place already in the abelian background) while
the 
 $\ZZ_{2N}\to \ZZ_2$ breaking is an IR effect which appears only  when one
considers the full non-abelian regular background.
This same  breaking pattern was observed in the case of SQCD with  massive
flavors. Indeed,  as showed in \cite{Affleck:1983mk}, 
the theory with massive flavors has a non-anomalous discrete  
$\ZZ_{2N}$ R-symmetry, given by 
(component fields are used, $\lambda$ is the gluino and $\Phi$   and $\bar
\Phi$ are the  squarks):

\beq
\lambda \to e^{- i \pi n/N}\lambda\,\,, ~
\,\,\,\,\,\,\,\, \Phi \to e^{- i \pi n/ N} \Phi\,\,,
\,\,\,\,\,\,\,\, \bar\Phi \to e^{- i \pi n/ N} \bar\Phi\,\,,
\eeq
with $n= 1, ...2N$\ .

As shown in 
\cite{Affleck:1983mk}, this $\ZZ_{2N}$ symmetry is broken  down to $\ZZ_2$
by the formation of a squark condensate.
Indeed, one can see that $<\Phi \bar{\Phi}>$ transforms as 
$<\Phi \bar{\Phi}> \to e^{-2i\pi n/N}<\Phi \bar{\Phi}>$  leaving us with a
$\ZZ_2$. Besides, the squark condensate is consistent with supersymmetry,
because the $F$-term equation of motion $<\Phi \bar{\Phi}> - m=0$
is satisfied. Notice that this preservation of symmetry  when $m\not=0$ is
in agreement with the kappa symmetry of our brane probes.

Apart from all this, there is a vectorial $U(1)$ symmetry that remains 
unbroken in our brane probe analysis and that we have  associated with the
invariance under translations in  $\tilde\varphi$. On the field theory
side, this symmetry can be identified with a phase change of the full
chiral/antichiral multiplet 
$\Phi \to e^{i\alpha} \Phi,\,\,\bar\Phi \to e^{-i   \alpha} \bar\Phi$, 
which is nothing but the  $U(1)_B$ baryonic number symmetry of the theory.
The two possible assignations
$\pm 1$ of the baryonic charge are in correspondence with the two possible
identifications of the $S^1$ described by $\tilde\varphi$ with the $S^1$
parametrized by $\varphi$. The spectrum we have found  is independent of
this identification.

We would like also to comment briefly on the possibility of taking the 
parameter $r_*=0$. Given that this parameter can be associated with the 
mass of the quark (since it is the characteristic distance between the 
probe brane and the background), one would like to study the case in
which  this parameter is taken to be zero. Nevertheless, the approach
implemented  here seems to break down for this particular value of the
parameter.  Indeed, taking $r_*=0$ will imply that $\theta_0=0$, so, a
fluctuation of  it can lead to negative values of $\theta$ taking us out
of the range of  this coordinate. The fact that our approach apparently
does not work for the case of massless quarks seems to be in agreement
with the  fact that SQCD with massless flavors, 
has some special properties like spontaneous breaking of  supersymmetry,
the existence of a runaway potential (Affleck-Dine-Seiberg potential)  and
the non-existence of a  well-defined vacuum state. Notice that, when
$r_*\not= 0$,  our approach, by 
construction, deals with massive quarks that preserve supersymmetry, since 
our probes were constructed by that requirement.

\setcounter{equation}{0}
\section{Discussion}

In this chapter, a search for surfaces where a D5-brane can
be placed in the Maldacena-N\'u\~nez background without
spoiling supersymmetry has been performed.

Solutions where the 
probes are at a fixed distance from the "background branes" are shown to 
break supersymmetry. This phenomenon is in agreement with the non-existence 
of moduli space of  ${\cal N}=1$ SYM. On the other hand, allowing
the distance to the brane probe to vary,
a wide variety of kappa symmetric solutions was found and a rich 
mathematical structure was pointed out. Typically, these probes 
extend to infinity. On the spirit of AdS/CFT correspondence, when
the gravity setup is changed at infinity, some new ingredient is
introduced in the associated gauge theory. In our case, we interpret
the configurations found (at least some of them) as the
introduction of flavor in the dual ${\cal N}=1$ SYM.

First, configurations in the abelian background (the  large $r$ regime
of the full background) have been studied. These abelian solutions are 
shown to have a very interesting analytic structure (harmonic functions
and Cauchy-Riemann equations show up)  allowing interesting  explicit
solutions. In extending these studies to the full background, we  found
several classes of non-abelian solitons. It might be possible that these
solutions show  themselves useful when studying other aspects of the
model not addressed in  this work.

Then, in section (\ref{secmesons}), the kappa symmetric
solutions which are argued to introduce fundamental  matter in the dual
to SYM theory (those  we called
unit-winding) have been explored in detail. The surfaces where the
flavor branes are placed are the equivalent of the non-compact
holomorphic 2-cycles considered in ref. \cite{civ} to add
chiral superfields in a similar scenario.

Given that the brane probes do not backreact on the background, these 
flavors are introduced in the so-called quenched approximation. 
Nevertheless, many qualitative features and quantitative predictions 
for the strong coupling regime of  ${\cal N}=1$ SQCD can 
be addressed. Among them, the qualitative  difference between the massless
and  massive flavors is clear in this picture. Indeed, by construction,
our  approach deals with the massive-flavor case, so, the problems or 
peculiarities of the massless case can be seen in this approach under a 
different geometrical perspective.

Other characteristic feature of SQCD with few flavors is the breaking 
pattern of R-symmetry. The $\ZZ_{2N} \to \ZZ_{2}$ breaking  is
geometrically very  clear from the  brane probe perspective. Also, the
preserved vectorial symmetry $U(1)_B$ is  geometrically realized, as
explained in the previous section, by arbitrary  changes in the coordinate
$\tilde{\varphi}$.

A mass spectrum for the low energy excitations (mesons) of massive SQCD
was found and this may be a very interesting and quantitative 
prediction of this 
approach.
In fact, a nice formula for the masses is derived that exhibits a BPS-like 
behavior with the level ($n$) and  with the Kaluza-Klein  quantum number
($l$) of the meson
(\ref{Fspectrum}), (\ref{Fparabola}).
  Basically, our formula gives the meson masses in terms
of the mass of the fundamentals and the Casimir energy due to finite size 
of the meson,  and shows explicitly the  contamination of the meson
spectrum due to the  Kaluza-Klein modes. It would be very 
interesting if lattice calculations could validate or invalidate the 
formula  found here.

Our results are in perfect agreement with two general assertions about
vector-like gauge theories
proved by Vafa and Witten \cite{VafaWitten}: massive quarks cannot
constitute massless mesons and a vectorial symmetry 
 cannot be spontaneously broken.

It would be really interesting to find a backreacted gravity
solution in order to go beyond the quenched approximation of
the gauge theory. Among other aspects, the Affleck-Dine-Seiberg
potential, the corrections to the $\beta$-function due to
flavors or Seiberg dualities might be found. The techniques developed
will  hopefully be useful to perform similar analysis in different
gravity setups.

\setcounter{equation}{0}
\section{Appendix: asymptotic  behavior of the fluctuations}
\label{appendixB}

This appendix is devoted to the determination of the  asymptotic form of
the fluctuations around the kappa-symmetric static configurations of the
D5-brane probe. In subsection \ref{B.1} we will consider the large and 
small
$r$ behavior of the solutions of the equation of motion corresponding to
the lagrangian (\ref{Flagrafluct}), which describes the small oscillations
around the unit-winding embedding (\ref{Fthetazero}).  In section \ref{B.2}
we will study the asymptotic form of the fluctuations of the
$n$-winding embedding in the abelian background. Following  our general
arguments, the UV behavior of the abelian and non-abelian fluctuations
must be the same and, thus,  from this analysis we can get an idea of the
nature of the small oscillations around the general non-abelian $n$-winding
configurations (whose analytical form we have not determined) for large
values of the radial coordinate.

\subsection{Non-abelian unit-winding embedding}
\label{B.1}

For large $r$ the  lagrangian (\ref{Flagrafluct}) takes the form:
\beq
{\cal L}\,=\,-{1\over 2}\,e^{2\phi}\,\big[\,
r\,(\partial_r\chi)^2\,+\,2r\,\chi\,\partial_r\chi\,+\,r\chi^2\,+\,r\,
(\partial_x\chi)^2\,+\,r\,(\partial_{\varphi}\chi)^2\,\big]\,\,,
\eeq
where we have not expanded the dilaton and we have  eliminated the
exponentially suppressed terms. Using the asymptotic value of
$\partial_r\phi$:
\beq
\partial_r\phi\,\approx\,1\,-{1\over 4r-1}\,\,,
\eeq
we obtain the following equation for $\xi$:
\beq
\partial^2_r\,\xi\,+\,{8r^2-1\over 4r^2-r}\,
\partial_r\,\,\xi\,+\,
\Bigg(M^2-l^2+1\,+\,{2r-1\over 4r^2-r}
\Bigg)\,\,\xi\,=\,0\,\,.
\eeq
To study the UV behavior of the solutions of this  differential equation,
it is interesting to rewrite it with the different coefficient functions
expanded in powers of 
$1/r$ as:
\beq
\partial^2_r\,\xi\,+\,\big(\,a_0+
{a_1\over r}+{a_2\over r^2}\,+\,\cdots\big)\,\partial_r\,\xi\,+\,
\big(\,b_0+{b_1\over r}+{b_2\over r^2}\,+\,\cdots\big)\,\xi\,=\,0\,\,,
\label{FUVexpand}
\eeq
where:
\beq
a_0\,=\,2\,,\,\,\,\,\,\,\,\,
b_0\,=\,M^2-l^2+1\,,\,\,\,\,\,\,\,\,\,\,\,\,\,\,\,\,\,\,\,\,
a_1\,=\,b_1\,=\,{1\over 2}\,,\,\,\,\,\,\,\,\,\,\,\,\,\,\,\,
a_2\,=\,b_2\,=\,-{1\over 8}\,,\cdots\rc\rc
\eeq
We want to solve  eq. (\ref{FUVexpand}) by means  of an  asymptotic
Frobenius expansion of the type 
$\xi=r^{\rho}\,(c_0+c_1/r+\cdots)$ for some  exponent $\rho$. By
substituting this expansion on the right-hand side of  eq.
(\ref{FUVexpand}) and comparing the terms with the different powers of
$r$, we notice that there is only one term with
$r^{\rho}$, namely $b_0c_0r^{\rho}$, which  cannot  be canceled. In order
to get rid of this term, let us define a new function $w$ as:
\beq
\xi\,=\,e^{\alpha r}\,w\,\,,
\eeq
with $\alpha$ being a number to be determined.  The equation satisfied by
$w$ is the same as that of $\xi$ with the changes:
\bear
&&a_0\rightarrow a_0+2\alpha\,,\,\,\,\,\,\,\,\,\,\,\,\,\,\,\,\,
b_0\rightarrow \alpha^2+\alpha a_0+b_0\,\,,\rc\rc
&&a_i\rightarrow a_i\,\,,\,\,\,\,\,\,\,\,\,\,\,\,
\,\,\,\,\,\,\,\,\,\,\,\,\,\,\, b_i\rightarrow b_i+\alpha
a_i\,\,,\,\,\,\,\,\,\,\, (i=1,2,\cdots)\,\,.
\eear
It is clear that we must impose the condition:
\beq
\alpha^2+\alpha a_0+b_0\,=\,0\,\,,
\label{FUVindicial}
\eeq
which  determines the values of  $\alpha$. Writing now 
$w=r^{\rho}\,(c_0+c_1/r+\cdots)$ and looking at the  highest power of $r$
in the equation of $w$ (\ie\ $\rho-1$), we immediately obtain the value of
$\rho$, namely:
\beq
\rho=-{\alpha a_1+b_1\over 2\alpha+a_0}\,\,,
\label{FUVfrobenius}
\eeq
and, clearly, as $r\rightarrow\infty$ the asymptotic  behavior of $\xi(r)$
is:
\beq
\xi(r)\approx e^{\alpha r}\,r^{\rho}\,(\,1\,+\,o({1\over r})\,)\,\,.
\eeq
In our case, it is easy to verify that the values of  $\alpha$ and $\rho$
are:
\beq
\alpha=-1\pm i\sqrt{M^2-l^2}\,\,,\,\,\,\,\,\,\,\,\,\,\,\,
\rho =-{1\over 4}\,\,.
\eeq
Then, it is clear that we have two independent  behaviors for the real
function
$\xi(r)$, namely:
\beq
\xi(r)\sim {e^{-r}\over r^{{1\over 4}}}\cos 
\Big(\,\sqrt{M^2-l^2}\,\,r\,\Big)\,,
\,\,\,\,\,\,
{e^{-r}\over r^{{1\over 4}}}\sin\Big(\,\sqrt{M^2-l^2}\,\,r\,\Big)\,\,.
\label{FUVasymp}
\eeq
Notice that all solutions decrease exponentially when $r\rightarrow\infty$. 

Let us now turn to the analysis of the fluctuations  for small values of
the radial coordinate. Recall that $r_*\le r\le \infty$. 
Near $r_*$, one can expand $\sin\theta_0$ and $\cos\theta_0$ as follows:
\bear
\sin\theta_0&\approx& 1\,-\,\coth r_*\,(r-r_*)\,+\,{1\over 2}\,
{1+\cosh^2r_*\over \sinh^2r_*}\,(r-r_*)^2\,+\,\cdots\,\,,\rc\rc
\cos\theta_0&\approx&\sqrt{2\coth r_* (r-r_*)}\,\bigg[\,
1\,-{\coth r_*\over 2}\,(1+{1\over 2\cosh^2 r_*})\, (r-r_*)
\,+\,\cdots\bigg]\,\,.
\label{Fsincos}
\eear
Using these expressions it is straightforward to show that 
the lagrangian density of the quadratic fluctuations is given by:
\beq
{\cal L}\,=\,-{1\over 2}\,e^{2\phi_*}\,\,\big[\,
A(r)\,(\,\partial_r\chi)^2\,+\,B(r)2\chi\partial_r\chi\,+\,
C(r)\,\chi^2\,+\,D(r)\,(\,\partial_x\chi)^2\,+\,
E(r)\,(\,\partial_{\varphi}\chi)^2
\,\,\big]\,\,,
\label{FIRL}
\eeq
where $\phi_*=\phi(r_*)$ and the functions $A, B, C, D$  and $E$ are of the
form:
\bear
A(r)&=&\big[\,8\tanh r_*\,(r-r_*)^3\,\big]^{{1\over 2}}\,\,
\big(\,1\,+\,{\cal A}(r)\,\big)\,\,,\rc\rc
B(r)&=&{\sqrt{2(r-r_*)}\over \sqrt{\coth r_*}}\,\,
\big(\,1\,+\,{\cal B}(r)\,\big)\,\,,\rc\rc
C(r)&=&{1\over \sqrt{2\coth r_*\,(r-r_*)}}\,\,
\big(\,1\,+\,{\cal C}(r)\,\big)\,\,,\rc\rc
D(r)&=&{r_*\over \sqrt{\coth r_*}}\,\sqrt{2(r-r_*)}\,\,
\big(\,1\,+\,{\cal D}(r)\,\big)\,\,,\rc\rc
E(r)&=&\big[\,\tanh r_*\,\big]^{{3\over 2}}\,\,
\sqrt{2(r-r_*)}\,\,\big(\,1\,+\,{\cal E}(r)\,\big)\,\,.
\eear
The functions ${\cal A},{\cal B},{\cal C},{\cal D}, {\cal E}$, satisfy:
\beq
{\cal A}(r),{\cal B}(r), {\cal C}(r),{\cal D}(r)\,{\cal E}(r)\,
\,\sim\,\,o(r-r_*)\,\,.
\eeq
Notice that, remarkably, after integrating by parts the 
$\chi\partial_r\chi$ term in the lagrangian, the singular term of $C(r)$
cancels against the leading term of $B(r)$. The equation of motion for
$\xi$ near $r_*$ is:
\beq
\partial^2_r\xi\,+\,{A'(r)\over A(r)}\,\,\partial_r\xi\,+\,
{B'(r)-C(r)+M^2D(r)-l^2E(r)\over A(r)}\,\,\xi\,=\,0\,\,.
\label{FIReq}
\eeq
In order to solve this equation in a power series expansion  in $r-r_*$, it
is important to understand the singularities of the different coefficients
near
$r_*$. It is immediate that:
\beq
{A'(r)\over A(r)}\,=\,{3\over 2}\,\,{1\over r-r_*}\,+\,
{{\cal A}'(r)\over 1+{\cal A}(r)}\,=\, 
{3\over 2}\,\,{1\over r-r_*}\,+\,{\rm regular}\,\,.
\eeq
Similarly, the coefficient of $\xi$ has a simple pole near $r_*$:
\bear
{B'(r)-C(r)+M^2D(r)-l^2E(r)\over A(r)}\,=\,
{3{\cal B}\,'(r_*)-{\cal C}\,'(r_*)+2M^2r_*-2l^2\tanh r_*\over
4}\,{1\over r-r_*}\,+\,{\rm regular}\,\,.\rc
\eear
It follows that the point $r=r_*$ is a singular regular point.  The
corresponding Frobenius expansion reads:
\beq
\xi(r)\,=\,(r-r_*)^{\lambda}\,
\sum_{n=0}^{\infty}\,c_n\,(r-r_*)^n\,\,,
\eeq
where $\lambda$ satisfies the indicial equation, which can be obtained by
plugging the expansion in the equation and looking at the term with lowest
power of $r-r_*$ (\ie\ $\lambda-2$). In our case, $\lambda$ must be a root
of the quadratic equation:
\beq
\lambda(\lambda-1)+{3\over 2}\,\lambda\,=\,0\,\,,
\label{FIRindicial}
\eeq
\ie:
\beq
\lambda=0, -{1\over 2}\,\,.
\eeq
This means that there are two independent solutions of  the differential
equation which can be represented by a Frobenius series around $r=r_*$, one
of then is regular as
$r\rightarrow r_*$ (the one corresponding to $\lambda=0$),  whereas the 
other  diverges as $(r-r_*)^{-{1\over 2}}$. 

\subsection{ $n$-Winding embeddings in the abelian background}
\label{B.2}

Let us consider the abelian background and an embedding with winding number
$n$, for which $\partial_{\varphi}\tilde\varphi=n$,
$\sin\theta\partial_{\theta}\tilde\theta\,=\,n\sin\tilde\theta$ and
$\psi=\psi_0={\rm constant}$. Let us define:
\bear
V&\equiv&{n\over 2}{\cos\tilde\theta\over \sin\theta}\,+\,{1\over
2}\cot\theta\,=\, {n\over 2\sin\theta}\,\,
{(1+\cos\theta)^n\,-\,(1-\cos\theta)^n\over
(1+\cos\theta)^n\,+\,(1-\cos\theta)^n}\,+\,{1\over 2}\cot\theta\,\,,\rc\rc
W&\equiv&{n\over 2}{\sin\tilde\theta\over \sin\theta}\,=\,
{n(\sin\theta)^{(n-1)}\over 
(1+\cos\theta)^n\,+\,(1-\cos\theta)^n}\,\,.
\eear
Choosing $r$ as worldvolume coordinate and considering
embeddings in which $\theta$ depends
on both $r$ and on the unwrapped coordinates $x$, we obtain the following
lagrangian:
\bear
{\cal L}=-e^{2\phi}\,\sin\theta&\Bigg[
&\sqrt{\big(\,e^{2h}\,+\,V^2\,+\,W^2\,\big)\,\big(\,1\,+\,
(e^{2h}\,+\,W^2)((\partial_r\theta)^2\,+\,(\partial_x\theta)^2)\big)}\,
+\cr\cr &&+\,(e^{2h}\,+\,W^2)\partial_r\theta\,-\,V\,\Bigg]\,\,.
\eear
We shall expand this lagrangian around a configuration $\theta_0(r)$  such
that:
\beq
e^{2(r-r_*)}\,=\,{1\over 2}\,
{(1+\cos\theta_0(r))^n\,+\,(1-\cos\theta_0(r))^n\over 
(\sin\theta_0(r))^{n+1}}\,\,.
\label{Fnthetazero}
\eeq
Notice that $\theta_0(r)$ remains unchanged under the transformation
$n\to -n$. Thus, without loss  of generality we shall restrict ourselves 
 from now on to the case $n\ge 0$. Let us now define:
\beq
V_0(r)\equiv V|_{\theta(r)=\theta_0(r)}\,\,,
\,\,\,\,\,\,\,\,\,\,\,\,\,\,\,\,
W_0(r)\equiv W|_{\theta(r)=\theta_0(r)}\,\,,
\,\,\,\,\,\,\,\,\,\,\,\,\,\,\,\,
V_{0\theta}(r)\equiv{\partial V\over \partial\theta}
\Big|_{\theta(r)=\theta_0(r)}\,\,.
\eeq
Using:
\beq
\partial_{r}\theta_0\,=\,-{1\over V_0}\,\,,
\eeq
we obtain the following quadratic lagrangian:
\bear
{\cal L}\,=\,-{e^{2\phi}\over 2}\,\sin\theta_0\,(e^{2h}\,+\,W_0^2\,)\,
&\Bigg[& {1\over e^{2h}\,+\,V_0^2\,+\,W_0^2}\,
\Big(\,V_0^3\,(\partial_r\chi)^2\,-\,2V_0\,V_{0\theta}\,\chi
\partial_r\chi\,+
\rc\rc
&&+\,{V_{0\theta}^2\over V_0}\,\chi^2\,\Big)\,+\,V_0\,(\partial_x\chi)^2
\,\Bigg]\,\,.
\eear
Keeping the leading terms for large $r$, the lagrangian becomes:
\beq
{\cal L}\,=\,-{r\over 2}e^{2\phi}\,\big[\,
{n+1\over 2}\,(\partial_r\chi)^2\,+\,2\chi\partial_r\chi\,+\,
{2\over n+1}\,\chi^2\,+\,{n+1\over 2}\,(\partial_x\chi)^2\,
\big]\,\,,
\eeq
and, if we represent $\chi$ as in eq. (\ref{Fchiansatz}) with $l=0$, 
the equation of motion for $\xi$ becomes:
\beq
\partial_r^2\xi\,+\,\big(2+{1\over 2r}\big)\,
\partial_r\xi\,+\,\bigg(\,M^2\,+\,{4n\over (n+1)^2}\,+\,{1\over n+1}\,
{1\over r}\,\bigg)\,\xi\,=\,0\,\,.
\eeq
Now we have the following coefficients in eq. (\ref{FUVexpand}):
\beq
a_0\,=\,2\,\,,
\,\,\,\,\,\,\,\,\,\,\,\,\,\,\,\,\,\,
b_0\,=\,M^2+{4n\over (n+1)^2}\,\,,
\,\,\,\,\,\,\,\,\,\,\,\,\,\,\,\,\,\,
a_1\,=\,{1\over 2}\,\,,
\,\,\,\,\,\,\,\,\,\,\,\,\,\,\,\,\,\,
b_1\,=\,{1\over n+1}\,\,.
\eeq
By plugging these values in eq. (\ref{FUVindicial}), we obtain the
following result for the coefficient $\alpha$ of the exponential:
\beq
\alpha\,=\,-1\,\pm\,
\sqrt{\Big({n-1\over n+1}\Big)^2\,-\,M^2}\,\,.
\eeq
Let us distinguish two cases, depending on the sign inside the square root.
Suppose first that $M^2\ge \big({n-1\over n+1}\big)^2$ and define:
\beq
\tilde M^2\,=\,M^2\,-\,\Big({n-1\over n+1}\Big)^2\,\,.
\eeq
In this case the values of the exponent $\rho$ obtained from 
(\ref{FUVfrobenius}) are:
\beq
\rho\,=\,-{1\over 4}\,\mp {n-1\over 2(n+1)\tilde M}\,\,i\,\,.
\eeq
It follows that the two real asymptotic solutions are:
\beq
\xi(r)\sim {e^{-r}\over r^{{1\over 4}}}\,
\cos\big[\tilde M r\,-\,{n-1\over 2(n+1)\tilde M}\,\log r\big]\,\,,
\,\,\,\,\,\,\,\,\,\,\,\,
{e^{-r}\over r^{{1\over 4}}}\,
\sin\big[\tilde M r\,-\,{n-1\over 2(n+1)\tilde M}\,\log r\big]\,\,.
\label{FnUV}
\eeq
Both solutions vanish exponentially when $r\rightarrow\infty$. 

If $\tilde M^2<0$, let us define $\bar M^2=-\tilde M^2$. In this  case
$\alpha$ is real, namely $\alpha=-1\pm\bar M$. Notice that $\bar M<1$ and
thus
$\alpha<0$. The independent asymptotic solutions are:
\beq
\xi(r)\sim e^{(\bar M-1)r}\,\,
r^{-{1\over 4}\,\big(1\,-\, {n-1\over (n+1)\bar M}\big)}\,\,,
\,\,\,\,\,\,\,\,\,\,\,\,
e^{-(\bar M+1)r}\,\,
r^{-{1\over 4}\,\big(1\,+\, {n-1\over (n+1)\bar M}\big)}\,\,,
\label{FnUVotro}
\eeq
and both decrease exponentially, without oscillations,  when
$r\rightarrow\infty$. This non-oscillatory character of the functions in
eq. (\ref{FnUV}) make them inadequate for the type of boundary conditions we
are imposing and, therefore, we shall discard them.

Notice that, for $n=1$, the large $r$ asymptotic solutions  (\ref{FUVasymp})
and (\ref{FnUV}) coincide. This is of course to be expected since the
abelian and non-abelian configurations coincide in the UV. It is also
interesting to compare the magnitude of the fluctuation with that of the
unperturbed configuration for large $r$. By inspecting eq. 
(\ref{Fnthetazero}) one readily concludes that:
\beq
\theta_0(r)\,\sim\,e^{-{2\over n+1}r}\,\,,
\,\,\,\,\,\,\,\,\,
(r\to \infty)\,\,.
\eeq
By comparing this behavior with eq. (\ref{FnUV}) one finds:
\beq
{\xi(r)\over \theta_0(r)}\sim {e^{-{n-1\over n+1}r}\over  r^{{1\over
4}}}\,\,,
\,\,\,\,\,\,\,\,\,
(r\to \infty)\,\,.
\eeq
Thus, for $n\ge 1$ one has that ${\xi(r)\over \theta_0(r)}\to 0$  as $r\to
\infty$. On the contrary for $n= 0$, both in the abelian and non-abelian
case, the ratio 
${\xi(r)\over \theta_0(r)}$ diverges in the UV and the first order 
expansion breaks down.

Let us now consider the IR behavior of the fluctuations. 
Near $r_*$ one has to leading order that $\sin\theta_0\approx 1$ and:
\bear
&&\cos\theta_0\approx {2\over \sqrt{1+n^2}}\,\sqrt{r-r_*}\,\,,
\,\,\,\,\,\,\,\,\,\,\,\,\,\,\,\,\,\,\,\,\,\,\,\,
V_0\approx \sqrt{n^2+1}\,\sqrt{r-r_*}\,\,,\rc\rc
&&W_0\approx{n\over 2}\,\,,
\,\,\,\,\,\,\,\,\,\,\,\,\,\,\,\,\,\,\,\,\,\,\,\,
\,\,\,\,\,\,\,\,\,\,\,\,\,\,\,\,\,\,\,\,\,\,\,\,
\,\,\,\,\,\,\,\,\,\,\,\,\,\,\,\,\,\,\,\,\,\,\,\,
V_{0\theta}\approx-{n^2+1\over 2}\,\,.
\eear
The IR lagrangian is of the same form as in eq. (\ref{FIRL})  (with $E=0$
since we are now considering the case in which  $\chi$ is independent of
$\varphi$).  The functions $A(r)$ to $D(r)$ are now of the form:
\bear
A(r)&=&\Big[\,(n^2+1)(r-r_*)\Big]^{{3\over 2}}\,\,
\Big(\,1+o(r-r_*)\Big)\,\,,\rc\rc
B(r)&=&{(n^2+1)^{{3\over 2}}\over 2}\,\sqrt{r-r_*}\,\,
\Big(\,1+o(r-r_*)\Big)\,\,,\rc\rc
C(r)&=&{1\over 4}\,{(n^2+1)^{{3\over 2}}\over\sqrt{r-r_*}}\,\,
\Big(\,1+o(r-r_*)\Big)\,\,,\rc\rc
D(r)&=&\Big(r_*+{n^2-1\over 4}\,\Big)\sqrt{n^2+1}\,\sqrt{r-r_*}\,\,
\Big(\,1+o(r-r_*)\Big)\,\,.
\eear
Notice that, also in this case, the coefficients of the functions above are
such that, after a partial integration, the singular term of $C(r)$ cancels
against the leading term of $B(r)$. The differential equation  that follows
for 
$\xi$ in the IR is:
\beq
\partial_r^2\xi\,+\,\Big({3\over 2}\,{1\over r-r_*}\,+\,o
(r-r_*)\,\Big)\,
\partial_r\xi\,+\,o({1\over r-r_*})\,\xi\,=\,0\,\,,
\eeq
and, therefore, the indicial equation is the same as for eq.  (\ref{FIReq})
(\ie\ eq. (\ref{FIRindicial})). It follows that also in this case there
exists an independent solution which does not diverge when $r\rightarrow
r_*$.

%*************************************************************

\chapter{Other solutions from supersymmetry}
\label{chothers}

\medskip

This miscellaneous chapter is devoted to the description of a few
supergravity backgrounds that have not been discussed elsewhere
in this thesis. Except for section \ref{maldanassec} \cite{flavoring},
the rest is based on unpublished work. In section \ref{maldanassec}, 
the supersymmetry analysis and Killing spinors of the supergravity
dual of three dimensional ${\cal N}=1$ gauge theory are shown.
Section \ref{hypersec} is devoted to the study of solutions
arising from branes wrapped in hyperbolic spaces. We will find
in all cases that the metric runs into a bad singularity.
Then, in section \ref{secspin7}, we review the possibility of
obtaining eight dimensional $Spin(7)$ holonomy metrics from
gauged supergravity. We will see that the procedure of allowing
a rotation of the Killing spinor in a simple ansatz, which leads to new
metrics in the conifold and $G_2$ cases, only yields trivial solutions.
Finally, some D=7, $SO(4)$ gauged supergravity 
setups are discussed in section \ref{so4sec}. This is a truncation
of the theory of section \ref{PPvN}. In
particular, we show how to obtain the Maldacena-N\'u\~nez solution from
this viewpoint and discuss a (singular) background dual to 
${\cal N}=2$ SYM. We show that the procedure that desingularizes
the gravity solution dual to ${\cal N}=1$  SYM does not work
in this case.

\setcounter{equation}{0}
\section{Fivebranes wrapped on a three-cycle}
\label{maldanassec}

In this section, we are going to analyze the case of D5-branes
wrapping an associative three-cycle inside a $G_2$ holonomy manifold in
type IIB theory, first studied in \cite{acharya}\footnote{The case in which
D5-branes
 wrap
a SLag three-cycle inside a Calabi-Yau three-fold \cite{kim} is also interesting,
but will not be presented here.}. This leaves 1+2 unwrapped flat dimensions
where the dual gauge theory lives. As usual, the normal bundle must be twisted in
order to preserve some supersymmetry. It can be shown that 1/16 of the maximal
supersymmetry is preserved, so there are two supercharges, which amounts to ${\cal
N}=1$ in three dimensions. We will check this  explicitly by finding the four
independent projections that must be imposed on the Killing spinor. The bosonic
degrees of freedom of the dual gauge theory are just a gauge
boson and the action is Yang-Mills with a Chern-Simons term.
This gravity solution and its dual gauge theory were studied in
\cite{acharya,MaldaNasta}.

We will follow a similar procedure to section
\ref{secMNsol}, where the solution for D5 wrapping $S^2$
was found. The natural framework is the D=7 supergravity of
section \ref{TVN}. The seven dimensional ansatz will be given,
but, for the sake of brevity, the supersymmetry analysis
will  be only performed in ten dimensions for the uplifted 
ansatz \cite{flavoring}. A difference with the case of section
\ref{secMNsol} is that the BPS first order equations can be found, but
the general analytical solution is not known.

The ansatz for the metric in seven dimensions is (string frame):
\beq
ds_7^2\,=\,
\,dx_{1,2}^{2}\,+\,{1\over 4}\,R(r)^2\,(\,\unbrw^i\,)^2\,+\,
dr^2\,\,,
\eeq
where $R(r)$ is a function to be determined. The
$\unbrw^i$ ($i=1,2,3$) are a new set of left invariant
one-forms defined as in (\ref{otherom}),
$d\unbrw^i\,=\,-\,{1\over 2}\,\epsilon_{ijk}\,\unbrw^j\,
\wedge\,\unbrw^k$, such that $d\Omega_3^2={1 \over 4}(\unbrw^i)^2$.
The ansatz for the gauge field is\footnote{The same ansatz with
$\omega(r)=0$ was introduced in \cite{acharya}. It leads to
a consistent system of BPS equations, but the resulting
gravity solution is singular at $r=0$. The function $\omega(r)$
plays the same smoothing role as the function $a(r)$ in section
\ref{secMNsol}. This ansatz for the gauge field resembles
(\ref{seis}) just as (\ref{Foneform}) resembled (\ref{conigauge}).}:
\beq
A^i\,=\,{1+\omega(r)\over 2}\,\,\unbrw^i\,\,.
\label{MNasA}
\eeq
It is immediate to get the gauge field strength by inserting
(\ref{MNasA}) in (\ref{gaugefstr}):
\beq
F^i\,=\,{\omega'\over 2}\,dr\wedge \unbrw^i\,+\,
{\omega^2-1\over 8}\,\epsilon_{ijk}\,\unbrw^j\wedge\unbrw^k\,\,.
\eeq
From the lagrangian (\ref{lagrTVN}), we see that $F\wedge F$ acts
as a source for the 3-form $B$. Upon uplifting, this leads
to an additional term in the RR 3-form $F_{(3)}$ (see below).
By explicit calculation, one finds that $F\wedge F$ is not
zero for this ansatz:
\beq
\sum_i\,F^i\wedge F^i\,=\,{1\over 8}\,
(\,\omega^2\,-\,1\,)\,\omega'\,\,
\epsilon_{jkm}\,dr\wedge\unbrw^j\wedge\unbrw^k\wedge\unbrw^m\,\,.
\label{MNasFF}
\eeq
The corresponding type IIB Einstein frame metric, describing D5 branes
wrapped in $S^3$, reads:
\beq
ds_{10}^2\,=\,e^{{\phi\over 2}}\,
\big[\,dx_{1,2}^{2}\,+\,{1\over 4}\,R(r)^2\,(\,\unbrw^i\,)^2\,+\,
dr^2\,+\,{1\over 4}\,(\,\om^i\,-\,A^i\,)^2\,\big]\,\,,
\eeq
while the Ramond-Ramond 3-form $F_{(3)}$ is:
\beq
F_{(3)}\,=\,-{1\over 4}\,
(\om^1-A^1)\wedge (\om^2-A^2) \wedge (\om^3-A^3)\,+\,
{1\over 4}\,\sum_i\,F^i \wedge (\om^i-A^i)\,+\,h\,\,,
\eeq
where $h$ comes from the uplifting of the non-zero 3-form $B$ of seven
dimensional sugra and is
determined by requiring that
$F_{(3)}$ satisfies the Bianchi identity $dF_{(3)}=0$. 
This is equivalent to requiring the form $B$ to solve the
equations of motion of 7d sugra. 
One easily
verifies that $h$ must satisfy the following equation:
\beq
dh\,=\,{1\over 4}\,\sum_i\,F^i\wedge F^i\,\,.
\eeq
Thus, taking (\ref{MNasFF}) into account, the equation for $h$ can be
solved as:
\beq
h\,=\,{1\over 32}\,{1\over 3!}\,V(r)\,
\epsilon_{ijk}\,\unbrw^i\wedge\unbrw^j\wedge\unbrw^k\,\,,
\eeq
where:
\beq
V(r)\,=\,2\omega^3\,-\,6\omega\,+\,8k\,\,,
\eeq
with $k$ being a constant. 
Let us study the supersymmetry preserved by this ansatz in the
frame:
\bear
&&e^{x^i}\,=\,e^{{\phi\over 4}}\,dx^i\,\,,
\,\,\,\,\,\,\,\,\,\,\,\,\,\,
(i=0,1,2)\,\,,
\,\,\,\,\,\,\,\,\,\,\,\,\,\,\,\,\,\,\,\,
e^r\,=\,e^{{\phi\over 4}}\,dr\,\,,\rc\rc
&&e^{i}\,=\,{1\over 2}\,\,e^{{\phi\over 4}}\,\,R
\,\unbrw^i\,\,,
\,\,\,\,\,\,\,\,\,\,\,\,\,\,\,\,\,\,\,\,\,\,\,\,\,\,\,\,
\,\,\,\,\,\,\,\,
e^{\hat i}\,=\,{1\over 2}\,\,e^{{\phi\over 4}}\,\,
(\om^i-A^i)\,\,,
\,\,\,\,\,\,\,\,\,\,
(i=1,2,3)\,\,,
\eear
We want to have $\delta\la=\delta\psi_\mu=0$
in the expressions (\ref{SUSYIIB}).
We  start by imposing the following projections on the spinors:
\bear
&&\Gamma_1\hat\Gamma_1\,\epsilon\,=\,
\Gamma_2\hat\Gamma_2\,\epsilon\,=\,
\Gamma_3\hat\Gamma_3\,\epsilon\,\,,\rc\rc
&&\epsilon\,=\,i\epsilon^*\,\,.
\label{MNasproj1}
\eear
Then, the condition $\delta\lambda\,=\,0$ becomes:
\beq
\phi'\,\epsilon\,-\,\bigg(\,1\,+\,3\,{\omega^2\,-\,1\over 4R^2}
\bigg)\,\Gamma_r\,\hat\Gamma_{123}\,\epsilon\,+\,
{3\omega'\over 4R}\,\Gamma_1\hat\Gamma_1\epsilon\,-\,
{V\over 8R^3}\,\Gamma_1\hat\Gamma_1\,
\Gamma_r\,\hat\Gamma_{123}\,\epsilon\,=\,0\,\,.
\label{MNasdilatino}
\eeq
Moreover,
from $\delta\psi_i=0$, one gets:
\beq
R'\epsilon\,-\,{\omega'\over 2}\,
\Gamma_1\hat\Gamma_1\epsilon\,+\,
{\omega^2-1\over R}\,\Gamma_r\,\hat\Gamma_{123}\,\epsilon\,+\,
\Big(\,{V\over 8R^2}\,-\,\omega\,\Big)\,
\Gamma_1\hat\Gamma_1\,
\Gamma_r\,\hat\Gamma_{123}\,\epsilon\,=\,0\,\,.
\label{MNasgravitino}
\eeq
The vanishing of the supersymmetric variation of the radial component
of the gravitino gives rise to:
\beq
\partial_r\epsilon\,-\,{3\omega'\over 4R}\,
\Gamma_1\hat\Gamma_1\epsilon\,-\,{1\over 8}\,\phi'\,\epsilon\,=\,0\,\,,
\label{MNasgravir}
\eeq
where we have used eq. (\ref{MNasdilatino}). 
Let us solve these equations by taking the projection:
\beq
\Gamma_r\,\hat\Gamma_{123}\,\epsilon\,=\,\big(\,
\beta\,+\,\tilde\beta\Gamma_1\hat\Gamma_1\,\big)\,\epsilon\,\,.
\label{MNasrad}
\eeq
As usual, from 
$(\Gamma_r\,\hat\Gamma_{123})^2\,\epsilon\,=\,\epsilon$ and 
$\{\Gamma_r\,\hat\Gamma_{123}, \Gamma_1\hat\Gamma_1\}=0$ 
it follows  that
$\beta^2+\tilde\beta^2\,=\,1$ and thus we can take $\beta=\cos\alpha$
and $\tilde\beta=\sin\alpha$.

Let us substitute our ansatz for
$\Gamma_r\,\hat\Gamma_{123}\,\epsilon$ (eq. (\ref{MNasrad}))
on the equations coming from
the dilatino and gravitino (eqs. (\ref{MNasdilatino}) and  
(\ref{MNasgravitino})). From the terms containing the unit matrix, we
obtain equations for $\phi'$ and $R'$:
\bear
\phi'&=&\Big(\,1\,+\,3\,{\omega^2-1\over 4R^2}\,\Big)\,\beta\,-\,
{V\over 8R^3}\,\tilde\beta\,\,,\rc\rc
R'&=&{1-\omega^2\over R}\,\beta\,+\,\Big(\,{V\over 8R^2}\,-\,\omega
\,\Big)\,\tilde\beta\,\,.
\eear
Moreover, by considering the terms with $\Gamma_1\hat\Gamma_1$, we
obtain two expressions for $\omega'$ :
\bear
3\omega'&=&{V\over 2R^2}\,\beta\,+\,\Big(\,4R\,+\,3\,
{\omega^2-1\over R}\,\Big)\,\tilde\beta\,\,,\rc\rc
\omega'&=&\Big(\,{V\over 4R^2}\,-\,2\omega\,\Big)\,\beta\,+\,
2\,{\omega^2-1\over R}\,\tilde\beta\,\,.
\eear
By combining these last two equations we get:
\beq
\Big(\,{V\over 24R^2}\,-\,w\,\Big)\,\beta\,=\,
\Big(\,{1-w^2\over 2R}\,+\,{2R\over 3}\,\Big)\,\tilde\beta\,\,.
\eeq
By plugging this last relation in the condition 
$\beta^2+\tilde\beta^2\,=\,1$, one can easily obtain the expression of
$\beta$ and $\tilde\beta$. Indeed, let us define $M$  as follows:
\beq
M\,\equiv\,\Big(\,{V\over 24R^2}\,-\,w\,\Big)^2\,+\,
\Big(\,{1-w^2\over 2R}\,+\,{2R\over 3}\,\Big)^2\,\,.
\eeq
In terms of this new quantity $M$, the coefficients $\beta$ and
$\tilde\beta$ are given by:
\beq
\beta\,=\,\cos\alpha\,=\,{1\over\sqrt{M}}\,\,
\Bigg(\,
{2R\over 3}\,+\,{1-\omega^2\over 2R}\,\Bigg)
\,\,,
\,\,\,\,\,\,\,\,\,\,\,\,\,\,\,\,\,\,
\tilde\beta\,=\,\sin\alpha\,=\,{1\over\sqrt{M}}\,\,
\Bigg(\,
{V\over 24R^2}\,-\,\omega
\,\Bigg)\,\,.
\eeq
By using these values of $\beta$ and
$\tilde\beta$ in the equations which determine $R\,'$, $\omega\,'$ and
$\phi'$, we obtain a system of first-order
BPS equations which are identical  to those written in refs. 
\cite{MaldaNasta}. They are:
\bear
R\,'&=&{1\over 3\sqrt{M}}\,\Big[\,{V^2\over 64R^4}\,+\,
{1\over
2R^2}\,\Big(\,3(1-\omega^2)^2\,-\,V\omega\,\Big)
\,+\,\omega^2\,+\,2\,\Big]\,\,,\rc\rc
\omega\,'&=&{4R\over 3\sqrt{M}}\,\Big[\,{V\over
32R^4}\,(\,1-\omega^2\,)\,+\, {2k\,-\,\omega^3\over
2R^2}\,-\,\omega\,\Big]\,\,,\rc\rc
\phi'&=&-{3\over 2}\,\big(\,\log R\,\big)'\,+\,{3\over 2}\,
{\sqrt{M}\over R}\,\,.
\label{MNasDE}
\eear
The radial projection condition (\ref{MNasrad}) can be written as:
\beq
\Gamma_r\,\hat\Gamma_{123}\,\epsilon\,=\,
e^{\alpha\Gamma_1\hat\Gamma_1}\,\epsilon\,\,.
\eeq
Since $\{\Gamma_r\,\hat\Gamma_{123}, \Gamma_1\hat\Gamma_1\}=0$, this
equation can be solved as: 
\beq
\epsilon\,=\,
e^{-{\alpha\over 2}\,\Gamma_1\hat\Gamma_1}\,\epsilon_0\,\,,
\,\,\,\,\,\,\,\,\,\,\,\,\,\,\,
\Gamma_r\,\hat\Gamma_{123}\,\epsilon_0\,=\,\epsilon_0\,\,.
\eeq
Plugging now this parametrization of $\epsilon$ into the equation
obtained from the variation of the radial component of the gravitino
(eq. (\ref{MNasgravir}) ), we arrive at the  following two equations:
\beq
\partial_r\,\epsilon_0\,=\,{\phi'\over 8}\,\epsilon_0\,\,,
\,\,\,\,\,\,\,\,\,\,\,\,\,\,\,\,\,\,
\alpha'\,=\,-{3\over 2}\,\omega'\,R\,\,.
\eeq
The equation for $\epsilon_0$ can be solved immediately:
\beq
\epsilon_0\,=\,e^{{\phi\over 8}}\,\eta\,\,,
\eeq
where $\eta$ is a constant spinor. Moreover, one can verify that 
the equation for $\alpha$ is a consequence of the first-order BPS
equations (\ref{MNasDE}). Therefore, the Killing spinors for this 
geometry are of the form:
\beq
\epsilon\,=\,e^{-{\alpha\over 2}\,\Gamma_1\hat\Gamma_1}\,\,
e^{{\phi\over 8}}\,\eta\,\,,
\eeq
where $\eta$ is constant and satisfies the following conditions:
\bear
&&\Gamma_{x^0\cdots x^2}\,\Gamma_{123}\,\eta\,=\,\eta\,\,,\rc\rc
&&\Gamma_1\hat\Gamma_1\,\eta\,=\,\Gamma_2\hat\Gamma_2\,\eta\,=\,
\Gamma_3\hat\Gamma_3\,\eta\,\,,\rc\rc
&&\eta\,=\,i\eta^*\,\,.
\eear
Notice that, for the spinor $\epsilon$,  the first of these projections
can be rewritten as:
\beq
\Gamma_{x^0\cdots x^2}\,\Big(\,\cos\,\alpha\Gamma_{123}\,-\,
\sin\alpha\hat\Gamma_{123}\,\Big)\,\epsilon\,=\,\epsilon\,\,.
\eeq
As in section \ref{secMNsol}, the susy analysis directly
performed in seven dimensional sugra is  analogous to the one
performed above in ten dimensions. In the seven dimensional language, the
projections (\ref{MNasproj1}) and (\ref{MNasrad}) read: $\G_1 \sigma^1
\epsilon=\G_2 \sigma^2\epsilon=\G_3 \sigma^3
\epsilon$ and $\G_r \epsilon = (\beta+\tilde\beta \G_1\,i\,\sigma^1)
\epsilon$ (one must also take into account $\G^{x_0x_1x_2r123}=1$).

%*******************************************************

\setcounter{equation}{0}
\section{Branes wrapping hyperbolic spaces}
\label{hypersec}

In this section, some cases where branes are wrapped on
spaces with constant negative curvature will be analyzed. We will
follow the strategy of previous chapters by looking for
a solution of some low dimensional gauged supergravity in
order to uplift it
to  ten or eleven dimensions.

Hyperbolic  spaces are, in principle, non-compact. However,
once the solution is obtained, we can quotient the hyperbolic
space by some discrete (infinite) group, so we end up with finite volume.
If the Killing spinors do not depend on the coordinates of
the hyperbolic space  where the quotienting is made, the
solution will continue to be supersymmetric.
Moreover, this procedure gives rise to Riemann surfaces of different
genus. This is interesting because it can introduce adjoint matter
in the associated gauge theory. Indeed, by an
index theorem, the genus of the surface is related to the
number of zero modes on the submanifold in which the brane is
wrapped (see, for example, \cite{MNfirst}). Therefore, if the genus is
bigger than one, there are  zero modes that
survive in the limit in which the volume of the compact space
is taken to be small and, thus, they must appear somehow in the
gauge theory. Because of supersymmetry, they must reorganize
themselves in supermultiplets (which may be different depending
on the case). Note that this kind of zero modes are absent
when the brane is wrapping a sphere $S^2$ or $S^3$, so this
adjoint matter is not present in the cases studied in
previous chapters\footnote{The quark-like matter introduced in chapter
\ref{flavoring}, unlike this case, transforms in the fundamental
representation of the gauge group, as it comes from strings stretching
from  the flavor brane to the gauge theory brane.}.

However, we will see in different cases that this leads
(for a general gauge field in the ansatz) to pathological
supergravity backgrounds: the factor multiplying the
hyperbolic space first grows but then collapses (and gets
negative, changing the signature of space-time) at some
finite value of the distance $r$. Maybe the reason is that
there is something in the gauge theory that makes the
gauge-gravity duality fail, so one should only trust it
far away from this singular point, but further work is
required to clarify this idea.

It would also be interesting to look for similar solutions
when the reduction from ten or eleven dimensions to gauged
supergravity is made on a hyperbolic space and not on a 
sphere (see \cite{ortin} for Scherk-Schwarz reductions in
general three dimensional group manifolds).

In the following subsections, we study the hyperbolic 
counterpart of the spaces described on chapters
\ref{conichapter}, \ref{g2chapter} and \ref{MNchapter} respectively.

%*******************************************************

\subsection{D6-branes wrapping $H_2$}

Let us consider the  ansatz  obtained by substituting the $S^2$
in eq. (\ref{d6s2metric}) by the hyperbolic space $H_2$.
The subsequent analysis will be very similar to that in sections
\ref{wrappingS2}, \ref{rescon}, so some details will
be skipped. The metric in the 8d
Salam-Sezgin  gauged sugra (section \ref{ss8d}) is:
\beq
ds^2_8=e^{{2\phi\over 3}}\,dx_{1,4}^2\,+\,
{e^{2h}\over y^2}\,\big(dz^2+dy^2\big)+dr^2\,\,.
\eeq
The ansatz for the scalars is again (\ref{coniscalars}):
$L_{\alpha}^i\,=\,{\rm diag}\,\Big(\,e^{\lambda}\,,\,e^{\lambda}\,,\,
e^{-2\lambda}\,\Big)$, while
 for the gauge field we write:
\beq
A^1\,=\,{b\over y}\,dy\,\,,
\,\,\,\,\,\,\,\,\,\,\,\,\,\,
A^2\,=\,-{b\over y}\,dz\,\,,
\,\,\,\,\,\,\,\,\,\,\,\,\,\
A^3\,=\,{1\over y}\,dz\,\,.
\eeq
The $P$ and $Q$ matrices are again given by (\ref{conipq}),
and the field strength reads:
\beq
F^1\,=\,\,{b'\over y}\,dr\wedge dy\,\,,
\,\,\,\,\,\,\,\,\,\,\,\,\,\,
F^2\,=\,-\,{b'\over y}\,dr\wedge dz\,\,,
\,\,\,\,\,\,\,\,\,\,\,\,\,\,
F_{zy}^3\,=\,\,\,{1+b^2\over y^2}\,dz\wedge dy\,\,.
\eeq
We will use the following angular projection:
\beq
\Gamma_{zy}\epsilon\,=\,-\hat\Gamma_{12}\epsilon\,\,.
\eeq
By combining the equations $\delta\chi_1=\delta\chi_2=0$ and
$\delta\chi_3=0$, we get an equation for $\phi'$:
\bear
\phi'\,\epsilon\,-\,
e^{\phi+\lambda-h}\,b'\,\hat\Gamma_2\Gamma_z\epsilon\,+\,
\Big[\,{1\over 2}\,
e^{\phi-2\lambda-2h}\,\big(\,1+b^2\,\big)\,+\,
{1\over 8}\,e^{-\phi}\,\big(\,e^{-4\lambda}\,+\,2e^{2\lambda}\,\big)\,
\Big]\,
\Gamma_r\hat\Gamma_{123}\,\epsilon\,=\,0\,\,,\rc
\label{hypphi}
\eear
and an equation for $\lambda'$:
\bear
&&\lambda'\,\epsilon\,+\,
be^{-h}\sinh
3\lambda\,\hat\Gamma_2\Gamma_z\Gamma_r\hat\Gamma_{123}\,\epsilon
\,-\,{1\over
3}\,e^{\phi+\lambda-h}\,b'\,\hat\Gamma_2\Gamma_z\epsilon\,+\,\rc
&&\,+\,
\Big[\,-{1\over 3}\,
e^{\phi-2\lambda-2h}\,\big(\,1+b^2\,\big)\,+\,
{1\over 6}\,e^{-\phi}\,\big(\,e^{-4\lambda}\,-\,e^{2\lambda}\,\big)\,
\Big]\,
\Gamma_r\hat\Gamma_{123}\,\epsilon\,=\,0\,\,.
\eear
From the gravitino variation we get a new equation:
\bear
&&h'\epsilon\,-\,
be^{-h}\cosh
3\lambda\,\hat\Gamma_2\Gamma_z\Gamma_r\hat\Gamma_{123}\,\epsilon
\,+\,{2\over 3}\,e^{\phi+\lambda-h}\,b'\,\hat\Gamma_2\Gamma_z\,\epsilon
\,+\,\rc\rc
&&+\,\Big[\,-{5\over 6}\,
e^{\phi-2\lambda-2h}\,\big(\,1+b^2\,\big)\,+\,
{1\over 24}\,e^{-\phi}\,\big(\,e^{-4\lambda}\,+\,2e^{2\lambda}\,\big)\,
\Big]\,
\Gamma_r\hat\Gamma_{123}\,\epsilon\,=\,0\,\,.
\eear
The equation for $\phi'$ is of the form:
\beq
\Gamma_r\hat\Gamma_{123}\,\epsilon\,=\,-(\,\beta\,+\,
\tilde\beta\,\hat\Gamma_2\Gamma_z\,)\,\epsilon\,\,,
\label{hyprot}
\eeq
where $\beta$ and $\tilde\beta$ can be directly read from
(\ref{hypphi}).
As in other cases, $\beta$ and $\tilde\beta$ should satisfy 
that $\beta^2+\tilde\beta^2=1$. Plugging (\ref{hyprot}) in the 
BPS equations we get two algebraic constraints
analogue to (\ref{constr}), (\ref{constr2}).
 After eliminating $\beta$ and $\tilde\beta$, they imply the following
constraint for the functions of the ansatz:
\beq
b\,\Big[\,1\,+\,4\,e^{2\phi+2\lambda-2h}\,(\,1\,+\,b^2\,)\,\Big]\,=\,0
\,\,.
\eeq
Notice that the only solution of this equation is $b=0$, since the other
solution corresponds to imaginary $b$. Thus, there is no analogue of the
deformed conifold and $\beta=1$, $\tilde\beta=0$. The resulting BPS 
equations are:
\bear
\phi'&=&{1\over 2}\,e^{\phi-2\lambda-2h}\,+\,
{1\over 8}\,e^{-\phi}\,(\,2e^{2\lambda}\,+\,e^{-4\lambda}\,)\,\,,\rc\rc
h'&=&-{5\over 6}\,e^{\phi-2\lambda-2h}\,+\,{1\over 24}\,e^{-\phi}\,
(\,2e^{2\lambda}\,+\,e^{-4\lambda}\,)\,\,,\rc\rc
\lambda'&=&-{1\over 3}\,e^{\phi-2\lambda-2h}\,+\,{1\over 6}\,
e^{-\phi}\,(\,e^{-4\lambda}\,-\,e^{2\lambda}\,)\,\,.
\eear
These eqs. can be integrated by the usual method. First of all, we 
define a new function $x$ and a new variable $t$:
\beq
x\,\equiv\, 4e^{2\phi-2h+2\lambda}\,\,,
\,\,\,\,\,\,\,\,\,\,\,\,\,
{dr\over dt}\,=\,e^{\phi+4\lambda}\,\,.
\eeq
Then, one gets a decoupled differential equation for $x$:
$\frac{dx}{dt}=\half x(x+1)$. Knowing $x$, the integration of
the full system is easy. The result is:
\beq
e^{\phi}={1 \over 12}\,\rho^{{3\over 2}}
\kappa(\rho)^{{1\over 4}}\,\,,\,\,\,\,\,\,\,\,\,\,\,\,
e^{\lambda}=\left({3\over 2\,\kappa(\rho)}\right)^{{1\over 6}}\,
\,\,,\,\,\,\,\,\,\,\,\,\,\,\,
e^{2h}={1 \over 6\,(12)^{\frac{2}{3}}}\,
\rho(6a^2-\rho^2)
\kappa(\rho)^{{1\over 6}}\,\,,
\eeq	
where the new radial variable $\rho$ and the function
$\kappa(\rho)$ are defined as follows:
\beq
\rho^2\,=\,6\,(96)^{\frac{1}{9}}\, e^{{t\over
2}}\,\,,\,\,\,\,\,\,\,\,\,\,\,\,
\,\,\,\,\,\,
\kappa(\rho)\,=\,{\rho^6-9a^2\rho^4+A\over \rho^6-6a^2\rho^4}\,\,,
\eeq
$a$, $A$ being integration constants.
The eleven dimensional metric takes the form:
\bear
&&ds^2_{11}\,=\,dx^2_{1,4}\,+\,\big(\kappa(\rho)\big)^{-1}\,d\rho^2\,+\,
{6a^2-\rho^2\over 6}\,dH_2\,+\,{\rho^2\over 6}\,
\Big(\tilde{w}_1^2+\tilde{w}_2^2\Big)\,+\,
{\rho^2\kappa(\rho)\over 9}\,\Big(\tilde{w}_3\,+\,{dz\over
y}\Big)^2\,\,,\rc\rc
\eear
where $\tilde{w}_i$ are left-invariant $SU(2)$ one-forms for the
external $S^3$ and $dH_2=y^{-2}(dz^2+dy^2)$ is the metric of the
hyperbolic space.

Let us study the behavior of this metric.  If we want to keep the same
$(1,10)$ signature, we must require that $6a^2-\rho^2>0$, \ie:
\beq
\rho< \rho_{max}\,\,,
\,\,\,\,\,\,\,\,\,\,\,\,\,\,\,\,
\rho^2_{max}\,=\,6a^2\,\,.
\eeq
Moreover, if we want to keep $\kappa(\rho)>0$ in the range $\rho^2<
\rho^2_{max}$, we should require $f(\rho)\equiv \rho^6-9a^2\rho^4+A<0$. 
Notice that 
$f'(\rho)=6\rho^3(\rho^2-6a^2)<0$ if $\rho< \rho_{max}$.  There are two
possible cases. If $A\le 0$, the minimal value of $\rho$ is $\rho=0$,
\ie\ in this case
$0\le\rho<\rho_{\max}$. If $A>0$ we must consider two cases.  If $0<A<108
a^6$, then $\rho_{min}\le\rho<\rho_{\max}$, where $\rho_{min}$ is the
solution of the equation $f(\rho)=0$. If $A\ge 108 a^6$, then
$\rho_{min}\ge \rho_{\max}$ and the solution makes no sense. 

The conclusion is that, in all the cases, there is a maximum
value of the radial coordinate $\rho_{max}$, where the metric
has a singularity that cannot be removed and, hence, the
solution becomes pathological. As a matter of fact, this
supergravity solution cannot be used as the dual of a gauge
theory, at least for values of $\rho$ near the singular point.
In any case, if one wants to use the small $\rho$ regime
of the solution as a gauge theory dual, the singularity
should be somehow interpreted.

\subsection{D6-branes wrapping $H_3$}

The following analysis is analogue to that of section
\ref{roundg2}, changing the sphere $S^3$ by the hyperbolic 
space $H_3$.
The natural ansatz for the  8d metric is\footnote{The reader
should not be confused by the fact that one of the directions
of the $H_3$ is denoted by $x$, unrelated to the Minkowski-like
directions $x_i$ .}:
\beq
ds^2_8=e^{{2\phi\over 3}}\,dx_{1,3}^2\,+\,
{e^{2h}\over z^2}\,\big(dx^2+dy^2+dz^2\big)+dr^2\,\,.
\eeq
The ansatz for the gauge field is:
\beq
A^1\,=\,{b\over z}\,dz\,\,,
\,\,\,\,\,\,\,\,\,\,\,\,\,\,
A^2\,=\,{dy\over z}\,+\,{b\over z}\,dx\,\,,
\,\,\,\,\,\,\,\,\,\,\,\,\,\
A^3\,=\,-{dx\over z}\,+\,{b\over z}\,dy\,\,,
\eeq
where, $b$ is a function of $r$.
We will not excite any scalar field. The  components
of the field strength are:
\bear
&&F_{xy}^1\,=\,F_{yz}^2\,=\,F_{zx}^3\,=\,{1+b^2\over z^2}\,\,,\rc\rc
&&F_{rz}^1\,=\,F_{rx}^2\,=\,F_{ry}^3\,=\,{b'\over z}\,\,.
\eear
We will now use the following angular projections (note that
only two are independent):
\beq
\Gamma_{zx}\epsilon\,=\,-\hat\Gamma_{12}\epsilon\,\,,
\,\,\,\,\,\,\,\,\,\,\,\,\,\,
\Gamma_{zy}\epsilon\,=\,-\hat\Gamma_{13}\epsilon\,\,,
\,\,\,\,\,\,\,\,\,\,\,\,\,\,
\Gamma_{xy}\epsilon\,=\,-\hat\Gamma_{23}\epsilon\,\,.
\eeq
The following equations are obtained after imposing the vanishing of the
variation of the fermionic fields (\ref{susy8}):
\bear
&&\phi'\epsilon\,+\,{3\over 2}\,e^{\phi-h}\,b'\,
\hat\Gamma_1\,\Gamma_z\,\epsilon\,+\,
\Big(\,{3\over 2}\,e^{\phi-2h}\,(1+b^2)\,+\,{3\over 8}\,e^{-\phi}\,\Big)\,
\Gamma_r\hat\Gamma_{123}\epsilon\,=\,0\,\,,\rc\rc
&&h'\epsilon\,-{1\over 2}\,b'\,e^{\phi-h}\,\,\hat\Gamma_1\Gamma_z
\epsilon\,+
\,be^{-h}\,\hat\Gamma_1\,\Gamma_z\Gamma_r\hat\Gamma_{123}\epsilon\,
+\,\rc\rc &&\,\,\,\,\,\,\,\,\,\,\,\,\,\,\,\,\,\,\,\,\,\,\,\,\,\,\,\,
\,+\,\Big(\,-{3\over 2}\,e^{\phi-2h}\,(1+b^2)\,+\,{1\over
8}\,e^{-\phi}\,\Big)\,
\Gamma_r\hat\Gamma_{123}\epsilon\,=\,0\,\,,\rc\rc
&&\partial_r\epsilon\,=\,{\phi'\over 6}\,\epsilon\,
+\,{3\over 2}\,e^{\phi-h}\,b'\,\hat\Gamma_1\Gamma_z\epsilon\,\,.
\label{h3eqs}
\eear
As usual, from the first of these equations, one can directly
read
$\Gamma_r\hat\Gamma_{123}\,\epsilon\,=\,-(\,\beta\,+\,
\tilde\beta\hat\Gamma_1\Gamma_z\,)\,\epsilon$, which
implies the consistency condition
$\beta^2+\tilde\beta^2=1$.
Proceeding as in section \ref{roundg2},
we get the following BPS equations:
\bear
\phi'&=&{3\over 8}\,{e^{-2h-\phi}\over K}\,\,\Big[\,e^{4h}\,-\,16\,
(\,1+b^2\,)^2\,e^{4\phi}\,\Big]\,\,,\rc\rc
h'&=&{e^{-2h-\phi}\over 8K}\,\,\Big[\,e^{4h}\,+\,16\,(\,b^2\,-\,1\,)\,
e^{2\phi+2h}\,+\,48\,(\,1\,+\,b^2\,)^2\,e^{4\phi}\,\Big]\,\,,\rc\rc
b'&=&-{be^{-\phi}\over K}\,\,\Big[\,4\,(\,1+b^2\,)\,e^{2\phi}\,+\,
e^{2h}\,\Big]\,\,,
\label{H3BPS}
\eear
with
\beq
K\,\equiv\,\sqrt{16\,(\,1+b^2\,)^2\,e^{4\phi}\,+
\,8\,(\,b^2\,-\,1\,)\,
e^{2\phi+2h}\,+\,e^{4h}}\,\,.
\eeq
Representing $\beta=\cos\alpha$, $\tilde\beta=\sin\alpha$, we have:
\beq
\tan\alpha\,=\,4b\,{e^{\phi+h}\over 4\,(\,1+b^2\,)\,e^{2\phi}\,-\,e^{2h}}
\,\,.
\eeq
Moreover, by taking the spinor $\epsilon$ as:
\beq
\epsilon\,=\,e^{-{1\over 2}\alpha\,\hat\Gamma_1\Gamma_z}\,\epsilon_0\,\,,
\,\,\,\,\,\,\,\,\,\,\,\,\,\,\,\,\,\,\,\,\,\,\,\,\,\,\,\,
\Gamma_r\,\hat\Gamma_{123}\,\epsilon_0\,=\,-\epsilon_0\,\,,
\eeq
we obtain from the last eq. in (\ref{h3eqs}) that
$\alpha'=-3e^{\phi-h}b'$
and $\epsilon_0\,=\,e^{\phi/6}\,\eta$, where $\eta$ is, as usual, a
constant spinor satisfying the same projections as $\epsilon_0$. It can be
checked that  this equation for
$\alpha'$ is satisfied as a consequence of the BPS equations for $\phi$,
$h$ and
$b$.  The BPS equations (\ref{H3BPS}) can be integrated as in the case of
the round 
$S^3$. Let us define:
\bear
Y(\rho)&=&\rho^2-2m\rho+m^2(1+\lambda^2)\,\,,\rc
F(\rho)&=&-3\rho^4+8m\rho^3-6\,(\,1\,+\,\lambda^2\,)\,m^2\,\rho^2\,+\,
m^4\,(\,1\,+\,\lambda^2\,)^2\,\,,
\eear
where $m$ and $\lambda$ are constants of integration and  $\rho$ is a new
radial variable. Then, the uplifting to eleven dimensions yields:
\beq
ds^2_{11}\,=\,dx^2_{1,3}\,+\, F^{-{1\over 3}}\,d\rho^2\,
+\, F^{{2\over 3}}
\,Y^{-1}\,dH_3\,+\,F^{-{1\over
3}}\,Y\,(\, \tilde{w}_i+A_i\,)^2\,\,,
\eeq
where $dH_3\equiv {1\over z^2}(dx^2+dy^2+dz^2)$ and the
components of the gauge field are determined  in terms of
$b$, which is given by:
\beq
b(\rho)\,=\,{2\lambda m\rho\over Y(\rho)}\,\,.
\eeq

Notice that $F(\rho)$ is positive near $\rho=0$  (for $m\not=0$) and
negative for large values of $\rho$. Thus, there is a limiting value of
$\rho$ also in this case. Again, there is a singularity of the
metric at some $\rho_{max}$ that
renders the solution pathological.

%*******************************************************

\subsection{D5-branes wrapping $H_2$}

\subsubsection{BPS equations for D5-branes wrapped on a $H_2$}

We look here for a solution in the case in which the $H_2$ space is
considered in the place of the $S^2$ of the Maldacena-N\'u\~nez model
\footnote{This setup was studied in \cite{radu} using four dimensional gauged
supergravity.}. Thus, the analysis below is similar to section \ref{FMN7dim}.
This gravity solution should, in principle, be also dual to
${\cal N}=1$ SYM, although, as discussed above, adjoint
matter is incorporated in the model. However, we will again
find that the gravity solution becomes pathological for
large values of the radial coordinate.
Let us use the gauged supergravity of section \ref{TVN} and consider the
seven dimensional metric:
\beq
ds^2_7=dx_{1,3}^2+dr^2+{e^{2h}\over y^2}\,\big(dz^2+dy^2\big)\,\,,
\eeq
where $h=h(r)$. 
Let us consider the following gauge field:
\beq
A^1\,=\,-\,\frac{b}{y}\,dy\,\,,\,\,\,\,\,\,\,\,\,\,\,\,
A^2\,=\,-\,\frac{b}{y}\,dz\,\,,\,\,\,\,\,\,\,\,\,\,\,\,
A^3\,=\,-\,\frac{1}{y}\,dz\,\,,
\eeq
where $b=b(r)$. The corresponding field strength is:
\beq
F^1\,=\,-\,\frac{b'}{y}\,dr\wedge dy\,\,,\,\,\,\,\,\,\,\,\,\,
F^2\,=\,-\,\frac{b'}{y}\,dr\wedge dz\,\,,\,\,\,\,\,\,\,\,\,\,
F^3\,=\,\frac{1+b^2}{y^2}\,dy\wedge dz\,\,.
\eeq
Now we turn to solve the equations for the vanishing of the
variation of the fermion fields (\ref{susyTVN}). From $\delta
\psi_z=0$ we find the angular projection needed to perform the
twisting:
\beq
\Gamma_{zy}\,\epsilon\,=\,i\,\sigma^3\,\epsilon
\,=\,\sigma^1\,\sigma^2\,\epsilon\,\,.
\eeq
Using it, the conditions $\delta\psi_z=\delta\psi_y=0$ yield:
\beq
 h'\epsilon\,+\,b \,e^{-h}\,\Gamma_{r y}\,i\sigma^1\,\epsilon\,+\,
{e^{-2h}\over 2}\,(\,1+b^2\,)\,\Gamma_{r}\epsilon\,-\,
{b'\over 2}\,e^{-h}\,\Gamma_{y}\,i\sigma^1\,\epsilon=0\,\,.
\label{gravMNH2}
\eeq
From $\delta\lambda=0$ we obtain:
\beq
\phi'\epsilon\,+\,\Big(\,1\,+\,{1\over 4 }\,e^{-2h}\,(1+b^2)\,\Big)\,
\Gamma_{r}\epsilon\,-\,
{e^{-h}\over 2}\,b'\,\Gamma_{y}\,i\,\sigma_1\,\epsilon\,=\,0\,\,.
\label{phiMNH2}
\eeq
Finally, from the variation of the radial component
of the gravitino, we arrive at:
\beq
\partial_{r}\epsilon\,=\,{e^{-h}\,b'\over 2}\,
\Gamma_{y}\,i\,\sigma_1\,\epsilon\,\,.
\eeq
Taking (\ref{phiMNH2}) into account, we obtain a
rotated radial projection:
\beq
\Gamma_{r}\epsilon\,=\,\big(\,\beta\,+\,\tilde\beta\,\Gamma_{y}
\,i\,\sigma_1\,\big)\,\epsilon\,\,,
\label{radialMNH2}
\eeq
where:
\beq
\beta\,=\,-\,{\phi'\over 1+{e^{-2h}\over 4}\,(1+b^2)}\,\,,
\,\,\,\,\,\,\,\,\,\,\,
\tilde\beta\,=\,{1\over 2}\,
{e^{-h}b'\over 1+{e^{-2h}\over 4}\,(1+b^2)}\,\,.
\eeq
Since $(\Gamma_{r})^2\epsilon=\epsilon$, we get 
$\beta^2+\tilde\beta^2=1$ and use the usual parametrization:
$\beta=\cos\alpha$,
$\tilde\beta=\sin\alpha$
in order to write (\ref{radialMNH2}) as:
\beq
\Gamma_{r}\epsilon\,=\,e^{\alpha\,\Gamma_{y}\,i\,\sigma_1}\,
\epsilon\,\,,
\eeq
which is solved by:
\beq
\epsilon=e^{-{\alpha\over 2}\Gamma_{y}i\sigma_1}\,
\epsilon_0\,\,,\,\,\,\,\,\,\,\,\,\,\,
\Gamma_{r}\,\epsilon_0=\epsilon_0\,\,.
\eeq
By inserting the projection in (\ref{gravMNH2}), we find   the radial
derivatives  of $h$ and $b$ In terms of
$\beta$ and $\tilde\beta$:
\beq
h'\,=\,-{1\over 2}\,e^{-2h}\,(1+b^2)\,\beta-b \,e^{-h}\,\tilde\beta\,\,,
\,\,\,\,\,\,\,\,\,\,\,
b'\,=\,-2\,b\,\beta\,+\,e^{-h}\,(1+b^2)\,\tilde\beta\,\,.
\eeq
Moreover, $\phi'$ is given in terms of $b'$ as:
\beq
\phi'\,=\,{b'\over 2b}\,\Big[\,1\,-\,{1\over 4}\,e^{-2h}\,
(1+b^2)\,\Big]\,\,.
\eeq
Define:
\beq
K\equiv \sqrt{1+{1\over 2}\,e^{-2h}\,(\,b^2-1\,)\,+\,
{1\over 16}\,e^{- 4h}\,(\,1+b^2\,)^2}\,\,.
\eeq
Then, we have the following first-order differential 
equations\footnote{The sign of all these equations
can be changed by changing the sign in the radial
projection, what amounts to taking $r\to -r$ in the final
solution. This is general for all the solutions treated in
this thesis, although in most of them, the sign is clear
when one wants the final solution to make sense. Here, both
possibilities are pathological.}:
\bear
\phi'&=&{1\over K}\,
\Big[\,1\,-\,{e^{- 4h}\over 16}\,(\,1+b^2\,)^2\,\Big]\,\,,\rc\rc
h'&=&-{e^{- 2h}\over 2K}\,
\Big[\,b^2-1+{e^{- 2h}\over 4}\,(\,1+b^2\,)^2\,\Big]\,\,,\rc\rc
b'&=&{2b\over K}\,
\Big[\,1+{e^{- 2h}\over 4}\,(\,1+b^2\,)\,\Big]\,\,.
\label{difeq}
\eear 
By substituting our ansatz for the spinor in the equation for 
$\partial_{r}\epsilon$, we get:
\beq
\alpha'\,=\,-e^{-h}\,b'\,\,.
\eeq
Moreover:
\beq
\tan\alpha\,=\,{4b\over (1+b^2)e^{-h}-4e^{h}}\,\,.
\eeq
One can show using this last equation that the equation for $\alpha'$ is a
consequence of (\ref{difeq}). The radial projection can be written as:
\beq
\Gamma_{x^0\cdots
x^3\,z}\,(\,\cos\alpha\,\Gamma_{y}+\,\sin\alpha\,i\sigma_{1}
\,)\epsilon\,=\,\epsilon\,\,.
\eeq

\subsubsection{Solution of the BPS equations}

It is clear that $b=0$ is a solution of the last BPS equation 
(\ref{difeq}). Let us start by solving this particular case,
in which the radial projection on the spinor is not rotated.
The expression of $K$ gets simplified:
\beq
K\,=\,1-{1\over 4}\,e^{-2h}\,\,,
\,\,\,\,\,\,\,\,\,\,\,\,\,\,\,\,\,\,\,\,\,\,\,\,\,\,
(b=0)\,\,.
\eeq
(Notice that there is no square root). The remaining BPS equations are:
\beq
h'\,=\,{1\over 2}\,\,e^{-2h}\,\,,
\,\,\,\,\,\,\,\,\,\,\,\,\,\,\,\,\,\,\,\,\,\,\,\,\,\,
\phi'\,=\,1\,+\,{1\over 4}\,\,e^{-2h}\,\,,
\eeq
whose general solution is:
\beq
e^{2h}\,=\,r\,+\,C\,\,,
\,\,\,\,\,\,\,\,\,\,\,\,\,\,\,\,\,\,\,\,\,\,\,\,\,\,
e^{2\phi}\,=\,2e^{2\phi_0}\,e^{2r}\,(\,r\,+\,C\,)^{{1\over 2}}\,\,,
\eeq
where $C$ and $\phi_0$ are constants of integration. We will see that the
matching with the general solution fixes the  value of $C$ to
be
$1/4$. For this value of $C$, the constant $\phi_0$ is the  value of
$\phi$ at
$r=0$ (which is finite). Notice that, in general,  $\phi(r=0)$ is
finite for any positive $C$, whereas $\phi(r=0)\rightarrow
-\infty$ for $C=0$ .

Let us now search for the general solution of the system (\ref{difeq}).
 We will follow closely
the Chamseddine-Volkov method (see section \ref{FMNsolve}). Let us define
first the quantities:
\beq
x\equiv b^2\,\,,
\,\,\,\,\,\,\,\,\,\,\,\,\,\,\,\,\,\,\,
R^2\equiv 4e^{2h}\,\,.
\label{MNH2defs}
\eeq
From the equations of $h$ and $b$ we get:
\beq
x(R^2+x+1)\,{dR^2\over dx}\,+\,(x-1)\,R^2\,+\,(x+1)^2\,=\,0\,\,.
\label{difeqR}
\eeq
To solve this equation we introduce a parametrization in terms  of an
auxiliary variable $\rho$ and an auxiliary function $\xi(\rho)$, related
to $x$ and $R$ as:
\beq
x\,=\,\rho^2\,e^{\xi(\rho)}\,\,,
\,\,\,\,\,\,\,\,\,\,\,\,\,\,\,\,\,\,\,
R^2\,=\,\rho{d\xi(\rho)\over d\rho}\,-\,\rho^2\,e^{\xi(\rho)}+1\,\,.
\label{MNH2defs2}
\eeq
Then, the differential equation (\ref{difeqR}) reduces to the  following
simple equation for $\xi$:
\beq
{d^2 \xi(\rho)\over d\rho^2}\,=\,-2\,e^{\xi(\rho)}\,\,.
\eeq
This is the equation of motion of a particle whose position is the 
variable
$\xi(\rho)$ moving under the force $-2e^{\xi(\rho)}$. The conservation of
energy  for this auxiliary mechanical problem gives
${1\over 2}\,\left(\,{d \xi\over d\rho}\,\right)^2\,+\,2\,e^{\xi}\,=\,E$,
and the general solution is:
\beq
e^{\xi(\rho)}\,=\,{E\over 2\cosh^2\big[\sqrt{{E\over
2}}\,(\,\rho-\rho_0\,)\,\big]}\,\,.
\eeq
By inserting this result in (\ref{MNH2defs2}) and
(\ref{MNH2defs}), and going back to the original radial
variable, the values of $h$ and $b$ are straightforwardly 
obtained. With them, it is easy to integrate the equation
for $\phi$. Finally, the  functions of
the ansatz are (notice that $E$ disappears from the expressions):
\bear
b(r)&=&{2r\over \cosh (2r-2r_0)}\,,\rc\rc
e^{2h}&=&-r\tanh(2r-2r_0)\,-\,
{r^2\over \cosh^2(2r-2r_0)}\,+\,{1\over 4}\,\,,\rc\rc
e^{2\phi}&=&e^{2\phi_0}\,\,{2e^{h}\cosh(2r_0)\over 
\cosh(2r-2r_0)}\,\,.
\eear

For $r$ close to zero and $r_0=0$, $e^{2h}$ behaves as 
$e^{2h}\approx {1\over 4}\,-\,3r^2$. Moreover, $e^{2h}$ becomes zero
for some value of $r=r_{max}$ and, thus, one should restrict  to the
region
$r< r_{max}$ if we want $e^{2h}>0$. The $b=0$ case is
recovered in the limit $r_0 \to \infty$.

Once more, a bad singularity is reached at some maximum value of 
the radial variable. For all we have seen in this section, this
seems to be a common property for solutions arising from branes
wrapped in hyperbolic cycles. The singularity
can be taken to infinity by choosing $r_0=\infty$, what amounts to having
an unrotated radial projection on the spinor.

\setcounter{equation}{0}
\section{$Spin(7)$ holonomy metrics from gauged 
supergravity}
\label{secspin7}

In this section\footnote{I would specially like to thank
R. Hern\'andez and K. Sfetsos for collaboration on these
topics.},
 we are going to consider D6-branes wrapping
an $S^4$ in eight dimensional Salam-Sezgin supergravity.
Upon uplifting to eleven dimensions, one gets a direct
product of 1+2 Minkowski
space and an eight-manifold with $Spin(7)$ holonomy 
\cite{rafa}, see also \cite{hsiii}. In ten dimensions, it corresponds to
D6-branes wrapping a coassociative four-cycle inside
a seven dimensional manifold of $G_2$ holonomy \cite{jaume}.

First, it will be shown how to obtain from gauged sugra (\cite{rafa}) a
complete,
$Spin(7)$ holonomy metric, first found by Bryant and Salamon in \cite{bs}.
Then, in the spirit of chapters
\ref{conichapter} and \ref{g2chapter}, a new ansatz will be proposed, by
allowing the gauge fields to depend on the radial coordinate and
the radial projection of the spinor to be rotated. Unlike the
previous cases, no new non-trivial metric can be found by 
this procedure, the only alternative to Bryant-Salamon metric
being flat eight dimensional space. This is in agreement
with the fact that the only consistent supersymmetric deformation of the 
Hopf fibration that gives $S^7$
is the squashed $S^7$ (with the $S^7$, one can construct the
metric of flat eight dimensional space and with the squashed
$S^7$, the large $r$ limit of the Bryant-Salamon metric).

Furthermore, it will be shown how the BPS equations can also
be obtained by imposing a self-duality condition to the spin connection,
 the duality being performed with  the
octonionic structure constants \cite{bakas,ads}. This  directly
proves  that the holonomy group is
contained in $Spin(7)$.

It would be interesting to clarify if other $Spin(7)$ 
holonomy metrics \cite{cglp} can be understood in the context of
gauged supergravity.

\subsubsection{The Bryant-Salamon $Spin(7)$ holonomy metric}

Following \cite{rafa,hsiii}, let us consider the following ansatz
for the metric in 8d sugra:
\beq
ds^2_8\,=\,e^{{2\phi \over 3}}\,dx_{1,2}^2\,+\,
dr^2\,+\,e^{2h}\,ds_4^2\,\,,
\eeq
where $ds_4^2$ is the de Sitter metric on the  $S^4$ with unit diameter
that will be written as:
\beq
ds_4^2\,=\,{1 \over \left(1+\xi^2\right)^2}\,
\left(d\xi^2\,+\,{\xi^2 \over
4}\Big((w^1)^2+(w^2)^2+(w^3)^2\Big)\right)\,\,,
\eeq
where $\xi$  ranges form $0$ to $\infty$ and the $w^i$ are a set
of $SU(2)$ left invariant one-forms defined as in
(\ref{omegas}). We will not consider any coset scalars excited:
$L_\alpha^i=\delta_\alpha^i$ so that $P_{ij}=0$ and 
$Q_{ij}=-\epsilon_{ijk}A^k$. The gauge field needed for the
twisting is that of the $SU(2)$ instanton on $S^4$:
\beq
A^i\,=\,-\,{1 \over 1+\xi^2}\,w^i\,\,\,,
\,\,\,\,\,\,\,\,\,\,\,\,\,\,\,\,\,\,
(i=1,2,3)\,\,\,,
\label{spin7gauge}
\eeq
and its field strength (self-dual on the $S^4$) reads:
\beq
F^i\,=\,4\,a(\xi)\,b(\xi)\,d\xi\wedge w^i\,-\,
2\,a(\xi)^2\,\epsilon_{ijk}\,w^j\wedge w^k\,\,,
\eeq
where the definitions:
\beq
a(\xi) \equiv \frac {\xi/2}{1+\xi^2} \ ,\qquad\quad \: \: \: \: b(\xi)
\equiv
\frac {1}{1+\xi^2} \ ,
\label{abdefs}
\eeq
have been made. Then, by imposing the following set of projections
on the Killing spinor:
\bear
\G_{\xi r}\,\epsilon\,=\,\G_{\hat 1 1}\,\epsilon\,\,,
\,\,\,\,\,\,\,\,\,\,\,\,\,\,\,\,\,\,
\G_{1 2}\,\epsilon&=&-\hat\G_{12}\,\epsilon\,\,,
\,\,\,\,\,\,\,\,\,\,\,\,\,\,\,\,\,\,
\G_{2 3}\,\epsilon\,=\,-\hat\G_{23}\,\epsilon\,\,,
\label{spin7proj}\\
\G_r\hat\G_{123}\,\epsilon&=&-\epsilon\,\,,
\label{spin7proj2}
\eear
the vanishing of the fermion field variations yields the system
of BPS equations:
\bear
\phi'&=&-12\,e^{\phi-2h}\,+\,{3 \over 8}\,e^{-\phi}\,\,,\rc
h'&=&8\,e^{\phi-2h}\,+\,{1 \over 8}\,e^{-\phi}\,\,.
\label{spin7eqs}
\eear
It is not hard to solve this system proceeding similarly to 
previous cases. Then, by using (\ref{uplift8d}), the solution
leads to the eleven dimensional metric:
\beq
ds_{11}^2=dx_{1,2}^2+{d\tau^2 \over \left(
1-\frac{l^{10/3}}{\tau^{10/3}}\right)}+
{9 \,\tau^2\over 5\left(1+\xi^2\right)^2}
\left(d\xi^2+{\xi^2 \over
4}(w^i)^2\right)+{9\over 100}
\tau^2\left(1-\frac{l^{10/3}}{\tau^{10/3}}\right)
\left(\tilde{w}^i-{w^i \over 1+\xi^2}\right)^2,
\label{bss7met}
\eeq
where $\tau$ is a new radial variable $d\tau =e^{\phi/3} dr$.
One can read the Bryant-Salamon metric by discarding
the three dimensional Minkowski space. Note that for $l=0$
(or large $\tau$), the fibration of $S^3$ on $S^4$ gives
the metric of the squashed seven-sphere.

\subsubsection{Extending the ansatz}

Let us consider an extension of the ansatz where the gauge
field can depend on $r$ and one of the projections on
the Killing spinor is rotated.
We saw in chapters \ref{conichapter} and \ref{g2chapter} how
such a generalization leads to new metrics for the cases of 
D6 wrapping $S^2$ or $S^3$. We therefore take, for the gauge
field:
\beq
A^i\,=\,A\,w^i\,\,,
\eeq
where $A=A(r,\xi)$. The field strength is then:
\beq
F^i=\partial_r A\,dr\wedge w^i + \partial_\xi A\, d\xi\wedge
w^i+
\half F\epsilon_{ijk}w^j\wedge w^k\,\,,
\eeq
 with:
\beq
F\equiv A(1+A)\,\,.
\eeq
As in previous chapters, we permit a rotation in the 
Killing spinor, leaving the projections
(\ref{spin7proj}) unchanged while rotating  (\ref{spin7proj2}):
\beq
\G_r \hat{\G}_{123} \, \epsilon =  - (\cos\alpha + \sin \alpha \,\G_1
\hat{\G}_1) \, \epsilon \ \ \Rightarrow\ \
\epsilon = e^{- {\alpha \over 2} \, \G_1 \hat{\G}_1} \epsilon_0 \ ,
\label{s7rot}
\eeq
with $\epsilon_0$ satisfying (\ref{spin7proj2}). Now it is
straightforward to compute the supersymmetry variation of the
fermion fields (\ref{susy8}) and to get the system of
BPS equations. From the dilatino variations one obtains:
\bear
\frac {d \phi}{d r} & = & \frac {3}{8}  e^{-\phi} \cos \alpha - \frac
{3}{2ab} \frac {\partial A}{\partial \xi}
e^{\phi -2 h} + \frac {3}{2} \frac {F}{a^2} e^{\phi-2h} \cos \alpha  \ ,
\rc
\frac {\partial A}{\partial r} \frac {e^{\phi-h}}{a} & = &  -\frac {1}{4}
 e^{-\phi} \sin \alpha -
\frac {F}{a^2} e^{\phi-2h} \sin \alpha \ ,
\label{ults71}
\eear
where the definitions of $a$ and $b$ (\ref{abdefs}) have been used.
From $\delta\psi_1=\delta\psi_2=\delta\psi_3=0$ one arrives at:
\bear
a \,e^h \frac {dh}{dr} & = &  A \sin \alpha + \frac {1}{2} \sin \alpha +
\frac {1}{2b} \frac{\partial A}{\partial \xi}
e^{\phi-h} + \frac {1}{8} a\, e^{-\phi+h} \cos \alpha - \frac {3}{2} \frac
{F}{a} e^{\phi-h} \cos \alpha \ , \rc
\frac {1}{2} \frac {\partial A}{\partial r} e^{\phi} & = & -  A \cos
\alpha -
\frac {1}{2} \cos \alpha -\half \left({1-\xi^2 \over 1+\xi^2} \right)+
\frac {1}{8} a \,e^{-\phi+h} \sin \alpha - \frac {3}{2} \frac {F}{a}
e^{\phi-h}
\sin \alpha \ .\rc
\eear
Finally, $\delta\psi_\xi=0$ and $\delta\psi_r=0$ yield:
\bear
\xi \frac {\partial \alpha}{\partial \xi}&=&-2 e^h a \frac {dh}{dr} +
\frac {F}{a} e^{\phi-h} \cos \alpha +
\frac {5}{b} \frac {\partial A}{\partial \xi} e^{\phi-h} + \frac {1}{4} a
\,e^{-\phi+h} \cos \alpha \ ,\rc
a \frac {\partial \alpha}{\partial r} &=& - \frac {F}{2a} e^{\phi-2h}
\sin \alpha + {5 \over 2} \frac {\partial A}{\partial r} e^{\phi-h} -
\frac {1}{8} a\, e^{-\phi} \sin \alpha \ .
\label{ults72}
\eear
Notice that by taking $\alpha=0$ and the gauge field of
(\ref{spin7gauge}), the system of BPS equations (\ref{spin7eqs}) is
recovered. Once the system (\ref{ults71})-(\ref{ults72}) is solved, it is
easy to get the uplifted eleven dimensional solution. It can be written:
\beq
ds_{11}^2 = dx_{1,2}^2 + d\rho^2  + 4P^2 \sum_{i=1}^3 \left( 
\tilde{w}^i +   A\, w_i \right)^2
+ Q^2 ds_4^2 \ ,
\label{s11}
\eeq
where the redefinition in terms of the eight dimensional 
variables reads:
\beq
d\rho \equiv e^{-\phi/3} d r \ , \quad\quad P \equiv e^{2\phi/3} \ ,
\quad\quad Q\equiv e^{-\phi/3 + h} \ .
\eeq
The Killing spinor equations  become, for the new set of functions:
\bear
\frac {dP}{d\rho} & = &  -\frac {P^2}{Q^2} \frac {\partial A}{\partial
\xi}
\frac {1}{ab}
+ \frac {F}{a^2} \frac {P^2}{Q^2} \cos \alpha + \frac {1}{4} \cos \alpha
\ ,
\label{seq1}
\\
\frac {1}{a} \frac {\partial A}{\partial \rho} \frac {P}{Q} & = & - \frac
{F}{a^2} \frac {P}{Q^2} \sin \alpha
- \frac {1}{4P} \sin \alpha \ , 
\label{seq2}\\
P \frac {\partial A}{\partial \rho} & = & -  A \cos \alpha - \frac {1}{2}
\cos \alpha - \half \left({1-\xi^2 \over 1+\xi^2} \right) - 2 \frac {F}{a}
\frac {P}{Q} \sin \alpha \ , 
\label{seq3}\\
a \frac {dQ}{d\rho} & = &  A \sin \alpha + \frac {1}{2} \sin \alpha +
\frac {P}{Q}
\frac {1}{b}
\frac {\partial A}{\partial \xi} - 2 \frac {F}{a} \frac {P}{Q} \cos
\alpha , 
\label{seq4}\\
\frac {\partial \alpha}{\partial \xi} & = &- b \frac {dQ}{d\rho} + 3 \frac
{P}{Q}
\frac {\partial A}{\partial \xi}
\frac {1}{a}  , 
\label{seq5}\\
a \frac {d\alpha}{d\rho} & = &  \frac {3P}{Q} \frac {\partial A}{\partial
\rho}
\  .
\label{seq6}
\eear

\subsubsection{Solving the BPS system}

Despite the terrifying appearance of (\ref{seq1})-(\ref{seq6}),
we will show that  there only exist two inequivalent
solutions, one leading to (\ref{bss7met}) and the other to
flat space. First of all, notice that an algebraic constraint
can be deduced from (\ref{seq2}) and (\ref{seq3}):

\beq
A \cos \alpha + \frac {1}{2} \cos \alpha +\half\left({1-\xi^2 \over
1+\xi^2}
\right) + \frac {F}{a}
\frac {P}{Q} \sin \alpha - \frac {Qa}{4P} \sin \alpha = 0 \ .
\eeq
By performing the radial derivative of this expression and
using (\ref{seq1}), (\ref{seq2}), (\ref{seq4}) and (\ref{seq6}),
one gets a new, simpler,  algebraic constraint:
\beq
\sin \alpha \,\frac {\partial A}{\partial \xi} \left( \frac{F P^2}{Q^2
a^2} +
\frac{1}{4} \right)
= 0 \ .
\eeq
There are three possibilities to fulfil this constraint. 
Clearly, $\sin \alpha=0$ leads to the system (\ref{spin7eqs})
and therefore to the metric (\ref{bss7met}). On the other
hand, $\frac {\partial A}{\partial \xi}=0$ is inconsistent
with the set of equations (\ref{seq1})-(\ref{seq6}).
Finally, 
$ \frac{F P^2}{Q^2 a^2} + \frac{1}{4}  = 0$ yields two
equivalent solutions (one corresponding to the instanton and
the other one to the antiinstanton on $S^4$). One of them is:
\bear
P & = & \frac{r}{4},  \: \: \: \: \: \: Q=r,  \: \: \: \: \: \:
 A=-\frac{1}{1+\xi^2},  \rc
 \cos \alpha & = & - \frac{\xi^4 -6 \xi^2 +1}{\xi^4 +2 \xi^2 +1},  \: \:
\:
\:
\sin \alpha = - \frac{4\xi (1 - \xi^2)}{\xi^4 +2 \xi^2 +1},
\eear
so the metric obtained by inserting this in (\ref{s11}) is:
\beq
ds_{8}^2 = dr^2  + r^2 ds_4^2 + \frac{1}{4} \,r^2
\sum_{i=1}^3 \left( \tilde{w}_i - \frac{1}{1+\xi^2} \, w_i \right)^2 \ .
\eeq
This is just a flat space metric, as can be checked by direct calculation
of the Riemann tensor (in fact, the angular part of this metric is just
the sphere $S^7$ written as a Hopf fibration). The holonomy group of flat
space is trivial, so it is contained in
$Spin(7)$ as it should. This has been quite a long way to obtain
just the Minkowski metric, but it is a nice result in the sense that
we obtain the sphere $S^7$ and its only supersymmetric deformation,
the squashed $S^7$ from the same system in gauged supergravity.

\subsubsection{Octonions and BPS equations from the spin
connection theorem}

The so-called spin connection theorem asserts that if an
eight dimensional manifold satisfies the condition:
\beq
\omega^{\alpha\beta} = \frac {1}{2} 
\Psi_{\alpha\beta\gamma\delta} \omega^{\gamma\delta} \, ,
\label{s7self}
\eeq
then its holonomy group is contained in $Spin(7)$ \cite{ads,bakas}.
 $\omega^{\alpha\beta}$
is the spin connection defined in some frame, and 
$\Psi_{\alpha\beta\gamma\delta}$ is an antisymmetric four-form constructed
from the octonionic structure constants (its precise definition will be
given below), and is invariant under the action of the
 $Spin(7)$  subgroup of $SO(8)$. Then, the system 
(\ref{seq1})-(\ref{seq6}), should  be obtainable by this method
by directly imposing this condition on the metric (\ref{s11}).
We will see that the key point is a rotation of the frame,
which is related to the rotation (\ref{s7rot}) on the Killing
spinor. As pointed out in section \ref{secgeng2}, an analogous procedure
can be followed in order to obtain the $G_2$ holonomy metrics studied
in chapter \ref{g2chapter}. Let us start with the constants that define
the octonion algebra:
\beq
\psi_{7i\hat j}=\delta_{ij}\ ,\quad\quad
 \psi_{ijk}=-\psi_{i\hat j\hat k}= \epsilon_{ijk}\ ,
\eeq
Using the definition $\psi_{abcd}={(1/6)}\epsilon_{abcdefg}\psi_{efg}$
\ , one finds that:
\beq
\psi_{7ij\hat k}= -\psi_{7\hat i\hat j\hat k}  = \epsilon_{ijk}\ ,\quad
\psi_{ij\hat m\hat n}=\delta_{im}\delta_{jn}-\delta_{in}\delta_{jm}\ .
\eeq
By splitting the eight dimensional index $\alpha$ as $\alpha=(a,8)$,
so that $a=1,2,\ldots,7$ runs over the seven imaginary octonions,
we can define the  totally antisymmetric 4-index tensor that is
invariant under $Spin(7)$:
\beq
\Psi_{abc8}\equiv\psi_{abc}\,\,,\quad\quad
\Psi_{abcd}\equiv\psi_{abcd}\,\,.
\label{psidef}
\eeq
Then, the self-duality condition (\ref{s7self}), explicitly
written in components, amounts to:
\bear
\omega^{8i}_0 & = &-\half \epsilon_{ijk} (\omega^{jk}_0 -\omega^{\hat
j\hat k}_0) +
\omega^{7\hat i}_0\ ,
\nonumber\\
\omega^{8\hat i}_0 & = & \epsilon_{ijk}\, \omega^{j\hat k}_0
-\omega^{7i}_0\ ,
\nonumber\\
\omega^{87}_0 & = & -\omega^{i\hat i}_0\ .
\label{selff0}
\eear
The subindex 0 means that the spin connection is referred to some 
vielbein frame $e^\alpha_0$. Now, consider a new frame $e^\alpha$
related to the former as:
\beq
e_0=\Lambda^{-1} e\ ,\quad\quad
 \Lambda=\pmatrix{\cos\alpha&-\sin\alpha\cr \sin\alpha
&\cos\alpha}\ ,
\eeq
where $\Lambda$ is a rotation matrix acting in the $(1-\hat 1)$ 
plane. Then, the relation of the spin connections is, as usual:
\beq
\omega_0=\Lambda^{-1}\, \omega\,\Lambda\, +\, \Lambda^{-1} d\Lambda\ ,
\label{hdg}
\eeq
so the self-duality equations (\ref{selff0}) in this frame  read:
\bear
&& \omega^{8i}=\half \cos\alpha \,\epsilon_{ijk}\,(\omega^{\hat j\hat
k}-\omega^{jk}) -\sin\alpha \,\epsilon_{ijk}\,\omega^{j\hat k}
+\omega^{7\hat i}\ ,
\label{8i} \rc
 &&\omega^{8\hat i}=\half \sin\alpha\,
\epsilon_{ijk}(\omega^{\hat j\hat k}-\omega^{jk}) +\cos\alpha\,
\epsilon_{ijk}\,\omega^{j\hat k} -\omega^{7i}\ , \label{8hi} \rc
 && \omega^{87} 
=  -\omega^{i\hat i}  + d
\alpha \ .
\label{87}
\eear
Notice that, although the rotation has been performed in the
$(1-\hat 1)$ plane, it would be exactly the same if it had
been done in $(2-\hat 2)$ or $(3-\hat 3)$. This is due to the
invariance of the 4-form $\Psi$ under cyclic permutations of (1,2,3).

In order to impose this condition on a metric of the form
(\ref{s11}), let us define the following eight dimensional
vielbein:
\beq
e^i=Q a w_i\ ,\quad\quad e^{\hat i} 
= 2P(\tilde{w}_i + A \,w_i)\ ,\quad\quad e^7 =d\rho\ ,\quad\quad e^8=Q b
d\xi\ ,
\eeq
where $P(\rho)$, $Q(\rho)$, $A(\rho,\xi)$, $a(\xi)$, $b(\xi)$.
We need the spin connection, which can be obtained from the structure
equations
$de^a+w^a_{\ b}\wedge e^b=0$. This computation  yields:
\bear
\omega^{87}={\partial_\rho Q \over Q}\,e^8\ ,\
\quad\quad\quad\quad\quad\quad\ \
&&\omega^{\hat i8}={P \partial_\xi A \over Q^2\,a\,b} e^i \ ,
\rc\rc
\omega^{i8}={\partial_\xi a\over Q\,a\,b} e^i +
{P \partial_\xi A \over Q^2\,a\,b} e^{\hat i} \ ,\
\quad\quad
&&\omega^{i7}={\partial_\rho Q \over Q} e^i +{P\,\partial_\rho A 
\over Q\,a}e^{\hat i}
\ ,
\rc\rc
\omega^{\hat i 7}={\partial_\rho P \over P} e^{\hat i} +
{P\,\partial_\rho A \over Q\,a} e^{i} \ ,\ \,
\quad\quad
&&\omega^{ij}=\epsilon_{ijk} \left({1\over 2Qa} e^k - {P\,F \over
Q^2a^2} e^{\hat k}\right)\,\,,
\rc\rc
\omega^{\hat i\hat j}=\epsilon_{ijk} \left({1 \over 4P} e^{\hat k} 
-{A\over Qa}
e^{k}\right)\ ,
\quad
&&\omega^{i \hat j }=\epsilon_{ijk} {P\,F \over Q^2a^2} e^k 
+ \delta_{ij} \left({P \partial_\xi A \over Q^2\,a\,b} e^8 + 
{P\,\partial_\rho A \over Q\,a} e^7\right).\ \ \ \ \
\eear
Now, one can substitute the value of the spin connection in
(\ref{87}) and check that the system (\ref{seq1})-(\ref{seq6})
is recovered.

Finally, let us see how the projection on the spinor can be
described in terms of $\Psi_{(4)}$. The conditions for a
$Spin(7)$ invariant spinor read
$\Bigl( \G_{\alpha\beta}+{1\over
6}\Psi_{\alpha\beta\gamma\delta}\G_{\gamma\delta}\Bigr)\epsilon_0=0$,
which are just (\ref{spin7proj}), (\ref{spin7proj2}). In order
to include the rotation of the Killing spinor, one can
make a rotation on $\Psi$ (this is what was
done in section \ref{calibrating} for the $G_2$ case):
\bear
&&\psi^{(\alpha)}_{7ij\hat k}= -\psi^{(\alpha)}_{7\hat i\hat j\hat k}  =
\cos\alpha\,
\epsilon_{ijk}\ ,
\quad
\psi^{(\alpha)}_{7ijk}= -\psi^{(\alpha)}_{7i\hat j\hat k}  = -\sin\alpha
\,\epsilon_{ijk}\ ,
\nonumber\\
&&
\psi^{(\alpha)}_{ij\hat m\hat n}=\psi_{ij\hat m\hat n}=
\delta_{im}\delta_{jn}-\delta_{in}\delta_{jm}\,\,,\nonumber
\eear
\beq
\psi^{(\alpha)}_{7i\hat j}=\psi_{7i\hat j}=\delta_{ij}\ ,
\quad \psi^{(\alpha)}_{ijk}=
-\psi^{(\alpha)}_{i\hat j\hat k}=\cos\alpha \,\epsilon_{ijk}\ ,
\quad \psi^{(\alpha)}_{\hat i \hat j\hat k}=-\psi^{(\alpha)}_{ij\hat
k}=-\sin\alpha\,
\epsilon_{ijk}\ .
\eeq
Then, the Killing
spinor satisfies the condition:
\beq
\Bigl( \G_{\alpha\beta}+{1\over
6}\Psi^{(\alpha)}_{\alpha\beta\gamma\delta}\G_{\gamma\delta}\Bigr)
\epsilon=0\,\,,
\eeq
which is just (\ref{spin7proj}), together with (\ref{s7rot}).

\setcounter{equation}{0}
\section{$SO(4)$ twistings in D=7 gauged supergravity}
\label{so4sec}

The gauge group of the maximal supergravity in seven
 dimensions is $SO(5)$ (see section \ref{PPvN}).
This gauge group comes from the symmetries of the $S^4$ sphere,
where the reduction from $D=11$ to $D=7$ was performed.
However, by taking an appropriate scaling limit for the fields
and the gauge coupling constant,
one can make a consistent truncation of the gauge group to 
$SO(4)$. This process was described in detail in ref. \cite{tran}.
It is equivalent to first reducing $D=11$ sugra to $D=10$ IIA
theory and then compactifying it in  $S^3$, so the $SO(4)$ is
the isometry group of the $S^3$ sphere.

In the following, the 3-form $C$ of section \ref{PPvN} will not
be considered. Looking at the lagrangian (\ref{PPvNlagr}), one
concludes that this is consistent as long as $[F,F]=0$, what
 indeed happens in the cases addressed below. 
Then, the
variation of the fermionic fields is 
\cite{liumin,apreda} (the notation used in this section is mainly
borrowed from \cite{apreda}):
\bear
\delta(\G^i\la_i)&=&\Big[\,\half\,(T_{ij}\,-\,\frac{1}{5}\,T\,\delta_{ij})
\,\G^i\,\G^j\,+\,\half\,\gamma_\mu\, P^\mu_{ij}\,\G^i\,\G^j\,+\rc
&&+\,\frac{1}{16}\,\gamma^{\mu\nu}\,(\G^i\,\G^{kl}\,\G^i
\,-\,\frac{1}{5}\,\G^{kl})\,
F_{\mu\nu}^{kl}\,\Big]\,\epsilon\,,\qquad\quad \textrm{(no sum in $i$)},
\rc\rc
\delta\hat\psi_\mu&=&\left[{\cal D}_\mu-\frac{1}{4}\gamma_\mu\gamma^\nu
(V^{-1})^I_i\partial_\nu V_I^i+\frac{1}{4}\G^{ij}F_{\mu\la}^{ij}
\gamma^\la\right]\epsilon\,\,,
\label{so4susy}
\eear
where now $i$, $j$, $I$, $J$ range from 1 to 4. These transformations
can be read from (\ref{PPvNsusy}), taking into account
that the shifted gravitino $\hat\psi$ has been defined as
a combination of the original gravitino and the spin-$\half$ fermions:
$\hat\psi_\mu=\psi_\mu-\half\gamma_\mu\sum_{i=1}^4\G^i\la_i$.

Following \cite{apreda}, we write $SO(4)=SU(2)^+\times SU(2)^-$
and express the two sets of independent $SU(2)$ generators
in $SO(4)$ notation:
\beq
\eta_1^\pm=\half\pmatrix{0&1&0&0\cr -1&0&0&0\cr
0&0&0&\pm 1\cr 0&0&\mp 1&0}\,,\,\,
\eta_2^\pm=\half\pmatrix{0&0&\mp 1&0\cr 0&0&0&1\cr
\pm 1&0&0&0\cr 0&-1&0&0}\,,\,\,
\eta_3^\pm=\half\pmatrix{0&0&0&1\cr 0&0&\pm 1&0\cr
0&\mp 1&0&0\cr -1&0&0&0}\,.
\eeq
These matrices satisfy the commutation algebra:
\beq
[\eta_a^\pm ,\eta_b^\pm]=\epsilon_{abc}\eta_c^\pm\,\,,
\qquad\quad
[\eta_a^+,\eta_b^-]=0\,\,.
\eeq
It is also convenient to write the Dirac matrices on the 
$SO(4)$ in an $SU(2)^+\times SU(2)^-$ form. Define:
\beq
\sigma_1^\pm\,=\,{1 \over 2i}\left(\G^{24}\pm\G^{31}\right)\,\,,
\qquad
\sigma_2^\pm\,=\,-\,{1 \over 2i}\left(\G^{14}\pm\G^{23}\right)\,\,,
\qquad
\sigma_3^\pm\,=\,{1 \over 2i}\left(\G^{12}\pm\G^{34}\right)\,\,.
\eeq
These matrices satisfy the following relations:
\beq
\sigma_a^\pm\sigma_b^\pm\,=\,i\,\epsilon_{abc}\,\sigma_c^\pm\,\,,
\label{sigmarel}
\eeq
but they are not really Pauli matrices as they do not square to
unity.

In the following, we will see a few different possibilities for the
$SO(4)$ gauge field in this general framework, \ie\ different
twistings of the normal bundle, leading to different supergravity setups
and therefore  to different dual gauge theories. Concretely, we will
only consider branes wrapping two-spheres, so the ansatz for
the seven dimensional metric is:
\beq
ds_7^2\,=\,e^{2f}\,\left(dx_{1,3}^3\,+\,d\rho^2\right)\,+\,
e^{2g}\,\left(d\theta^2\,+\,\sin^2\theta\,d\varphi^2\right)\,\,,
\label{so4met}
\eeq
where $f\equiv f(\rho)$, $g\equiv g(\rho)$. First, we will
see how to recover the supergravity dual of ${\cal N}=1$ SYM
in this setup. Then, a singular sugra solution dual to 
${\cal N}=2$ SYM is described. To finish the chapter, it will
be proved that this solution cannot be desingularized along
the lines followed to resolve the singularity in the 
 ${\cal N}=1$ case.

\subsubsection{The Maldacena-N\'u\~nez model}

It was shown in section \ref{FMN7dim} that the Chamseddine-Volkov
background used in the Maldacena-N\'u\~nez model can be 
obtained in $SU(2)$ gauged supergravity. Certainly, it is
 possible to get it in this formalism by just switching on
the gauge field $SU(2)^+ \subset SO(4)$ \cite{apreda}:
\beq
A\,=\,\half\,\left[\,\cos\theta\, d\varphi \,\eta_1^+\,+\,
a(\rho)\, d\theta \,\eta_2^+\,+\,a(\rho) \,
\sin\theta\, d\varphi \,\eta_3^+\,\right]\,\,.
\eeq
There is only one scalar: $V_I^i=\textrm{diag}(
e^{-\la},e^{-\la},e^{-\la},e^{-\la})$. Then, 
after an appropriate identification of the functions
in the ansatz, one obtains
the differential equations of sec. \ref{secMNsol}. The projections
needed in this formalism are (\ref{FMNangproj7})
and (\ref{Fradial7}). Furthermore, one has to impose
$\sigma^-\epsilon=0$ (notice that if one the $\sigma^-$ 
annihilates the spinor, all of them do because of 
(\ref{sigmarel})). Basically, this projection (which halves
the number of supercharges) kills the $SU(2)^-$ so it
leads from $SO(4)$ sugra to the Townsend-van Nieuwenhuizen
$SU(2)$ gauged sugra.

\subsubsection{The ${\cal N}=2$ singular solution}

This supergravity dual of ${\cal N}=2$ Yang-Mills was first
considered in \cite{gaunN2}, see also \cite{bigazzi}. The idea is similar
to that of the Maldacena-N\'u\~nez model: the low energy description
of a D5-brane wrapping a two-cycle will be Yang-Mills in
3+1 dimensions. The difference is the way in which the
twisting is performed. From a geometrical perspective, we now want to have
the brane wrapping a two-cycle inside a Calabi-Yau two-fold (instead of
being embedded in a Calabi-Yau three-fold). Then, it can be
seen that the total number of supercharges is eight. To achieve
this, we take the gauge field to live in $U(1)^+\times U(1)^-
\subset SO(4)$:
\beq
A\,=\,\half\,\cos\theta\,d\varphi\,(\,\eta_1^+\,+\,\eta_1^-\,)\,\,.
\label{gaungf}
\eeq
This breaks the symmetry group as
$SO(1,3)\times SO(2)\times SO(4)\rightarrow 
SO(1,3)\times SO(2)\times SO(2)_1\times SO(2)_2$, where
$SO(2)_1$ rotates the \{1,2\} directions (where the gauge
field lives)
inside $SO(4)$ and
$SO(2)_2$ rotates the \{3,4\} ones. The spin connection of
$S^2$ is identified with $SO(2)_1$. Geometrically, the 
normal bundle is split into $SO(2)_1$ describing normal 
directions to the brane within the Calabi-Yau manifold
and $SO(2)_2$ describing the rest of the normal directions.
To preserve these symmetries, we take the matrix of 
scalars\footnote{Actually, one must keep the symmetry
of the $SO(2)_1$ ($V_1^1=V^2_2$), but it is not necessary
in the untwisted plane, so the matrix of scalars
$V_I^i\,=\,\textrm{diag}\left(e^{-\la_1},e^{-\la_1},
e^{-\la_2},e^{-\la_3}\right)$
 is possible. This was considered in \cite{bigazzi}.}:
\beq
V_I^i\,=\,\textrm{diag}\left(e^{-\la_1},e^{-\la_1},
e^{-\la_2},e^{-\la_2}\right)\,\,.
\label{gaunsc}
\eeq
Now, by plugging the ansatz (\ref{so4met}), (\ref{gaungf}),
(\ref{gaunsc}) in (\ref{so4susy}), a system of BPS equations
is found. As expected, only two projections on the 
Killing spinor are needed:
\beq
\gamma^\rho \epsilon\,=\,\epsilon\,\,,\qquad\quad
\gamma^{\theta\varphi}\epsilon\,=\,\G^{12}\epsilon\,\equiv\,
i\,(\sigma_3^++\sigma_3^-)\epsilon\,\,.
\label{n2projs}
\eeq
Let us make the following definitions:
\beq
p\equiv \la_1+\la_2\,\,,\qquad
y\equiv \la_1-\la_2\,\,,\qquad
h\equiv g-f\,\,.
\label{xyhdef}
\eeq
Then, it is immediate to check that $\delta\hat\psi_x=0$
implies $f=-\la_1-\la_2$. From 
$\delta\hat\psi_\theta=\delta\hat\psi_\varphi=0$, one finds:
\beq
\partial_\rho h\,=\,-\,\half \,e^{-2h-y}\,\,,
\eeq
and by combining the spin-$\half$ fermion variations one arrives at:
\bear
\partial_\rho p&=&\frac{2}{5}\,\cosh
y\,-\,\frac{1}{10}\,e^{-2h-y}\,\,,\rc 
\partial_\rho y&=&2\,\sinh
y\,-\,\half\,e^{-2h-y}\,\,.
\eear
Finally, from $\delta\hat\psi_\rho=0$, one gets $\partial_\rho
\epsilon=(\partial_\rho f/2) \epsilon$ and, therefore
$\epsilon=e^{\frac{f}{2}}\eta$, where $\eta$ is a constant spinor
that satisfies the same projections as $\epsilon$.

This system was found in \cite{gaunN2} by applying the superpotential 
method in seven dimensions. Then, the supersymmetry
of the corresponding ten dimensional solution was studied
and it turns out that a rather non-obvious frame is needed
 (unlike what happens in $S^3$ upliftings coming from
$SU(2)$ gaugings). The ten dimensional gravity solution is
dual to ${\cal N}=2$ SYM in four dimensions. It
is singular in the IR. In the string frame, it reads \cite{gaunN2}:
\bear
ds_{10}^2&=&dx_{1,3}^2\,+\,z\,(d\tilde\theta^2\,+\,\sin^2
\tilde\theta^2\,d\tilde\phi^2)\,+ \,e^{2x}\,dz^2\,+\,d\theta^2\,+\rc 
&&+\,{1 \over \Omega}\,e^{-x}\,\cos^2\theta\,(d\phi_1\,+\,
\cos\tilde\theta\, d\tilde\phi)^2\,+\,{1 \over \Omega}\,e^x\,\sin^2\theta
\,d\phi_2^2\,\,,
\eear
with the definitions:
\beq
e^{-2x}\,\equiv\,1\,-\,{1+k\,e^{-2z} \over 2z}\,\,,\qquad\quad
\Omega\,\equiv\,e^x \cos^2\theta+e^{-x}\sin^2\theta\,\,,
\eeq
where $k$ is a constant. The dilaton and NS 3-form are:
\bear
&&\,\,e^{-2\phi}\,=\,e^{-2\phi_0}\,e^{2z}\,\left(1-\sin^2\theta
{1+k\,e^{-2z} \over 2z}\right)\,\,,\rc\rc
H_{(3)}&=&{2\sin\theta\cos\theta \over \Omega^2}
\left(\sin\theta\cos\theta {dx\over dz}dz-d\theta\right)
\wedge(d\phi_1+\cos\tilde\theta d\tilde\phi)\wedge d\phi_2+\rc
&&+\,{e^{-x}\sin^2\theta \over \Omega}\sin\tilde\theta
d\tilde\theta\wedge d\tilde\phi\wedge d\phi_2\,\,.
\eear

\subsubsection{Trying (unsuccessfully) to desingularize
the ${\cal N}=2$ solution}

In the ${\cal N}=1$ case, we saw how generalizing the
$U(1)$ gauge field to $SU(2)$
(in a 't Hooft-Polyakov way), the gravity solution gets
desingularized. It is natural to try here to look for
an $SU(2)\times SU(2)$ gauge field that generalizes the
$U(1)\times U(1)$ showed above. This was proposed
in \cite{russo}. By studying the relevant system of
first order
equations for this ansatz,  it will be proved below
that the only possible solution is the one already described.

So, let us consider the gauge field:
\beq
A\,=\,\half\,\left[\cos\theta\,d\varphi\, (\eta_1^++\eta_1^-)+
a(\rho)d\theta\, (\eta_2^++\eta_2^-)+a(\rho)
\sin\theta d\varphi\, (\eta_3^++\eta_3^-)\right]\,\,,
\label{n2gf}
\eeq
whose gauge field strength reads:
\bear
F&=&\half\,\Big[\,(a^2-1)\,\sin\theta\, d\theta\wedge d\varphi
\,(\eta_1^++\eta_1^-)\,+\,\partial_\rho a \,d\rho\wedge d\theta
\,(\eta_2^++\eta_2^-)+\rc
&&+\,\partial_\rho a\,\sin\theta\,
d\rho\wedge d\varphi\, (\eta_3^++\eta_3^-)\,
\Big]\,\,.
\eear
The ansatz for the scalar fields is:
\beq
V_I^i\,=\,\textrm{diag}\left(e^{-\la_1},e^{-\la_1},
e^{-\la_2},e^{-\la_3}\right)\,\,.
\eeq
It is useful to use the same definition of $h$ as in
(\ref{xyhdef}) and also to take:
\beq
x\equiv \la_1+{\la_2+\la_3 \over 2}\,\,,\qquad\quad
y\equiv \la_1-{\la_2+\la_3 \over 2}\,\,,\qquad\quad
z\equiv {\la_2-\la_3 \over 2}\,\,.
\eeq
Following the reasoning used in previous cases, we maintain
the angular projection:
\beq
\gamma^{\theta\varphi}\epsilon\,=\,\G^{12}\epsilon\,\equiv\,
i\,(\sigma_3^++\sigma_3^-)\epsilon\,\,,
\label{n2angproj}
\eeq
which is necessary for the twisting but do not make any
{\it a priori} choice for the radial projection, that,
in principle, may be rotated. As usual, we must now
put the ansatz (\ref{n2gf})-(\ref{n2angproj}), (\ref{so4met})
into the supersymmetry variation of the fermions. 
$\delta\hat\psi_x=0$ gives $f=-x$ and the 
equations $\delta\hat\psi_\theta=\delta\hat\psi_\varphi=0$ yield:
\beq
\left[e^h\,(\partial_\rho h) +a\cosh (y+z)\, \gamma^{\rho\theta}\G^{24}+
\half e^z\, (\partial_\rho a) \,\,\gamma^{\theta}\G^{24}
-\half e^{-h-y}\,(a^2-1) \,\gamma^{\rho}\right]\epsilon=0\,\,,
\label{n2h}
\eeq
whereas $\delta\hat\psi_\rho=0$ leads to:
\beq
\left(\partial_\rho + {\partial_\rho x \over 2}+
\half (\partial_\rho a) \,\,e^{-h+z}\,\gamma^{\theta}\G^{24}
\right)\epsilon=0\,\,.
\label{n2rad}
\eeq
Finally, by appropriately combining the equations $\delta\la_i=0$,  one
finds:
\bear
\left(- \partial_\rho z\,\gamma^\rho+e^{-y}\sinh 2z + e^{-h}
a\,\sinh(y+z)\,\gamma^{\theta}\G^{24}+
\half (\partial_\rho a)e^{-h+z}\gamma^{\rho\theta}\G^{24}\right)
\epsilon=0\,\,,\,\,\,\,\,\,\,\,\,\,\,\,\,\,
\label{la1}\\
\left(-\partial_\rho x\,\gamma^\rho+
\frac{1}{5}(e^y+e^{-y}\cosh
2z)+\frac{1}{10}e^{-2h-y}(a^2-1)-\frac{1}{5}(\partial_\rho a)e^{-h+z}
\gamma^{\rho\theta}\G^{24}\right)\epsilon=0\,,\,\,\,\,\,\ \
\label{la2}\\
\left(-\partial_\rho y\,\gamma^\rho+(e^y-e^{-y}\cosh 2z)+\half
e^{-2h-y}(a^2-1)+2 a \,e^{-h}\sinh(y+z)\gamma^{\theta}\G^{24}
\right)\epsilon=0\,\,.\,\,\,\,\,\,\,\,
\label{la3}
\eear
Comparing with previous cases, it seems clear that, in order to 
solve this system, one should consider a Killing spinor of the form:
\beq
\epsilon= e^{-\frac{\alpha(\rho)}{2}\, \gamma^\theta
\G^{24}}\epsilon_0\,\,,
\eeq
 where $\epsilon_0$ satisfies the unrotated radial
projection (\ref{n2projs}). Then, $\epsilon$ would satisfy the rotated
projection:
$\gamma^\rho\epsilon=e^{\alpha(\rho)\gamma^\theta\G^{24}}\epsilon$.
However, by adding (\ref{la1})+$\frac{5}{2}$(\ref{la2})-$\half$
(\ref{la3}), one gets:
\beq
\partial_\rho(z+\frac{5}{2}x-\half y)\ \gamma^\rho\epsilon\,=\,
e^{-y+2z}\epsilon\,\,,
\eeq
and, as the right hand side cannot vanish, for consistency, the radial
projection must be unrotated $\gamma^\rho\epsilon=\epsilon$. Thus,
we have
$\alpha(\rho)=0$. Then, by looking at the matrix structure 
of eq. (\ref{n2rad}), it
is immediate to conclude
$\partial_\rho a =0$, and hence, $a=0$ because of (\ref{n2h}). This
takes us back to the $U(1)\times U(1)$ singular solution studied above.

The presence of IR singularities is common to all ${\cal N}=2$ 
supergravity duals. The resolution of these singularities is
believed to be the enhan\c{c}on mechanism. The enhan\c{c}on is a locus
where symmetry gets enhanced and extra string states become
important. Therefore, one cannot trust the supergravity approach
beyond it, and, hence, one never sees the IR singularity in the dual
gauge theory. For a review on this topic and further references, see 
\cite{n1n2}.

\vspace{3cm}

\centerline{{\large \textbf{Acknowledgements}}}

\vspace{1cm}

First of all, I would like to thank my supervisor
Alfonso V\'azquez Ramallo for giving me the opportunity
to carry out this Ph. D. thesis work at the University of
Santiago de Compostela. I am also indebted to Jos\'e
Camino, Pepe Barb\'on, Jos\'e Edelstein and Carlos
N\'u\~nez. I have really enjoyed collaborating with them
during this years. Finally, I also wish to thank 
Kostas Sfetsos, Rafael Hern\'andez and Javier Mas for
stimulating discussions.

%*******************************************

\end{document}